\documentclass[3p,sort&compress]{elsarticle}

\usepackage{graphicx}        
\usepackage{bbm}
\usepackage{amsmath,amssymb,units}
\usepackage{url}
\usepackage[colorlinks=true,hyperindex,breaklinks]{hyperref}

    \makeatletter
    \def\ps@pprintTitle{%
       \let\@oddhead\@empty
       \let\@evenhead\@empty
       \def\@oddfoot{\reset@font\hfil\thepage\hfil}
       \let\@evenfoot\@oddfoot
    }
    \makeatother

\newcommand{\bra}[1]{\left\langle{#1}\right\vert}
\newcommand{\ket}[1]{\left\vert{#1}\right\rangle}
\newcommand{\braket}[2]{\ensuremath{{\langle #1}|{#2 \rangle}}}
\newcommand{\ketbra}[2]{\ensuremath{|{#1 \rangle}{\langle #2}|}}
\providecommand{\abs}[1]{\left\lvert#1\right\rvert}
\newcommand{\D}{\ensuremath{\mathrm{d}}}
\newcommand{\E}{\ensuremath{e}}
\newcommand{\I}{\ensuremath{i}}
\newcommand{\op}[1]{#1}

\renewcommand{\vec}[1]{\ensuremath{\mathbf{#1}}}

\begin{document}

\begin{frontmatter}

\title{Quantum Decoherence}
\author{Maximilian Schlosshauer} 
\address{Department of Physics, University of Portland,\\5000 North Willamette Boulevard, Portland, OR 97203, USA}
\ead{schlossh@up.edu}

\begin{abstract}
Quantum decoherence plays a pivotal role in the dynamical description of the quantum-to-classical transition and is the main impediment to the realization of devices for quantum information processing. This paper gives an overview of the theory and experimental observation of the decoherence mechanism. We introduce the essential concepts and the mathematical formalism of decoherence, focusing on the picture of the decoherence process as a continuous monitoring of a quantum system by its environment. We review several classes of decoherence models and discuss the description of the decoherence dynamics in terms of master equations. We survey methods for avoiding and mitigating decoherence and give an overview of several experiments that have studied decoherence processes. We also comment on the role decoherence may play in interpretations of quantum mechanics and in addressing foundational questions.\\[.2cm]
\emph{Journal reference:} \emph{Phys.\ Rep.\ }{\bf 831}, 1--57 (2019), \href{https://doi.org/10.1016/j.physrep.2019.10.001}{\texttt{doi.org/10.1016/j.physrep.2019.10.001}}
\end{abstract}


\end{frontmatter}

\vspace{.4cm}

\begin{center}
\emph{In memory of H.\,Dieter Zeh (1932--2018)}
\end{center}

\tableofcontents

\section{Introduction}

Hilbert space is a vast and seemingly egalitarian place. If $\ket{\psi_1}$ and $\ket{\psi_2}$ represent two possible physical states of a quantum system, then quantum mechanics postulates that an arbitrary superposition $\alpha\ket{\psi_1} + \beta \ket{\psi_2}$ constitutes another possible physical state. The question, then, is why most such states, especially for mesoscopic and macroscopic systems, are found to be very difficult to prepare and observe, often prohibitively so. For example, it turns out to be extremely challenging to prepare a macroscopic quantum system in a spatial superposition of two macroscopically separated, narrow wave packets, with each individual wave packet approximately representing the kind of spatial localization familiar from the classical world of our experience. Even if one succeeded in generating such a superposition and confirming its existence---for example, by measuring fringes arising from interference between the wave-packet components---one would find that it becomes very rapidly unobservable. Thus, we arrive at the dynamical \emph{problem of the quantum-to-classical transition}: Why are certain ``nonclassical'' quantum states so fragile and easily degraded? The question is of immense importance not only from a fundamental point of view, but also because quantum information processing and quantum technologies crucially depend on our ability to generate, maintain, and manipulate such nonclassical superposition states. 

The key insight in addressing the problem of the quantum-to-classical transition was first spelled out almost fifty years ago by Zeh \cite{Zeh:1970:yt}, and it gave birth to the theory of \emph{quantum decoherence}, sometimes also called \emph{dynamical decoherence} or \emph{environment-induced decoherence} \cite{Zeh:1970:yt,Zurek:1981:dd,Zurek:1982:tv,Paz:2001:aa,Zurek:2002:ii,Schlosshauer:2003:tv,Bacciagaluppi:2003:yz,Joos:2003:jh,Schlosshauer:2007:un}. The insight is that realistic quantum systems are never completely isolated from their environment, and that when a quantum system interacts with its environment, it will in general become rapidly  and strongly entangled with a large number of environmental degrees of freedom. This entanglement dramatically influences what we can locally observe upon measuring the system, even when from a classical point of view the influence of the environment on the system (in terms of dissipation, perturbations, noise, etc.) is negligibly small. In particular, quantum interference effects with respect to certain physical quantities (most notably, ``classical'' quantities such as position) become effectively suppressed, making them prohibitively difficult to observe in most cases of practical interest.

This, in a nutshell, is the process of decoherence \cite{Zeh:1970:yt,Zurek:1981:dd,Zurek:1982:tv,Paz:2001:aa,Zurek:2002:ii,Schlosshauer:2003:tv,Bacciagaluppi:2003:yz,Joos:2003:jh,Schlosshauer:2007:un}. Stated in general and interpretation-neutral terms, decoherence describes how entangling interactions with the environment influence the statistics of future measurements on the system. Formally, decoherence can be viewed as a dynamical filter on the space of quantum states, singling out those states that, for a given system, can be stably prepared and maintained, while effectively excluding most other states, in particular, nonclassical superposition states of the kind epitomized by Schr\"odinger's cat \cite{Schrodinger:1935:gs}. In this way, decoherence lies at the heart of the quantum-to-classical transition. It ensures consistency between quantum and classical predictions for systems observed to behave classically. It provides a quantitative, dynamical account of the boundary between quantum and classical physics. In any concrete experimental situation, decoherence theory specifies the physical requirements, both qualitatively and quantitatively, for pushing the quantum--classical boundary toward the quantum realm. Decoherence is a genuinely quantum-mechanical effect, to be carefully distinguished from classical dissipation and stochastic fluctuations.

One of the most surprising aspects of the decoherence process is its extreme efficiency, especially for mesoscopic and macroscopic quantum systems. Furthermore, due to the many uncontrollable degrees of freedom of the environment, the dynamically created entanglement between system and environment is usually irreversible for all practical purposes; indeed, this effective irreversibility is a hallmark of decoherence. Increasingly realistic models of decoherence processes have been developed, progressing from toy models to complex models tailored to specific experiments (see Sec.~\ref{sec:decmodels}). Advances in experimental techniques have made it possible to observe the gradual action of decoherence in experiments such as cavity QED \cite{Raimond:2001:aa}, matter-wave interferometry \cite{Hornberger:2012:ii}, superconducting systems \cite{Leggett:2002:uy}, and ion traps \cite{Leibfried:2003:om,Haffner:2008:pp} (see Sec.~\ref{sec:exper-observ-decoh}). 

The superposition states necessary for quantum information processing are typically also those most susceptible to decoherence. Thus, decoherence is a major barrier to the implementation of devices for quantum information processing such as quantum computers. Qubit systems must be engineered to minimize environmental interactions detrimental to the preparation and longevity of the desired superposition states. At the same time, these systems must remain sufficiently open to allow for their control. Strategies for combatting the adverse effects of decoherence include decoherence avoidance, such as the encoding of information in decoherence-free subspaces (see Sec.~\ref{sec:dfs}), and quantum error correction \cite{Lidar:2013:pp}, which can undo the decoherence-induced degradation of the superposition state (see Sec.~\ref{sec:corr-decoh-induc}). Such strategies will be an integral part of quantum computers. Not only is decoherence relevant to quantum information, but also vice versa. An information-centric view of quantum mechanics proves helpful in conveying the essence of the decoherence process and is also used in recent explorations of the role of the environment as an information channel (see Secs.~\ref{sec:envir-monit-inform} and \ref{sec:prol-inform-quant}).

Decoherence is a technical result concerning the dynamics and measurement statistics of open quantum systems. From this view, decoherence merely addresses a \emph{consistency problem}, by explaining how and when the quantum probability distributions approach the classically expected distributions. Since decoherence follows directly from an application of the quantum formalism to interacting quantum systems, it is not tied to any particular interpretation of quantum mechanics, and it neither supplies such an interpretation nor amounts to a theory that could make predictions beyond those of standard quantum mechanics.  However, the bearing decoherence has on the problem of the relation between quantum and classical has been frequently invoked to assess or support various interpretations of quantum mechanics, and the implications of decoherence for the so-called quantum measurement problem have been analyzed extensively (see Sec.~\ref{sec:impl-found-quant}). Indeed, historically decoherence theory arose in the context of Zeh's independent formulation of an Everett-style interpretation \cite{Zeh:1970:yt}; see Ref.~\cite{Camilleri:2009:aq} for an analysis of the connections between the roots of decoherence and matters of interpretation.

It is a curious ``historical accident'' (Joos's term \cite[p.~13]{Joos:1999:po}) that the implications of environmental entanglement were appreciated only relatively late. While one can find---for example, in Heisenberg's writings (see Sec.~\ref{sec:niels-bohrs-views} and Ref.~\cite{Camilleri:2015:oo})---a few early anticipatory remarks about the role of environmental interactions in the quantum-mechanical description of physical systems, it was not until the 1970s that the ubiquity and implications of environmental entanglement were realized by Zeh \cite{Zeh:1970:yt,Kubler:1973:ux}. In the 1980s, the formalism of decoherence was further developed, chiefly by Zurek \cite{Zurek:1981:dd,Zurek:1982:tv}, and the first concrete decoherence models and numerical estimates of decoherence rates were worked out by Joos and Zeh \cite{Joos:1985:iu} and Zurek \cite{Zurek:1986:uz} (see also Refs~\cite{Walls:1985:pp,Walls:1985:lm,Caldeira:1985:tt}). Zurek's 1991 \emph{Physics Today} article \cite{Zurek:1991:vv} was an important factor in introducing a broader audience of physicists to decoherence theory. Such dissemination and maturing of decoherence theory came at a perfect time, as the 1990s also saw the blossoming of quantum information \cite{Feynman:1982:yy,Deutsch:1985:ym,Deutsch:1992:tv,Berthiaume:1992:lk,Berthiaume:1992:lm,Bernstein:1993:yy,Simon:1994:lk,Shor:1994:om,Shor:1997:tt,Grover:1996:rr,Grover:1997:mm}, as well as experimental advances in the creation of superpositions of mesoscopically and macroscopically distinct states \cite{Brune:1996:om,Arndt:1999:rc,Friedman:2000:rr,Wal:2000:om}. The quantum states relevant to quantum information processing and Schr\"odinger-cat-type experiments required the insights of decoherence theory, and conversely the new experiments served as a fertile ground for testing the predictions of decoherence theory. Accordingly, these developments led to a rapid rise in interest and research activity in the field of decoherence. Today, decoherence has become a central topic of modern quantum mechanics and is studied intensely both theoretically and experimentally.

Existing reviews of decoherence include the papers by Zurek \cite{Zurek:2002:ii}, Paz and Zurek \cite{Paz:2001:aa}, and Hornberger \cite{Hornberger:2009:aq}. Two books dedicated to decoherence are presently available: a volume by Joos et al.\ \cite{Joos:2003:jh} (a collection of chapters written by different authors), and a monograph by this author \cite{Schlosshauer:2007:un}, which offers, among other material, a detailed treatment of the topics surveyed in this paper. Textbooks on open quantum systems, such as Ref.~\cite{Breuer:2002:oq}, also contain a substantial amount of material on decoherence, especially in the context of quantum master equations. 

This article is organized as follows. Section~\ref{sec:form-basic-conc} introduces the theory, formalism, and fundamental concepts of decoherence. Section~\ref{sec:mastereqs} discusses the description of decoherence dynamics in terms of master equations. Section~\ref{sec:decmodels} reviews several classes of important decoherence models. Section~\ref{sec:decoh-errcorr} describes methods for avoiding and mitigating the influence of decoherence. Section~\ref{sec:exper-observ-decoh} gives an overview of several experiments that have demonstrated the gradual, controlled action of decoherence. Section~\ref{sec:impl-found-quant} comments on the implications of decoherence for foundational issues in quantum mechanics and for the different interpretations of quantum mechanics. Section~\ref{sec:concluding-remarks} offers concluding remarks.

\section{\label{sec:form-basic-conc}Basic formalism and concepts}

\begin{figure}
{\footnotesize \emph{(a)} \hspace{5cm} \emph{(b)} \hspace{5cm} \emph{(c)} }

\centering
\includegraphics[scale=.33]{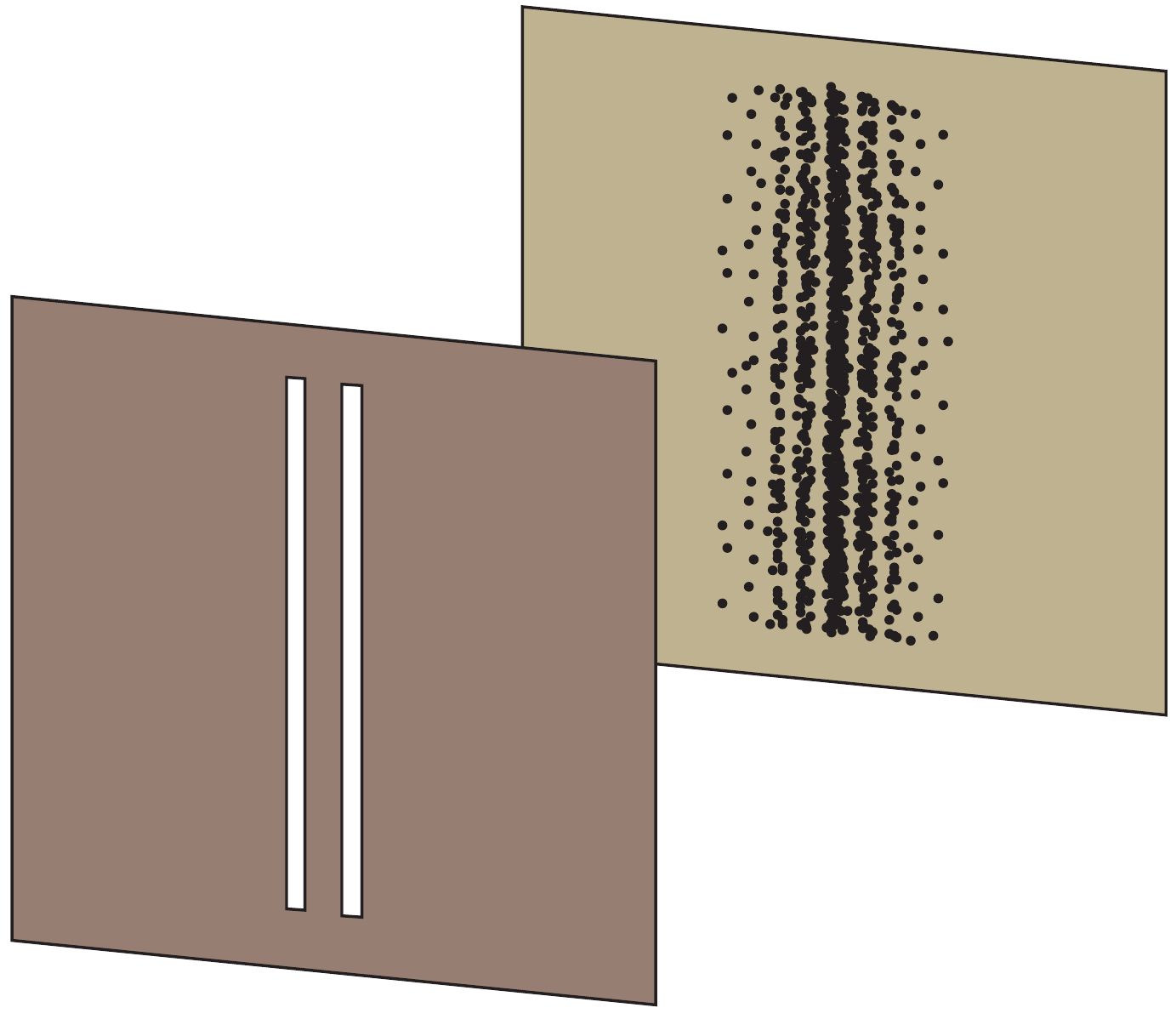} \hspace{.7cm}
\includegraphics[scale=.33]{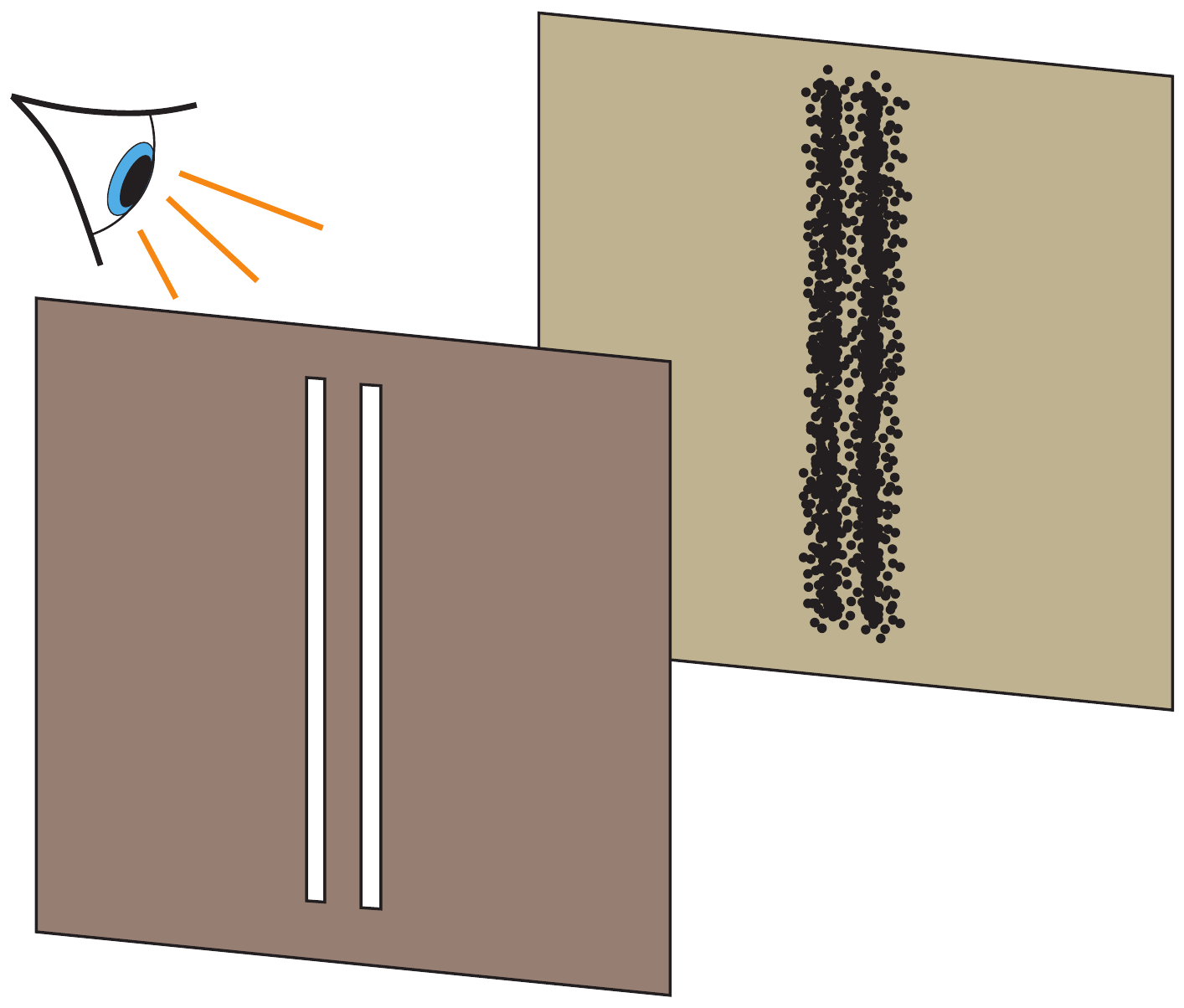}\hspace{.7cm}
\includegraphics[scale=.33]{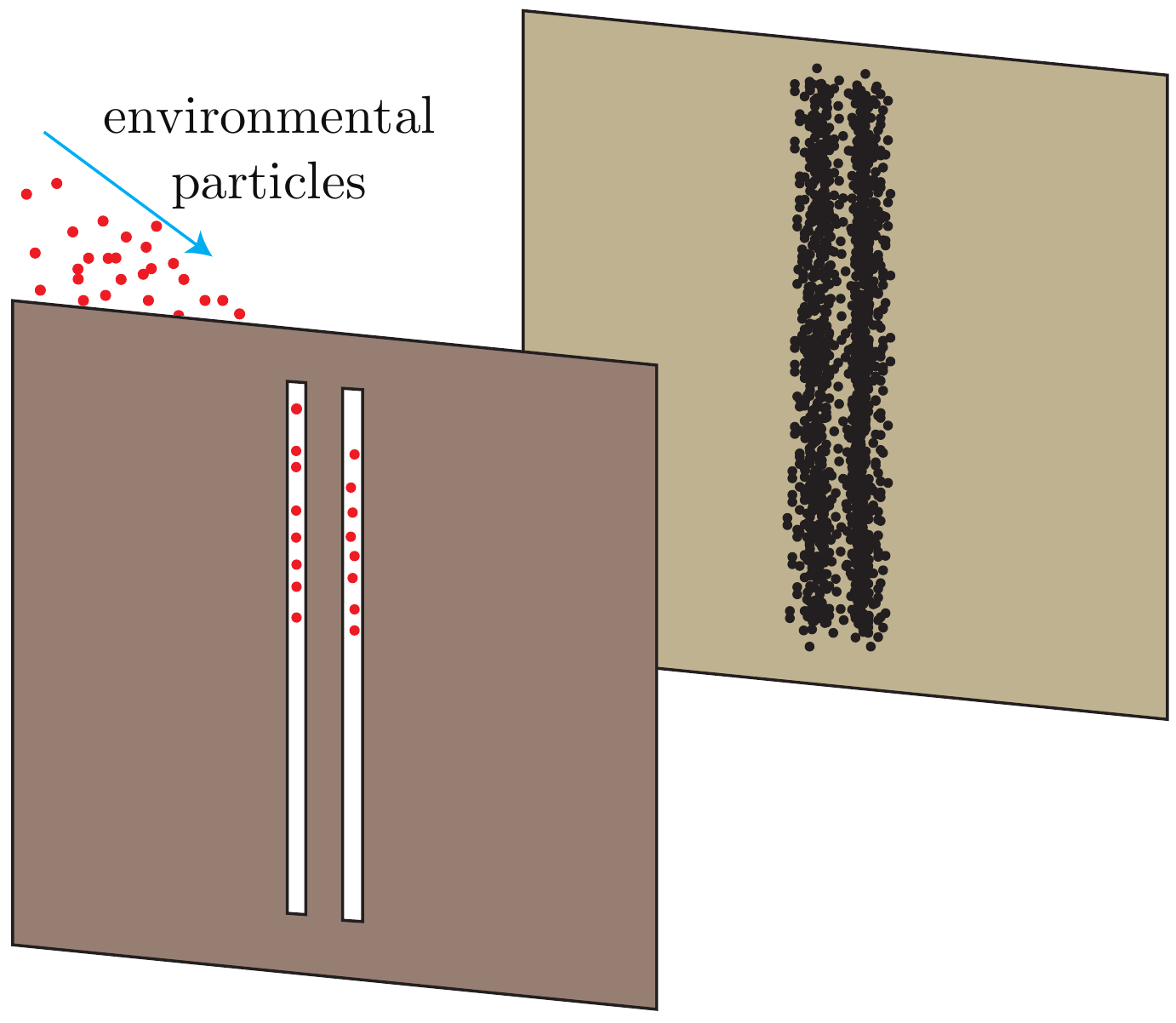}
\caption{\label{fig:bi}Basic idea of the decoherence process, illustrated in the context of a quantum double-slit experiment. \emph{(a)} Particles passing through a double slit create an interference pattern on a distant screen. \emph{(b)} If one monitors which slit each particle passes through, the interference pattern vanishes. \emph{(c)} The monitoring may arise from any measurement-like interaction, such as the scattering of environmental particles. The motional states of the environmental particles will then encode information about the path of the particle through the slits, resulting in the disappearance of the interference pattern. If the environment obtains only partial (rather than complete) which-path information, an interference pattern with reduced visibility obtains.}
\end{figure}

In the double-slit experiment, we cannot observe an interference pattern if we also measure which slit the particle passes through, that is, if we obtain perfect \emph{which-path information} (Fig.~\ref{fig:bi}). In fact, there is a continuous tradeoff between interference (phase information) and which-path information: the better we can distinguish the two possible paths, the less visible the interference pattern becomes \cite{Wooters:1979:az,Englert:1996:km}. What is more, for a decrease in interference visibility to occur it suffices that there are degrees of freedom \emph{somewhere in the world} that, \emph{if they were measured}, would allow us to make, with a certain degree of confidence, a statement about the path of the particle through the slits. While we cannot say that prior to their measurement, these degrees of freedom have encoded information about a particular, definitive path of the particle---instead, we have merely \emph{correlations} involving both possible paths---no actual measurement is required to bring about the decrease in interference visibility. It is enough that, \emph{in principle}, we could make such a measurement to obtain which-path information. 

This is somewhat loose talk, and conceptual caveats lurk. But it captures quite well the essence of what is happening in decoherence, where those ``degrees of freedom somewhere in the world'' are degrees of freedom of the system's environment that interact with the system, leading to the creation of quantum correlations (entanglement) between system and environment. Decoherence can thus be thought of as a process arising from the \emph{continuous monitoring of the system by the environment} \cite{Zurek:1981:dd}; effectively, the environment is performing nondemolition measurements on the system (see Sec.~\ref{sec:envir-monit-inform}). We now give a formal quantum-mechanical account of what we have just tried to convey in words, and then flesh out the consequences and details.

\subsection{Decoherence and interference damping}

Consider again the double-slit experiment and denote the quantum states of the particle (call it $S$, for ``system'') corresponding to passage through slit 1 and 2 by $\ket{s_1}$ and $\ket{s_2}$, respectively. Suppose that the particle interacts with another system $E$---for example, a detector or an environment---such that if the quantum state of the particle before the interaction is $\ket{s_1}$, then the quantum state of $E$ will become $\ket{E_1}$ (and similarly for $\ket{s_2}$), resulting in the final composite states $\ket{s_1}\ket{E_1}$ and $\ket{s_2}\ket{E_2}$, respectively. Owing to the linearity of the Schr\"odinger time evolution, for an initial superposition state $\alpha\ket{s_1}+\beta\ket{s_2}$ the final composite state will be entangled,
\begin{equation}
\label{eq:1dlkf}
\ket{\Psi} = \alpha \ket{s_1} \ket{E_1} + \beta \ket{s_2} \ket{E_2}.
\end{equation}
Consider now the \emph{reduced density matrix} $\op{\rho}_S$ for the system \cite{Landau:1927:uy,Neumann:1932:gq,Furry:1936:pp}, which is obtained by tracing out (i.e., averaging over) the degrees of freedom of the environment in the composite system--environment density matrix $\op{\rho}_{SE}$,
\begin{align}
  \label{eq:aa12rm}
 \op{\rho}_S &= \text{Tr}_E(\op{\rho}_{SE}).
\end{align}
The reduced density matrix exhaustively encodes the statistics of all possible local measurements on the system $S$. That is to say, for any observable that pertains only to the Hilbert space of the system, $\op{O} = \op{O}_S\otimes \op{I}_E$, where $\op{I}_E$ is the identity operator in the Hilbert space of the environment, the reduced density matrix $\op{\rho}_S$ will be sufficient to calculate the expectation value of $\op{O}$. To see this, let $\{ \ket{\psi_k} \}$ and $\{ \ket{\phi_l} \}$ be orthonormal bases of the Hilbert spaces of the system and environment, respectively. Then the expectation value of $\op{O}$ is
\begin{align}
\label{eq:rcae}
\langle \op{O} \rangle &= \text{Tr} \, (\op{\rho}_{SE}\op{O})  \notag \\
&= \sum_{kl} \bra{\phi_l} \bra{\psi_k} \op{\rho}_{SE}
\left(\op{O}_S \otimes
\op{I}_E\right) \ket{\psi_k}  \ket{\phi_l} \notag \\
&= \sum_{k} \bra{\psi_k} \left( \sum_l \bra{\phi_l}
  \op{\rho}_{SE}
  \ket{\phi_l}  \right) \op{O}_S \ket{\psi_k} \notag \\
&= \sum_{k} \bra{\psi_k} \left( \text{Tr}_E \, \op{\rho}_{SE}
\right) \op{O}_S \ket{\psi_k} \notag \\
&= \sum_{k} \bra{\psi_k} \op{\rho}_S\op{O}_S
\ket{\psi_k} \notag \\
&= \text{Tr}_S \left( \op{\rho}_S
  \op{O}_S \right),
\end{align}
showing that indeed only the reduced density matrix, rather than the full composite density matrix $\op{\rho}_{SE}$, is needed to calculate the expectation value. Since in the context of decoherence we are chiefly concerned with the effects of the environment on the measurable properties of the system, the reduced density matrix  plays an essential role in decoherence theory for describing the quantum state of a system in the presence of environmental entanglement \cite{Zurek:1981:dd,Zurek:1982:tv,Schlosshauer:2007:un}. 

For the composite state vector described by Eq.~\eqref{eq:1dlkf}, the reduced density matrix is \cite{Schlosshauer:2007:un}
\begin{align}
  \label{eq:aa12}
 \op{\rho}_S &= \text{Tr}_E(\op{\rho}_{SE}) =\text{Tr}_E \ketbra{\Psi}{\Psi} \nonumber \\
 &= \abs{\alpha}^2
    \ketbra{s_1}{s_1} + \abs{\beta}^2 \ketbra{s_2}{s_2} + \alpha \beta^*\ketbra{s_1}{s_2}
      \braket{E_2}{E_1} +  \alpha^*\beta\ketbra{s_2}{s_1}
      \braket{E_1}{E_2}.
\end{align}
Now suppose, for example, that we measure the particle's position by letting the particle impinge on a distant detection screen. Statistically, the resulting particle probability density $P(x)$ will be given by
\begin{align}\label{eq:fdljksjk2}
P(x) & = \text{Tr}_S(\op{\rho}_S x) = \notag \\ &= \abs{\alpha}^2
\abs{\psi_1(x)}^2 +\abs{\beta}^2 \abs{\psi_2(x)}^2 + 2 \,\text{Re} \left\{\alpha \beta^* \psi_1(x) \psi_2^*(x)\braket{E_2}{E_1} \right\},
\end{align}
where $\psi_i(x) \equiv \braket{x}{s_i}$. The last term represents the interference contribution. Thus, the visibility of the interference pattern is quantified by the overlap $\braket{E_2}{E_1}$, i.e., by the distinguishability of $\ket{E_1}$ and $\ket{E_2}$. In the limiting case of perfect distinguishability ($\braket{E_2}{E_1} = 0$), no interference pattern will be observable and we obtain the classical prediction. Phase relations have become \emph{locally} (i.e., with respect to $S$) inaccessible, and there is no measurement on $S$ that can reveal coherence between $\ket{s_1}$ and $\ket{s_2}$. The coherence is now between the states  $\ket{s_1} \ket{E_1}$ and $\ket{s_2} \ket{E_2}$, requiring an appropriate \emph{global} measurement (acting jointly on $S$ and $E$) for it to be revealed. Conversely, if the interaction between $S$ and $E$ is such that $E$ is completely unable to resolve the path of the particle, then $\ket{E_1}$ and $\ket{E_2}$ are indistinguishable and full coherence is retained at the level of $S$, as is also directly obvious from Eq.~\eqref{eq:1dlkf}. 

Here is another way of putting the matter. Looking back at Eq.~\eqref{eq:1dlkf}, we see that $E$ encodes which-way information about $S$ in the same ``relative-state'' sense \cite{Everett:1957:rw} in which EPR correlations \cite{Einstein:1935:dr,Bell:1964:ep,Bell:1966:ph} may be said to encode ``information.'' That is, if $\braket{E_2}{E_1} = 0$ and we were to measure $E$ and found it to be in state $\ket{E_1}$, we could, in EPR's words \cite[p.~777]{Einstein:1935:dr}, ``predict with certainty'' that we will find $S$ in $\ket{s_1}$.\footnote{Of course, this must not be read as saying that $S$ was already in $\ket{s_1}$ (i.e., ``went through slit 1'') prior to the measurement of $E$. Nor does it mean that the result of a subsequent path measurement on $S$ is necessarily determined, by virtue of the measurement on $E$, prior to this $S$-measurement's actually being carried out. After all, as Peres \cite{Peres:1978:aa} has cautioned us, unperformed measurements have no outcomes. So while the picture of $E$ as ``encoding which-path information'' about $S$ is certainly suggestive and helpful, it should be used with an understanding of its conceptual pitfalls.} Whenever such a prediction is possible were we to measure $E$, no interference effects between the components $\ket{s_1}$ and $\ket{s_2}$ can be measured at $S$, even if $E$ is never actually measured. In the intermediary regime where $0 < \abs{\braket{E_2}{E_1}} < 1$, $E$ encodes only \emph{partial} which-way information about $S$, in the sense that a measurement of $E$ could not reliably distinguish between $\ket{E_1}$ and $\ket{E_2}$; instead, sometimes the measurement will result in an outcome compatible with both $\ket{E_1}$ and $\ket{E_2}$. Consequently, an interference experiment carried out on $S$ would find reduced visibility, representing diminished local coherence between the components $\ket{s_1}$ and $\ket{s_2}$. Equation~\eqref{eq:fdljksjk2} shows that the reduction in visibility increases as $\ket{E_1}$ and $\ket{E_2}$ become more distinguishable.

As hinted above, the description developed so far describes the essence of the decoherence process if we identify the particle $S$ more generally with an arbitrary quantum system and the second system $E$ with the environment of $S$. Then an idealized account of the decoherence interaction has the (von Neumann \cite{Neumann:1932:gq}) form 
\begin{equation}
\label{eq:d1}
\biggl( \sum_i c_i \ket{s_i} \biggr) \ket{E_0} \quad \longrightarrow \quad \sum_i c_i \ket{s_i} \ket{E_i(t)}.
\end{equation}
Here we have introduced a time parameter $t$, where $t=0$ corresponds to the onset of the environmental interaction, with $\ket{E_i(t=0)} \equiv \ket{E_0}$ for all $i$.\footnote{In cases where the environment does not start out in a pure state, we can always purify it through the introduction of an additional (fictitious) environment. Without loss of generality, we can therefore always take the environment to be in a pure state before its interaction with the system.} At $t<0$ the system and environment are assumed to be uncorrelated (an assumption common to most decoherence models). 

A single environmental particle interacting with the system will typically only insufficiently resolve the components $\ket{s_i}$ in the system's superposition state. But because of the large number of such particles (and, hence, degrees of freedom), the overlap between their different joint states $\ket{E_i(t)}$ will rapidly decrease as a result of the buildup of many interaction events. Specifically, in many decoherence models an exponential decay of overlap is found \cite{Zurek:1982:tv,Joos:1985:iu,Paz:1993:ta,Leggett:1987:pm,Mokarzel:2002:za,Hornberger:2003:un,Zurek:2002:ii,Schlosshauer:2007:un,Breuer:2002:oq},
\begin{equation}
  \label{eq:jkHjkfhjhPJHyudf615}
  \braket{E_i(t)}{E_j(t)} \, \propto \, \E^{-t/\tau_\text{d}} \qquad \text{for $i \not= j$}.
\end{equation}
Here, $\tau_\text{d}$ is the characteristic decoherence timescale, which can be evaluated for particular choices of the parameters in each model (see Sec.~\ref{sec:decmodels}). Because the overlap of the environmental states quantifies the observability of interference effects between the corresponding system states that are correlated with the environmental states, Eq.~\eqref{eq:jkHjkfhjhPJHyudf615} describes an exponentially fast suppression of local interference.

\subsection{\label{sec:envir-monit-inform}Environmental monitoring and information transfer}

We will now motivate, in a different and more rigorous way, the picture of decoherence as a process of environmental monitoring. First, we express the influence of the environment in a completely general way. We assume that at $t=0$ there are no correlations between system $S$ and environment $E$, $\op{\rho}_{SE}(0) = \op{\rho}_S(0) \otimes \op{\rho}_E(0)$. We write $\op{\rho}_E(0)$ in its diagonal decomposition, $\op{\rho}_E(0) = \sum_i p_i \ketbra{E_i}{E_i}$, where $\sum_i p_i =1$ and the states $\ket{E_i}$ form an orthonormal basis of the Hilbert space of $E$. If $H$ denotes the Hamiltonian (here assumed to be time-independent) of $SE$ and $U(t) = \E^{-\I H t/\hbar}$ represents the unitary time evolution operator, then the density matrix of $S$ evolves according to 
\begin{align}
  \label{eq:1slvjhvkjfkjvsj0}
  \op{\rho}_S(t) &= \text{Tr}_E \left\{ U(t) \left[ \op{\rho}_S(0) \otimes \left( \sum_i p_i \ketbra{E_i}{E_i} \right) \right] U^\dagger(t) \right\}\nonumber \\
&= \sum_{ij} p_i \bra{E_j} U(t) \ket{E_i} \op{\rho}_S(0)\bra{E_i} U^\dagger(t) \ket{E_j}.
\end{align}
Introducing the \emph{Kraus operators} \cite{Kraus:1971:ii,Kraus:1983:ee} defined by $E_{ij}(t) = \sqrt{p_i} \bra{E_j} U(t) \ket{E_i}$, we obtain
\begin{equation}
 \op{\rho}_S(t) = \sum_{ij} E_{ij}(t) \op{\rho}_S(0) E^\dagger_{ij}(t).
\end{equation}
It is customary to combine the two indices $i$ and $j$ into a single index and write the Kraus operators as
\begin{equation}
  \label{eq:worihfvsjvttrafs2}
  W_k(t) \equiv \sqrt{p_{i_k}} \bra{E_{j_k}} U(t) \ket{E_{i_k}},
\end{equation}
such that 
\begin{equation}
\label{eq:dfjsb1}
 \op{\rho}_S(t) = \sum_k W_k(t) \op{\rho}_S(0) W^\dagger_k(t).
\end{equation}
Unitarity of the evolution of $SE$ implies that the Kraus operators satisfy the completeness constraint
\begin{equation}
  \label{eq:19sirhgvksjvbkjsb}
\sum_k W_k(t) W^\dagger_k(t) = I_S,
\end{equation}
where $I_S$ is the identity operator in the Hilbert space of $S$.\footnote{Conversely, Eq.~\eqref{eq:19sirhgvksjvbkjsb} may also be used as an indicator of unitarity; if it were not obeyed, then one would need to conclude that $SE$ is evolving nonunitarily due to the presence of an additional environment $E'$.} The Kraus-operator formalism (also called \emph{operator-sum formalism}) represents the effect of the environment as a sequence of (in general nonunitary) transformations of $\op{\rho}_S$ generated by the operators $W_k$ \cite{Breuer:2002:oq,Alicki:2007:uu}. The Kraus operators exhaustively encode information about the initial state of the environment and about the dynamics of the joint $SE$ system, and they play the role of generators of so-called dynamical maps (see Sec.~\ref{sec:mastereqs}). 

Following the treatment given by Hornberger in Ref.~\cite{Hornberger:2009:aq}, we will now use Eq.~\eqref{eq:dfjsb1} to formally motivate the view that decoherence corresponds to an indirect measurement of the system by the environment, and that it thus results from a transfer of information from the system to the environment. In such an indirect measurement, we let the system $S$ interact with a probe---here the environment $E$---followed by a projective measurement on $E$. The probe is treated as a quantum system. This procedure aims to yield information about $S$ without performing a projective (and thus destructive) direct measurement on $S$. To model such an indirect measurement, consider again an initial composite density operator $\op{\rho}_{SE}(0) = \op{\rho}_S(0) \otimes \op{\rho}_E(0)$ evolving under the action of $U(t) = \E^{-\I H t/\hbar}$, where $H$ is the total Hamiltonian. Consider a projective measurement $M$ on $E$ with eigenvalues $\alpha$ and corresponding projectors $P_\alpha = \ketbra{\alpha}{\alpha}$, with $P_\alpha^2=P_\alpha^\dagger=P_\alpha$. The probability of obtaining outcome $\alpha$ in this measurement when $S$ is described by the density operator $\op{\rho}_S(t)$ is
\begin{equation}
 \text{Prob}\left(\alpha \mid \op{\rho}_S(t) \right) = \text{Tr}_E \left( P_\alpha \op{\rho}_E(t) \right) = \text{Tr}_E \left\{ P_\alpha \text{Tr}_S \left[ U(t) \left( \op{\rho}_S(0) \otimes \op{\rho}_E(0) \right) U^\dagger(t)\right] \right\}.
\end{equation}
The density matrix of $S$ conditioned on the particular outcome $\alpha$ is
\begin{align}
\op{\rho}_S^{(\alpha)}(t) &= 
\frac{ \text{Tr}_E \left\{ \left[I \otimes P_\alpha \right] \op{\rho}_{SE}(t) \left[I \otimes P_\alpha \right] \right\}}{\text{Prob}\left(\alpha \mid \op{\rho}_S(t)\right)}\nonumber\\
&=\frac{ \text{Tr}_E \left\{ \left[I \otimes P_\alpha \right] U(t) \left[ \op{\rho}_S(0) \otimes \op{\rho}_E(0) \right] U^\dagger(t)  \left[I \otimes P_\alpha \right] \right\}}{\text{Prob}\left(\alpha \mid \op{\rho}_S(t)\right)}.\label{eq:hiuvb}
\end{align}
Inserting the diagonal decomposition $\op{\rho}_E(0) = \sum_k p_k \ketbra{E_k}{E_k}$ and carrying out the trace gives \cite{Hornberger:2009:aq}
\begin{equation}\label{eq:hiuvbxxxxx}
\op{\rho}_S^{(\alpha)}(t) = \sum_k \frac{ M_{\alpha,k}(t)  \op{\rho}_S(0) M^\dagger_{\alpha,k}(t)}{\text{Prob}\left(\alpha \mid \op{\rho}_S(t)\right)},
\end{equation}
where we have introduced the measurement operators
\begin{equation}
M_{\alpha,k}(t) = \sqrt{p_k} \bra{\alpha} U(t) \ket{E_k},
\end{equation}
which obey the completeness constraint $\sum_{\alpha,k}M_{\alpha,k}(t)M_{\alpha,k}^\dagger(t)=I_S$. Equation~\eqref{eq:hiuvbxxxxx} describes the effect of the indirect measurement on the state of the system. If, however, we do not actually inquire about the result of this measurement, we must assign to the system a density operator that is a sum over all the possible conditional states $\op{\rho}_S^{(\alpha)}(t)$ weighted by their probabilities $\text{Prob}\left(\alpha \mid \op{\rho}_S(t)\right)$,
\begin{equation}
\op{\rho}_S(t) = \sum_\alpha \text{Prob}\left(\alpha \mid \op{\rho}_S(t)\right) \op{\rho}_S^{(\alpha)}(t) = \sum_{\alpha,k} M_{\alpha,k}(t)  \op{\rho}_S(0) M^\dagger_{\alpha,k}(t).
\end{equation}
Note that this expression is formally analogous to the Kraus-operator expression of Eq.~\eqref{eq:dfjsb1}, which described the effect of a general environmental interaction on the state of the system. Recall, further, that the situation we encounter in decoherence is precisely one in which we do not actually read out the environment---or, in the present picture, in which we do not inquire about the result of the indirect measurement. This suggests that decoherence can indeed be understood as an indirect measurement---a monitoring---of the system by its environment.

\subsection{\label{sec:meas}Measures and visualization of decoherence}

Given the reduced density matrix $\op{\rho}_S(t)$ for a system interacting with an environment, there exist several measures for quantifying the amount of decoherence introduced into the system by the environmental interaction. Two commonly used measures are the \emph{purity},
\begin{equation}
\label{eq:puri}
\varsigma(\op{\rho}_S) = \text{Tr} \op{\rho}_S^2,
\end{equation}
and the \emph{von Neumann entropy} \cite{VonNeumann:1926:tv},
\begin{equation}
\label{eq:ent}
S(\op{\rho}_S) = - \text{Tr}\left( \op{\rho}_S \log_2 \op{\rho}_S \right).
\end{equation}
Both are based on the fact that the entanglement with the environment causes an initially pure quantum state of the system to become progressively mixed.

Consider first the purity, $\varsigma(\op{\rho}_S) = \text{Tr} \,\op{\rho}_S^2$. If $S$ is in a pure state, i.e., if its density matrix can be written as a single projector $\op{\rho}_S=\ketbra{\psi}{\psi}$ on a pure state $\ket{\psi}$, then $\op{\rho}_S^2=\op{\rho}_S$ and therefore $\varsigma(\op{\op{\rho}}_S) =1$. In the opposite limit of a maximally mixed state of an $N$-dimensional system,
\begin{equation}
\label{eq:msd}
  \op{\rho}_S = \frac{1}{N} \sum_{i}  \ketbra{\psi_i}{\psi_i},
\end{equation}
where the states $\{\ket{\psi_i}\}$ form an orthonormal basis of the system's Hilbert space, the purity attains its lower bound, $\varsigma(\op{\rho}_S) =1/N$. 

Similarly, the von Neumann entropy $S(\op{\rho}_S) = - \text{Tr}\left( \op{\rho}_S \log_2 \op{\rho}_S \right)$ is equal to zero for a pure state and increases for nonpure states, up to a value of $\log_2 (N)$ for a maximally mixed state. This can be seen explicitly by writing out the trace in the expression for the von Neumann entropy, which for an arbitrary density matrix $\op{\rho}$ with eigenvalues $\lambda_i$ yields
\begin{equation}\label{eq:ojibef00011}
  S(\op{\rho}) = - \text{Tr} \, ( \op{\rho} \log_2 \op{\rho} ) =
  - \sum_i \lambda_i \log_2 \lambda_i,
\end{equation}
where any eigenvalues $\lambda_i$ that are equal to zero (representing states not contained in the mixture) are by convention excluded from the sum. For a pure state, there will be only a single nonzero eigenvalue $\lambda_i$, which must be equal to 1, and therefore $S(\op{\rho}) = 0$. For a maximally mixed state, $\lambda_i=1/N$ for all $i$, and thus $S(\op{\rho}) = \log_2 (N)$, its largest possible value. 

A conceptual note on mixed density matrices is in order. Such density matrices may arise in two fundamentally different ways. In the first, a state-preparation procedure produces different possible pure states for the system; the mixture then reflects an observer's ignorance of which (pure) state was prepared in a particular run, which connects with the statistical distribution of (pure) states in the limit of many runs of the experiment. In this case, the probabilities associated with the pure states in the mixture can be thought of as classical entities: they represent either subjective ignorance in a situation when a single pure state was actually prepared, or they describe relative frequencies of pure states (for a physical \emph{ensemble} of systems). Such mixtures are also called \emph{proper} \cite{Espagnat:1966:mf,Espagnat:1988:cf,Espagnat:1995:ma,Schlosshauer:2003:tv,Schlosshauer:2007:un}. The second, distinct way in which a mixture may obtain is for a system entangled with an environment. Now, however, the reduced density matrix describing the mixture is ``improper,'' in the sense that no pure state can be ascribed to the system because of the presence of entanglement. The ``mixedness'' of the reduced state---reflecting a loss of information about a particular pure state arising from the environmental information transfer described in Sec.~\ref{sec:envir-monit-inform}---is purely quantum-mechanical in nature. Therefore, mixed reduced density matrices for systems entangled with an environment are not \emph{ignorance-interpretable} \cite{Espagnat:1966:mf,Espagnat:1988:cf,Espagnat:1995:ma,Schlosshauer:2003:tv,Schlosshauer:2007:un}, i.e., they do not describe a situation in which the system is in a pure state but one does not know which.\footnote{The degree to which this distinction between proper and improper ensembles is considered fundamental, or even relevant, depends in some measure on one's interpretation of quantum states. For example, in the interpretation known as QBism \cite{Fuchs:2014:pp}, all quantum states are purely subjective entities that encode an observer's probabilistic expectations associated with his future measurement interactions. On this view, the notion of the system's \emph{being} in a particular quantum state is not applicable to begin with, and the above cautionary note against interpreting decohered reduced density matrices as proper mixtures would consequently appear unnecessary. The use of a mixed reduced density matrix would simply reflect an observer's adjustment of his subjective probabilistic expectations on account of the presence of an environment.}

\begin{figure}
{\footnotesize \hspace{1.1cm} \emph{(a)} \hspace{6.9cm} \emph{(b)}}

\centering
\includegraphics[scale=.77]{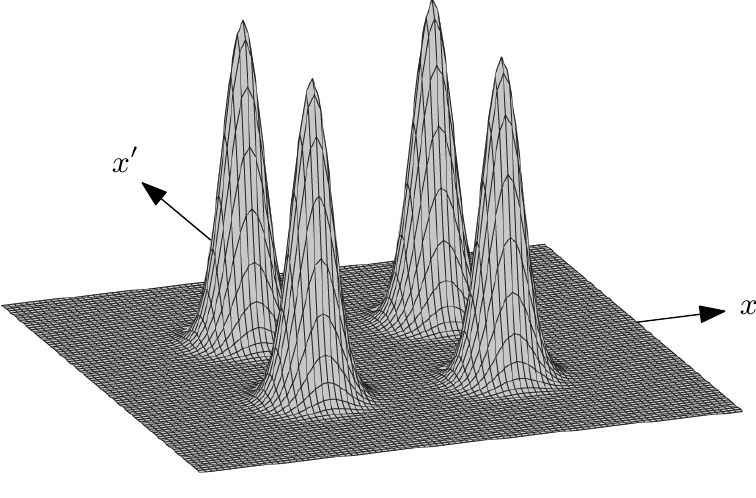} \hspace{1.5cm}
\includegraphics[scale=.77]{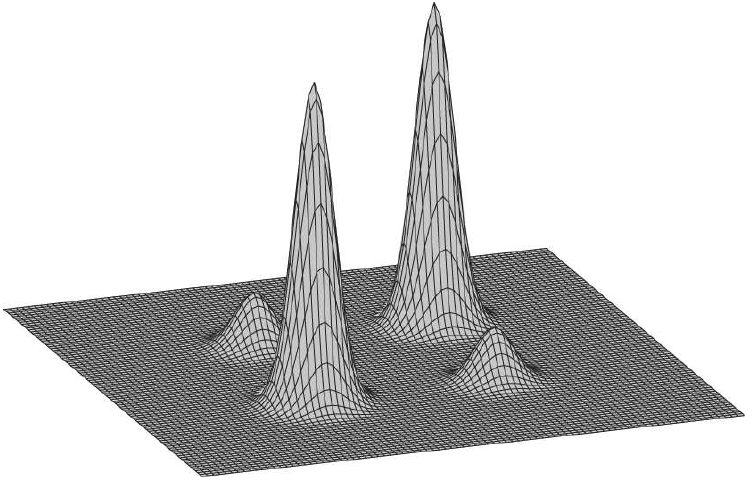}
\caption{\label{fig:r12}Visualization of the decoherence dynamics in one dimension, showing the reduced density matrix representing a superposition of two Gaussian wavepackets separated in position space. \emph{(a)} The initial density matrix before the onset of decoherence, exhibiting large off-diagonal terms that represent spatial coherence. \emph{(b)} Decoherence arising from entanglement with an environment diminishes the size of the off-diagonal terms over time. The direct peaks along the diagonal represent the position-space probability density $P(x)=\rho(x,x)$ and are, in the absence of dissipation, not affected by the decoherence process.}
\end{figure}

To visualize the decoherence of a quantum state, one may display the decay of the off-diagonal elements in the reduced density matrix as a function of time. As an example, Fig.~\ref{fig:r12} shows the reduced density matrix for a particle that moves in one spatial dimension and is described by a superposition of two position-space Gaussian wave packets. The interaction with the environment progressively reduces the size of the off-diagonal terms, while in the absence of dissipation the direct peaks (representing the probability distribution of finding the different possible position values in a measurement) remain unchanged. 

\begin{figure}
{\footnotesize \hspace{1.1cm}\emph{(a)} \hspace{6.9cm} \emph{(b)}}

\centering
\includegraphics[scale=.65]{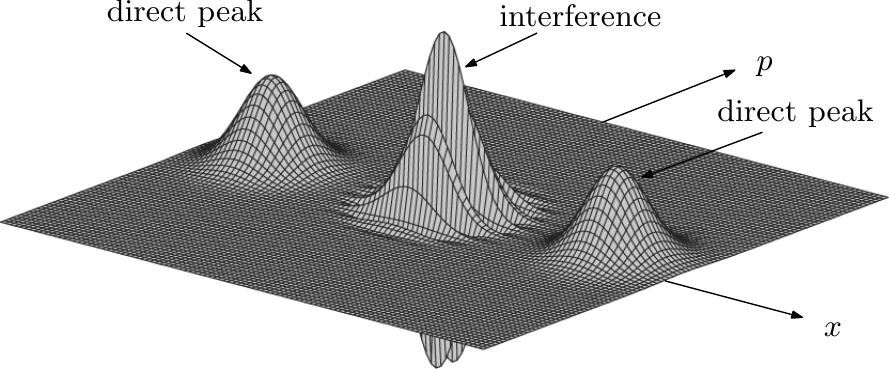} \hspace{1.5cm}
\includegraphics[scale=.65]{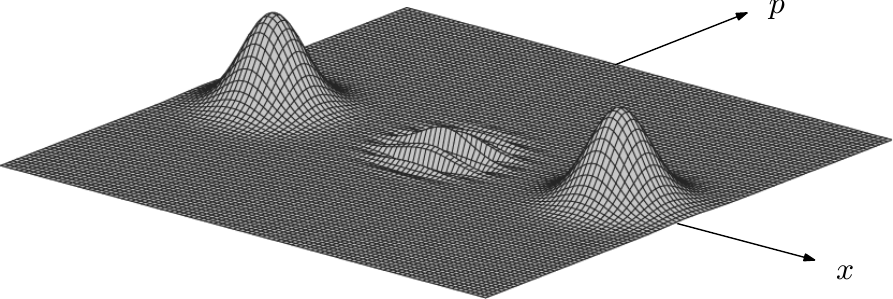}
\caption{Wigner representation of a decohering superposition of
  two position-space Gaussian wave packets in one spatial dimension. \emph{(a)} Interference is represented by an oscillatory, ridge-like pattern between the direct peaks. \emph{(b)} Decoherence manifests itself as a progressive damping of the oscillatory pattern.}
\label{fig:wig}
\end{figure}

An alternative and commonly used approach to representing the decoherence of a system represented by a continuous degree of freedom (such as position) is the \emph{Wigner function} \cite{Wigner:1932:un,Hillery:1984:tv}. Using the example of a position degree of freedom, the Wigner function representing a position-space density matrix $\rho(x,x') \equiv \bra{x}\op{\rho}\ket{x'}$ is defined by 
\begin{equation}
  \label{eq:fsoifhwddfs6611a}
W(x,p) = \frac{1}{2\pi\hbar} \int_{-\infty}^{+\infty} \D y \, \exp\left( \frac{\I p
  y}{\hbar}\right) \rho(x+y/2,x-y/2),
\end{equation}
where $p$ is the conjugate momentum variable. The Wigner function is attractive because it resembles a phase-space probability distribution: it is real-valued and normalized, $\int \D x \int \D p \,W(x,p) = 1$, and position and momentum distributions may be obtained from the marginals $P(x) = \int \D p \, W(x,p) $ and $P(p) = \int \D x \, W(x,p)$. Of course, owing to the uncertainty principle, no proper quantum phase-space probability distribution is admissible, a fact that is reflected in the observation that the Wigner function (with the notable exception of Gaussians \cite{Hudson:1974:ra}) may be negative in certain regions. In the Wigner representation, interference terms appear as an oscillatory, ridge-like pattern between the direct peaks, as shown in Fig.~\ref{fig:wig} for a superposition of two Gaussian wave packets separated in position space. The wavelength $\lambda$ of the oscillations is inversely proportional to the spatial separation $\Delta x$ of the wave packets, $\lambda =2\pi\hbar/\Delta x$, which implies that the oscillations become more rapid as the superposition becomes more nonclassical (i.e., as $\Delta X$ increases) \cite{Zurek:2002:ii,Schlosshauer:2007:un}. Decoherence then manifests itself as a progressive damping of these oscillations (see also Sec.~\ref{sec:quant-brown-moti} and Fig.~\ref{fig:gaussmov}). 

\subsection{\label{sec:envir-induc-supers}Environment-induced superselection}

As we have seen, decoherence results when a quantum system becomes entangled with its environment. How much the system becomes entangled---and thus how strong the effect of decoherence is---depends on how its initial quantum state relates to the Hamiltonian that governs the interaction between system and environment. In particular, as we will elaborate below, the specific structure of a given interaction Hamiltonian implies a set of quantum states that will become least entangled with the environment and are therefore most immune to the decohering influence of the environment. The states that are dynamically chosen through this \emph{stability criterion} \cite{Zurek:1981:dd,Zurek:1982:tv} are commonly referred to as \emph{preferred states} or \emph{pointer states}. In situations where the pointer states form a proper basis of the Hilbert space of the system---as is often the case for low-dimensional systems---one may also speak of a \emph{preferred basis} or \emph{pointer basis}. In this sense, the interaction with the environment imposes a dynamical filter on the state space, selecting those states that can be stably prepared and observed even in the presence of the environmental interaction \cite{Zeh:1970:yt,Kubler:1973:ux,Zurek:1981:dd,Zurek:1982:tv,Walls:1985:lm}. Zurek, who studied the process of state selection via environmental interactions in two influential papers in the 1980s \cite{Zurek:1981:dd,Zurek:1982:tv}, called it \emph{environment-induced superselection}. 

To find the pointer states, we decompose the total system--environment Hamiltonian $\op{H}$ into the self-Hamiltonians $\op{H}_S$ and $\op{H}_E$ of the system $S$ and environment $E$ (describing the intrinsic dynamics), and a part $\op{H}_\text{int}$ representing the interaction between system and environment, 
\begin{equation}
H = \op{H}_S + \op{H}_E + \op{H}_\text{int}. 
\end{equation}
In many cases of practical interest, $\op{H}_\text{int}$ dominates the evolution of the system, such that $\op{H} \approx \op{H}_\text{int}$; this situation is referred to as the \emph{quantum-measurement limit} of decoherence. Let us first consider this case and determine the corresponding pointer states. In the spirit of the stability criterion, the idea is to find a set of system states $\{\ket{s_i}\}$ that remain unchanged and do not get entangled with the environment under the evolution generated by $\op{H}_\text{int}$. This condition is met for the eigenstates $\{\ket{s_i}\}$ (with eigenvalues $\{\lambda_i\}$) of the part of the interaction Hamiltonian $\op{H}_\text{int}$ that addresses the Hilbert space of the system---i.e., for the states of the system that are stationary under $\op{H}_\text{int}$ \cite{Zurek:1981:dd}. In this case, a system--environment product state $\ket{s_i}\ket{E_0}$ at $t=0$ (when the interaction with the environment is turned on) will evolve according to 
\begin{equation}
  \label{eq:gxlknn98ygya24}
  \E^{-\I \op{H}_\text{int} t/\hbar} \ket{s_i}\ket{E_0}=
  \lambda_i \ket{s_i}\E^{-\I \op{H}_\text{int} t/\hbar} \ket{E_0} \equiv  \ket{s_i}\ket{E_i(t)},
\end{equation}
where we have assumed that $\op{H}_\text{int}$ is not explicitly time-dependent. Since the state remains a product state for all subsequent times $t>0$, there is no entanglement or decoherence. Note that \emph{superpositions} of pointer states are in general not immune to decoherence, since the environmental states $\ket{E_i(t)}$ tend to become rapidly distinguishable and therefore lead to an entangled system--environment state. 

Thus, in the quantum-measurement limit the pointer states $\ket{s_i}$ are obtained by diagonalizing the interaction Hamiltonian in the subspace of the system. We may also define a \emph{pointer observable} $\op{O}_S = \sum_i o_i \ketbra{s_i}{s_i}$ of the system as a linear combination of pointer-state projectors $\op{\Pi}_i=\ketbra{s_i}{s_i}$.  Because each $\ket{s_i}$ is an eigenstate of $\op{H}_\text{int}$, it follows that $\op{O}_S$ commutes with $\op{H}_\text{int}$,
\begin{equation}
  \label{eq:dhvvsdnbbfvs27}
  \bigl[ \op{O}_S, \op{H}_\text{int} \bigr] = 0.
\end{equation}
This \emph{commutativity criterion} \cite{Zurek:1981:dd,Zurek:1982:tv} is particularly easy to apply when $\op{H}_\text{int}$ takes the (commonly encountered) tensor-product form $\op{H}_\text{int} = \op{S} \otimes \op{E}$, in which case the pointer observables will be those observables that commute with the system part $\op{S}$ of the interaction Hamiltonian. 

If the operator $\op{S}$ appearing in $\op{H}_\text{int}$ is Hermitian and thus could represent a physical observable, it will describe the quantity monitored by the environment, in the spirit of the discussion in Sec.~\ref{sec:envir-monit-inform}. For example, often position dynamically emerges as the environment-selected quantity because many interaction Hamiltonians describe scattering processes governed by force laws that depend on some power of particle distance. Then the Hamiltonian will commute with the position operator, and the corresponding eigenstates---the pointer states---are approximate eigenstates of position, represented by narrow position-space wave packets. These states are dynamically robust, thus accounting for the fact that position is a preferred, stable quantity in our everyday world. Their superpositions, however, are typically rapidly decohered, especially if they refer to mesoscopically or macroscopically distinct positions. A ubiquitous source of decoherence of such spatial superpositions is the scattering of environmental particles, a process known as \emph{collisional decoherence} (see Sec.~\ref{sec:collisionaldecoherence}). This explains why mesoscopic and macroscopic spatial superpositions tend to be prohibitively difficult to observe for larger systems \cite{Zurek:1981:dd,Zurek:1982:tv,Joos:1985:iu,Zurek:1991:vv,Gallis:1990:un,Diosi:1995:um,Hornberger:2003:un,Hornberger:2006:tb,Hornberger:2008:ii,Busse:2009:aa,Busse:2010:aa}. 

But collisional decoherence may also be significant in microscopic systems. For instance, chiral molecules such as sugar occur in two distinct spatial configurations: left-handed and right-handed. When these molecules are immersed into a medium, the scattering of environmental particles resolves these two configurations, and thus the left-handed and right-handed chirality eigenstates dynamically emerge as the preferred states. Energy eigenstates of such molecules, on the other hand, are represented by superpositions of chirality eigenstates and are therefore subject to immediate decoherence. This explains why chiral molecules are found not in energy eigenstates but in chirality eigenstates \cite{Harris:1981:rc,Zeh:1999:qr,Trost:2009:ll,Bahrami:2012:oo}. 

To give another example, the fact that we do not observe superpositions of different electric charges can be explained as a consequence of the coupling of a charge to its own Coulomb far-field acting as an environment \cite{Zeh:1970:yt,Giulini:1995:zh,Giulini:2000:ry}. The interaction leads to decoherence of charge superpositions and therefore to the environment-induced superselection of eigenstates of the charge operator. This role of the environment was already spelled out by Zeh \cite{Zeh:1970:yt} in his 1970 paper marking the birth of decoherence theory:
\begin{quote}
This interpretation of measurement may also explain certain ``superselection rules'' which state, for example, that superpositions of states with different charge cannot occur. \dots [Such states] cannot be dynamically stable because of the significantly different interaction of their components with their environment, in analogy with the different handedness components of a sugar molecule.
\end{quote}

In general, any interaction Hamiltonian $\op{H}_\text{int}$ can be written as a diagonal decomposition of system and environment operators $\op{S}_\alpha$ and $\op{E}_\alpha$, $\op{H}_\text{int} =  \sum_\alpha \op{S}_\alpha \otimes \op{E}_\alpha$. For Hermitian operators $\op{S}_\alpha$, such a Hamiltonian represents the simultaneous environmental monitoring of different observables $\op{S}_\alpha$ of the system. Then the pointer states will be simultaneous eigenstates of the operators $\op{S}_\alpha$:
\begin{equation}
  \label{eq:OIbvsrhjkbv9}
  \op{S}_\alpha \ket{s_i} = \lambda_i^{(\alpha)}\ket{s_i} \qquad
  \text{for all $\alpha$ and $i$}. 
\end{equation}

The \emph{quantum limit of decoherence} \cite{Paz:1999:vv} applies in situation where the self-Hamiltonian of the system dominates over the interaction Hamiltonian. This represents a situation in which the frequencies of the environment are small compared with the frequencies of the system. Then the environment will be able to monitor only quantities that are constants of motion. In the case of nondegeneracy, this will be the energy of the system, leading to the environment-induced superselection of energy eigenstates for the system \cite{Paz:1999:vv}.\footnote{Energy eigenstates are given a special role in textbooks because of their stationarity. Note, however, that for closed systems superpositions of energy eigenstates are equally viable. It is only through the inclusion of an environment and consideration of the resulting decoherence that such superpositions become dynamically suppressed, leading to the emergence of energy eigenstates as the preferred states of the system. Therefore, decoherence can be used to justify the special status commonly attributed to energy eigenstates.}

For more realistic models of decoherence, the stability criterion, Eq.~\eqref{eq:dhvvsdnbbfvs27}, often cannot be fulfilled exactly. Furthermore, in many situations the self-Hamiltonian of the system and the interaction Hamiltonian are of approximately equal strength, which means that neither of the two limiting cases discussed above---the quantum-measurement limit of negligible intrinsic dynamics and the quantum limit of decoherence of a slow environment---are appropriate. To deal with such situations, a more general, operational method known as a \emph{predictability sieve} \cite{Zurek:1993:pu,Zurek:1993:qq,Zurek:1998:re} can be used to identify classes of approximate pointer states. Here one computes the amount of decoherence introduced into the system over time for a large set of initial, pure states of the system evolving under the total system--environment Hamiltonian. Typically, this decoherence is measured as a decrease in purity $\text{Tr} \op{\rho}_S^2$ or an increase in  von Neumann entropy $S(\op{\rho}_S) = - \text{Tr}\left( \op{\rho}_S \log_2 \op{\rho}_S \right)$ of the reduced density matrix $\op{\rho}_S$ over time (see Sec.~\ref{sec:meas} for a description of these measures). This allows one to rank the states according to their susceptibility to decoherence, and in this way the states most robust to the environmental interaction---the (approximate) pointer states---can be identified \cite{Zurek:1993:pu,Zurek:1993:qq,Zurek:1998:re,Zurek:2002:ii}. For example, in the model for quantum Brownian motion (see Sec.~\ref{sec:quant-brown-moti}), different measures of decoherence all lead to the selection of minimum-uncertainty wave packets in phase space as the most robust states \cite{Kubler:1973:ux,Zurek:1993:pu,Zurek:2002:ii,Diosi:2000:yn,Joos:2003:jh,Eisert:2003:ib}.

We note that the term ``predictability sieve'' is motivated by the connection between the purity of a state of a system and our knowledge of this state. A pure state encodes perfect knowledge (one assigns precisely one state vector to the system) and therefore maximum ``predictability.''  On the other hand, decoherence caused by entanglement with (and thus information transfer to) the environment renders the reduced density matrix progressively impure, which introduces an additional, purely quantum-mechanical probabilistic element and diminishes the degree of predictability.  In this sense, the states most robust to decoherence---the (exact or approximate) pointer states  whose purity is least affected by the presence of the environmental interactions---are also the most predictable \cite{Zurek:1993:pu,Zurek:1993:qq,Zurek:1998:re}.

Subspaces of a system's Hilbert space spanned by pointer states that couple to the environment in the same way are known as \emph{decoherence-free subspaces}. Because any state in such a subspace will be immune to decoherence, decoherence-free subspaces are a valuable tool for encoding quantum information in a manner that avoids decoherence. Decoherence-free subspaces will be discussed in Sec.~\ref{sec:dfs}.

\subsection{\label{sec:prol-inform-quant}Proliferation of information and quantum Darwinism}

Decoherence theory focuses on the effect that entanglement with an environment has on a quantum system. As discussed in Sec.~\ref{sec:envir-monit-inform}, decoherence represents a process in which the environment monitors the system and information is transferred from the system to the environment. In this spirit, \emph{quantum Darwinism} \cite{Zurek:2003:pl,Ollivier:2003:za,Ollivier:2004:im,Blume:2004:oo,Blume:2005:oo,Zurek:2009:om,Riedel:2010:un,Riedel:2011:un,Riedel:2012:un,Zurek:2014:xx, Zwolak:2016:zz,Zwolak:2017:mm,Zurek:2018:on,Unden:2018:ia} turns the focus from the system to the environment and considers the information that the environment encodes about the system. Building on the ideas of decoherence and environmental encoding of information, quantum Darwinism broadens the role of the environment to that of a communication and amplification channel. It studies how interactions between the system and its environment lead to the \emph{redundant} storage of selected information about the system in many fragments of the environment. Hence the name ``quantum Darwinism'': certain states of the system are \emph{fitter} than others in the sense that they are able to imprint their information robustly and redundantly across the environment.  By measuring some of these environmental fragments, observers can indirectly obtain information about the system without appreciably disturbing the system itself. Indeed, this represents how we typically observe objects. For example, we see an object not by directly interacting with it, but by intercepting scattered photons that encode information about the object's spatial structure \cite{Riedel:2010:un,Riedel:2011:un}. 

In this sense, quantum Darwinism provides a dynamical explanation for the robustness of states to observation, especially for macroscopic objects. It has been shown that the observable of the system that can be imprinted most completely and redundantly in many distinct fragments of the environment coincides with the pointer observable selected by the system--environment interaction  \cite{Ollivier:2003:za,Ollivier:2004:im,Blume:2004:oo,Blume:2005:oo}; conversely, most other states do not seem to be redundantly storable. Indeed, the redundant proliferation of information regarding pointer states may be as inevitable as decoherence itself \cite{Zwolak:2014:tt}. 

Quantum Darwinism has been studied in several concrete models, including spin environments \cite{Blume:2004:oo,Zwolak:2016:zz}, quantum Brownian motion \cite{Blume:2007:oo}, and photon and photon-like environments \cite{Riedel:2010:un,Riedel:2011:un,Zwolak:2014:tt}. The efficiency of the amplification process described by quantum Darwinism can be expressed in terms of the quantum Chernoff information \cite{Zwolak:2014:tt}, and the ability of this information measure to appropriately capture the rich dynamics of amplification has been confirmed in the context of realistic spin models \cite{Zwolak:2016:zz}. 

The structure and amount of information that the environment encodes about the system can be quantified using the measure of \emph{mutual information}, either in its classical \cite{Ollivier:2003:za,Ollivier:2004:im} or quantum \cite{Zurek:2002:ii,Blume:2004:oo,Blume:2005:oo} definition. Classical mutual information measures how well one can predict the outcome of a measurement of a given observable of the system $S$ by measuring an observable on a fraction of the environment $E$ \cite{Ollivier:2003:za,Ollivier:2004:im}. Quantum mutual information generalizes this concept and is defined as \cite{Zurek:2002:ii,Blume:2004:oo,Blume:2005:oo}
\begin{equation}
\mathcal{I}_{S:E} =S(\op{\rho}_S) + S(\op{\rho}_E) - S(\op{\rho}_{SE}),
\end{equation}
where $\op{\rho}_S$ is the density matrix of the system $S$, $\op{\rho}_E$ is the density matrix of the environment $E$, $\op{\rho}_{SE}$ is the density matrix of the composite system $SE$, and $S(\op{\rho})$ is the von Neumann entropy, Eq.~\eqref{eq:ent}, associated with $\op{\rho}$. Quantum mutual information represents the amount of entropy that would be created if all quantum correlations between $S$ and $E$ were destroyed; in other words, it quantifies how strongly system and environment are correlated. Classical and quantum mutual information give similar results \cite{Ollivier:2003:za,Ollivier:2004:im,Zurek:2002:ii,Blume:2004:oo,Blume:2005:oo} because the difference between the two measures, known as the \emph{quantum discord} \cite{Ollivier:2001:az}, vanishes when decoherence is effective enough to select a pointer basis \cite{Ollivier:2001:az}. We note here that the measure of quantum discord has also been applied to an analysis of Bohr's suggestion that the classicality of a measurement outcome is related to its communicability by classical means \cite{Streltsov:2013:oo}. 

Recently, many of the more subtle details of quantum Darwinism have begun to be investigated. For example, Zwolak and Zurek \cite{Zwolak:2017:mm} have shown that environmental imprints left by quantum systems other than the system of interest do not appreciably affect the redundancy of the environmental information about the system of interest. The influence of factors such as initial correlations, interactions between subenvironments, and non-Markovian dynamics that may hinder the redundant encoding of information in the environment has been studied by several authors \cite{Riedel:2012:un,Galve:2016:oo, Pleasance:2017:oo, Ciampini:2018:ii}. Among such studies, Pleasance and Garraway \cite{Pleasance:2017:oo} used the model of a single qubit interacting with a collection of bosonic environments to investigate environmental encoding of information in the presence of many subenvironments. Ciampini et al.\ \cite{Ciampini:2018:ii} employed photonic cluster states to explore, both theoretically and experimentally, the influence that correlations between parts of the environment have on the redundancy and objectivity of environmental information. It has also been shown \cite{Galve:2016:oo, Pleasance:2017:oo, Ciampini:2018:ii} that non-Markovian dynamics and the resulting memory effects can result in a backflow of information from the environment to the system in a manner that impedes the creation of robust, classical, redundant environmental records. In addition to the aforementioned photonic experiment by Ciampini et al.\ \cite{Ciampini:2018:ii}, experimental studies of the ideas of quantum Darwinism have been reported by Unden et al.\ \cite{Unden:2018:ia}, who used a controlled interaction between a nitrogen vacancy center (the system) and several nuclear spins (the environment). 

\subsection{\label{sec:decoh-vers-diss}Decoherence versus dissipation and noise}

Dissipation is always accompanied by decoherence (see, for example, the early studies by Walls and Milburn \cite{Walls:1985:pp} and by Caldeira and Leggett \cite{Caldeira:1985:tt}, who investigated  the influence of damping on the coherence of superpositions of macroscopically different states). The converse, however, is not necessarily true. In fact, one of the earliest models of decoherence due to random spin environments \cite{Zurek:1982:tv, Cucchietti:2005:om} clearly demonstrated that the system may rapidly decohere without any loss of energy from the system. When dissipation and decoherence are both present, the loss of coherence is usually many orders of magnitude faster than any relaxation processes induced by dissipation. For example, a classic paper by Zurek \cite{Zurek:1986:uz} gave a ballpark estimate for the ratio of the relaxation timescale $\tau_\text{r}$ to the decoherence timescale $\tau_\text{d}$ for a massive object represented by a coherent superposition of two positions separated by $\Delta x$:
\begin{equation}
  \label{eq:daf12}
  \frac{\tau_\text{r}}{\tau_\text{d}} \sim \left( \frac{\Delta
      x}{\lambda_\text{th}} \right)^2,
\end{equation}
where 
\begin{equation}
  \label{eq:daf12thermal}
\lambda_\text{th}=\frac{\hbar}{\sqrt{2mk_BT}}
\end{equation}
is the thermal de Broglie wavelength of the object. Applied to a macroscopic object of mass $m = \unit[1]{g}$ at $T=\unit[300]{K}$ with a macroscopic separation $\Delta x =\unit[1]{cm}$, Eq.~\eqref{eq:daf12} gives $\tau_\text{r}/\tau_\text{d} \sim10^{40}$ \cite{Zurek:1986:uz}. Thus, for macroscopic objects described by such nonclassical superposition states, dissipation is typically negligible over the timescale relevant to the decoherence process.

Decoherence is a consequence of environmental entanglement, and as such is a purely quantum-mechanical effect. In the literature (especially in the area of quantum information processing), the term ``decoherence'' is often used more broadly to encompass any process, quantum or classical, that detrimentally affects the desired superposition states. An example would be classical noise processes arising, for instance, from experimental fluctuations and imperfections, such as variations in laser intensities in ion-trap experiments \cite{Schneider:1998:yz,Miquel:1997:zz}, bias fluctuations in superconducting qubits \cite{Martinis:2003:bz}, and inhomogeneities in the magnetic fields used in NMR quantum processing \cite{Vandersypen:2004:ra}. When averaged over many different realizations of such noise processes, the density matrix of the system then shows a decay of off-diagonal terms, representing a loss of interference similar to what would result from environmental entanglement. But it is important to realize that for an individual instance of the noise process applied to an individual system, the evolution is completely unitary---there is no ``washing-out'' of phase information, no loss of information from the system, and no creation of entanglement between the system and an environment. Hence the consequences of the noise process could in principle be undone through a local operation acting on the system alone (in fact, this is the basis of the spin-echo method for reversing collective spin dephasing in NMR experiments). Such a local reversal is not possible for decoherence resulting from environmental entanglement; ``undoing'' decoherence to restore an (unknown) pre-decoherence state of the system will require appropriate measurements on the environment to gather information that has leaked from the system \cite{Myatt:2000:yy,Zurek:2002:ii}.\footnote{Incidentally, this is reminiscent of the situation in a quantum eraser experiment \cite{Jaynes:1980:lm,Peres:1980:im,Scully:1982:yb,Scully:1991:yb}, where interference fringes can be extracted from the no-fringes data only once the outcomes of measurements on one particle are correlated with the outcomes of measurements on the other, entangled partner \cite{Englert:1999:aq,Ashby:2016:pp}.}

We note that the loss of phase coherence due to environmental entanglement has sometimes been \emph{simulated} (with the above caveats) by classical fluctuations introduced through the addition of time-dependent perturbations to the self-Hamiltonian of the system; see, for example,  Refs.~\cite{Schneider:1998:yz,Schneider:1999:tt,Turchette:2000:aa,Myatt:2000:yy} and Sec.~\ref{sec:trapped} for applications of this approach to decoherence in ion traps.

\section{\label{sec:mastereqs}Master equations}

To calculate the time-evolved reduced density matrix of a decohering system, the route we have discussed so far consists of determining the time evolution $\ket{\psi}\ket{E_0} \longrightarrow \ket{\Psi_{SE}(t)}$ of the joint quantum state of the system and environment, and then obtaining the reduced density matrix of the system by tracing out the degrees of freedom of the environment in the composite density matrix $\op{\rho}_{SE}(t) = \ketbra{\Psi_{SE}(t)}{\Psi_{SE}(t)}=  U(t) \op{\rho}_{SE}(0) U^\dagger(t)$, where $U(t) = \E^{-\I \op{H} t/\hbar}$ is the time-evolution operator for the composite system $SE$ evolving under the total Hamiltonian $\op{H}$.  This is the procedure formally represented by Eq.~\eqref{eq:1slvjhvkjfkjvsj0}:
\begin{equation}
\label{eq:dmm2}
  \op{\rho}_S(t) = \text{Tr}_E \, \op{\rho}_{SE}(t) \equiv \text{Tr}_{E} \left\{ U(t) \op{\rho}_{SE}(0) U^\dagger(t) \right\}.
\end{equation}
Alternatively, we can start from the Liouville--von Neumann equation for the composite density matrix $\op{\rho}_{SE}(t)$,
\begin{equation}
\label{eq:dmm2xdfvb}
  \frac{\partial}{\partial t} \op{\rho}_{SE}(t) = -\frac{\I}{\hbar} \left[ \op{H}, \op{\rho}_{SE}(t)\right],
\end{equation}
and then take the trace over the environment, which yields a differential equation for the evolution of the reduced density operator,
\begin{equation}
\label{eq:dmsdlm2}
  \frac{\partial}{\partial t} \op{\rho}_S(t) = -\frac{\I}{\hbar} \text{Tr}_{E} \left\{ \left[ \op{H}, \op{\rho}_{SE}(t)\right]\right\}.
\end{equation}
The evolution equations \eqref{eq:dmm2} and \eqref{eq:dmsdlm2} require calculating the exact dynamics of the system and environment, as the reduced density matrix at some time $t$ (or, equivalently its differential change) depends on the full system--environment state and its entire past history. Typically, solving such equations presents an intractable problem both analytically and numerically, and therefore the exact description is generally not useful in practice. Furthermore, since we are usually not interested in the dynamics of the environment (unless we explicitly inquire about, say, the storage of information in the environment, as in the program of quantum Darwinism described in Sec.~\ref{sec:prol-inform-quant}), calculating the full composite system--environment state also provides unnecessary detail. 

Master equations offer a shortcut. To provide a reduction in computational effort over Eqs.~\eqref{eq:dmm2} and \eqref{eq:dmsdlm2}, such master equations are typically based on certain assumptions and simplifications that lead to an approximate (but in practice sufficiently accurate) description of the decoherence process. The reduced density matrix is calculated directly from an (in general nonunitary) evolution equation that depends only on the reduced, not the global, density matrix. A \emph{generalized master equation} for the reduced density matrix is of the form 
\begin{equation}
\label{eq:dggamsdlm2}
  \frac{\partial}{\partial t} \op{\rho}_S(t) = \mathcal{K} \left[ \op{\rho}_S(t'), t' < t \right ],
\end{equation}
where $\mathcal{K}$ is a superoperator that takes the history of the reduced density matrix as input. Of particular interest to the description of decoherence processes are \emph{Markovian master equations}, 
\begin{equation}
\label{eq:dggamsdlm58672}
  \frac{\partial}{\partial t} \op{\rho}_S(t) = \mathcal{L}\op{\rho}_S(t).
\end{equation}
Such master equations are local in time (the right-hand side of the equation does not depend on the history of the density matrix), and the superoperator $\mathcal{L}$ does not depend on time or the initial preparation. 

Markovian master equations are widely used in the description of decoherence dynamics. They enable a relatively easy calculation of the reduced dynamics while still providing, in many cases of practical interest, a good approximation to the exact dynamics and the experimentally observed data. We will now discuss their derivation and underlying assumptions. We start by introducing the concept of dynamical maps (Sec.~\ref{sec:dynamical-maps}), followed by the discussion of the approach to Markovian master equations via the formalism of quantum dynamical semigroups and their generators (Sec.~\ref{sec:semigr-deriv-mark}). Separately, we describe the derivation of Markovian master equations from microscopic considerations (Sec.~\ref{sec:micr-deriv-mark}). We also comment on the formalism of quantum trajectories (Sec.~\ref{sec:quantum-trajectories}) and on the treatment of non-Markovian decoherence (Sec.~\ref{sec:non-mark-decoh}).

\subsection{\label{sec:dynamical-maps}Dynamical maps} 

In what follows, we will make the usual assumption of an initially uncorrelated system--environment state, $\op{\rho}_{SE}(0)=\op{\rho}_{S}(0)\otimes \op{\rho}_{E}(0)$. Equation~\eqref{eq:dmm2} defines a state transformation
\begin{equation}
\label{eq:dmmetf2}
\op{\rho}_{S}(0) \mapsto \op{\rho}_S(t) = \mathcal{V}_t \op{\rho}_{S}(0),
\end{equation}
with 
\begin{equation}
  \label{eq:dmssmetf2}
\mathcal{V}_t \op{\rho}_{S}(0) \equiv \text{Tr}_{E} \left\{ U(t) \left[\op{\rho}_{S}(0)\otimes \op{\rho}_{E}(0) \right]U^\dagger(t) \right\}.
\end{equation}
The transformation $\mathcal{V}_t$ given in Eqs.~\eqref{eq:dmmetf2} and \eqref{eq:dmssmetf2} is an instance of a \emph{dynamical map} \cite{Breuer:2002:oq,Alicki:2007:uu}. A dynamical map $\mathcal{V}_t : \op{\rho}(0) \mapsto \op{\rho}(t)$ is a transformation that takes an arbitrary initial quantum state $\op{\rho}(0)$ to a final quantum state $\op{\rho}(t)$ at some fixed time $t$ in accordance with the rules of quantum mechanics. Since the map defined in Eq.~\eqref{eq:dmssmetf2} was solely derived from the Schr\"odinger equation and the trace operation, it will automatically obey the correct quantum rules. In Sec.~\ref{sec:envir-monit-inform}, we already showed [see Eq.~\eqref{eq:dfjsb1}] that the right-hand side of Eq.~\eqref{eq:dmssmetf2} can be expressed in terms of Kraus operators $\op{W}_k(t)$ \cite{Kraus:1971:ii,Kraus:1983:ee}, and therefore the dynamical map $\mathcal{V}_t$ defined by Eq.~\eqref{eq:dmssmetf2} can be written as
\begin{equation}
  \label{eq:asdmssmetf2}
\mathcal{V}_t \op{\rho}_{SE}(0) = \sum_k W_k(t) \op{\rho}_S(0) W^\dagger_k(t),
\end{equation}
where the $\op{W}_k(t)$ are operators in the Hilbert space $\mathcal{H}_S$ of $S$ obeying the completeness constraint~\eqref{eq:19sirhgvksjvbkjsb}, i.e.,
$\sum_k W_k(t) W^\dagger_k(t) = I_S$. For a Hilbert space $\mathcal{H}_S$ of finite dimension $N$, the number of Kraus operators required to represent a dynamical map is bounded by $N^2$ \cite{Alicki:2007:uu}.\footnote{For a (separable) Hilbert space of infinite dimension, a countable set of Kraus operators is needed.} Since Eq.~\eqref{eq:dmssmetf2} represents the most general way in which the state of the open quantum system $S$ may change, it follows from Eq.~\eqref{eq:asdmssmetf2} that any dynamical map can be completely characterized in terms of a set of Kraus operators $\op{W}_k \in \mathcal{H}_S$ with $\sum_k W_k W^\dagger_k = I_S$ \cite{Kraus:1983:ee,Lindblad:1976:um,Breuer:2002:oq,Alicki:2007:uu}. Therefore, the Kraus operators play the role of \emph{generators} of dynamical maps.

Alternatively, and equivalently \cite{Kraus:1983:ee}, a dynamical map $\mathcal{V}_t$ may be characterized by requiring it to obey the following three mathematical conditions \cite{Breuer:2002:oq,Alicki:2007:uu}: complete positivity, convex linearity, and trace preservation. Let us describe these conditions in turn.

\begin{enumerate} 

\item \emph{Complete positivity.} It is well known that a valid density operator must be positive semidefinite, i.e., its eigenvalues must be nonnegative, because these eigenvalues have the physical interpretation of probabilities. Therefore, a dynamical map $\mathcal{V}_t$ must be \emph{positive} in the sense that it takes positive semidefinite operators to positive semidefinite operators. But this is not sufficient. Instead, we must require \emph{complete positivity} \cite{Kraus:1971:ii,Gorini:1976:tt,Lindblad:1976:um,Breuer:2002:oq,Alicki:2001:aa,Alicki:2007:uu,Benatti:2005:ii}, a much stronger condition. It means that also all extensions $\mathcal{V}_t \otimes \text{id}_n$ of $\mathcal{V}_t$ to a composite Hilbert space $\mathcal{H}_\text{ext} = \mathcal{H}_S \otimes \tilde{\mathcal{H}}_n \equiv \mathcal{H}_S \otimes \mathbb{C}^n$ for all integer $n$ must be positive, where $\text{id}_n$ denotes the identity map on the space $\tilde{\mathcal{H}}_n$ that leaves all operators in that space unchanged. That is, we require that $\mathcal{V}_t \otimes \text{id}_n$ maps any $\op{\rho} \in \mathcal{H}_\text{ext}$ onto another valid density operator. 

The physical motivation behind this requirement is as follows. Imagine an ancillary system $A$ (represented by the Hilbert space $\tilde{\mathcal{H}}_n$), which is assumed to have no intrinsic dynamics (i.e., the self-Hamiltonian is $H_A=0$), and which is placed at a large distance from the system $S$ of interest such that it does not interact with $S$. Then the dynamical map for the composite system $SA$ will be given by $\mathcal{V}_t \otimes \text{id}_n$, which should again be positive for any state of the composite system $SA$; this is the condition of complete positivity. If this condition is not met, then it can be shown (see, for example, Refs.~\cite{Horodecki:1995:oo,Peres:1996:oo,Alicki:2001:aa,Benatti:2005:ii}) that, if $S$ and $A$ start out entangled,\footnote{We can imagine that the quantum correlations between $S$ and $A$ arose from some past interaction prior to the initial time point when the map  $\mathcal{V}_t \otimes \text{id}_n$ is applied.} the linear map $\mathcal{V}_t \otimes \text{id}_n$ may give rise to negative probabilities.\footnote{A simple example is the 2D transposition map $\mathcal{T} : \left(\begin{smallmatrix}a&b\\c&d\end{smallmatrix}\right) \mapsto \left(\begin{smallmatrix}a&c\\b&d\end{smallmatrix}\right)$, which is linear and positive. However, when its extension $\mathcal{T}\otimes\text{id}_2$ is applied to the density matrix $\op{\rho}=\ketbra{\Psi^+}{\Psi^+}$ representing the maximally entangled bipartite state $\ket{\Psi^+}=2^{-1/2}\left(\ket{0}\ket{0}+\ket{1}\ket{1}\right)$, the resulting matrix is no longer positive.} Thus, complete positivity ensures that $\mathcal{V}_t$ generates physically consistent dynamics even when the system $S$ initially has correlations with another system. 

A related motivation of completely positivity that is especially pertinent to open quantum systems can be given by considering two identical, noninteracting $N$-level systems $S_1$ and $S_2$ immersed into the same environment \cite{Benatti:2002:oo,Benatti:2005:ii}. Suppose that $\mathcal{V}_t$ generates the reduced evolution of $S_1$ so that, to first approximation, $\mathcal{V}_t\otimes \mathcal{V}_t$ generates the evolution of the joint system $S_1S_2$. Then one can show that for entangled states of $S_1S_2$, $\mathcal{V}_t\otimes \mathcal{V}_t$ preserves positivity if and only if $\mathcal{V}_t$ is completely positive \cite{Benatti:2002:oo}. See Ref.~\cite{Benatti:2005:ii} for a detailed discussion of the requirement of completely positivity in the context of open quantum systems and quantum master equations.

\item \emph{Convex linearity.} Consider a convex-linear combination of density operators, $\op{\rho} = \lambda \op{\rho}_1 + (1-\lambda) \op{\rho}_2$ with $0 < \lambda < 1$. This represents an ignorance-interpretable mixture of the two ensembles $\op{\rho}_1$ and $\op{\rho}_2$, with (classical) probability weights $\lambda$ and $1-\lambda$. Thus, we require that it should be possible to represent the time-evolved mixture again as an ignorance-interpretable mixture of the two ensembles evolved \emph{individually} under the action of the dynamical map:
\begin{equation}
  \label{eq:cl}
\mathcal{V}_t \op{\rho} = \mathcal{V}_t \left\{ \lambda \op{\rho}_1 + (1-\lambda) \op{\rho}_2\right\} = \lambda \mathcal{V}_t\op{\rho}_1 + (1-\lambda) \mathcal{V}_t\op{\rho}_2.
\end{equation}

\item \emph{Trace preservation.} We demand that the time-evolved density matrix remains a trace-one operator:
\begin{equation}
  \label{eq:t78sccl}
\text{Tr} \left\{ \mathcal{V}_t \op{\rho} \right\} =1.
\end{equation}

\end{enumerate}

We emphasize again that this characterization of dynamical maps in terms of completely positive, convex-linear, trace-preserving maps is equivalent to the characterization in terms of maps generated by a complete set of Kraus operators [see Eq.~\eqref{eq:asdmssmetf2}] \cite{Kraus:1983:ee}.

\subsection{Markovian master equations}

In Eq.~\eqref{eq:dggamsdlm58672}, we introduced the notion of a Markovian master equation, given by $\partial_t \op{\rho}_S(t) = \mathcal{L}\op{\rho}_S(t)$ with a time-independent superoperator $\mathcal{L}$. There exist several approaches to deriving the dynamical maps representing Markovian master equations: an axiomatic approach based on the theory of quantum dynamical semigroups (see Sec.~\ref{sec:semigr-deriv-mark}) \cite{Lindblad:1976:um,Gorini:1976:tt,Gorini:1978:uf,Davies:1974:tw,Kossakowski:1972:tf,Alicki:2007:uu}; a microscopic approach proceeding from a consideration of the relevant Hamiltonians and the evolution generated by them (see Sec.~\ref{sec:micr-deriv-mark}); and a monitoring approach (see Refs.~\cite{Hornberger:2006:tc,Hornberger:2008:ii}).

The common key assumption underlying these approaches is known as the \emph{Markov approximation}. Here one considers two timescales: (i) the typical relaxation time $\tau_r$ of the open quantum system, describing the timescale on which the environment affects the evolution of the system; and (ii) the typical coherence time $\tau_c$ of the environment, representing the characteristic timescale for the decay of correlations between the degrees of the environment that are being generated by the interaction with the system. In the Markov approximation, one assumes that $\tau_r \gg \tau_c$, i.e., the environmental self-correlations are assumed to decay rapidly compared to the timescale on which the open system evolves. Then, on this coarse-grained timescale defined by $\tau_r$, the environment may be considered memoryless, meaning that it does not appreciably retain information about its interaction with the system between time points much farther apart than the timescale set by environmental self-correlations. 

\subsubsection{\label{sec:semigr-deriv-mark}Semigroup approach to Markovian master equations and the Lindblad form}

In terms of a family $\{ \mathcal{V}_t \mid t \ge 0 \}$ of dynamical maps parametrized by $t$, the Markov property can be rigorously stated in terms of the \emph{semigroup condition} \cite{Lindblad:1976:um,Gorini:1976:tt,Gorini:1978:uf,Davies:1974:tw,Kossakowski:1972:tf,Alicki:2007:uu},
\begin{equation}
\label{eq:d4488m58672}
\mathcal{V}_{t_2}\mathcal{V}_{t_1}=\mathcal{V}_{t_1+t_2}.
\end{equation}
If this relation is fulfilled, then $\{ \mathcal{V}_t \mid t \ge 0 \}$ is said to form a \emph{quantum dynamical semigroup} \cite{Alicki:2007:uu}. In this case, it can be shown that  (given mild assumptions) there exists a superoperator $\mathcal{L}$ such that \cite{Alicki:2007:uu} 
\begin{equation}\label{eq:767n8m58672}
\mathcal{V}_t = \exp (\mathcal{L}t),
\end{equation}
which implies the quantum Markovian master equation~\eqref{eq:dggamsdlm58672}, $\partial_t \op{\rho}_S(t) = \mathcal{L}\op{\rho}_S(t)$.
The superoperator $\mathcal{L}$ is a linear map known as the \emph{generator of the dynamical semigroup}; it is also often referred to as the \emph{Liouville superoperator}.

Gorini, Kossakowski, and Sudarshan \cite{Gorini:1976:tt} first showed that for a finite-dimensional Hilbert space $\mathcal{H}_S$ of the system, the most general form of the generator $\mathcal{L}$ is \cite{Breuer:2002:oq,Alicki:2007:uu}
\begin{equation}
  \label{eq:sdfkhwr69}
\mathcal{L}\op{\rho}_S = \underbrace{-\frac{\I}{\hbar} \bigl[
  \op{H}'_S, \op{\rho}_S \bigr]}_{\text{unitary part}} + \underbrace{\sum_{\alpha,\beta=1}^{N^2-1} \gamma_{\alpha\beta} \left\{
\op{F}_\alpha \op{\rho}_S \op{F}^\dagger_\beta - \frac{1}{2} \op{F}^\dagger_\beta\op{F}_\alpha \op{\rho}_S
- \frac{1}{2} \op{\rho}_S\op{F}^\dagger_\beta\op{F}_\alpha \right\}}_{\text{nonunitary part (``dissipator'')}}.
\end{equation}
Here, $N=\dim(\mathcal{H}_S)$, and the $\op{F}_\alpha$ are a set of $N^2$ linear operators forming an orthonormal\footnote{Orthonormality is here defined in terms of the Hilbert--Schmidt scalar product $(\op{F}_\alpha, \op{F}_\beta) = \text{tr}_S (\op{F}_\alpha^\dagger \op{F}_\beta)$.}  basis in the Liouville space of linear and bounded operators in $\mathcal{H}_S$, with $\op{F}_{N^2}$ chosen to be proportional to the identity (see Refs.~\cite{Breuer:2002:oq,Alicki:2007:uu} for mathematical details). The coefficients $\gamma_{\alpha\beta}$ define a matrix, and one can show that this matrix is positive, i.e., that all its eigenvalues $\kappa_\mu$ are nonnegative. Conversely, if a master equation can be brought into the form~\eqref{eq:sdfkhwr69} with a positive coefficient matrix, it will represent the generator of a quantum dynamical semigroup and hence ensure complete positivity. Equation~\eqref{eq:sdfkhwr69} is known as the \emph{first standard form} of the generator. 

The first term on the right-hand side of Eq.~\eqref{eq:sdfkhwr69} describes the unitary evolution of the system under a Hamiltonian $\op{H}'_S$. This Hamiltonian will, in general, differ from the self-Hamiltonian $\op{H}_S$ of the system because of the presence of the environment, which renormalizes the energy levels of the system. Accordingly, $\op{H}'_S$ is often referred to as the \emph{environment-renormalized} (or \emph{Lamb-shifted}) Hamiltonian (one can show that it commutes with $\op{H}_S$).  The second term on the right-hand side of Eq.~\eqref{eq:sdfkhwr69} reflects the nonunitary influence of the environment, which changes the coherence of the system and may also lead to a loss of energy from the system (i.e., dissipation). It is sometimes referred to as the \emph{dissipator} \cite{Breuer:2002:oq} (but note that it may generate decoherence without dissipation, so the terminology is not always apt). In the context of applications to decoherence models, the time-independent coefficients $\gamma_{\alpha\beta}$ encapsulate all relevant information about the physical parameters of the decoherence (and possibly dissipation) processes. If the $\op{F}_\alpha$ are chosen to be dimensionless, then the $\gamma_{\alpha\beta}$ have units of frequency (i.e., inverse time).

Because the coefficient matrix $\gamma_{\alpha\beta}$ is positive, we may diagonalize it and rewrite Eq.~\eqref{eq:sdfkhwr69} as
\begin{equation}\label{eq:lindblad}
\mathcal{L}\op{\rho}_S = -\frac{\I}{\hbar} \bigl[
  \op{H}'_S, \op{\rho}_S \bigr] + \sum_{\mu=1}^{N^2-1} \kappa_\mu \left\{
\op{L}_\mu \op{\rho}_S \op{L}^\dagger_\mu - \frac{1}{2} \op{L}^\dagger_\mu\op{L}_\mu \op{\rho}_S
- \frac{1}{2} \op{\rho}_S\op{L}^\dagger_\mu\op{L}_\mu \right\},
\end{equation}
where the \emph{Lindblad operators} $\op{L}_\mu$ are linear combinations of the operators $\op{F}_\alpha$. Lindblad \cite{Lindblad:1976:um} (see also Ref.~\cite{Gorini:1978:uf}) showed that Eq.~\eqref{eq:lindblad} is the most general form for a bounded generator in any separable Hilbert space for a countable set of indices $\{\mu\}$.\footnote{The assumption of boundedness does not hold in many physical applications: both the Hamiltonian $\op{H}_S$ of the system and the Lindblad operators $\op{L}_\mu$ will in general be unbounded. It turns out, however, that one can define quantum dynamical semigroups using expressions of the Lindblad form~\eqref{eq:lindblad} even for unbounded Lindblad operators \cite{Davies:1976:uu,Holevo:1996:ll}, and conversely one finds that all known instances of generators of quantum dynamical semigroups are of Lindblad form (or can be readily adapted to it) \cite{Breuer:2002:oq,Alicki:2007:uu}.}  Equation~\eqref{eq:lindblad} is known as the \emph{second standard form}, the \emph{diagonal standard form}, or the \emph{Lindblad form} of the generator \cite{Breuer:2002:oq,Alicki:2007:uu}. When the generator~\eqref{eq:lindblad} is inserted into Eq.~\eqref{eq:dggamsdlm58672}, the resulting master equation is referred to as the \emph{Gorini--Kossakowski--Sudarshan--Lindblad master equation}, or \emph{Lindblad master equation} for short. If one chooses the Lindblad operators to be dimensionless, then the quantities $\kappa_\mu$ (i.e., the eigenvalues of the coefficient matrix) have units of inverse time and may be interpreted directly as decoherence rates. We note that a given Lindblad generator $\mathcal{L}$ does not uniquely determine the Lindblad operators or the Hamiltonian $\op{H}'_S$ \cite{Breuer:2002:oq}.

In applications to decoherence models, the Lindblad operators $\op{L}_\mu$ are constructed from linear combinations of the system operators $\op{S}_\alpha$ appearing in the diagonal decomposition of the interaction Hamiltonian, $\op{H}_\text{int} = \sum_\alpha \op{S}_\alpha \otimes \op{E}_\alpha$. When the system operators $\op{S}_\alpha$ represent physical observables monitored by the environment, they will be Hermitian and therefore the Lindblad operators, being linear combinations of the $\op{S}_\alpha$, will be Hermitian as well. In this case, the Lindblad generator~\eqref{eq:lindblad} may be further simplified by writing it in double-commutator form, resulting in the master equation 
\begin{equation}\label{eq:lindbladc}
\frac{\partial}{\partial t} \op{\rho}_S(t) = - \frac{\I}{\hbar} \left[ \op{H}'_S, \op{\rho}_S(t) \right] - \frac{1}{2} \sum_{\mu=1}^{N^2-1} \kappa_\mu \left[ \op{L}_\mu, \left[ \op{L}_\mu, \op{\rho}_S(t) \right]
\right].
\end{equation}
Note that the second, nonunitary term on the right-hand side vanishes if 
\begin{equation}
  \label{eq:7fskjhytw10}
  \left[ \op{L}_\mu, \op{\rho}_S(t) \right] = 0 \qquad \text{for all $\mu,t$},
\end{equation}
in which case $\op{\rho}_S(t)$ will evolve unitarily. Incidentally, this leads to a connection with the concept of pointer states. Recall that the Lindblad operators $\op{L}_\mu$ are linear combinations of the operators $\op{S}_\alpha$ in the diagonal decomposition of the interaction Hamiltonian. Thus Eq.~\eqref{eq:7fskjhytw10} implies (disregarding very specific linear combinations of $\op{S}_\alpha$) that we must also have $\left[ \op{S}_\alpha, \op{\rho}_S(t) \right] = 0$ for all $\alpha, t$. This, however, is nothing but the pointer-state criterion of Eq.~\eqref{eq:OIbvsrhjkbv9}, which says that quantum states that are simultaneous eigenstates of all operators $\op{S}_\alpha$ will not decohere, and therefore will evolve unitarily.

\subsubsection{\label{sec:two-simple-examples}Two simple examples of Lindblad master equations}

A particularly important and simple case is that of a single system observable monitored by the environment, corresponding to $\op{H}_\text{int} = \op{S} \otimes \op{E}$. Let us mention two such basic examples. 

\paragraph{Pure decoherence in the spin--boson model} Consider a qubit whose $\op{\sigma}_z$ spin coordinate is coupled to an environment of harmonic oscillators (this is the spin--boson model discussed in Sec.~\ref{sec:spin-boson-models}). In the absence of intrinsic tunneling dynamics, the qubit evolution can be described by a Lindblad master equation with a single Lindblad operator $\op{L}=\op{\sigma}_z$, 
\begin{equation}
\label{eq:vjp32gbntrkh22max}
\frac{\partial}{\partial t} \op{\rho}_\mathcal{S}(t) = -\frac{\I}{\hbar} \bigl[
  \op{H}_\mathcal{S}, \op{\rho}_\mathcal{S}(t) \bigr] -
D \left[ \op{\sigma}_z, \left[ \op{\sigma}_z,
    \op{\rho}_\mathcal{S}(t) \right]\right].
\end{equation}
This equation may be derived from the relevant Hamiltonians using a microscopic approach; see Eq.~\eqref{eq:vjp32gbntrkh22}  and Ref.~\cite{Schlosshauer:2007:un} for details. It describes the environmental monitoring and resulting decoherence in the $\{\ket{0},\ket{1}\}$ eigenbasis of $\op{\sigma}_z$, with $D$ playing the role of a decoherence rate.  To see this explicitly, we write the Lindblad double commutator on the right-hand side of Eq.~\eqref{eq:vjp32gbntrkh22max} in matrix form in the $\{\ket{0},\ket{1}\}$ basis, 
\begin{equation}
  \label{eq:fsigj98gre42}
  D \left[ \op{\sigma}_z, \left[ \op{\sigma}_z,
    \op{\rho}_\mathcal{S}(t) \right]\right] = D \left(
  \frac{1}{2} \op{\rho}_\mathcal{S}(t) - 2\op{\sigma}_z
  \op{\rho}_\mathcal{S}(t) \op{\sigma}_z \right) \dot{=}
\,\,D \left( \begin{array}{cc} 0 & \rho_\mathcal{S}^{(01)}(t) \\
    \rho_\mathcal{S}^{(10)}(t) & 0\end{array} \right),
\end{equation}
where $\rho_\mathcal{S}^{(ij)}(t)$ denotes the matrix element $\bra{i} \op{\rho}_\mathcal{S}(t) \ket{j}$, $i \in \{0,1\}$. It then follows from Eq.~\eqref{eq:vjp32gbntrkh22max} that the evolution of the off-diagonal matrix elements of the reduced density matrix (expressed in the eigenbasis of $\op{\sigma}_z$) governed by the $D$ term alone is
\begin{align}
  \label{eq:44}
  \frac{\partial}{\partial t} \rho_\mathcal{S}^{(01)}(t) = -
  D\rho_\mathcal{S}^{(01)}(t), \qquad
  \frac{\partial}{\partial t} \rho_\mathcal{S}^{(10)}(t) = - D\rho_\mathcal{S}^{(10)}(t).
\end{align}
This shows that the off-diagonal elements decay exponentially at a rate given by $D$, while the diagonal elements (the occupation probabilities) are not affected. Thus Eq.~\eqref{eq:fsigj98gre42} generates pure decoherence in the $\{\ket{0},\ket{1}\}$ basis without dissipation.

\paragraph{Spatial decoherence} As another example, consider a free particle in one dimension, subject to environmental monitoring of its position. The most simple way in which we may represent this environmental interaction is in terms of a single Lindblad operator $L \propto \op{x}$. With $\op{H}'_S = \op{H}_S = p^2/2m$, Eq.~\eqref{eq:lindbladc} reads
\begin{equation}\label{eq:lifsfdndbladc}
\frac{\partial}{\partial t} \op{\rho}_S(t) =  -\frac{\I}{2m\hbar}\left[p^2, \op{\rho}_S(t) \right] - \Lambda \left[ x, \left[ x, \op{\rho}_S(t) \right]\right],
\end{equation}
where the coefficient $\Lambda$ has dimensions of $\text{(time)}^{-1} \times \text{(length)}^{-2}$. Writing this equation in the position representation, one obtains
\begin{equation}\label{eq:lifsshvgvvvxayhcgiefdndbladc}
  \frac{\partial \op{\rho}_S(x,x',t)}{\partial t} = - \frac{\I}{2m\hbar} \left(\frac{ \partial^2}{\partial x'^2}- \frac{ \partial^2}{\partial x^2} \right) \op{\rho}_S(x,x',t) -  \Lambda
  \left(x-x'\right)^2   \op{\rho}_S(x,x',t),
\end{equation}
which is the classic equation of motion for spatial decoherence due to environmental scattering first derived in Ref.~\cite{Joos:1985:iu} (see Sec.~\ref{sec:collisionaldecoherence}). The second term on the right-hand side of Eq.~\eqref{eq:lifsshvgvvvxayhcgiefdndbladc} generates exponential decay of spatial coherences (represented by the off-diagonal elements $x\not= x'$) at a rate given by $ \Lambda \left(x-x'\right)^2$, 
\begin{equation}
\rho_S(x,x',t) =\rho_S(x,x',0) \exp\left[-\Lambda (x-x')^2 t\right],
\end{equation}
where we have neglected the intrinsic dynamics. We see that the localization rate depends on the square of the separation $\abs{x-x'}$. We have already encountered this dependence in Eq.~\eqref{eq:daf12}, and we will find it again below in the context of an explicit scattering model [see Eqs.~\eqref{eq:scwer2}
 and \eqref{eq:scwer6565}] and the Caldeira--Leggett master equation for quantum Brownian motion [see Eqs.~\eqref{eq:fsdojgdj1} and \eqref{eq:odijsvuhfsw21}]. In fact, if we add a harmonic potential to the system, Eq.~\eqref{eq:lifsfdndbladc} represents the high-temperature limit of the master equation for quantum Brownian motion with the dissipative term neglected, as given in Eq.~\eqref {eq:vfnbcclasclnd9s27} of Sec.~\ref{sec:cald-legg-mast} (see also Sec.~5.2.4 of Ref.~\cite{Schlosshauer:2007:un} for details). When dissipative effects are included, the appropriate Lindblad operator is a linear combination of the position and momentum operators of the system [see Eq.~\eqref{eq:dkvnkl1}].

\subsubsection{\label{sec:micr-deriv-mark}Microscopic derivation of Markovian master equations}

In the previous section, we obtained the most general form of a Markovian (and time-homogeneous) quantum master equation from the formalism of quantum dynamical semigroups and their generators. This approach is mathematically elegant, leads to a very general result, and automatically ensures the complete positivity of the evolution. Still, from a physical point of view, it would also be desirable to directly derive Markovian master equations from the underlying Hamiltonian description of the system and its environment. To do so, one proceeds from the total Hamiltonian, $\op{H}=\op{H}_S+\op{H}_E+\op{H}_\text{int}$, and an initially uncorrelated system--environment state, $\op{\rho}_{SE}(0)=\op{\rho}_{S}(0)\otimes \op{\rho}_{E}(0)$, and then imposes the following two main assumptions. 

The first assumption is the \emph{Born approximation}, which takes the coupling between system and environment to be sufficiently weak and the environment to be sufficiently large such that, to second order in the interaction Hamiltonian $\op{H}_\text{int}$, changes of the density operator of the environment may be neglected and the composite system--environment state remains in an approximate product state over time. That is, one assumes that $\op{\rho}_{SE}(t) \approx \op{\rho}_S(t) \otimes \op{\rho}_E$, where $\op{\rho}_E$ is the stationary state of $E$ (i.e., $\bigl[ \op{H}_E, \op{\rho}_E \bigr] = 0$). The second assumption is the Markov approximation mentioned above, i.e., the assumption of a memoryless environment on the coarse-grained relaxation timescale $\tau_r$ defined by the evolution of the open system. Comparisons of the predictions of master equations derived from the Born and Markov approximations with experimental data indicate that these approximations are reasonable in many physical situations (but see Sec.~\ref{sec:non-mark-decoh} for comments on exceptions and non-Markovian models). 

We will now give a brief sketch of the derivation (see, e.g., Refs.~\cite{Breuer:2002:oq,Schlosshauer:2007:un} for details). We start from the Liouville--von Neumann equation for the total density operator $\op{\rho}^{(I)}(t)$ in the interaction picture,
\begin{equation}
  \label{eq:pexp}
  \frac{\partial}{\partial t} \op{\rho}^{(I)}(t) = \frac{1}{\I \hbar} \bigl[
    \op{H}_\text{int}(t),  \op{\rho}^{(I)}(t) \bigr],
\end{equation}
where $\op{H}_\text{int}(t)$ is the interaction Hamiltonian in the interaction picture. (From here on, operators bearing explicit time arguments shall be understood as interaction-picture operators, while for interaction-picture density operators we use the superscript ``$(I)$'' to distinguish them from time-dependent Schr\"odinger-picture density operators.)  We formally integrate Eq.~\eqref{eq:pexp}, insert the resulting expression for $\op{\rho}^{(I)}(t)$ into the right side of Eq.~\eqref{eq:pexp}, and trace over the environment. This gives
\begin{equation} 
\label{eq:pexps2} 
\frac{\partial}{\partial t}
  \op{\rho}^{(I)}_S(t) = \frac{1}{\I \hbar}\text{Tr}_E \bigl[ \op{H}_\text{int}(t), \op{\rho}(0) \bigr] + \left(\frac{1}{\I \hbar}\right)^2 \int_0^t \D t'
  \, \text{Tr}_E \bigl[ \op{H}_\text{int}(t), \bigl[
      \op{H}_\text{int}(t'), \op{\rho}^{(I)}(t') \bigr]
  \bigr].
\end{equation}
Without loss of generality, the first commutator can be made to vanish by redefining $\op{H}_0$ and $\op{H}_\text{int}$. Imposing the Born approximation allows us to replace the total density operator $\op{\rho}^{(I)}(t')$ by $\op{\rho}^{(I)}_S(t') \otimes  \op{\rho}_E$,
\begin{equation} 
\label{eq:pexaa2} 
\frac{\partial}{\partial t}
\op{\rho}^{(I)}_S(t) = -\frac{1}{\hbar^2} \int_0^t \D t'
\, \text{Tr}_E \bigl[ \op{H}_\text{int}(t), \bigl[
    \op{H}_\text{int}(t'),  \op{\rho}^{(I)}_S(t') \otimes
    \op{\rho}_E \bigr]
\bigr].
\end{equation}
Now the master equation is expressed entirely in  terms of the reduced state of the system and the initial state of the environment. Inserting into Eq.~\eqref{eq:pexaa2}  the diagonal decomposition of the interaction Hamiltonian in the interaction picture, $\op{H}_\text{int}(t) =\sum_\alpha \op{S}_\alpha(t) \otimes \op{E}_\alpha(t)$,  and writing out the double commutator, one finds
\begin{multline} 
\label{eq:pexa51} 
\frac{\partial}{\partial t}
\op{\rho}^{(I)}_S(t) =  - \frac{1}{\hbar^2}\int_0^t \D t'
\, \sum_{\alpha\beta} \, \left\{ \mathcal{C}_{\alpha\beta}(t-t') \left[
    \op{S}_\alpha(t) \op{S}_\beta(t') \op{\rho}^{(I)}_S(t')
    - \op{S}_\beta(t') \op{\rho}^{(I)}_S(t') \op{S}_\alpha(t)
  \right] \right. \\  \left. + \, \mathcal{C}_{\beta\alpha}(t'-t) \left[
    \op{\rho}^{(I)}_S(t')\op{S}_\beta(t')  \op{S}_\alpha(t)
    - \op{S}_\alpha(t)\op{\rho}^{(I)}_S(t')\op{S}_\beta(t')
  \right] \right\},
\end{multline}
with
\begin{equation}
  \label{eq:xbab20}
\mathcal{C}_{\alpha\beta}(t-t') = \text{Tr}_E \left\{ \op{E}_\alpha(t-t')
    \op{E}_\beta \op{\rho}_E\right\} = \left\langle \op{E}_\alpha(t-t') \op{E}_\beta
  \right\rangle_{\op{\rho}_E}.
\end{equation}
The quantities $C_{\alpha\beta}(t-t')$ are referred to as \emph{environment self-correlation functions}. This term is motivated by the following observation. Each operator $\op{E}_\alpha$ (provided it is Hermitian) may be thought of as an observable measured on the environment through its coupling to the system. Equation~\eqref{eq:xbab20} then quantifies how much the result of such a measurement is correlated with the same measurement carried out at a different instant a time $\tau=t-t'$ apart. Thus, it measures how much information the environment retains over time about its interaction with the system. 

Imposing the Markov approximation corresponds to assuming that the environment self-correlation functions $C_{\alpha\beta}(\tau)$ are sharply peaked around $\tau = 0$ and decay rapidly (on a timescale $\tau_c$) compared to the system relaxation timescale $\tau_r$ that measures the change of $\op{\rho}^{(I)}_S(t)$ due to the interaction with the environment. This leads to two immediate consequences. First, because the change of $\op{\rho}^{(I)}_S(t)$ is negligibly small over the time interval for which the $C_{\alpha\beta}(\tau)$ have appreciable magnitude, we may replace the retarded-time density operator $\op{\rho}^{(I)}_S(t')$ on the right-hand side of Eq.~\eqref{eq:pexa51} by the current-time density operator $\op{\rho}^{(I)}_S(t)$. The resulting master equation is time-local, but the dependence of the integral limit on $t$ in Eq.~\eqref{eq:pexa51} means that the equation is not yet Markovian. However, we can replace the lower limit of the integral  in Eq.~\eqref{eq:pexa51} by $-\infty$, because the functions $C_{\alpha\beta}(t-t')$ vanish for $t' \ll t$. After making the substitution $t' \longrightarrow \tau = t-t'$, we arrive at the Markovian master equation
\begin{multline} 
\label{eq:pexarr} 
\frac{\partial}{\partial t}
\op{\rho}^{(I)}_S(t) =  - \frac{1}{\hbar^2}\int_0^\infty \D \tau
\, \sum_{\alpha\beta} \, \left\{ \mathcal{C}_{\alpha\beta}(\tau) \left[
    \op{S}_\alpha(t) \op{S}_\beta(t-\tau) \op{\rho}^{(I)}_S(t)
    - \op{S}_\beta(t-\tau) \op{\rho}^{(I)}_S(t) \op{S}_\alpha(t)
  \right]\right. \\ \left. + \mathcal{C}_{\beta\alpha}(-\tau) \left[
    \op{\rho}^{(I)}_S(t)\op{S}_\beta(t-\tau)  \op{S}_\alpha(t)
    - \op{S}_\alpha(t)\op{\rho}^{(I)}_S(t)\op{S}_\beta(t-\tau)
  \right] \right\}.
\end{multline}
We note again that imposing the Markov assumption means that the evolution is considered only on a coarse-grained timescale, since we are not resolving changes on the order of the environment self-correlation timescale $\tau_c$. 

Finally, transforming Eq.~\eqref{eq:pexarr} back to the Schr\"odinger picture yields the \emph{Born--Markov master equation} (sometimes also called 
\emph{Redfield equation} \cite{Redfield:1957:im,Blum:1981:qq}\footnote{In the literature, the term  ``Redfield equation'' is occasionally (see, e.g., Refs.~\cite{Breuer:2002:oq,Hornberger:2009:aq}) associated with the time-local but pre-Markovian master equation obtained just prior to Eq.~\eqref{eq:pexarr}, i.e., before the integration limit is extended to infinity. In his original paper \cite{Redfield:1957:im}, Redfield did, however,  include the step of extending the integration limit in this way (see his Eq.~2.14).})
\begin{equation}
\label{eq:born-markov-master}
\frac{\partial}{\partial t} \op{\rho}_S(t) = -\frac{\I}{\hbar} \left[ \op{H}_S, \op{\rho}_S(t) \right] - \frac{1}{\hbar^2}\sum_{\alpha} \left\{ \left[
\op{S}_\alpha, B_\alpha \op{\rho}_S(t) \right] + \left[ \op{\rho}_S(t) C_\alpha, \op{S}_\alpha \right] \right\},
\end{equation}
where the system operators $B_\alpha$ and $C_\alpha$ are defined as
\begin{subequations} \label{eq:hvdwg643r5gsxkjcvbsvnx20}
\begin{align}
\label{eq:iuwrgf8}
B_\alpha &=\int_0^\infty \D \tau \, \sum_{\beta}
C_{\alpha\beta}(\tau)  S^{(I)}_\beta(-\tau),
\\ \label{eq:112823jmn2} C_\alpha &= \int_0^\infty \D \tau \,
\sum_{\beta} C_{\beta\alpha}(-\tau)  S^{(I)}_\beta(-\tau).
\end{align}
\end{subequations}
In many situations of interest, the general form~\eqref{eq:born-markov-master} of the Born--Markov master equation simplifies considerably. For example, if only a single system observable $S$ is monitored by the environment, Eq.~\eqref{eq:born-markov-master} becomes
\begin{equation}
\label{eq:born-markov-mastersim}
\frac{\partial}{\partial t} \op{\rho}_S(t) = -\frac{\I}{\hbar} \left[ \op{H}_S, \op{\rho}_S(t) \right] - \frac{1}{\hbar^2}\left\{\left[
\op{S}, B \op{\rho}_S(t) \right] + \left[ \op{\rho}_S(t) C, \op{S} \right]\right\},
\end{equation}
with corresponding simplifications of Eqs.~\eqref{eq:xbab20}, \eqref{eq:iuwrgf8}, and \eqref{eq:112823jmn2}. Moreover, in many cases one finds a rather simple time dependence for the operators $\op{S}_\alpha(\tau)$ and $E_\alpha(\tau)$ appearing in Eqs.~\eqref{eq:xbab20}, \eqref{eq:iuwrgf8}, and \eqref{eq:112823jmn2}, which in turn makes calculating the quantities $B_\alpha$ and $C_\alpha$ relatively easy. Examples of applications of Born--Markov master equations to specific models are given in Sec.~\ref{sec:decmodels}.
 
While the Born--Markov master equation~\eqref{eq:born-markov-master} is both time-local and Markovian, it does \emph{not} guarantee the complete positivity of the evolution \cite{Alicki:2007:uu,Benatti:2005:ii}. Therefore, it may lead to unphysical states with negative populations and does not, in general, describe a quantum dynamical semigroup (see Ref.~\cite{Benatti:2005:ii} for a detailed discussion). A well-known case of a Born--Markov master equation that violates complete positivity is the Caldeira--Leggett master equation \cite{Caldeira:1983:on} discussed in Sec.~\ref{sec:cald-legg-mast}. In fact, there are situations in which a Born--Markov master equation does not even preserve the positivity of the time-evolved reduced density matrix \cite{Dumke:1979:ia,Benatti:2005:ii} (see, e.g., the model discussed in Example~3.4 of Ref.~\cite{Benatti:2005:ii}).

Complete positivity can be ensured, however, by imposing a third, secular approximation, known as the \emph{rotating-wave approximation}, which was analyzed in detail by Davies \cite{Davies:1974:tw,Davies:1976:oo,Davies:1976:uu,Davies:1978:uu} (see also Refs.~\cite{Dumke:1979:ia,Breuer:2002:oq,Alicki:2007:uu,Hornberger:2009:aq}). Its application requires that the system has a discrete and nondegenerate (or exactly degenerate) spectrum (see Ref.~\cite{Davies:1978:uu} for an analysis of the case of nearly degenerate spectra). It is justified when the relaxation timescale $\tau_r$ of the open quantum system $S$ is much larger than the timescale $\tau_S$ set by the typical energy differences $\hbar(\omega-\omega')$ of the system Hamiltonian $\op{H}_S$, i.e., if $\tau_r \gg \abs{\omega-\omega'}^{-1}$. This condition is fulfilled, for example, in many quantum-optical settings. One then proceeds by first inserting a decomposition of the interaction-picture interaction Hamiltonian in terms of eigenoperators of $\op{H}_S$ into the Markovian master equation~\eqref{eq:pexarr}, which leads to Fourier-type summations of the form $\sum_{\omega \omega'} \E^{\I (\omega-\omega')t} f(\omega, \omega')$. Since the exponentials $\E^{\I (\omega-\omega')t}$ oscillate rapidly over the relaxation timescale $\tau_r$, they will average out to zero unless $\omega \approx \omega'$. In the rotating-wave approximation, one therefore neglects all terms $\omega \not= \omega'$.  One can then show \cite{Davies:1974:tw,Davies:1976:oo,Davies:1976:uu,Breuer:2002:oq,Alicki:2007:uu,Hornberger:2009:aq} that this procedure transforms the Born--Markov master equation into the first standard form \eqref{eq:sdfkhwr69} for the generator of a quantum dynamical semigroup, with, as required, a positive coefficient matrix $\gamma_{\alpha\beta}$. Accordingly, the master equation ensures complete positivity, and can also be brought into the Lindblad form~\eqref{eq:lindblad}.

It is interesting to note that, while complete positivity is of course desirable (and a necessary feature of any exact, physically meaningful evolution), imposing the rotating-wave approximation may in turn obscure other relevant physical features \cite{Dodin:2018:zz}. Such observations serve as a reminder that master equations and their underlying approximations must be judiciously chosen and applied to ensure that they are appropriate to a given physical situation.

\subsection{\label{sec:quantum-trajectories}Quantum trajectories} 

In \emph{quantum-jump} and \emph{quantum-trajectory} approaches \cite{Barchielli:1991:fv,Belavkin:1989:an,Belavkin:1989:am,Belavkin:1989:um,Belavkin:1995:tt,Diosi:1988:wx,Diosi:1988:hn,Diosi:1988:bv,Gisin:1984:qs,Gisin:1989:jn,Wiseman:1994:qq,Goan:2001:rz,Plenio:1998:bb}, the evolution of the reduced density matrix is conditioned on the results of a sequence of measurements performed on the environment. In this way, one may consider an individual system evolving stochastically, conditioned on a particular measurement record. This evolution is described by a Lindblad master equation of the form \eqref{eq:lindbladc}, where now the reduced density matrix (denoted by $\op{\rho}^C_S$ below) is conditioned on the records of measurements of the Lindblad operators $\op{L}_\mu$,
\begin{equation} \label{eq:cme}
\D \op{\rho}^C_S = -\frac{\I}{\hbar} \left[\op{H}_S, \op{\rho}_S^C \right] \D t  - \frac{1}{2} \sum_\mu \kappa_\mu \left[\op{L}_\mu, \left[\op{L}_\mu, \op{\rho}_S^C\right] \right] \D t + \sum_\mu \sqrt{\kappa_\mu} \, \mathcal{W}[\op{L}_\mu] \op{\rho}_S^C \, \D W_\mu,  
\end{equation}
where $\mathcal{W}[L]\op{\rho} \equiv L \op{\rho} + \op{\rho} L^\dagger - \op{\rho} \, \text{Tr} \left\{ L\op{\rho} + \op{\rho} L^\dagger \right\}$, and the $\D W_\mu$ are so-called \emph{Wiener increments}. Equation~\eqref{eq:cme} represents what is known as a \emph{diffusive unraveling} of the Lindblad equation into individual quantum trajectories, which can then be expressed by means of a \emph{stochastic Schr\"odinger equation} \cite{Barchielli:1991:fv,Belavkin:1989:an,Belavkin:1989:am,Belavkin:1989:um,Belavkin:1995:tt,Diosi:1988:wx,Diosi:1988:hn,Diosi:1988:bv,Gisin:1984:qs,Gisin:1989:jn,Wiseman:1994:qq,Goan:2001:rz,Plenio:1998:bb}. Unraveling of a master equation has been used, for example, to characterize the dynamically emerging pointer states in collisional decoherence \cite{Busse:2009:aa,Busse:2010:aa} and quantum Brownian motion \cite{Sorgel:2015:pp}. 

\subsection{\label{sec:non-mark-decoh}Non-Markovian decoherence}

While Born--Markov master equations adequately capture the decoherence dynamics of many physically relevant systems, the underlying assumption of weak coupling to an essentially unchanging, memoryless environment is not always fulfilled in practice, and significantly non-Markovian dynamics may arise. An important example of such a breakdown of Markovian decoherence dynamics is encountered in the case of a superconducting qubit that interacts strongly with a low-temperature environment of other two-level systems \cite{Prokofev:2000:zz,Dube:2001:zz}. Another example is an experiment \cite{Groeblacher:2013:im} that has measured strongly non-Ohmic spectral densities for the environment of a quantum nanomechanical system; such densities lead to non-Markovian evolution. 

If memory effects in the environment are substantial, then the evolution of the reduced density matrix will depend on the past history of the system and the environment. In general, this may mean that time-local master equations are no longer  applicable and one has to instead solve integro-differential equations, which is typically a difficult task (see also the Nakajima--Zwanzig projection-operator technique
\cite{Nakajima:1958:im,Zwanzig:1960:om,Zwanzig:1960:mo,Joos:2003:jh}). It turns out, however, that in some cases time-local master equations can still provide a good representation even of non-Markovian processes. Such equations are  of the form
\begin{equation}
\label{eq:sfihvsfhv7}
  \frac{\partial}{\partial t} \op{\rho}_S(t) = \mathcal{K}(t) \op{\rho}_S(t),
\end{equation}
where the superoperator $\mathcal{K}(t)$ is now time dependent but evaluated at a single time $t$ only, as is the reduced density matrix. To give an example, a non-Markovian but time-local master equation  for quantum Brownian motion (see Sec.~\ref{sec:quant-brown-moti}) can be obtained through a formal modification of the Born--Markov master equation \cite{Paz:2001:aa,Zurek:2002:ii}. In general, non-Markovian, time-local master equations may be obtained using the so-called time-convolutionless projection operator technique \cite{Chaturvedi:1979:pm,Shibata:1980:ma,Royer:1972:um,Royer:2003:za}. 

\section{\label{sec:decmodels}Decoherence models}

Many physical systems can be represented either by a qubit (i.e., a spin-$\frac{1}{2}$ particle) if the state space of the system is discrete and effectively two-dimensional, or by a particle described by continuous phase-space coordinates. Similarly, a wide range of environments can be modeled as a collection of quantum harmonic oscillators (``oscillator environments,'' representing a quasicontinuum of delocalized bosonic modes) or qubits (``spin environments,'' representing a collection of localized, discrete modes). 

A harmonic-oscillator environment is a very general model at low energies. Many systems interacting with an environment can be effectively described by one or two degrees of freedom of the system linearly coupled to an environment of harmonic oscillators; indeed, it turns out that sufficiently weak interactions with an \emph{arbitrary} environment can be mapped onto a system linearly coupled to a harmonic-oscillator environment \cite{Feynman:1963:jj,Caldeira:1983:gv}.

Spin environments are particularly appropriate models in the low-temperature regime, where decoherence is typically dominated by interactions with localized modes, such as paramagnetic spins, paramagnetic electronic impurities, tunneling charges, defects, and nuclear spins \cite{Dube:2001:zz,Prokofev:2000:zz,Lounasmaa:1974:yb}. Each such localized mode may be described by a finite-dimensional Hilbert space with a finite energy cutoff, allowing one to model these modes as a set of discrete states. Since typically only two such states are relevant, the localized modes can be mapped onto an environment of spin-$\frac{1}{2}$ particles.

In the following, we will discuss four important standard models, namely, collisional decoherence (Sec.~\ref{sec:collisionaldecoherence}), quantum Brownian motion (Sec.~\ref{sec:quant-brown-moti}), the spin--boson model (Sec.~\ref{sec:spin-boson-models}), and the spin--spin model (Sec.~\ref{sec:spin-envir-models}). For details on these and other decoherence models, including derivations of the relevant master equations, see, e.g., Secs.~3 and 5 of Ref.~\cite{Schlosshauer:2007:un}. 

\subsection{\label{sec:collisionaldecoherence}Collisional decoherence}

\begin{figure}
\centering
\includegraphics[scale=0.7]{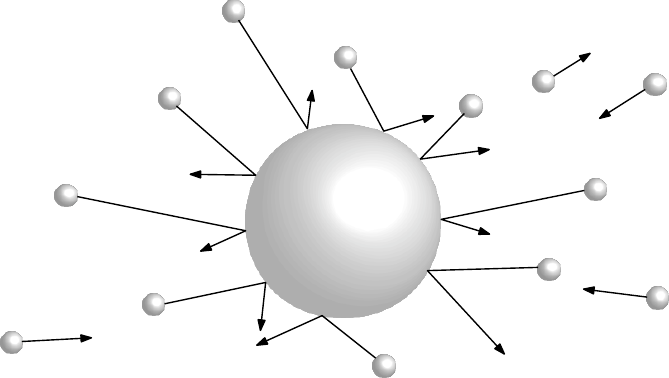}
\caption{Illustration of environmental scattering. Particles in the environment, such as
  photons or gas molecules, are scattered by a central particle. When they carry away which-path information, collisional decoherence results.}
\label{fig:scatmod}
\end{figure}

Collisional decoherence arises from the scattering of environmental particles by a massive free quantum particle, a process by which the scattered environmental particles obtain which-path information about the central particle (see Fig.~\ref{fig:scatmod}; compare also Fig.~\ref{fig:bi}\emph{c}). Models of collisional decoherence were first studied in the classic paper by Joos and Zeh \cite{Joos:1985:iu}. Subsequently, a more rigorous derivation of the master equation was given by Hornberger and Sipe \cite{Hornberger:2003:un}; it remedied a flaw in Joos and Zeh's original derivation that had resulted in decoherence rates that were too large by a factor of $2\pi$ (see also Refs.~\cite{Gallis:1990:un,Diosi:1995:um,Adler:2006:yb}). These treatments consider the case in which the mass $M$ of the central particle is much larger than the masses $m$ of the scattered environmental particles, such that the center-of-mass state of the central particle is not disturbed by the scattering events (i.e., no recoil). This situation is applicable, for example, to the decoherence of a macroscopic object due to scattering of microscopic or mesoscopic particles such as photons or air molecules, an ubiquitous process in nature; it also applies to scenarios such as the controlled decoherence of fullerene molecules due to collisions with a gaseous environment \cite{Hackermuller:2003:uu,Hornberger:2003:tv}. If the masses of the central particle and the environmental particles are similar, then a more general description is required that also includes the dissipative dynamics (see below) \cite{Diosi:1995:um,Hornberger:2006:tb,Hornberger:2006:tc,Hornberger:2008:ii, Vacchini:2009:pp,Busse:2009:aa,Busse:2010:aa,Busse:2010:oo}. 

\subsubsection{Master equation}

Assuming $M \gg m$ holds, the time evolution of the reduced density matrix is given by \cite{Joos:1985:iu,Gallis:1990:un,Diosi:1995:um,Hornberger:2003:un,Schlosshauer:2007:un,Hornberger:2009:aq}
\begin{equation} 
\label{eq:scatq} 
\frac{\partial \rho_S(\vec{x}, \vec{x}', t)}{\partial t} 
= - F(\vec{x} - \vec{x}') \rho_S(\vec{x}, \vec{x}', t).
\end{equation}
This master equation describes pure spatial decoherence without dissipation. The decoherence factor $F(\vec{x} - \vec{x}')$ represents the characteristic decoherence rate at which coherence between two positions $\vec{x}$ and $\vec{x}'$ becomes locally unobservable. It is given by
\begin{equation}
\label{eq:scatf}
  F(\vec{x} - \vec{x}') =  \int_0^\infty \D q \,
 \varrho(q) v(q) \int \frac{\D \hat{n}\,\D \hat{n}'}{4\pi}  \left(1- \E^{\I
    q\left(\vec{\hat{n}} - \vec{\hat{n}}'\right) \cdot \left( \vec{x} - \vec{x'}
    \right) /\hbar} \right) \abs{ f(q\vec{\hat{n}}, q\vec{\hat{n}}') }^2,
\end{equation}
where $\varrho(q)$ is the number density of incoming environmental particles with magnitude of momentum equal to $q=\abs{\vec{q}}$, $\vec{\hat{n}}$ and $\vec{\hat{n}}'$ are unit vectors (with $\D \hat{n}$ and $\D \hat{n}'$ representing the associated solid-angle differentials), and $v(q)$ is the speed of particles with momentum $q$. If the environmental particles are massive, we have $v(q) = q/m$, where $m$ is each particle's mass; for massless particles such as photons, $v(q)$ is equal to the speed of light. The quantity $\abs{ f(q\vec{\hat{n}},  q\vec{\hat{n}}')}^2$ is the differential cross-section for the scattering of an environmental particle from initial momentum $\vec{q}=q\vec{\hat{n}}$ to final momentum $\vec{q}'=q\vec{\hat{n}}'$.

To further evaluate the decoherence factor $F(\vec{x} - \vec{x}')$ [Eq.~\eqref{eq:scatf}], we distinguish two important limiting cases. In the \emph{short-wavelength limit}, the typical wavelength of the scattered environmental particles is much shorter than the coherent separation $\Delta x = \abs{\vec{x}-\vec{x}'}$ between the well-localized wave packets in the spatial superposition state of the system. Then a single scattering event will be able to fully resolve this separation and thus carry away complete which-path information, leading to maximum spatial decoherence per scattering event. In this limit, $F(\vec{x} - \vec{x}')$ turns out to be simply equal to the total scattering rate $\Gamma_\text{tot}$ \cite{Schlosshauer:2007:un}. This implies the existence of an upper limit (saturation) for the decoherence rate when increasing the separation $\Delta x$, in contrast with decoherence rates obtained from linear models [compare Eqs.~\eqref{eq:daf12} and \eqref{eq:odijsvuhfsw21}]. If we ignore the comparably slow internal dynamics of the system, Equation~\eqref{eq:scatq} then implies exponential decay of spatial interference terms at a rate given by $\Gamma_\text{tot}$,
\begin{equation}\label{eq:sees}
\rho_S(\vec{x},\vec{x}',t) =
\rho_S(\vec{x},\vec{x}',0) \E^{-\Gamma_\text{tot} t}.
\end{equation}
Such collisional decoherence in the short-wavelength regime has been observed, for example, for fullerene molecules interacting with an environment of background gas particles \cite{Hackermuller:2003:uu}, and good agreement of the measured decoherence rates with theoretical predictions obtained from Eq.~\eqref{eq:sees} has been found \cite{Hornberger:2003:tv} (see also Sec.~\ref{sec:matt-wave-interf}).

In the opposite \emph{long-wavelength limit}, the environmental wavelengths are much larger than the coherent separation $\Delta x = \abs{\vec{x}-\vec{x}'}$, which implies that an individual scattering event will reveal only incomplete which-path information. For this case, the change of the reduced density matrix imparted by environmental scattering is given by
\begin{equation}\label{eq:scwer1} 
  \frac{\partial\rho_S(\vec{x},\vec{x}',t)}{\partial t} = - \Lambda
  (\vec{x} -\vec{x'})^2   \rho_{S}(\vec{x},\vec{x}',t).
\end{equation}
Here, $\Lambda$ is a scattering constant that represents the physical properties of the system--environment interaction and is given by
\begin{equation}\label{eq:scatfls2} 
\Lambda =  \int \D q\,  \varrho(q) v(q) 
\frac{q^2}{\hbar^2} \sigma_\text{eff}(q),
\end{equation}
where
\begin{equation}\label{eq:sccs} 
  \sigma_\text{eff}(q) = \frac{2\pi}{3} \int  \D \cos\Theta \,
  \left(1 - \cos \Theta \right)  \abs{ f(q, \cos\Theta)}^2 
\end{equation}
is the effective cross-section for the scattering interaction, with $\Theta$ denoting the scattering angle (i.e., the angle between incoming and outgoing directions of a scattered environmental particle).

If we again neglect the internal dynamics, then Eq.~\eqref{eq:scwer1}  leads to
\begin{equation}\label{eq:scwer2}
\rho_S(\vec{x},\vec{x}',t) =
\rho_S(\vec{x},\vec{x}',0) \E^{-\Lambda (\Delta x)^2 t},
\end{equation}
showing that spatial coherences become exponentially suppressed at a rate that depends on the square of the separation $\Delta x$ \cite{Schlosshauer:2007:un}. We see that the quantity $\Lambda (\Delta x)^2$ plays the role of a decoherence rate, and therefore
\begin{equation}\label{eq:scwer6565}
  \tau_{\Delta x} = \frac{1}{\Lambda (\Delta x)^2}
\end{equation}
is the characteristic spatial decoherence time. The dependence on the coherent separation $\Delta x$ is reasonable: if the environmental wavelengths are much larger than $\Delta x$, a large number of scattering events will need to accumulate before an appreciable amount of which-path information has become encoded in the environment, and this amount will increase, for a constant number of scattering events, as $\Delta x$ becomes larger. Note that if $\Delta x$ is increased beyond the typical wavelength of the environment, the short-wavelength limit needs to be considered instead, for which the decoherence rate is independent of $\Delta x$ and attains its maximum possible value.

\subsubsection{Time evolution and decoherence rates}

To study the effect of the environment on a given initial wave function in one spatial dimension, let us consider the evolution \eqref{eq:scwer1} in the long-wavelength limit and also include the self-Hamiltonian $\op{H}_S=\op{p}^2/2M$ of the central particle. In the position representation, the evolution is then given by the master equation [compare Eq.~\eqref{eq:lifsshvgvvvxayhcgiefdndbladc}]
\begin{equation}\label{eq:scweraass1hallo} 
   \frac{\partial\rho_{S}(x,x',t)}{\partial t} =  -\frac{\I}{2m\hbar}
  \left(\frac{ \partial^2}{\partial x'^2} - \frac{ \partial^2}{\partial
      x^2} \right) \rho_{S}(x,x',t)  -  \Lambda
  (x-x')^2   \rho_{S}(x,x',t). 
\end{equation}

\begin{figure}
\centering
\includegraphics[scale=0.54]{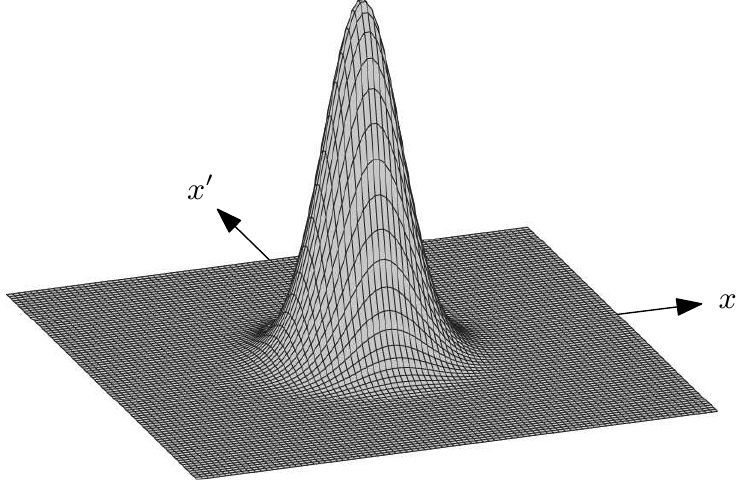} \hspace{1.2cm}\includegraphics[scale=0.54]{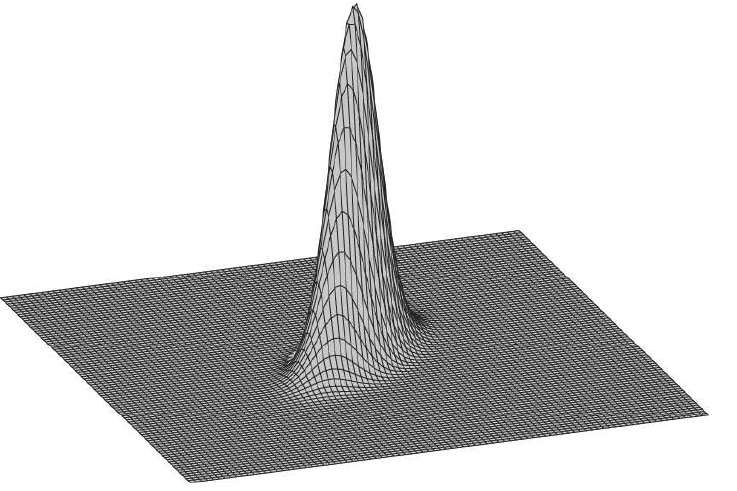} \hspace{1cm} \includegraphics[scale=0.54]{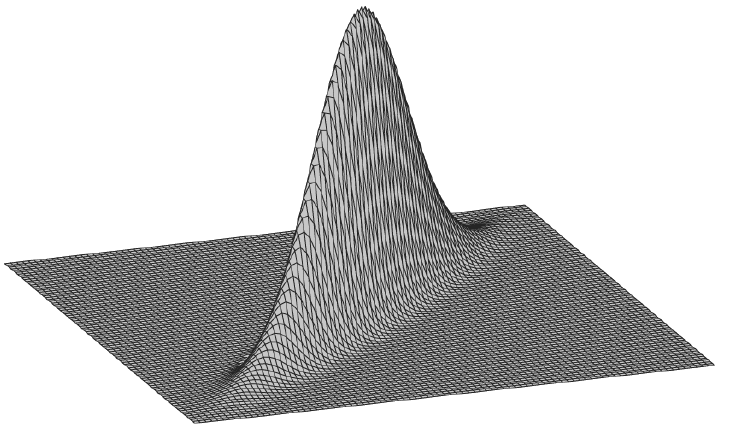} 
\caption{Collisional decoherence of a density matrix representing a Gaussian wave packet, as generated by the master equation \eqref{eq:scweraass1hallo}. Reading from left to right, the spatial coherence length, represented by the width of the Gaussian in the off-diagonal direction $x=-x'$, becomes progressively reduced by the environmental interaction.}
\label{fig:gaev}
\end{figure}

\begin{figure}
\centering
\includegraphics[scale=0.22]{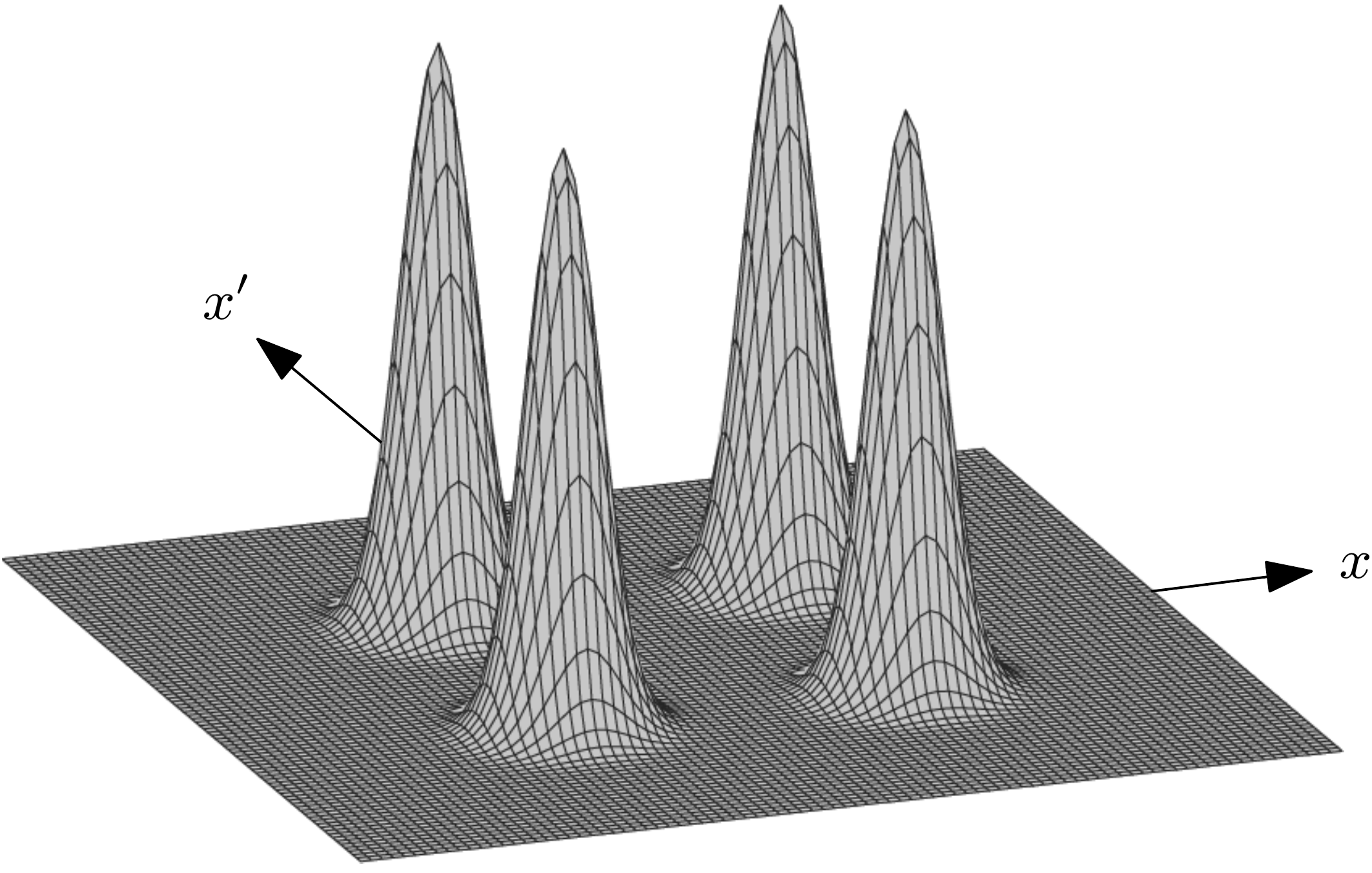} \hspace{.8cm} \includegraphics[scale=0.22]{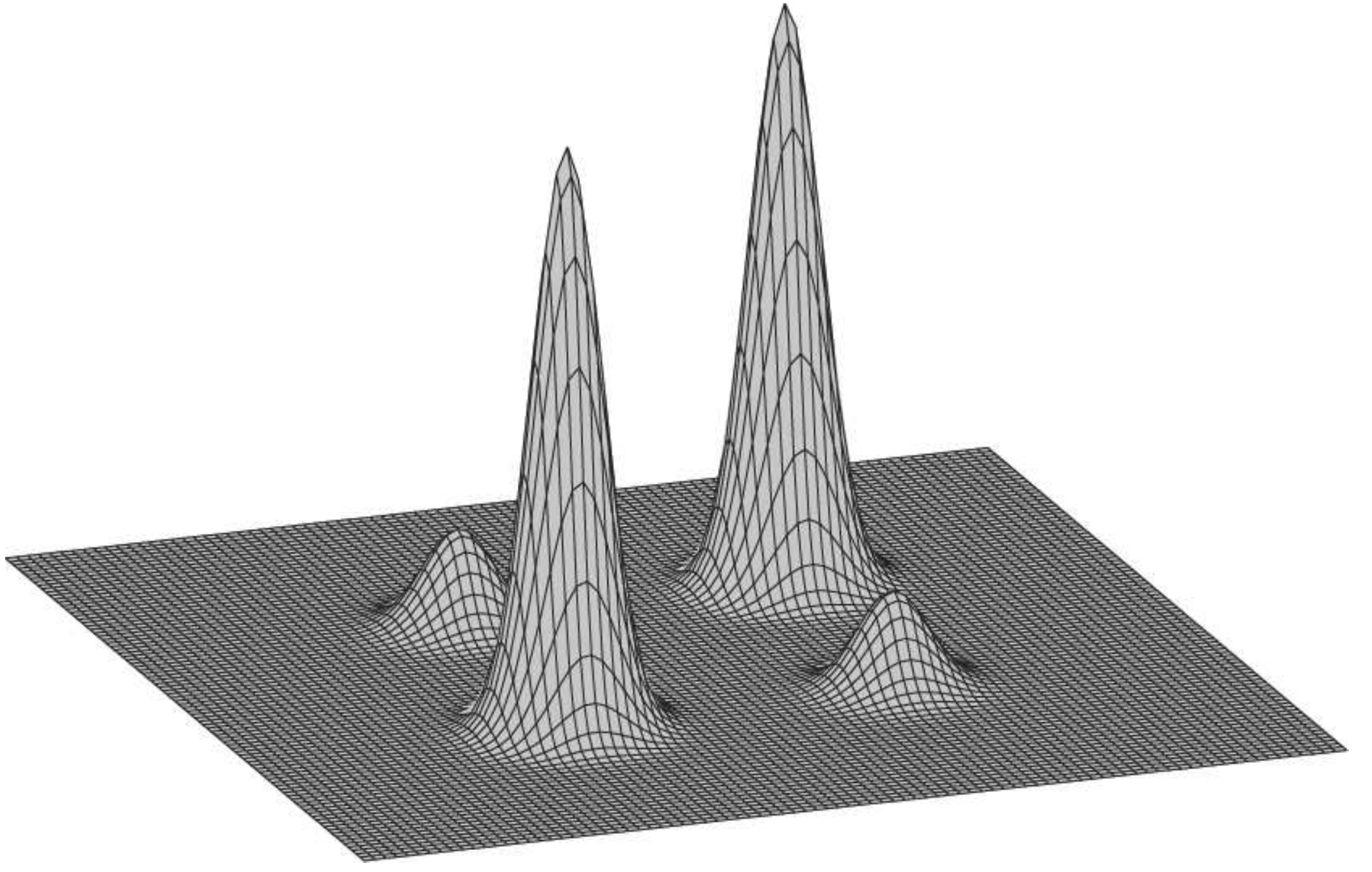} \hspace{.8cm} \includegraphics[scale=0.22]{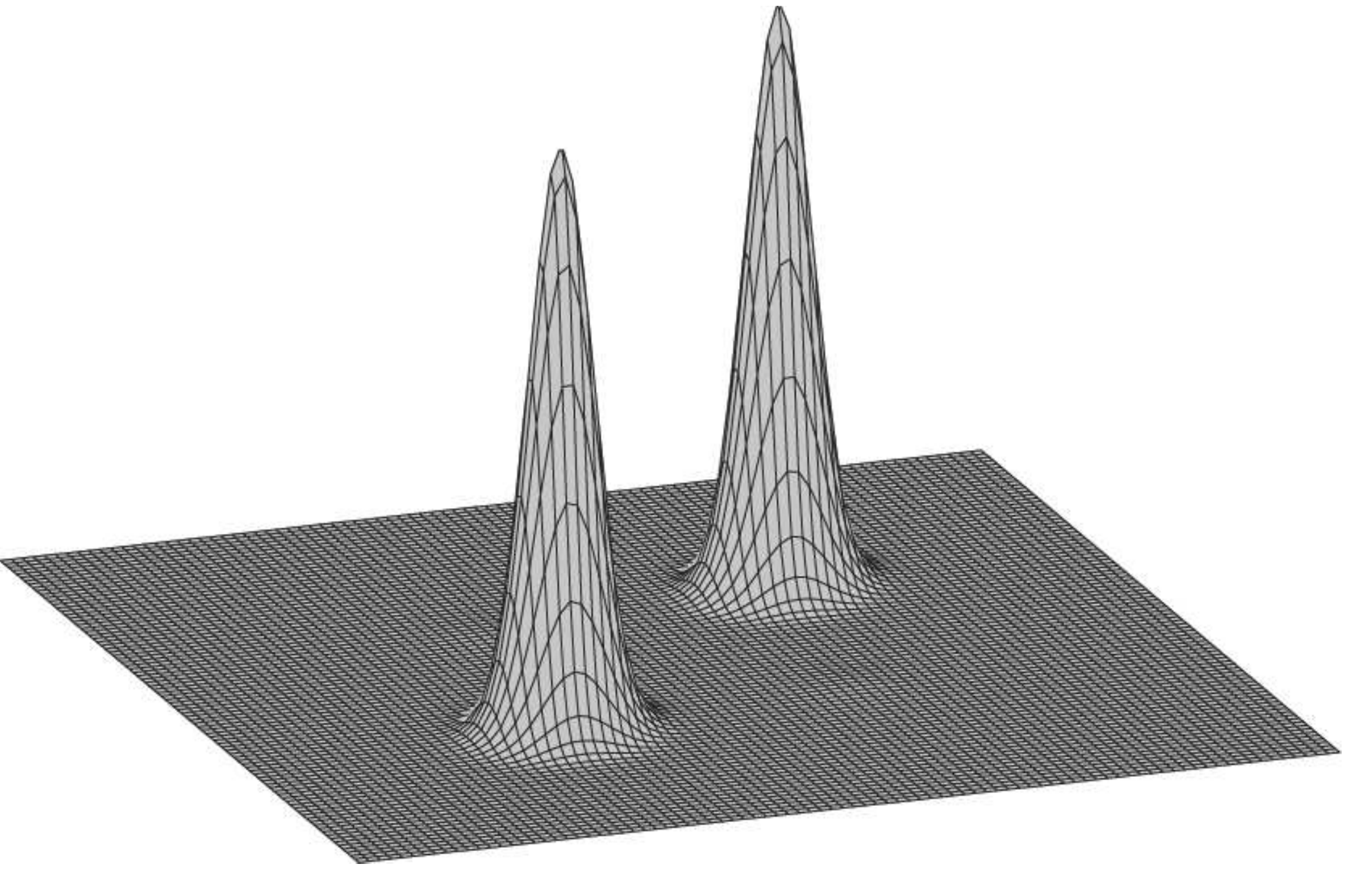}
\caption{Progressive decoherence (left to right) of a density matrix describing a spatial superposition of two Gaussian wave packets, as generated by the master equation \eqref{eq:scweraass1hallo}. Spatial coherence, represented by the peaks along the off-diagonal direction $x=-x'$, becomes damped by the environmental interaction.}
  \label{fig:twoevl}
\end{figure}

Let us start with an initial Gaussian wave packet centered at $x=0$ and apply the master equation~\eqref{eq:scweraass1hallo}. The resulting time evolution is shown in Fig.~\ref{fig:gaev}. We see that the coherence length (the width of the Gaussian in the off-diagonal direction $x=-x'$, representing spatial coherences) decreases over time, describing the collisional decoherence process. In this way, the density matrix approaches a quasiclassical probability distribution of positions clustered around the diagonal $x=x'$. Note that the width of the ensemble in the diagonal $x=x'$ direction---i.e., the size of the probability distribution $P(x,t) \equiv \rho_{S}(x,x,t)$ for different positions---increases in time. This is due to two influences: the free spreading of the wave packet (which is equally present in the absence of an environment), and an increase in the mean energy of the system due to the scattering interaction, rooted in the no-recoil assumption made in deriving the master equation \eqref{eq:scweraass1hallo}. Figure~\ref{fig:twoevl} shows the evolution generated by Eq.~\eqref{eq:scweraass1hallo} for a superposition of two Gaussian wave packets separated in position space. The off-diagonal peaks, which represent spatial coherence between the wave packets, become gradually suppressed due to the coupling to the environment. 

Numerical values of collisional decoherence rates obtained from Eq.~\eqref{eq:scwer6565}, with the physically relevant scattering parameters $\Gamma_\text{tot}$ and $\Lambda$ appropriately evaluated (see Ref.~\cite{Schlosshauer:2007:un} for details), have demonstrated the extreme efficiency of collisions with environmental particles in suppressing spatial interferences. Table~\ref{tab:decrate} lists a few classic order-of-magnitude estimates \cite{Joos:1985:iu,Joos:2003:jh,Schlosshauer:2007:un}. Carefully controlled decoherence experiments have shown excellent agreement between theory and experimental data, for example, for the decoherence of fullerenes due to collisions with background gas molecules in a Talbot--Lau interferometer \cite{Hackermuller:2003:uu,Hornberger:2003:tv,Hornberger:2003:un,Hornberger:2004:bb,Nimmrichter:2011:pr} (see Sec.~\ref{sec:matt-wave-interf}), and for the decoherence of sodium atoms in a Mach--Zehnder interferometer due to the scattering of photons \cite{Kokorowski:2001:ub} and gas molecules \cite{Uys:2005:yb}.

\begin{table}
\centering
\resizebox{0.5\textwidth}{!}{%
\begin{tabular}{lcc}
  \hline\noalign{\smallskip}
\small Environment & \,\,Dust grain\,\, &  Large molecule \\
  \noalign{\smallskip}\hline\noalign{\smallskip}
  Cosmic background radiation  & $1$  & $10^{24}$ \\
  Photons at room temperature  & $10^{-18}$  & $10^{6}$ \\
  Best laboratory vacuum  & $10^{-14}$ & $10^{-2}$\\
  Air at normal pressure  & $10^{-31}$ & $10^{-19}$\\
  \noalign{\smallskip}\hline
\end{tabular}}
\caption{\label{tab:decrate}Estimates of collisional decoherence timescales (in seconds) obtained from Eq.~\eqref{eq:scwer6565} for spatial coherences over a distance $\Delta x$ chosen to be equal to the size of the object ($\Delta x =  \unit[10^{-3}]{cm}$ for a dust grain and $\Delta x = \unit[10^{-6}]{cm}$ for a large molecule), calculated for four different environments. The first two entries represent photon environments, and the last two entries represent an environment of ambient air molecules at room temperature. See Ref.~\cite{Schlosshauer:2007:un} for details on the calculation of the shown values.}
\end{table}

\subsubsection{Generalizations and refinements}

Whenever the mass of the central particle becomes comparable to the mass of the environmental particles (as in the case of air molecules scattered by small molecules and free electrons \cite{Tegmark:1993:uz}), the no-recoil assumption does not hold and more general models for collisional decoherence and dissipation have to be considered. An important step in this direction was the master equation given by Di{\'o}si \cite{Diosi:1995:um}, though the derivation was based on a number of approximations that may be considered difficult to justify at the microscopic level \cite{Hornberger:2008:ii}. Later, a general, nonperturbative treatment based on the quantum linear Boltzmann equation was developed by Hornberger and collaborators \cite{Hornberger:2006:tb,Hornberger:2006:tc,Hornberger:2008:ii,Busse:2009:aa,Vacchini:2009:pp,Busse:2010:aa,Busse:2010:oo} (see Ref.~\cite{Vacchini:2009:pp} for a comprehensive review). The resulting framework properly accounts for the dynamical interplay between decoherence (in both position and momentum) and dissipation; previous results are recovered as limiting cases \cite{Hornberger:2006:tb,Hornberger:2006:tc,Hornberger:2008:ii,Vacchini:2009:pp,Busse:2010:oo}. The dynamically selected pointer states are found to be exponentially localized solitonic wave functions that follow the classical equations of motion \cite{Busse:2009:aa,Busse:2010:aa}. 

In all aforementioned models of collisional decoherence, the central particle is treated as a point-like particle with no orientational degrees of freedom---i.e., as an isotropic sphere with no rotational motion. Motivated by experiments involving molecular rotors and the observation that near-field interferometry with massive molecules is highly sensitive to molecular rotations \cite{Gring:2010:aa,Stickler:2015:zz}, recently the theoretical treatment of collisional decoherence has been extended to the derivation of Markovian master equations describing the spatio-orientational decoherence of rotating, anisotropic, nonspherical molecules due to scattering interactions with a gaseous environment \cite{Walter:2016:zz,Stickler:2016:yy,Papendell:2017:yy, Stickler:2018:oo,Stickler:2018:uu}. Specific cases considered include molecules with a dipole moment \cite{Walter:2016:zz} and molecules with a high rotation rate, known as superrotors \cite{Stickler:2018:oo}, and good agreement with experimental data has been found \cite{Stickler:2018:oo}. In this way, the development of increasingly refined models of collisional decoherence in response to experimental advances and insights speak nicely to the interplay between theory and experiment.

\subsection{\label{sec:quant-brown-moti}Quantum Brownian motion}

A classic and extensively studied model of decoherence and dissipation is the one-dimensional motion of a particle weakly coupled to a thermal bath of noninteracting harmonic oscillators, a model known as \emph{quantum Brownian motion} \cite{Kubler:1973:ux,Caldeira:1983:on,Hu:1992:om,Paz:1993:ta,Zurek:1993:pu,Weiss:1999:tv,Zurek:2002:ii,Breuer:2002:oq,Diosi:2000:yn,Joos:2003:jh,Eisert:2003:ib,Schlosshauer:2007:un}.  (By ``thermal bath'' we shall mean an environment in thermal equilibrium.) The self-Hamiltonian $\op{H}_E$ of the environment is given by
\begin{equation}
  \label{eq:sfsfjaa11}
  \op{H}_E = \sum_i \left( \frac{1}{2m_i}p_i^2 +
  \frac{1}{2}m_i\omega_i^2q_i^2 \right),  
\end{equation}
where $m_i$ and $\omega_i$ are the mass and natural frequency of the $i$th oscillator, and $q_i$ and $p_i$ denote the canonical position and momentum operators. The interaction Hamiltonian $\op{H}_\text{int}$ is chosen to be
\begin{equation}\label{eq:sfsfjaaaaaa11}
\op{H}_\text{int} = x \otimes \sum_i c_i q_i,
\end{equation}
which describes the bilinear coupling of the system's position $x$ to the positions $q_i$ of the environmental oscillators, with $c_i$ denoting the coupling strength between the system and the $i$th environmental oscillator. Note that the interaction Hamiltonian \eqref{eq:sfsfjaaaaaa11} describes a continuous monitoring of the position of the system by the environment. 

\subsubsection{Master equation} 

Given the Hamiltonians~\eqref{eq:sfsfjaa11} and \eqref{eq:sfsfjaaaaaa11}, one can derive the Born--Markov master equation for quantum Brownian motion. The result is (see, e.g., Refs.~\cite{Breuer:2002:oq,Schlosshauer:2007:un} for a derivation)
\begin{equation}
\label{eq:vjp32q22}
  \frac{\partial}{\partial t} \op{\rho}_S(t) 
  = -\frac{\I}{\hbar} \bigl[ \op{H}_S, \op{\rho}_S(t) \bigr]  -
  \frac{1}{\hbar} \int_0^\infty \D \tau \, \left\{ \nu(\tau) \bigl[ x, \bigl[
      x(-\tau), \op{\rho}_S(t) \bigr]\bigr] - \I \eta(\tau) \bigl[ x,
    \bigl\{ x(-\tau), \op{\rho}_S(t)
    \bigr\}\bigr] \right\}.
\end{equation}
Here, $x(\tau)$ denotes the system's position operator in the interaction picture, $x(\tau) = \E^{\I \op{H}_S\tau/\hbar} x\E^{-\I \op{H}_S\tau/\hbar}$. The curly brackets $\{\cdot , \cdot \}$ in the second line denote the anticommutator $\{ A, B \} \equiv AB + BA$. The functions
\begin{align}
  \nu(\tau) &= \int_0^\infty \D \omega
  \, J(\omega) \coth
  \left(\frac{\hbar\omega}{2k_B T}\right) \cos \left(\omega\tau\right), \label{eq:vdjpoo17} \\
  \eta(\tau) &= \int_0^\infty \D \omega\, J(\omega)
  \sin\left(\omega\tau\right), \label{eq:ponol218}
\end{align}
are known as the \emph{noise kernel} and \emph{dissipation kernel}, respectively. The function $J(\omega)$, called the \emph{spectral density} of the environment, is given by 
\begin{equation}
\label{eq:vdfpmdmv16}
  J(\omega) = \sum_i  \frac{c_i^2}{2m_i\omega_i} \delta(\omega-\omega_i).
\end{equation}
Spectral densities encode physical properties of the environment. One frequently replaces the collection of individual environmental oscillators by an (often phenomenologically motivated) continuous spectral-density function $J(\omega)$ of the environmental frequencies $\omega$. 

If we focus on the important case of a system represented by a harmonic oscillator with self-Hamiltonian
\begin{equation}
  \label{eq:sfsfsdfy7jaa11}
  \op{H}_S =  \frac{1}{2M}p^2 +
  \frac{1}{2}M\Omega^2x^2,
\end{equation}
the resulting Born--Markov master equation is (see, e.g., Refs.~\cite{Breuer:2002:oq,Zurek:2002:ii,Schlosshauer:2007:un})
\begin{equation}
\label{eq:vfoinbnd9s27}
  \frac{\partial}{\partial t} \op{\rho}_S(t) 
  = -\frac{\I}{\hbar} \bigl[ \op{H}_S + \frac{1}{2}M
    \widetilde{\Omega}^2 x^2, \op{\rho}_S(t) \bigr]
  - \frac{\I \gamma}{\hbar} \bigl[ x, \bigr\{ p,
      \op{\rho}_S(t) \bigr\} \bigr] 
 - D \bigl[ x, \bigl[ x, \op{\rho}_S(t) \bigr]
\bigr] 
- \frac{f}{\hbar} \bigl[ x, \bigl[ p, \op{\rho}_S(t) \bigr]
\bigr].
\end{equation}
The coefficients $\widetilde{\Omega}^2$, $\gamma$, $D$, and $f$ are given by\footnote{In the literature, the upper integral limit in the expressions for the coefficients is sometimes  considered explicitly time-dependent, rather than being extended to infinity (see, e.g., Refs.~\cite{Paz:2001:aa,Zurek:2002:ii}). This corresponds to not taking the final step in the microscopic derivation of the Born--Markov equation, namely, the replacement of the integral limit by infinity (see Sec.~\ref{sec:micr-deriv-mark}). This results in a partially pre-Markovian master equation that in certain cases  (for example, for low-temperature environments \cite{Unruh:1989:rc,Lombardo:2005:ia}) provides a more physically appropriate description than can be given using the Markovian coefficients~\eqref{eq:vfoinbnd9s27}.}
\begin{subequations}\label{eq:jcsfr09355378}
\begin{align}
  \widetilde{\Omega}^2 &= - \frac{2}{M} \int_0^\infty \D \tau \,
  \eta(\tau) \cos\left( \Omega \tau \right), \label{eq:caytcs1} \\
  \gamma &= \frac{2}{M\Omega} \int_0^\infty \D \tau \,
  \eta(\tau) \sin\left( \Omega \tau \right), \label{eq:caytcs2} \\
  D &=  \frac{1}{\hbar} \int_0^\infty \D \tau \,
  \nu(\tau) \cos\left( \Omega \tau \right), \label{eq:caytcs3}  \\
  f &= - \frac{1}{M\Omega} \int_0^\infty \D \tau \,
  \nu(\tau) \sin\left( \Omega \tau \right). \label{eq:caytcs4} 
\end{align}
\end{subequations}
The first term on the right-hand side of Eq.~\eqref{eq:vfoinbnd9s27} represents the unitary dynamics of a harmonic oscillator whose natural frequency is shifted by $\widetilde{\Omega}$. The second term describes momentum damping (dissipation) at a rate proportional to $\gamma$; it depends on the spectral density $J(\omega)$ of the environment but not on the temperature $T$. The third term has the Lindblad double-commutator form [see Eq.~\eqref{eq:lindbladc}] and describes decoherence of spatial coherences over a distance $\Delta X$ at a rate $D(\Delta X)^2$. Note that $D$ depends on both the spectral density $J(\omega)$ and the temperature $T$ of the environment. The fourth term also represents decoherence, but its influence on the dynamics of the system is usually negligible, especially at higher temperatures. In the long-time limit $\gamma t \gg 1$, the master equation \eqref{eq:vfoinbnd9s27} describes dispersion in position space given by
\begin{equation}
  \Delta X^2(t) = \frac{\hbar^2D}{2m^2 \gamma^2} t.
\end{equation}
We thus see that the ensemble width $\Delta X(t)$ in position space asymptotically scales as $\Delta X(t) \propto \sqrt{t}$. This is the same scaling behavior as in classical Brownian motion, thus motivating the term ``quantum Brownian motion.'' 

We note that it is possible to derive the \emph{exact}, non-Markovian master equation for quantum Brownian motion \cite{Hu:1992:om} (see also Refs.~\cite{Caldeira:1983:on,Caldeira:1985:tt,Haake:1932:tt,Grabert:1988:bf,Unruh:1989:rc} for preliminary results). Remarkably, this equation also turns out to be time-local. The exact master equation takes the same form as the Born--Markov equation~\eqref{eq:vjp32q22} presented above, but with the coefficients $\widetilde{\Omega}^2$, $\gamma$, $D$, and $f$ now being substantially more complex functions of time and other
time-dependent coefficients, which in turn are described in terms of integrals over the noise and dissipation kernels \eqref{eq:vdjpoo17} and
\eqref{eq:ponol218}. We refer the reader to Ref.~\cite{Hu:1992:om} for details. 

\subsubsection{Time evolution}

\begin{figure}
\centering
\includegraphics[scale=1]{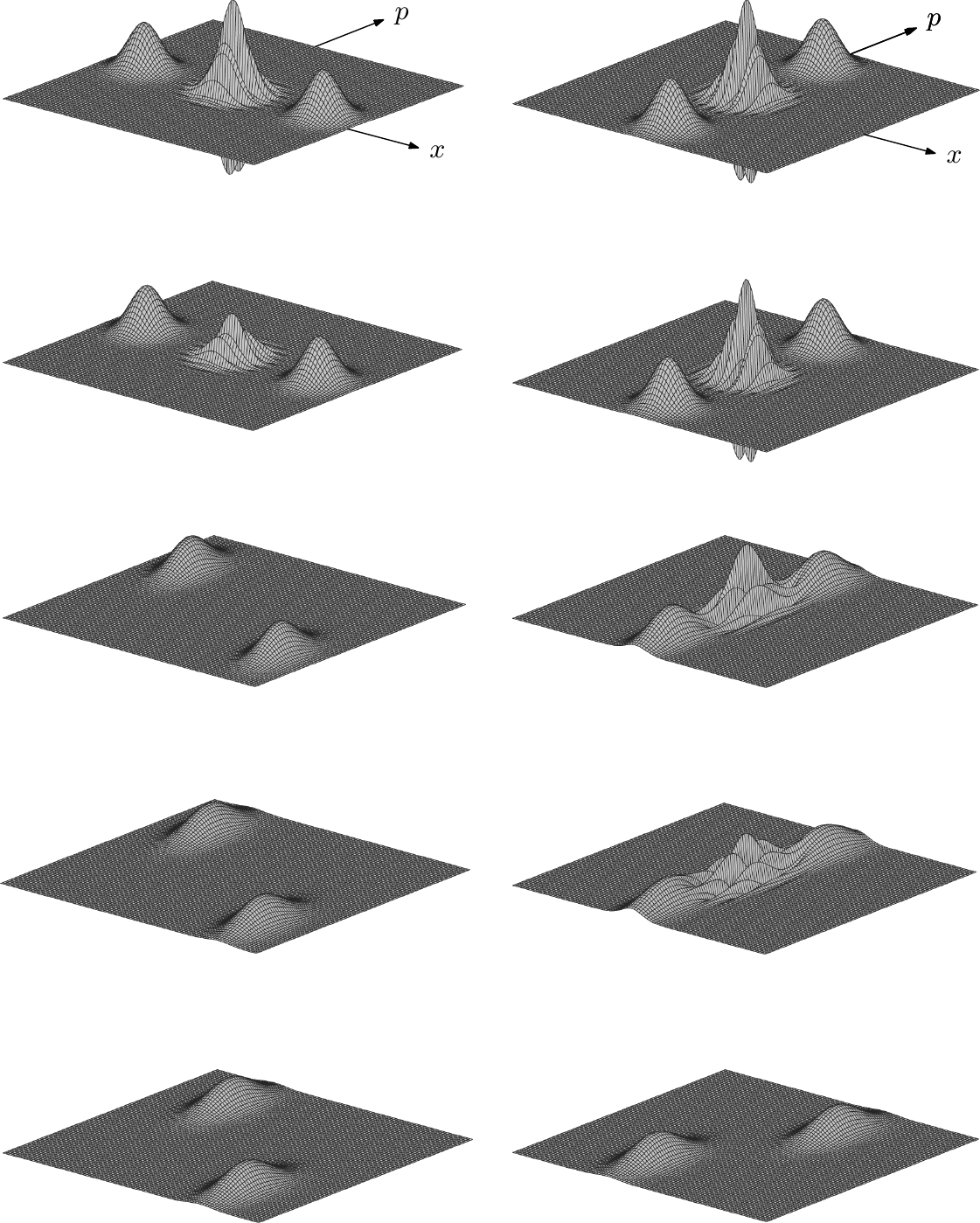}
\caption{Evolution of superpositions of Gaussian wave packets in quantum Brownian motion as studied in Ref.~\cite{Paz:1993:ta}, visualized in the Wigner representation \cite{Wigner:1932:un,Hillery:1984:tv}. Time increases from top to bottom. In the left column, the initial wave packets are separated in position; in the right column, the separation is in momentum. Interference between the two wave packets is represented by the oscillatory pattern between the two direct peaks.}
\label{fig:gaussmov}
\end{figure}

Figure~\ref{fig:gaussmov} shows the time evolution of position-space and momentum-space superpositions of two Gaussian wave
packets (represented in the Wigner picture \cite{Wigner:1932:un,Hillery:1984:tv}; see Sec.~\ref{sec:meas}), as described by Eq.~\eqref{eq:vfoinbnd9s27} an studied by Paz, Habib, and Zurek in Ref.~\cite{Paz:1993:ta}. The oscillations between the direct peaks represent interference between the two wave packets; as time goes on, these oscillations become progressively suppressed due to the interaction with the environment. The superposition of \emph{spatially} separated wave packets is decohered much more rapidly than the superposition of two distinct \emph{momentum} states. 

This can be explained in terms of the structure of the environmental monitoring. The interaction Hamiltonian couples only the position coordinate of the system to the environment, but not the momentum coordinate. Thus the environment monitors only position, and therefore one would expect that decoherence should occur only in position, not momentum. However, the intrinsic dynamics of the system are such that a superposition of two momenta will evolve into a superposition of positions, which in turn is sensitive to environmental monitoring. Thus, superpositions of momenta will also be decohered, albeit on a timescale associated with the intrinsic dynamics, which  is much longer than the timescale for the decoherence interaction described by the interaction Hamiltonian. 

This interplay of environmental monitoring and intrinsic dynamics leads to the emergence of pointer states that are minimum-uncertainty Gaussians (coherent states) well-localized in both position and momentum, thus approximating classical points in phase space \cite{Kubler:1973:ux,Paz:1993:ta,Zurek:1993:pu,Zurek:2002:ii,Diosi:2000:yn,Joos:2003:jh,Eisert:2003:ib,Sorgel:2015:pp}. Using a Poissonian unraveling of the master equation into individual quantum trajectories, the motion of these Gaussians may be represented by a stochastic differential equation that describes momentum damping as well as diffusion in position and momentum \cite{Sorgel:2015:pp}. 

\begin{figure}
\centering
\includegraphics[scale=0.27]{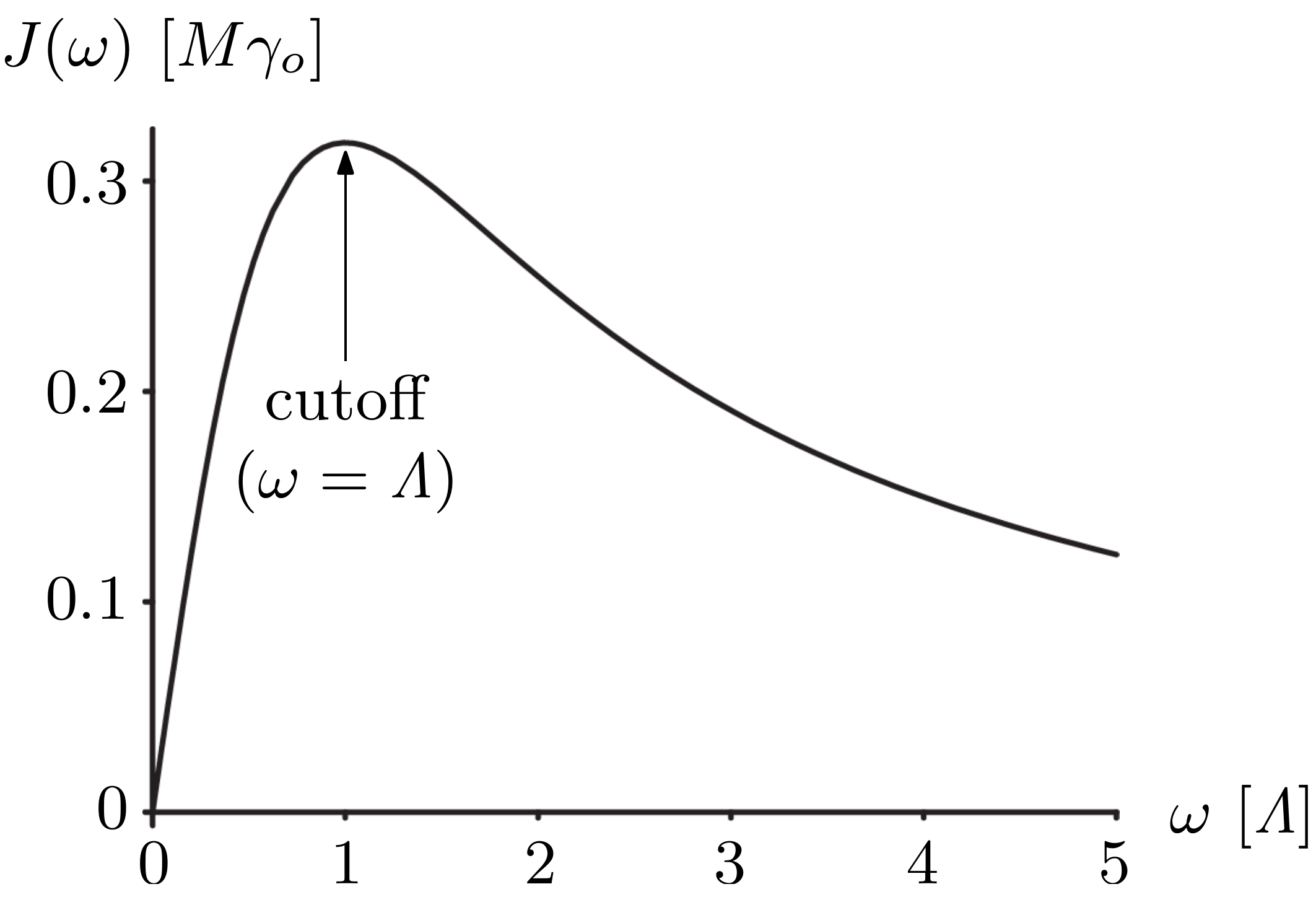}
\caption{Ohmic spectral density $J(\omega)$ with a high-frequency cutoff $\Lambda$ as defined by Eq.~\eqref{eq:pojsvsddsjldfv1}. The frequency axis is in units of $\Lambda$.}
\label{fig:spectraldensity}
\end{figure}

\subsubsection{\label{sec:cald-legg-mast}The Caldeira--Leggett master equation}

Let us now consider the frequently used case of an ohmic spectral density $J(\omega) \propto \omega$ with a high-frequency cutoff $\Lambda$,
\begin{equation}
  \label{eq:pojsvsddsjldfv1}
  J(\omega) = \frac{2M\gamma_0}{\pi} \omega
  \frac{\Lambda^2}{\Lambda^2 + \omega^2},
\end{equation}
where $\gamma_0$ is the effective coupling strength between  system and  environment. This function is shown in Fig.~\ref{fig:spectraldensity}. We evaluate the coefficients $\widetilde{\Omega}^2$, $\gamma$, $D$, and $f$ in Eq.~\eqref{eq:jcsfr09355378} for this spectral density in the limit of a high-temperature environment ($k_B T \gg \Omega$ and $k_B T \gg \Lambda$), and in the limit of the cutoff $\Lambda$ of environmental frequencies being much higher than the characteristic frequency $\Omega$ of the system. Then the master equation for quantum Brownian motion becomes (see, e.g., Refs.~\cite{Breuer:2002:oq,Schlosshauer:2007:un} for a derivation)

\begin{equation}
\label{eq:vfnbcclclnd6t48efg89s27}
  \frac{\partial}{\partial t} \op{\rho}_S(t) 
  = -\frac{\I}{\hbar} \bigl[ \op{H}'_S, \op{\rho}_S(t) \bigr]
- \frac{\I \gamma_0}{\hbar} \bigl[ x, \bigl\{ p,
      \op{\rho}_S(t) \bigr\} \bigr] 
 - \frac{2 M\gamma_0 k_B T}{\hbar^2} \bigl[ x, \bigl[ x, \op{\rho}_S(t) \bigr]\bigr]  + \frac{2\gamma_0k_B T}{\hbar^2\Lambda} \bigl[ x, \bigl[ p, \op{\rho}_S(t) \bigr]\bigr].
\end{equation}
where 
\begin{equation}
  \op{H}'_S = \op{H}_S + \frac{1}{2}M
  \widetilde{\Omega}^2 x^2 = \frac{1}{2M}p^2 +
  \frac{1}{2}M\left[ \Omega^2 - 2\gamma_0 \Lambda \right]x^2
\end{equation}
is the frequency-shifted Hamiltonian $\op{H}'_S$ of the system. The last term on the right-hand side of Eq.~\eqref{eq:vfnbcclclnd6t48efg89s27} is negligible in the limit $\Lambda \gg \Omega$ considered here and can therefore be omitted. Thus, we arrive at 
\begin{equation}
\label{eq:vfnbcclclnd9s27}
  \frac{\partial}{\partial t} \op{\rho}_S(t) 
  = -\frac{\I}{\hbar} \bigl[ \op{H}'_S, \op{\rho}_S(t) \bigr]
  - \frac{\I \gamma_0}{\hbar} \bigl[ x, \bigl\{ p,
      \op{\rho}_S(t) \bigr\} \bigr] 
 - \frac{2 M\gamma_0 k_B T}{\hbar^2} \bigl[ x, \bigl[ x, \op{\rho}_S(t) \bigr]\bigr],
\end{equation}
which is known as the \emph{Caldeira--Leggett master equation} \cite{Caldeira:1983:on} (here applied to the case of a system represented by harmonic oscillator). It has been used extensively to model decoherence and dissipation processes 
\cite{Gallis:1990:un,Gallis:1992:im,Anglin:1997:za}. It has been shown to sometimes provide an adequate representation even when the underlying assumptions are not strictly fulfilled, for example, in quantum-optical settings for which we often have $k_B T \lesssim \hbar\Lambda$ \cite{Walls:1985:lm}. Comparisons of the predictions from the Caldeira--Leggett master equation with those of more complicated, non-Markovian models often exhibit surprisingly good agreement \cite{Paz:1993:ta}.

In the position representation, the final term on the right-hand side of Eq.~\eqref{eq:vfnbcclclnd9s27} may be expressed as
\begin{equation}
  \label{eq:fsdojgdj1}
  -  \gamma_0 \left( \frac{x-x'}{\lambda_\text{th}} \right)^2 \rho_S(x,x',t),
\end{equation}
where $\lambda_\text{th}$ is the thermal de Broglie wavelength  defined in Eq.~\eqref{eq:daf12thermal}. This term describes spatial localization with a decoherence rate $\tau_{\abs{x-x'}}^{-1}$ given by \cite{Zurek:1986:uz}
\begin{equation}
\label{eq:odijsvuhfsw21}
  \tau_{\abs{x-x'}}^{-1} = \gamma_0 \left(
    \frac{x-x'}{\lambda_\text{th}} \right)^2.
\end{equation}
This is Eq.~\eqref{eq:daf12}, and as discussed there, given that $\lambda_\text{th}$ is extremely small for macroscopic and even mesoscopic objects, it follows that, typically, superpositions of macroscopically separated center-of-mass positions will be decohered on a timescale that is many orders of magnitude shorter than the dissipation (relaxation) timescale $\gamma^{-1}_0$. Therefore, over timescales on the order of the decoherence time, it is often safe to drop the second, dissipative term on the right-hand side of Eq.~\eqref{eq:vfnbcclclnd9s27}, yielding a master equation that describes pure decoherence,
\begin{equation}
\label{eq:vfnbcclasclnd9s27}
  \frac{\partial}{\partial t} \op{\rho}_S(t)  = -\frac{\I}{\hbar} \bigl[
  \op{H}'_S, \op{\rho}_S(t) \bigr] 
 - \frac{2 M\gamma_0 k_B T}{\hbar^2} \bigl[ x, \bigl[ x, \op{\rho}_S(t) \bigr]
\bigr],
\end{equation}
which is of the Lindblad double-commutator form~\eqref{eq:lindbladc}. Incidentally, this result provides a microscopic motivation for the simple Lindblad master equation~\eqref{eq:lifsfdndbladc} for spatial decoherence we had written down previously. 

Note that the decoherence rate given by Eq.~\eqref{eq:odijsvuhfsw21} grows without bounds as the separation $x-x'$ is increased, which is an obviously unphysical behavior. Just as in the case of collisional decoherence, we expect that the decoherence rate will saturate; in the present case, this should happen when the separation $x-x'$ grows to approach the maximum coherence length of the oscillator environment. The absence of such a saturation point indicates the limitations of the Caldeira--Leggett model. In fact, by considering the general model of a massive particle coupled to a massless scalar field, one can show \cite{Unruh:1989:rc,Anglin:1997:za,Paz:2001:aa} that not only the Caldeira--Leggett model but also the quantum Brownian master equation \eqref{eq:vfoinbnd9s27} is based on an implicit long-wavelength assumption (compare the discussion in Sec.~\ref{sec:collisionaldecoherence}). The absence of a saturation point for the decoherence rate given by Eq.~\eqref{eq:odijsvuhfsw21} is therefore nothing but an artifact of the underlying model, showing that one needs to be careful when extrapolating models to different parameter regimes \cite{Gallis:1990:un,Gallis:1992:im,Anglin:1997:za}. 

While we presented the Caldeira--Leggett master equation in the context of quantum Brownian motion, it is in fact far more general.  It may be applied to arbitrary system potentials, and instead of a system described by position and momentum coordinates, we may instead use a spin-$\frac{1}{2}$ particle represented by the Pauli spin operators $\op{\sigma}_x$ and $\op{\sigma}_z$, giving rise to models of the spin--boson kind (see Sec.~\ref{sec:spin-boson-models}). 

As it stands, the Caldeira--Leggett master equation~\eqref{eq:vfnbcclclnd9s27} cannot be expressed in Lindblad form and therefore cannot guarantee complete positivity \cite{Breuer:2002:oq}. However, it can be brought into Lindblad form through a minimal modification. This amounts to adding a term $-\gamma_0(8Mk_B T)^{-1} [p,[p, \op{\rho}_S(t)]]$, which is small in the relevant high-temperature limit \cite{Breuer:2002:oq}. The resulting Lindblad master equation then has the single Lindblad operator 
\begin{equation}\label{eq:dkvnkl1}
\op{L} = \sqrt{\frac{4Mk_B T}{\hbar^2}}\op{x} + \I \sqrt{\frac{1}{4Mk_B T}} \,\op{p}.
\end{equation}

\subsection{\label{sec:spin-boson-models}Spin--boson models}

In the spin--boson model (see, e.g., Refs.~\cite{Leggett:1987:pm,Breuer:2002:oq,Schlosshauer:2007:un} for reviews), a qubit interacts with an environment of harmonic oscillators. This model has been of strong interest in investigations of decoherence; it has been used, for example, in the first studies of qubit decoherence in the early years of quantum information \cite{Unruh:1995:uy,Palma:1996:yy}. Spin--boson models are exceptionally versatile because many quantum systems can be represented by a two-level system, and because harmonic-oscillator environments are, as mentioned before, of great generality \cite{Feynman:1963:jj,Caldeira:1983:gv}. 

Let us first consider a simplified spin--boson model where the  self-Hamiltonian of the system is taken to be  
\begin{equation}
\op{H}_S = \frac{1}{2} \hbar\omega_0 \sigma_z,
\end{equation}
 with eigenstates $\ket{0}$ and $\ket{1}$. In contrast with the more general case discussed below, this Hamiltonian does not include a tunneling term $\frac{1}{2}\hbar \Delta_0 \sigma_x$, and thus $\op{H}_S$ does not generate any nontrivial intrinsic dynamics. The self-Hamiltonian for the environment of harmonic oscillators is the same as in Eq.~\eqref{eq:sfsfjaa11},
\begin{equation}
  \op{H}_E = \sum_i \left( \frac{1}{2m_i}p_i^2 +
  \frac{1}{2}m_i\omega_i^2q_i^2 \right),  
\end{equation}
 and we choose the bilinear interaction Hamiltonian 
\begin{equation}
\op{H}_\text{int} =   \sigma_z \otimes \sum_i c_i q_i.
\end{equation}
Using the bosonic creation and annihilation (i.e., raising and lowering) operators $a_i^\dagger$ and $a_i$, we may recast the total Hamiltonian as
\begin{equation}\label{eq:h-ssb}
  H =  \frac{1}{2}\hbar \omega_0 \sigma_z 
  + \sum_i \hbar\omega_i a_i^\dagger a_i + \sigma_z \otimes  \sum_i 
  \left( g_ia_i^\dagger + g_i^* a_i \right),
\end{equation}
with $[a_i, a_j^\dagger]=\delta_{ij}$ (for simplicity, we have dropped the vacuum-energy term $\sum_i \frac{\hbar\omega_i}{2}$).

Note that since the total Hamiltonian $H=H_S+H_E+H_\text{int}$ commutes with $\sigma_z$, no transitions between the $\sigma_z$ eigenstates $\ket{0}$ and $\ket{1}$ can be induced by $H$. Because there is no energy exchange between the system and the environment, the model describes decoherence without dissipation. Such a model is a good representation of decoherence processes that occur on a timescale that is much shorter than the timescale for dissipation, as is often the case in physical applications. The resulting evolution can be solved exactly (see, e.g., Refs.~\cite{Schlosshauer:2007:un,Hornberger:2009:aq} for details). For an ohmic spectral density with a high-frequency cutoff, it is found that superpositions of the form $\alpha\ket{0}+\beta\ket{1}$ are exponentially decohered on a timescale set by the thermal correlation time $\tau_B=2\hbar (k_B T)^{-1}$ of the environment.

Inclusion of a tunneling term $\frac{1}{2} \hbar\Delta_0 \sigma_x$ yields the general spin--boson model defined by the
Hamiltonian
\begin{equation}\label{eq:h-sbhdcskgsf}
H = \frac{1}{2} \hbar\omega_0 \sigma_z + 
\frac{1}{2} \hbar\Delta_0 \sigma_x 
+  \sum_i \left( \frac{1}{2m_i} p_i^2 + \frac{1}{2} m_i \omega_i^2 q_i^2 
   \right) + \sigma_z \otimes \sum_i
  c_i q_i.
\end{equation}
One typically considers the unbiased case $\omega_0=0$, corresponding to a symmetric double-well potential (but see Sec.~VII of Ref.~\cite{Leggett:1987:pm} for a treatment of the biased case). The non-Markovian dynamics of this model have been studied in great detail in Refs.~\cite{Leggett:1987:pm,Weiss:1999:tv}. The particular dynamics strongly depend on the various parameters of the model, such as the temperature of the environment, the form of the spectral density (subohmic, ohmic, or supraohmic), and the system--environment coupling strength. For each parameter regime, a characteristic dynamical behavior emerges: localization, exponential or incoherent relaxation, exponential decay, and strongly or weakly damped coherent oscillations \cite{Leggett:1987:pm}.

In the weak-coupling limit, one can derive the Born--Markov master equation for the Hamiltonian~\eqref{eq:h-sbhdcskgsf} (where we shall again assume $\omega_0=0$ for simplicity). Because of the formal similarity between the spin--boson Hamiltonian and the Hamiltonian for quantum Brownian motion, the derivation proceeds in much the same way as for the quantum Brownian motion. The result is (see, e.g., Refs.~\cite{Paz:2001:aa,Schlosshauer:2007:un} for details)
\begin{equation}
\label{eq:vjp32gbntrkh22}
\frac{\partial}{\partial t} \op{\rho}_S(t) = -\frac{\I}{\hbar} \left(
  \op{H}'_S \op{\rho}_S(t) - {\op{\rho}}_S(t)
  H'^\dagger_S \right) - D \left[
  \sigma_z, \left[ \sigma_z, \op{\rho}_S(t)
  \right]\right] - \zeta \sigma_z
\op{\rho}_S(t)\sigma_y - \zeta^* \sigma_y
\op{\rho}_S(t)\sigma_z,
\end{equation}
where
\begin{subequations}
\begin{align}
  \op{H}'_\mathcal{S} &= \hbar\left(  \frac{1}{2}
  \Delta_0 + \zeta^* \right) \op{\sigma}_x. \\
  \zeta^* &= \int_0^\infty \D \tau \, \left[\nu(\tau) - \I \eta(\tau)\right] \sin\left( \Delta_0 \tau \right), \\
  D &= \int_0^\infty \D \tau \,
  \nu(\tau) \cos\left( \Delta_0 \tau \right),
\end{align}
\end{subequations}
with the noise and the dissipation kernels $\nu(\tau)$ and $\eta(\tau)$ taking the same form as in quantum Brownian motion [see Eqs.~\eqref{eq:vdjpoo17} and \eqref{eq:ponol218}]. The first term on the right-hand side of Eq.~\eqref{eq:vjp32gbntrkh22} represents the evolution under the environment-renormalized (and in general non-Hermitian) Hamiltonian $\op{H}'_S$. The second term is in the Lindblad double-commutator form \eqref{eq:lindbladc} and generates decoherence in the $\sigma_z$ eigenbasis at a rate given by $D$. The last two terms describe the decay of the two-level system. In the absence of tunneling ($\Delta_0=0$ and hence also $\zeta=0$), Eq.~\eqref{eq:vjp32gbntrkh22} reduces to the pure-decoherence Lindblad master equation~\eqref{eq:vjp32gbntrkh22max} discussed in Sec.~\ref{sec:two-simple-examples}.

\subsection{\label{sec:spin-envir-models}Spin-environment models}

A qubit linearly coupled to a collection of other qubits---also known as a \emph{spin--spin model}---is often a good model of a two-level system (for example, a superconducting qubit) that interacts strongly with a low-temperature environment \cite{Prokofev:2000:zz,Dube:2001:zz}. The model of a harmonic oscillator interacting with a spin environment may be relevant to the description of decoherence and dissipation in quantum-nanomechanical systems and micron-scale ion traps \cite{Schlosshauer:2008:os}. For details on the theory of spin-environment models, see Refs.~\cite{Dube:2001:zz,Stamp:1998:im,Prokofev:1995:ab,Prokofev:1993:aa}.

A basic version of a spin--spin model was studied by Zurek in his seminal paper of 1982 \cite{Zurek:1982:tv}. This model neglects the intrinsic dynamics of the system and environment, and the interaction Hamiltonian describes a bilinear coupling between the system and environment spins, 

\begin{equation}\label{eq:hse-zurek}
  \op{H} = \op{H}_\text{int} = \frac{1}{2} \op{\sigma}_z \otimes
   \sum_{i=1}^N g_i \op{\sigma}^{(i)}_z  \equiv
  \frac{1}{2} \op{\sigma}_z \otimes \op{E}.
\end{equation}
This Hamiltonian represents the environmental monitoring of the observable $\sigma_z$ and leads to decoherence in the $\{\ket{0},\ket{1}\}$ eigenbasis of $\op{\sigma}_z$. Specifically, one can show \cite{Zurek:1982:tv,Cucchietti:2005:om} that the decoherence rate increases exponentially with the number $N$ of environmental spins, and that for large $N$ and a broad class of distributions of the coupling coefficients $g_i$, the interference damping follows an approximately Gaussian time dependence
$\propto \exp(- \Gamma^2 t^2)$, where the decay constant $\Gamma$ is determined by the initial state of the environment and the distribution of the couplings $g_i$. 

If a tunneling term is added to the Hamiltonian (representing the intrinsic dynamics of the system), the Hamiltonian becomes
\begin{equation}\label{eq:jhdsgwuygfurwb}
  H = \op{H}_S + \op{H}_\text{int} =  \frac{1}{2}\hbar
  \Delta_0 \sigma_x + \frac{1}{2} \sigma_z \otimes
  \sum_{i=1}^N g_i
  \sigma_z^{(i)} 
  \equiv \frac{1}{2}\hbar \Delta_0 \sigma_x + \frac{1}{2}
  \sigma_z \otimes E.
\end{equation}
This model can be solved exactly \cite{Dobrovitski:2003:az,Cucchietti:2005:om}. The particular preferred (pointer) states selected by the dynamics depend on the relative strengths of the self-Hamiltonian $\op{H}_S$ of the system and the interaction Hamiltonian $\op{H}_\text{int}$. In general, the preferred states are those that are most robust under the action of the \emph{total} Hamiltonian. In the quantum-measurement limit, where the interaction Hamiltonian dominates the evolution [this is the model described by Eq.~\eqref{eq:hse-zurek}], the emerging pointer states are indeed found to be close to the eigenstates of the interaction Hamiltonian (i.e., the eigenstates $\{\ket{0},\ket{1}\}$ of $\op{\sigma}_z$) \cite{Cucchietti:2005:om}, as also predicted by the commutativity criterion, Eq.~\eqref{eq:dhvvsdnbbfvs27}. In the quantum limit of decoherence, where the modes of the environment are slow and the self-Hamiltonian $\op{H}_S$ of the system dominates, the pointer states are found to be close to the eigenstates of $\op{H}_S$, i.e., the eigenstates $\ket{\pm}=\left(\ket{0}\pm\ket{1}\right)/\sqrt{2}$ of  $\op{\sigma}_x$ \cite{Cucchietti:2005:om}.

In the weak-coupling limit, spin environments can be mapped onto oscillator environments \cite{Feynman:1963:jj,Caldeira:1993:bz}. Specifically, the reduced dynamics of a system weakly coupled to a spin environment can be described by the system coupled to an \emph{equivalent} oscillator environment with an explicitly temperature-dependent spectral density of the form
\begin{equation}
\label{eq:vslkfvfgyiJA2}
J_\text{eff}(\omega, T) = J(\omega)
\tanh\left(\frac{\hbar\omega}{2k_B T}\right),
\end{equation}
where $J(\omega)$ is the original spectral density of the spin environment. See Sec.~5.4.2 of Ref.~\cite{Schlosshauer:2007:un} for details and examples.

Many physical settings in which spin environments are the appropriate model are also those where low temperatures and strong system--environment interactions render the Born--Markov assumptions of weak coupling and negligible memory effects inapplicable. Therefore, one needs to look for non-Markovian solutions, which are typically difficult to calculate. In principle, techniques such as the instanton formalism \cite{Prokofev:2000:zz} allow for analytical calculations of relevant quantities such as spin expectation values. One challenging task is the tracing (averaging) over the degrees of freedom of a strongly coupled environment. Prokof'ev and Stamp \cite{Prokofev:2000:zz} have demonstrated that this task can be accomplished by considering four limiting cases of the general spin-environment model, averaging over the environment in each case, and then combining the four averages to represent the average for the general model; see Ref.~\cite{Prokofev:2000:zz} for details.

\section{\label{sec:decoh-errcorr}Decoherence avoidance and mitigation}

Combatting the detrimental effect of decoherence is of paramount importance whenever nonclassical quantum superposition states need to be generated and maintained, for example, in quantum information processing, quantum computing, and quantum technologies \cite{Dowling:2003:tv}. Accordingly, a number of methods have been developed to prevent quantum states from decohering in the first place (or, at least, to minimize their decoherence), and to undo (correct for) the effects of decoherence. In the terminology of quantum information processing where the effects of decoherence amount to processing errors, the first approach is often known as \emph{error avoidance}, whereas the second approach is referred to as \emph{error correction}. Here, we will discuss decoherence-free subspaces  (Sec.~\ref{sec:dfs}) as an instance of an error-avoidance scheme, and also comment on techniques such as reservoir engineering and dynamical decoupling (Sec.~\ref{sec:reserv-engin-quant}). We will then give a brief description of quantum error correction (Sec.~\ref{sec:corr-decoh-induc}).

\subsection{\label{sec:dfs}Decoherence-free subspaces}

Recall that while the pointer states themselves are robust against decoherence, superpositions of such states will generally be rapidly decohered. By contrast, in a \emph{pointer subspace} \cite{Zurek:1982:tv} or \emph{decoherence-free subspace} (DFS) \cite{Palma:1996:yy,Lidar:1998:uu,Zanardi:1997:yy,Zanardi:1997:tv,Zanardi:1998:oo,Lidar:1999:fa,Bacon:2000:yy,Duan:1998:yb,Zanardi:2001:oo,Knill:2000:aa} (see Refs.~\cite{Lidar:2003:aa,Lidar:2014:pp} for reviews), \emph{every possible state in that subspace} will be robust to decoherence. This, obviously, is a much stronger statement than the existence of individual pointer states, and it is therefore not surprising that the existence of a DFS, especially a high-dimensional one for larger systems, is nontrivial and that the conditions required for such a DFS to be present tend to be correspondingly difficult to meet in practice. 

\subsubsection{Condition for the existence of a decoherence-free subspace} 

For a given interaction Hamiltonian $\op{H}_\text{int} =  \sum_\alpha \op{S}_\alpha \otimes \op{E}_\alpha$, a DFS is spanned by a basis $\{\ket{s_i}\}$ of orthonormal pointer states with the added condition that the action of each system operator $\op{S}_\alpha$ in $\op{H}_\text{int}$ must be the same for each of the pointer-basis states $\ket{s_i}$. In other words, the result of the system--environment interaction applied to these pointer states must be trivial in the sense that it must not distinguish between these states. In this way, the existence of a DFS corresponds to the presence of a symmetry in the structure of the system--environment interaction (i.e., a \emph{dynamical symmetry}). Mathematically, this condition is implemented through strengthening the pointer-state condition of Eq.~\eqref{eq:OIbvsrhjkbv9}, by requiring that the pointer states are simultaneous \emph{degenerate} eigenstates of each $\op{S}_\alpha$,
\begin{equation}
  \label{eq:OIbvsrhjkbvsfljvh9}
  \op{S}_\alpha \ket{s_i} = \lambda^{(\alpha)} \ket{s_i} \qquad
  \text{for all $\alpha$ and $i$}. 
\end{equation}
If this condition is fulfilled, then the evolution generated by the interaction Hamiltonian $\op{H}_\text{int} =  \sum_\alpha \op{S}_\alpha \otimes \op{E}_\alpha$ for an arbitrary, initially pure state $\ket{\psi} = \sum_i c_i \ket{s_i}$ in the DFS is 
\begin{equation}
  \label{eq:OIbvsrhjkbv9zFFHGSVCxc}
  \E^{-\I \op{H}_\text{int}t} \ket{\psi}\ket{E_0} = \ket{\psi}\E^{-\I
    \left( \sum_\alpha \lambda^{(\alpha)} E_\alpha \right)t/\hbar}
  \ket{E_0} \equiv \ket{\psi} \ket{E_\psi(t)}.
\end{equation}
This shows that the system does not get entangled with the environment and therefore remains decoherence-free. Of course, in general the self-Hamiltonian of the system will also contribute, in which case one needs to additionally ensure that this Hamiltonian does not take the state outside the DFS, which would then make it vulnerable to decoherence. The concept of a DFS can be generalized and extended to the formalism of \emph{noiseless subsystems} or \emph{noiseless quantum codes} \cite{Knill:2000:aa,Kempe:2001:oo,Lidar:2003:aa,Choi:2006:tt,Beny:2007:pp,BlumeKohout:2010:pp, Lidar:2014:pp}; see Ref.~\cite{Lidar:2014:pp} for a review.

\subsubsection{Collective versus independent decoherence}

The condition \eqref{eq:OIbvsrhjkbvsfljvh9} for the basis states of a DFS is strong and therefore often difficult to fulfill in practice. To illustrate this point, let us consider a system consisting of $N$ two-level systems (qubits) interacting with an environment of harmonic oscillators (this is the spin--boson model discussed in Sec.~\ref{sec:spin-boson-models}). The interaction Hamiltonian is
\begin{equation}
  \label{eq:dadrestinpeacdh11}
  \op{H}_\text{int} =  \sum_{i=1}^N  \sigma_z^{(i)} \otimes \sum_j
  \left( g_{ij}a_j^\dagger + g_{ij}^* a_j \right) \equiv
  \sum_{i=1}^N  \sigma_z^{(i)} \otimes E_i,
\end{equation}
where the $g_{ij}$ are coupling coefficients, and $a^\dagger$ and $a$ are the raising and lowering operators for the harmonic oscillators of the environment. 

Recall that the existence of a DFS is related to the presence of a dynamical symmetry in the interaction Hamiltonian. A drastic way of creating such a symmetry is to require that each qubit operator $\op{\sigma}_z^{(i)}$ couples to the environment in exactly the same way. This means that the interaction with the environment is invariant under an exchange of any two qubits, and therefore the environment cannot distinguish between the qubits. This limiting case is commonly referred to as \emph{collective decoherence}, and in the example of the spin--boson interaction \eqref{eq:dadrestinpeacdh11} corresponds to dropping the dependence of the coupling coefficients $g_{ij}$ on the index $i$ labeling the particular qubit. Then the interaction Hamiltonian \eqref{eq:dadrestinpeacdh11} becomes 
\begin{equation}
\label{eq:dadrestinpeacfxndh22}
\op{H}_\text{int} =  \left( \sum_{i} \sigma_z^{(i)} \right)
\otimes E \equiv
\op{S}_z \otimes E.
\end{equation}
One sees that this Hamiltonian represents an interaction between the collective spin operator and a single environment operator. 

The condition \eqref{eq:OIbvsrhjkbvsfljvh9} for basis states of a DFS then tells us that the DFS will be a spanned by states that are simultaneous degenerate eigenstates of $\op{S}_z=\sum_i \op{\sigma}_z^{(i)}$. Consider a computational-basis state $\ket{m_1}\ket{m_2} \otimes \cdots \otimes \ket{m_N}$, where we let $m_i=0$ represent the eigenvalue $+1$ of $\op{\sigma}_z$, and $m_i=1$ the eigenvalue $-1$. Any such computational-basis state will be an eigenstate of $\op{S}_z$, with integer eigenvalues ranging from $M=-N$ (for the state $\ket{11\cdots 1}$) to $M=+N$ (for the state $\ket{00\cdots 0}$). Since we would like to span a subspace from a set of degenerate of eigenstates of $\op{S}_z$, and would like this subspace to be as large as possible so we can make the largest possible number of quantum states immune to decoherence, we need to look for the greatest number of computational-basis states with the same eigenvalue $M$ of $\op{S}_z$. This happens for $M=0$, corresponding to computational-basis states for which half of the qubits are in the state $\ket{0}$ and half in the state $\ket{1}$. For a system of $N$ qubits, there are $\binom{N}{N/2}$ such computational-basis states. These states will then span a DFS. For example, for four qubits ($N=4$), represented by a Hilbert space of dimension $2^4=16$, we have $\binom{4}{2}=6$ computational-basis states with the same eigenvalue $M=0$, namely, $\ket{0011}, \ket{0101}, \ket{0110}, \ket{1001}, \ket{1010}$, and $\ket{1100}$. These states span a six-dimensional DFS. For large $N$, Stirling's formula for approximating the binomial coefficient gives 
\begin{equation}
 \log_2 \binom{N}{N/2} \approx N - \frac{1}{2} \log_2 (\pi N/2) \,\,
 \xrightarrow{N \gg 1} \,\, N,
\end{equation}
which shows that the dimension of the DFS approaches the dimension of the Hilbert space of the system. Thus, in this limiting case of perfectly collective decoherence of a very large qubit system, essentially every state in the system's Hilbert space will be immune to decoherence. 

Using the spin--boson example just discussed, we can also see that no DFS exists when no two qubits couple to the environment in the same way, that is, if the interaction Hamiltonian does not exhibit any dynamical symmetry. In this limiting case, known as \emph{independent decoherence}, the couplings $g_{ij}$ appearing in the interaction Hamiltonian~\eqref{eq:dadrestinpeacdh11} will be different for each qubit, and thus the Hamiltonian retains its form \eqref{eq:dadrestinpeacdh11}, $\op{H}_\text{int}  =  \sum_{i=1}^N  \sigma_z^{(i)} \otimes E_i$, with the environment operators $E_i$ differing between any two qubits. The usual DFS condition \eqref{eq:OIbvsrhjkbvsfljvh9} would then require us to find an orthonormal set of simultaneous, degenerate $N$-qubit eigenstates of \emph{each} single-qubit operator $\sigma_z^{(i)}$, $i=1,\hdots,N$. The only computational-basis states that fulfill this condition are $\ket{00\cdots 0}$ and $\ket{11\cdots 1}$, albeit with different eigenvalues, and therefore no DFS can exist \cite{Lidar:1998:uu}. 

In practice, it would be challenging to find a system in which each qubit couples to exactly the same environment. Fortunately, one can show that a DFS is robust to small deviations from perfect dynamical symmetry. Specifically, one may investigate the consequences of perturbing a symmetric interaction Hamiltonian, such as the Hamiltonian given by Eq.~\eqref{eq:dadrestinpeacfxndh22}, by adding additional, small coupling terms $\widetilde{g}_{ij}$ that break the symmetry by distinguishing between the qubits. Such perturbations will introduce a tunable dependence of the environmental interaction on the qubit index $i$. This poses the question of the sensitivity of a DFS to such perturbations, and how the sensitivity scales with the system. Using the measure of \emph{dynamical fidelity} \cite{Lidar:1998:uu,Bacon:1999:aq}, which quantifies how the evolution of a given initial state differs in the presence of additional system--environment couplings, it has been shown \cite{Lidar:1998:uu,Bacon:1999:aq,Kattemolle:2018:ii} that, to first order in the strength of the symmetry-breaking perturbations, a DFS is robust to such perturbations. The influence of perturbations resulting in noncollective decoherence effects has also been studied experimentally; see the photonic experiment reported in Ref.~\cite{Altepeter:2004:ll} for an example. We note that strategies for quantum error correction \cite{Steane:1996:cd,Shor:1995:rx,Steane:2001:dx,Knill:2002:rx,Nielsen:2000:tt,Lidar:2013:pp} may be used to combat decoherence arising from subspaces that are not perfectly decoherence-free \cite{Lidar:1999:fa}.

\subsubsection{Experimental realizations of decoherence-free subspaces}

Starting with the proof-of-principle demonstration for two-photon states by Kwiat et al.\ \cite{Kwiat:2000:kv}, several experiments have realized DFSs. Among the first, in 2001 Kielpinski et al.\ \cite{Kielpinski:2001:uu} reported the creation of a two-dimensional (i.e., one-bit) DFS using a pair of trapped, interacting $^9$Be$^+$ ions, with the qubit states formed by two hyperfine levels and a decohering environment simulated by fluctuations of the laser intensity, and Viola et al.\ \cite{Viola:2001:ra} described generation of a one-bit DFS using three NMR qubits. Roos et al.\ \cite{Roos:204:pp} created a DFS using two trapped $^{40}$Ca$^+$ ions subject to a dephasing environment and achieved coherence times around \unit[1]{s}. Two-ion DFSs were also reported by H{\"a}ffner et al.\ \cite{Haffner:2005:zz} and Langer et al.\ \cite{Langer:2005:uu}. Coherence times of up to \unit[34]{s} were found, and long lifetimes of up to \unit[20]{s} were observed for the entanglement between ion pairs. DFSs for a photon pair were further investigated experimentally by Altepeter et al.\ \cite{Altepeter:2004:ll}, who also studied the sensitivity of a DFS to perturbations that introduce noncollective couplings to the environment. 

Mohseni et al.\ \cite{Mohseni:2003:pp} experimentally demonstrated how the performance of a photonic implementation of the Deutsch--Jozsa quantum algorithm \cite{Deutsch:1989:mm} can be substantially enhanced through the use of a DFS. DFSs have also been experimentally realized in quantum cryptography, for example, in the fault-tolerant quantum key distribution protocol proposed and implemented by Zhang et al.\ \cite{Zhang:2006:zz}. The usefulness of DFSs is not limited to quantum information processing, either. For instance, a DFS has been successfully realized to protect a neutron interferometer from unwanted noise arising from low-frequency mechanical vibrations \cite{Pushin:2011:zz}. 
 
\subsection{\label{sec:reserv-engin-quant}Reservoir engineering and dynamical decoupling}

Even approximate dynamical symmetries will often be absent in multi-qubit systems, and therefore one approach consists of actively creating such symmetries through a strategy known as \emph{environment engineering} or \emph{reservoir engineering}. Dalvit, Dziarmaga, and Zurek \cite{Dalvit:2000:bb} have shown how this idea, in principle, could lead to a DFS that is spanned by superposition states of Bose--Einstein condensates. In the context of ion traps, theoretical \cite{Poyatos:1996:um} and experimental \cite{Myatt:2000:yy,Turchette:2000:aa,Carvalho:2001:ua} studies have investigated how different DFSs for the trapped ion can be created through a laser-induced manipulation of the system--environment coupling. 

Beyond DFSs but related in spirit, engineering of the couplings between system and  environment has also been used to drive the system into particular quantum superposition states (``quantum state engineering''). Experimental implementations of this approach have been reported, for example, with trapped ions \cite{Barreiro:2011:oo,Lin:2013:pp,Kienzler:2015:oo}, superconducting circuits \cite{Shankar:2013:pp}, and atomic ensembles \cite{Krauter:2011:ll}. Reservoir engineering even opens up the possibility of an unorthodox implementation of universal quantum computation \cite{Verstraete:2009:ii}, in which the interaction with the environment is not an adversary but rather a resource for quantum information processing. It has also been shown \cite{Braun:2002:aa, Benatti:2003:aa,Kim:2002:oo,Jakobczyk:2002:oo} that for two quantum systems that do not interact with each other but are coupled to a common environment, the decoherence and dissipation produced by the environment can sometimes lead to the creation of entanglement between the two systems. This is a noteworthy result, given that the action of decoherence and dissipation is usually considered detrimental to the presence of entanglement between systems \cite{Zyczkowski:2001:ii,Lee:2004:uu,Barreiro:2010:aa}. Such dynamical entanglement generation through environmental interactions was demonstrated, for example, in an exactly solvable model of two qubits interacting with a heat bath of harmonic oscillators \cite{Braun:2002:aa}, as well as arising from Markovian dissipative dynamics \cite{Benatti:2003:aa}; see also Refs.~\cite{Kim:2002:oo,Jakobczyk:2002:oo}. 

Coherence of quantum states may be maintained also through a sequence of rapidly applied control pulses (or projective measurements) that average the coupling between system and environment to zero, an approach known as \emph{dynamical decoupling} or \emph{quantum bang-bang control} \cite{Viola:1998:uu,Viola:1999:zp,Zanardi:1999:oo,Viola:2000:pp,Wu:2002:aa,Wu:2002:bb}. In this way, a dynamically decoupled subspace is created, and coherence may be maintained to a large degree as long as the rate of the control pulses is higher than the rate at which entanglement with the environment is being produced. Dynamical decoupling can also be used to enhance the fidelity of quantum gates by several orders of magnitude \cite{Viola:1999:pp,West:2010:oo}.

\subsection{\label{sec:corr-decoh-induc}Quantum error correction}

Quantum error correction (see, e.g., Refs.~\cite{Steane:2001:dx,Knill:2002:rx,Nielsen:2000:tt,Gaitan:2008:uu,Lidar:2013:pp} for reviews) is the technique of undoing the change of the quantum state of a system induced by decoherence. The idea of quantum error correction, going back to Steane \cite{Steane:1996:cd} and Shor \cite{Shor:1995:rx}, is to couple the system to an ancilla in such a way that the original, pre-decoherence state can be reconstructed.

\subsubsection{Basic concepts}

To see how such state reconstruction is made possible, we observe that any changes to the state of a qubit resulting from its interaction with an environment $E$ can be reduced to the combined action of three different, discrete transformations. Specifically, for a single qubit in an initially pure state $\ket{\psi}$, the evolution of the combined system--environment state may always be written in the form \cite{Steane:2001:dx,Knill:2002:rx,Nielsen:2000:tt,Schlosshauer:2007:un}
\begin{equation} \label{eq:qcerrc} 
\ket{\psi} \ket{e_r}  \, \longrightarrow \, I
  \ket{\psi} \ket{e_I} + \sum_{s= x,y,z }
  \left( \sigma_s \ket{\psi} \right)
  \ket{e_s},
\end{equation}
where $\op{I}$ is the identity operator, the Pauli operators $\sigma_s$ act on the Hilbert space of $S$, $\ket{e_r}$ is the initial state of the environment, and $\ket{e_I}$ and $\{ \ket{e_s} \}$ denote states of the environment that need not be orthogonal or normalized. Equation~\eqref{eq:qcerrc} is simply a consequence of the fact that the Pauli operators, together with the identity, form a complete set of operators in the Hilbert space of the qubit. 

The effects of $\sigma_x$ and $\sigma_z$ on the qubit state are referred to as a \emph{bit-flip error} and \emph{phase-flip error}, respectively (since $\op{\sigma}_y= \I\op{\sigma}_x\op{\sigma}_z$, the operator $\op{\sigma}_y$ represents the simultaneous presence of a bit-flip error and phase-flip error). State changes resulting from environmental entanglement alone (i.e., decoherence) are fully captured by phase-flip errors. To see this explicitly, consider a qubit in an arbitrary state $\ket{\psi}=a\ket{0}+b\ket{1}$ undergoing an entangling interaction with an environment,
\begin{equation} 
\ket{\psi}\ket{e_r} \,\longrightarrow\, a\ket{0}\ket{e_0}+b\ket{1}\ket{e_1}.
\end{equation}
We may rewrite the right-hand side as
\begin{align} 
a\ket{0}\ket{e_0}+b\ket{1}\ket{e_1} &= \left( a\ket{0}+b\ket{1} \right) \frac{1}{2} \left( \ket{e_0}+\ket{e_1}\right) + \left( a\ket{0}-b\ket{1} \right) \frac{1}{2} \left( \ket{e_0}-\ket{e_1}\right) \notag \\
&= \left(\op{I} \ket{\psi} \right) \frac{1}{2} \left( \ket{e_0}+\ket{e_1}\right) + \left(\op{\sigma}_z \ket{\psi} \right) \frac{1}{2} \left( \ket{e_0}-\ket{e_1}\right).
\end{align}
Making the identifications $\ket{e_I} \equiv \frac{1}{2} \left( \ket{e_0}+\ket{e_1}\right)$ and $\ket{e_z} \equiv \frac{1}{2} \left( \ket{e_0}-\ket{e_1}\right)$, we recover Eq.~\eqref{eq:qcerrc} with only the $\op{\sigma}_z$ (phase-flip) term present,
\begin{equation} \label{eq:qcerrcaa} \ket{\psi}
  \ket{e_r}  \, \longrightarrow \, I
  \ket{\psi} \ket{e_I} + 
  \left( \sigma_z \ket{\psi} \right)
  \ket{e_z}.
\end{equation}
For $N$ qubits, this can be shown to generalize to 
\begin{equation}\label{eq:qcerrcN}
  \ket{\psi} \ket{e_r} \, \longrightarrow\,
  \sum_{i} 
  \left( E_i \ket{\psi} \right) \ket{e_i},
\end{equation}
where the so-called \emph{error operators} $E_i$ represent tensor products of $N$ operators involving identity and $\sigma_z$ operators; the number of $\sigma_z$ operators appearing in a given operator $E_i$ is refereed to as the \emph{weight} of the error operator. In many cases of interest, only a limited number $K<N$ of qubits become entangled with the environment  (``partial decoherence''), and hence only the $2^K$ different error operators up to weight $K$ will need to be considered. A further dramatic reduction in the number of error operators to be taken into account occurs in the case of independent qubit decoherence, where each qubit couples independently to an environment (see Sec.~\ref{sec:dfs}). In this case, only error operators of weight equal to one will need to be considered (corresponding to independent phase flip errors), and therefore no more than $N$ such operators in total will be needed to describe the result of the decoherence process. 

To bring about the actual correction of the quantum error imparted by decoherence, we start from the post-decoherence state $\sum_{i}   \left( E_i \ket{\psi} \right) \ket{e_i}$ of Eq.~\eqref{eq:qcerrcN} and couple an ancilla to the qubit system in such a way that the composite system evolves as 
\begin{equation}\label{eq:errfsyn}
  \ket{a_0} \left[ \sum_{i} \left( E_i   \ket{\psi} \right)
    \ket{e_i} \right] \, \longrightarrow \, 
  \sum_{i} \ket{a_i} \left( E_i \ket{\psi} \right)
  \ket{e_i}.
\end{equation}
Here $\ket{a_0}$ is the initial state of the ancilla, and $\ket{a_i}$ are ancilla states that we shall assume to be (approximately) orthogonal so that they can be distinguished by a subsequent measurement. This measurement of the ancilla, represented by an observable  $\op{O}_A = \sum_i a_i \ketbra{a_i}{a_i}$ with all eigenvalues $a_i$ being distinct, will project the system--ancilla combination onto one of the states $\ket{a_k} \left( E_k \ket{\psi} \right) \ket{e_k}$, with measurement outcome $a_k$. We have therefore isolated a single error operator $E_k$, and knowledge of the outcome $a_k$ provides the necessary information for applying a countertransformation $E_k^{-1}=E_k^\dagger$ to the system. This transformation changes the state from $\ket{a_k} \left( E_k \ket{\psi} \right) \ket{e_k}$ to $\ket{a_k} \ket{\psi} \ket{e_k}$, thereby restoring the original pre-decoherence state of the system. Note that no information about the state of the system is necessary to correct the error---as must be, for otherwise any such information gain would result in an uncontrollable disturbance of the system.

\subsubsection{Challenges}

We have sketched here only the bare essentials of quantum error correction, and in a highly simplified form. In practice, a number of challenges and complications arise. Let us mention just three. First, and perhaps most importantly, it is usually impossible to realize a system--ancilla evolution of the form \eqref{eq:errfsyn} such that all possible error operators are perfectly distinguishable, by measurement, via corresponding orthogonal ancilla states. We have  to settle for a limited scope, usually one in which error operators only up to a certain weight can be distinguished, and one in which the error correction only works for a subspace of the qubit's Hilbert space (referred to as the \emph{code space}). In this endeavor, an important strategy for realizing an error-correcting code is the \emph{redundant encoding} of the qubit state in multiple physical qubits through a successive application of \textsc{cnot} gates, 
\begin{equation}\label{eq:snba}
  \left(a\ket{0} + b\ket{1}\right)\ket{0 0 \cdots
    0 }  \,\longrightarrow\, a \ket{0 0 0 \cdots
    0 } + b \ket{1 1 1 \cdots 1 }.
\end{equation}
For example, to correct phase-flip errors of a single qubit without needing to know or restrict the state of the qubit, redundant encoding of the qubit state in three qubits is required; this is the so-called \emph{three-bit code} for phase errors (see, e.g., Sec.~7.4.5 of Ref.~\cite{Schlosshauer:2007:un} for details). Incidentally, the three-bit code was the basis of the first experimental demonstration of quantum error correction, reported in 1998 \cite{Cory:1998:uu}.

A second challenge in realizing quantum error correction, especially for larger numbers of qubits, is the implementation of the final countertransformations restoring the original quantum state. A third challenge comes from the fact that adding an ancilla increases the effective size of the system, thereby making it potentially more prone to decoherence (recall that the rate of decoherence typically scales exponentially with system size). 

Undoubtedly, quantum error correction will be an integral part of any viable quantum computer. It may be combined (or \emph{concatenated} \cite{Lidar:1999:fa}) with decoherence-free subspaces to achieve universal fault-tolerant quantum computation \cite{Bacon:2000:yy,Lidar:2001:oo}.

\section{\label{sec:exper-observ-decoh}Experimental studies of decoherence}

Decoherence happens all around us, and in this sense its consequences are readily observed. But what we would like to be able to do is experimentally study the \emph{gradual} and \emph{controlled} action of decoherence, preferably of superpositions of mesoscopically or macroscopically distinguishable states. Such experiments have many important applications and implications. They demonstrate the possibility of generating nonclassical superposition quantum states for mesoscopic and macroscopic systems, and they show that the quantum--classical boundary can be shifted by changing the relevant experimental parameters. They are useful for assessing the predictions of decoherence models, and for designing quantum devices---for example, those needed for quantum information processing---that are good at evading the detrimental influence of the environment. They can also be used to search for deviations from standard unitary quantum mechanics.

Realizing an experiment capable of measuring the progressive decoherence of a quantum state requires meeting several challenges. The first is the task of preparing a suitable quantum superposition state (see Ref.~\cite{Arndt:2014:oo} for an overview of the state of the art in generating superpostions of mesoscopically and macroscopically distinguishable states). Second, decoherence of this superposition must be sufficiently slow for its gradual action to be observed.  Third, one must be able to measure the decoherence introduced over time without imparting a significant amount of additional, unwanted decoherence. Ideally, one would also like to have sufficient control over the environment, so that one can tune the strength and form of its interaction with the system.

In what follows, we shall focus on four experimental areas that have played a key role in experimental studies of decoherence: atom--photon interactions in a cavity (Sec.~\ref{sec:atoms-cavity}), interferometry with mesoscopic molecules (Sec.~\ref{sec:matt-wave-interf}), superconducting systems such as SQUIDs and Cooper-pair boxes (Sec.~\ref{sec:superc-syst}), and ion traps (Sec.~\ref{sec:trapped}). Section~\ref{sec:other} briefly lists a few other experimental areas in which decoherence has been studied. Finally, in Sec.~\ref{sec:exper-tests-quant}, we comment on the use of experimental investigations of decoherence for testing quantum mechanics.

\subsection{\label{sec:atoms-cavity}Photon states in a cavity}

In decoherence experiments of the cavity-QED type \cite{Raimond:2001:aa,Haroche:2006:hh}, an atom interacts with a radiation field in a cavity in such a way that information about the atomic state is imprinted on the state of the field, resulting in an entangled atom--field state. Atom and field are then disentangled through a measurement of the atomic state, resulting in a superposition of two coherent field states whose decoherence is monitored over time. In 1996 Brune et al.\ generated a superposition of radiation fields with classically distinguishable phases involving several photons and observed the controlled decoherence of this state \cite{Brune:1996:om,Raimond:2001:aa,Kaiser:2001:tm} (see Refs.~\cite{Raimond:2001:aa,Haroche:2006:hh} for reviews). 

\begin{figure}
\centering
\includegraphics[scale=0.9]{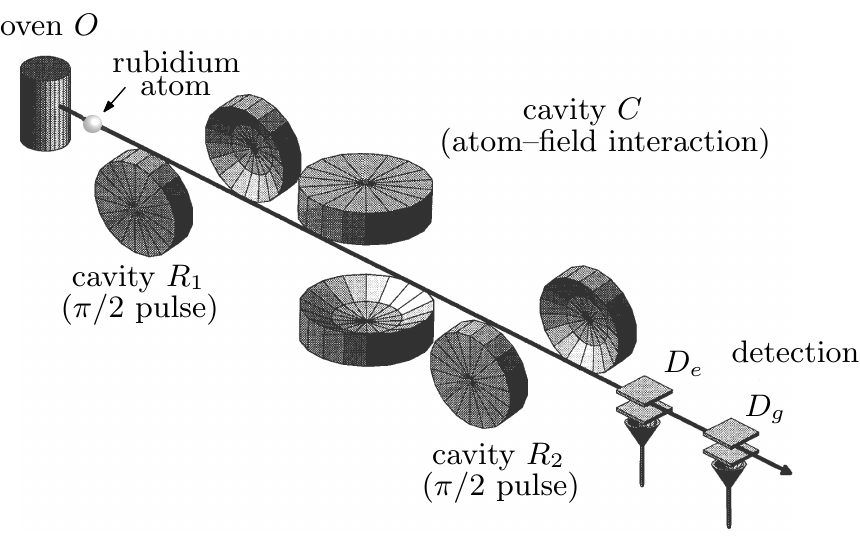}
\caption{Schematic illustration of the cavity experiment by Brune et al.\
  \cite{Brune:1996:om}. A rubidium atom emitted from an oven $O$ passes through a cavity $R_1$ that uses a microwave $\pi/2$ pulse to prepare the atom in a superposition of two circular Rydberg energy eigenstates $\ket{g}$ and $\ket{e}$. The atom then enters a cavity $C$ containing a photon field described by a coherent state $\ket{\alpha}$. The field state suffers a dispersive phase shift $\chi$ that depends on the energy state of the atom, leading to the creation of an entangled atom--field state of the form $\frac{1}{\sqrt{2}} \left( \ket{g} \ket{\alpha \E^{-\I \chi} } + \ket{e} \ket{\alpha \E^{\I \chi}} \right)$. Subsequently, another microwave $\pi/2$ pulse is applied in cavity $R_2$ to further transform the atomic state, and finally the atom's energy is measured by ionization chambers $D_e$ and $D_g$. This disentangles the atom from the field, leaving the latter in a superposition $\ket{\pm}=\frac{1}{\sqrt{2}} \left( \ket{\alpha \E^{\I \chi}} \pm \ket{\alpha \E^{-\I \chi}} \right)$ of two coherent field states with distinguishable phases.  Figure adapted with permission from Ref.~\cite{Davidovich:1996:sa}.}
\label{fig:brune-setup}
\end{figure}

In this experiment (see Fig.~\ref{fig:brune-setup}), a rubidium atom is prepared in a superposition of energy eigenstates $\ket{g}$ (the ``lower'' state) and $\ket{e}$ (the ``upper'' state) corresponding to two circular Rydberg states. This preparation is done by applying a $\pi/2$ pulse in a microwave   cavity $R_1$, at a frequency $\nu$ that is very close to the resonant frequency $\nu_{ge}$ of the atomic transition between the two Rydberg levels $g$ and $e$. The atom enters a cavity $C$ made of highly reflecting superconducting mirrors, with a long damping time $T_r$; more recent experiments have realized cavities with damping times in excess of a tenth of a second \cite{Kuhr:2007:aa,Deleglise:2008:oo}. The cavity contains a radiation field that consists of a few photons and is described by a coherent state $\ket{\alpha}$, 
\begin{equation}
\ket{\alpha} = \E^{- \abs{\alpha}^2 /2}\sum_{n=0}^\infty \frac{\alpha^n}{\sqrt{n!}}\ket{n},
\end{equation}
where $\alpha$ is a complex number. This state may be visualized as a vector in phase space whose squared length $\abs{\alpha}^2$ is equal to the  mean number $\bar{n}$ of photons. 

The interaction between the atom and field inside the cavity is tuned such that there is no energy transfer. The atom effectively acts as a transparent dielectric for the field, by imposing a state-dependent dispersive phase shift on the field. If the atom is in the state $\ket{e}$, then the coherent field state $\ket{\alpha}$ experiences a phase shift $\chi$ such that $\ket{\alpha}$ is transformed to $\ket{\E^{i\chi} \alpha}$; if the atom enters in the state $\ket{g}$ instead, the phase shift is $-\chi$ and $\ket{\alpha}$ is transformed to $\ket{\E^{-i\chi} \alpha}$. Thus, for an atom prepared in a coherent superposition $\frac{1}{\sqrt{2}}\left( \ket{g} + \ket{e} \right)$ of $\ket{g}$ and $\ket{e}$, the entangling evolution is
\begin{equation}\label{eq:cat-brune}
  \frac{1}{\sqrt{2}}\left( \ket{g} + \ket{e} \right)  \ket{\alpha} \,\longrightarrow\,
  \frac{1}{\sqrt{2}} \left( \ket{g} \ket{\alpha \E^{-\I \chi}} + \ket{e} \ket{\alpha
      \E^{\I \chi}} \right).
\end{equation}
The atom then passes through an additional microwave cavity $R_2$, which applies another $\pi/2$ pulse at the same frequency $\nu$ as cavity $R_1$. The pulse transforms the atomic states according to $\ket{g} \,\longrightarrow\,\frac{1}{\sqrt{2}} \left( \ket{g} - \ket{e} \right)$ and $\ket{e} \,\longrightarrow\, \frac{1}{\sqrt{2}} \left( \ket{g} + \ket{e} \right)$, resulting in the combined atom--field state
\begin{align}\label{eq:cat-brune4}
\ket{\Psi_\text{atom+field}} &=
\frac{1}{2} \left( \ket{g} \ket{\alpha \E^{-\I \chi}} - \ket{e} \ket{\alpha \E^{-\I \chi}} +
  \ket{g} \ket{\alpha \E^{\I \chi}} + \ket{e} \ket{\alpha \E^{\I \chi}} \right) \notag \\
&= \frac{1}{2} \left( \ket{\alpha \E^{-\I \chi}} + \ket{\alpha \E^{\I \chi}} \right) \ket{g} +
\frac{1}{2} \left(- \ket{\alpha \E^{-\I \chi}} + \ket{\alpha \E^{\I \chi}} \right) \ket{e}.
\end{align}
Finally, the energy of the atom is measured, collapsing the state \eqref{eq:cat-brune4} onto either one of the energy eigenstates $\ket{g}$ and $\ket{e}$. This measurement destroys the entanglement between atom and photon field, and the field is left in a superposition of the coherent-field states $\ket{\alpha \E^{\I \chi}}$ and $\ket{\alpha \E^{-\I \chi}}$ whose relative phase depends on the outcome of the measurement. If the outcome is the ground state $g$, the field state will be the ``even'' state
\begin{equation}\label{eq:cat-bhvlhvlrune}
\ket{+} = \frac{1}{\sqrt{2}} \left( \ket{\alpha \E^{\I \chi}} + \ket{\alpha \E^{-\I \chi}} \right).
\end{equation}
If the outcome is the excited state $e$, the field state will be the ``odd'' state
\begin{equation}\label{eq:cat-baahvlhvlrune}
\ket{-} = \frac{1}{\sqrt{2}} \left( \ket{\alpha \E^{\I \chi}} - \ket{\alpha \E^{-\I \chi}} \right).
\end{equation}
(The terminology ``even'' and ``odd'' is motivated by the observation that for a phase shift $\chi$ equal to $\pi/2$, these states contain, respectively, only even and odd photon numbers \cite{Deleglise:2008:oo}.) 

To quantify the ``catness'' of the superpositions $\ket{\pm} =\frac{1}{\sqrt{2}} \left(\ket{\alpha \E^{\I \chi}}\pm \ket{\alpha \E^{-\I \chi}} \right)$---i.e., to measure the degree to which the components $\ket{\alpha \E^{\I \chi}}$ and $\ket{\alpha \E^{-\I \chi}}$ represent mesoscopically or macroscopically distinguishable states---we consider the squared magnitude of the overlap between $\ket{\alpha \E^{\I \chi}}$ and $\ket{\alpha \E^{-\I \chi}}$, $\abs{\braket{\alpha \E^{\I \chi}}{\alpha \E^{-\I \chi}}}^2$. For two general coherent states $\ket{\alpha}$ and $\ket{\beta}$, we have
\begin{equation}\label{eq:hsvg}
\abs{\braket{\alpha}{\beta}}^2 = \E^{- \abs{\alpha-\beta}^2},
\end{equation}
and therefore
\begin{equation}\label{eq:hsvg11}
\abs{\braket{\alpha \E^{\I \chi}}{\alpha \E^{-\I \chi}}}^2 = \E^{- 2\abs{\alpha}^2(1-\cos2\chi)} =  \E^{- 4\abs{\alpha}^2\sin^2\chi}.
\end{equation}
We may therefore quantify the ``catness'' (or ``size'' of the cat-like state) by introducing a parameter $D^2$ equal to the argument of the exponential in Eq.~\eqref{eq:hsvg11} \cite{Brune:1996:om},
\begin{equation}\label{eq:hsADvg11}
D^2 = 4 \bar{n}\sin^2\chi,
\end{equation}
where we have used that $\bar{n}=\abs{\alpha}^2$. This also represents the squared distance between the (direct) peaks of the density matrices representing the superpositions $\ket{\pm} =\frac{1}{\sqrt{2}} \left(\ket{\alpha \E^{\I \chi}}\pm \ket{\alpha \E^{-\I \chi}} \right)$ of coherent states \cite{Deleglise:2008:oo}. We see that the overlap depends both on the mean photon number $\bar{n}=\abs{\alpha}^2$ and on the phase difference $\chi$, and decreases exponentially with $\bar{n}$. This makes sense: the ``catness'' increases if the size of the system and the phase difference are increased. For fixed mean photon number $\bar{n}$, Eq.~\eqref{eq:hsvg11} shows that the minimum overlap is obtained for $\chi=\pi/2$ (i.e., when the vectors representing the two coherent states point in opposite directions). 

Owing to the properties of the Rydberg atoms, the experiment by Brune et al.\ \cite{Brune:1996:om} achieved relatively large phase shifts of up to $\chi = 0.31\pi$, with mean photon number $\bar{n} \approx 10$. For these values, the overlap $\abs{\braket{\alpha \E^{\I \chi}}{\alpha \E^{-\I \chi}}}= \E^{- 2\abs{\alpha}^2\sin^2\chi}$ [see Eq.~\eqref{eq:hsvg11}] is less than $3 \times 10^{-5}$. Thus the states $\ket{\alpha \E^{\I \chi}}$ and $\ket{\alpha \E^{-\I \chi}}$ are very nearly orthogonal, and therefore the field effectively acts as a meter that encodes which-state information about the energy eigenstates $\ket{g}$ and $\ket{e}$ in the mesoscopically distinct states $\ket{\alpha \E^{\I \chi}}$ and $\ket{\alpha \E^{-\I \chi}}$. A subsequent experiment has realized mesoscopic superposition states involving $\bar{n}=29$ photons \cite{Auffeves:2003:za}. By coupling a superconducting qubit to a waveguide cavity resonator, a superposition of coherent states involving 111 photons has been generated \cite{Vlastakis:2013:pp} (see also Ref.~\cite{Hermann-Avigliano:2015:tt}). 

The experiment by Brune et al.\ \cite{Brune:1996:om,Maitre:1997:tv} then measured the progressive decoherence of the field superposition $\ket{\pm}=\frac{1}{\sqrt{2}} \left( \ket{\alpha \E^{\I \chi}} \pm \ket{\alpha \E^{-\I \chi}} \right)$ [see Eqs.~\eqref{eq:cat-bhvlhvlrune} and \eqref{eq:cat-bhvlhvlrune}] that is left behind in the cavity $C$ after the passage and detection of the atom. This measurement was accomplished by sending a second rubidium atom through the apparatus. One can show \cite{Davidovich:1996:sa, Maitre:1997:tv,Raimond:2001:aa,Kaiser:2001:tm,Schlosshauer:2007:un} that upon detection, this second atom will be found in the same state ($g$ or $e$) as the first atom provided the superposition has not been decohered.\footnote{For such a perfect correlation to obtain, we also need to require that the state components $\ket{\alpha \E^{\I \chi}}$ and $\ket{\alpha \E^{-\I \chi}}$ be orthogonal.  This holds to a good approximation even for the modest photon numbers used by Brune et al.; see the discussion following Eq.~\eqref{eq:hsADvg11}.} Thus, the conditional detection probability $P_{ee}$ for finding both the first and second atoms in the state $e$ after passage through the apparatus will be equal to one. If, however, the field state has started to decohere before the second atom has passed through the cavity $C$, $P_{ee}$ will decrease, approaching a value of $\frac{1}{2}$ in the limit of complete decoherence. The longer one waits before sending the second atom through $C$, the more the field state will have decohered. Thus, by adjusting the wait time $\tau$ between sending the first and second atoms through the apparatus and recording $P_{ee}$ as a function of $\tau$, the gradual decoherence of the field state can be measured \cite{Davidovich:1996:sa, Maitre:1997:tv,Raimond:2001:aa}. 

\begin{figure}
\centering
\includegraphics[scale=.7]{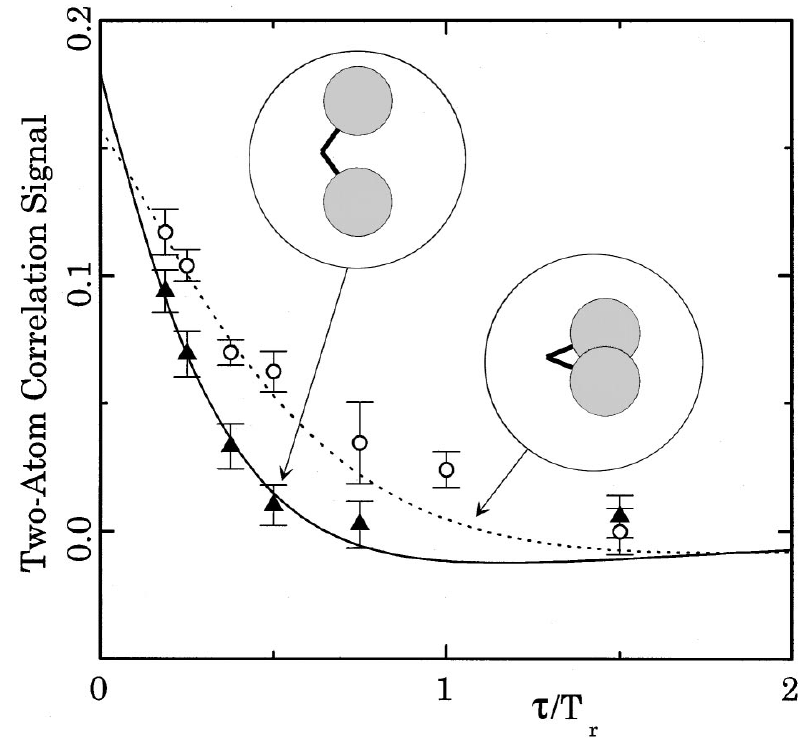}
\caption{Progressive decoherence of a superposition of two coherent field states as a function of the wait time $\tau$ (expressed in units of the cavity damping time $T_r$), as reported in Refs.~\cite{Brune:1996:om,Maitre:1997:tv}. The two-atom correlation signal measures the coherence of the superposition; a value of zero corresponds to a complete loss of decoherence. Decoherence of superpositions of the coherent states $\ket{\alpha \E^{\I \chi}}$ and $\ket{\alpha \E^{-\I \chi}}$ for two different values of the phase shift $\chi$ is shown: $\chi=0.13\pi$ (circles) and $\chi=0.31\pi$ (triangles). The phase difference between the coherent-state components is also visualized in the insets, showing the vector representations of the components. The mean photon number is $\bar{n}=3.3$. Solid lines are theoretical predictions \cite{Davidovich:1996:sa,Maitre:1997:tv}.  Figure reproduced with permission from Ref.~\cite{Raimond:2001:aa}.}
\label{fig:twoatom}
\end{figure}

The first experimental realizations \cite{Brune:1996:om,Maitre:1997:tv} used the observed data for $P_{ee}$ and $P_{eg}$ (the probability of finding the second atom in the state $e$ if the first atom was found in $g$) to define the two-atom correlation function $\eta(\tau)=P_{ee}(\tau)-P_{eg}(\tau)$. In the absence of decoherence and therefore perfect two-atom correlations, we have $P_{ee}=1$ and $P_{eg}=\frac{1}{2}$, and thus $\eta=\frac{1}{2}$; in the case of complete decoherence, the correlation is lost and we have $P_{ee}=P_{eg}=\frac{1}{2}$, and thus $\eta=0$. Figure~\ref{fig:twoatom} shows the experimental results for two different phase shifts $\chi=0.13\pi$ and $\chi=0.31\pi$. Good agreement is found with theoretical predictions obtained from a simple model \cite{Davidovich:1996:sa,Maitre:1997:tv}. It is clearly seen that, as expected, decoherence becomes more rapid as the phase shift $\chi$ between the coherent-state components $\ket{\alpha \E^{\I \chi}}$ and $\ket{\alpha \E^{-\I \chi}}$---and thus their distinguishability---is increased. 

\begin{figure}
\centering
\includegraphics[scale=0.9]{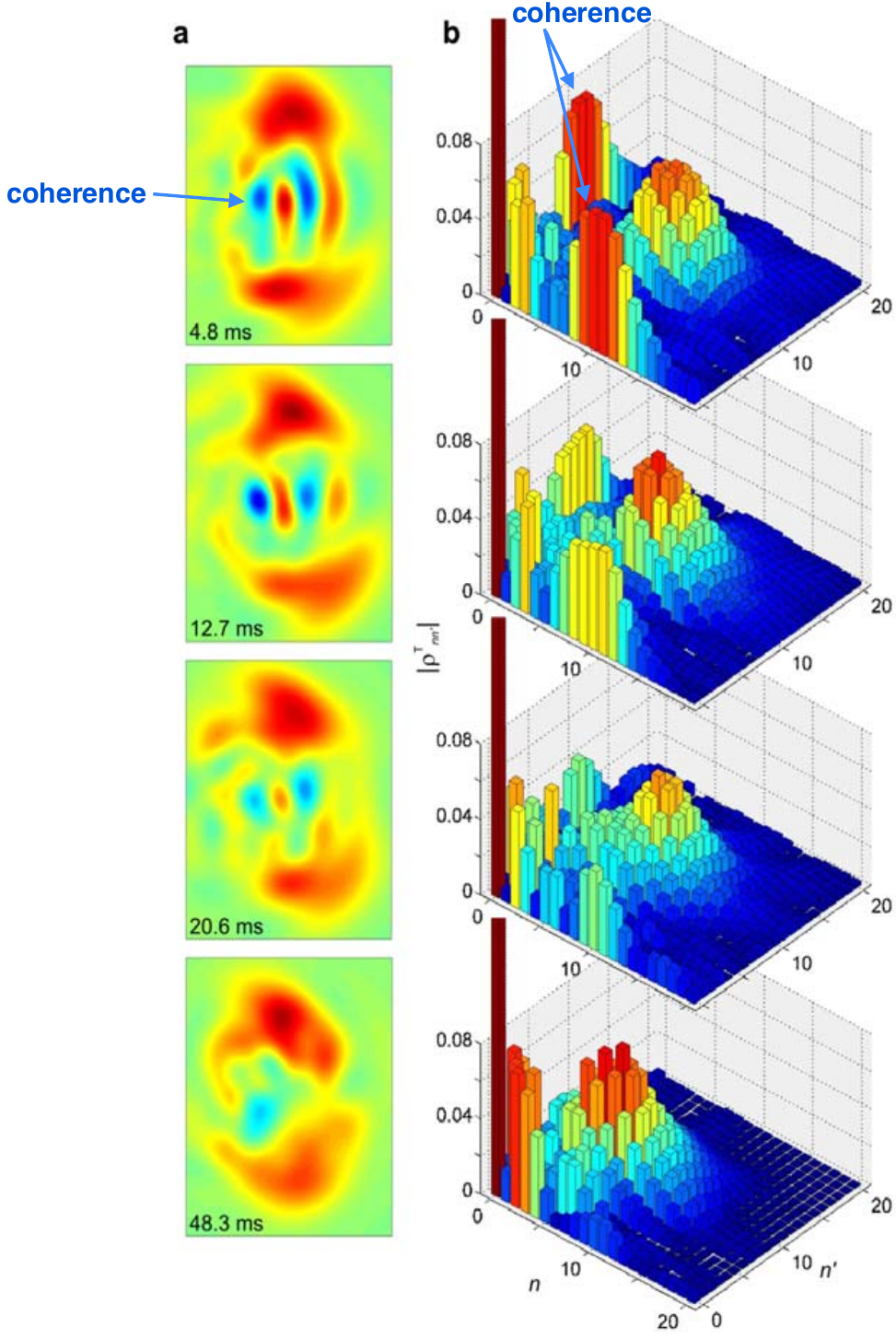} \,\,\, \includegraphics[scale=0.18]{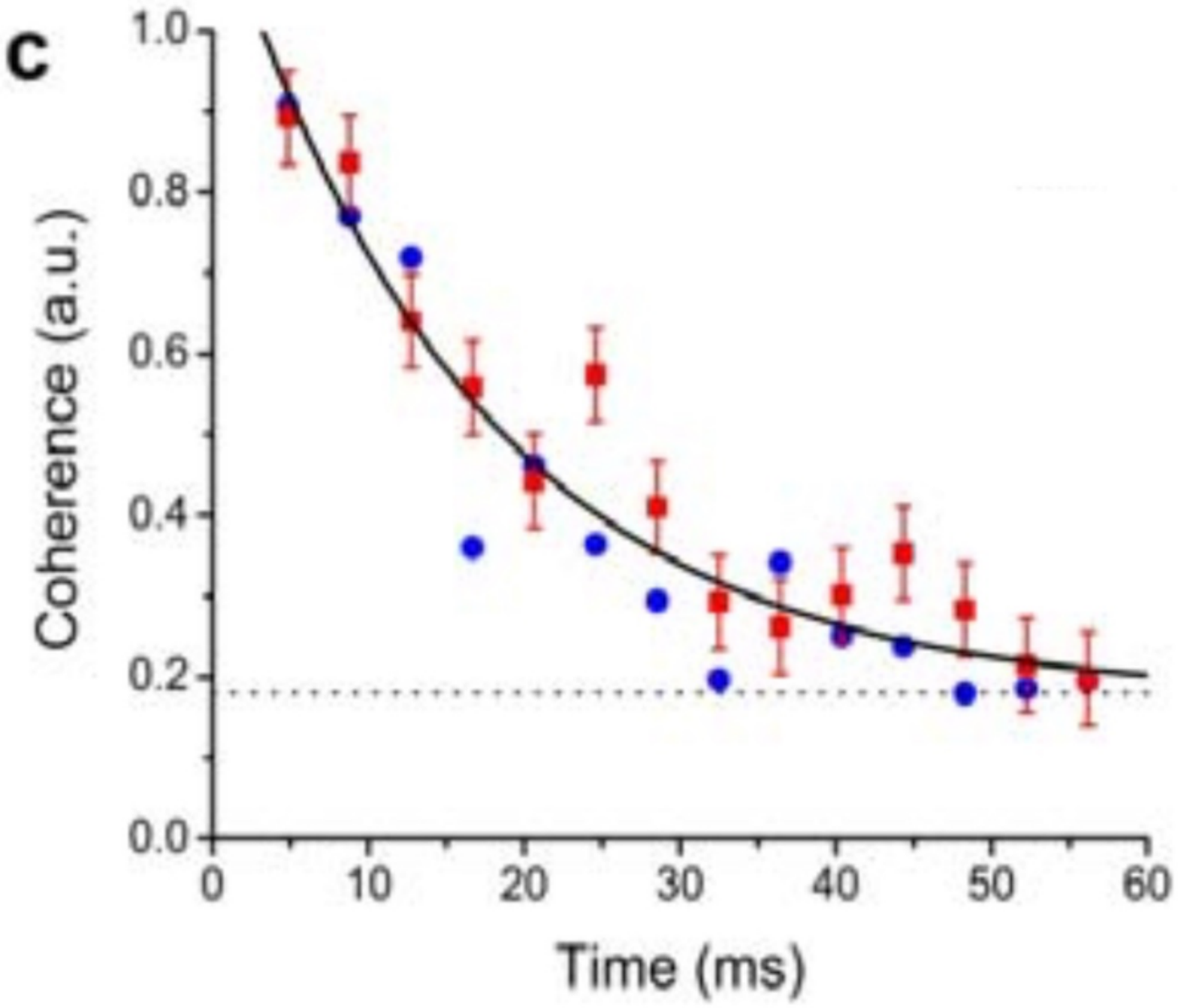} 
\caption{Experimental observation of the decoherence of the superposition $\ket{-} = \frac{1}{\sqrt{2}} \left( \ket{\alpha \E^{\I \chi}} - \ket{\alpha \E^{-\I \chi}} \right)$ [see Eq.~\eqref{eq:cat-baahvlhvlrune}] of two coherent photon fields, as reported by Del{\'e}glise et al.\ \cite{Deleglise:2008:oo}. The mean photon number is $\bar{n}=3.5$ and the phase difference is $\chi=0.37\pi$. \emph{(a)} Reconstructed wave functions as a function of the time elapsed since state preparation, shown in the Wigner representation projected onto the 2D plane \cite{Wigner:1932:un,Hillery:1984:tv} (see Ref.~\cite{Deleglise:2008:oo} for details on the state reconstruction procedure). Coherence between the state components is represented by oscillations in the indicated region between the top and bottom peaks. This coherence is seen to disappear as time progresses. \emph{(b)} Corresponding density matrices. Coherence is now represented by the off-diagonal terms in the first row and column. \emph{(c)} Measurement of coherence as a function of time, obtained from an analysis of the density matrices. The solid line represents an exponential fit with an offset to account for a noise background.  Figure adapted with permission from Ref.~\cite{Deleglise:2008:oo}.}
\label{fig:qeddecoh}
\end{figure}

In a subsequent experiment reported by Del{\'e}glise et al.\ \cite{Deleglise:2008:oo}, the field states inside the cavity were reconstructed at different stages of their gradual decoherence. Thus, the effect of the decoherence process on the quantum state could be visualized explicitly (see Fig.~\ref{fig:qeddecoh}); the authors even generated a movie of the decoherence process (available as supplementary information for Ref.~\cite{Deleglise:2008:oo}).  In the experiment, the mean photon number was $\bar{n}=3.5$ and the phase difference was $\chi=0.37\pi$. A simple model of decoherence \cite{Walls:1985:pp,Brune:1992:zz,Haroche:2006:hh} predicts a decoherence timescale on the order of $T_d=2T_r/D^2$, where $T_r$ is the damping time of the optical cavity. In the experiment by Del{\'e}glise et al.\ \cite{Deleglise:2008:oo}, this damping time was $T_r=\unit[0.13]{s}$, leading to a predicted decoherence time of $T_d=\unit[19.5]{ms}$ when adjusted for thermal background, which is in good agreement with the measured value $T_d=\unit[(17 \pm 3)]{ms}$. 

\begin{figure}
\centering
\includegraphics[scale=.75]{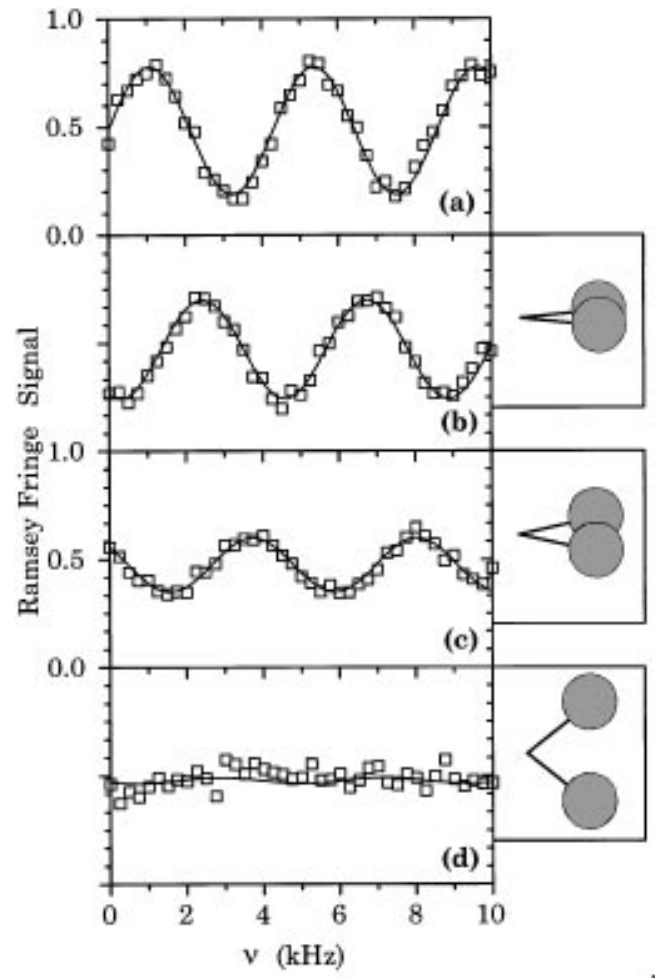}
\caption{Ramsey interference fringes measured in the experiment by Brune et al.\
  \cite{Brune:1996:om} for the probability of detecting the atom in the ground state $g$ after passage through the cavities $R_1$, $C$, and $R_2$ [see Fig.~\ref{fig:brune-setup}], shown as a function of the microwave frequency $\nu$ in $R_1$ and $R_2$. Solid lines represent sinusoidal fits. \emph{(a)} Fringes observed in the absence of a photon field in cavity $C$. The interference arises from the indistinguishability of the two quantum ``paths'' representing the atomic transition $g\rightarrow e$ occurring in either $R_1$ or $R_2$. \emph{(b)--(d)} When a photon field (mean photon number $\bar{n}=9.5$) is present in $C$, it obtains information about the two quantum paths, making them progressively distinguishable as the phase difference $\chi$ (and therefore the distinguishability) between the coherent states $\ket{\alpha \E^{\I \chi}}$ and $\ket{\alpha \E^{-\I \chi}}$ is increased (visualized by the insets on the right). This leads to a corresponding decrease in fringe visibility. The phase shifts are \emph{(a)} $\chi=0.03\pi$, \emph{(b)} $\chi=0.06\pi$, and \emph{(c)} $\chi=0.21\pi$.  Figure reproduced with permission from Ref.~\cite{Brune:1996:om}.}
\label{fig:ramsey}
\end{figure}

The experiment by Brune et al.\ \cite{Brune:1996:om} was also used to observe the decoherence of the \emph{atomic} state due to the photon field. The combination of the two microwave cavities $R_1$ and $R_2$ effectively forms a (spatially separated) Ramsey interferometer \cite{Ramsey:1950:pp,Haroche:2006:hh}. One may then measure the detection probability $P_g(\nu)$ of finding an atom, after passage through the apparatus, in the ground state $g$ as a function of the frequency $\nu$ in the cavities $R_1$ and $R_2$ (see Fig.~\ref{fig:ramsey}). One finds that, in the absence of a photon field in cavity $C$, $P_g(\nu)$ displays an oscillatory (fringe) pattern. This pattern arises because either cavity $R_1$ or $R_2$ may trigger a transition from state $g$ to state $e$, and it is impossible to distinguish in which cavity a given transition occurred. The indistinguishability of these two ``paths'' (transition in $R_1$ or in $R_2$) implies interference between the paths, which manifests itself in an interference pattern for the detection probability at the output. However, in the presence of a photon field in cavity $C$, the field will obtain information about the atomic state [compare Eq.~\eqref{eq:cat-brune}]. This which-path information diminishes the visibility of the interference pattern, with the visibility quantified by the overlap between $\ket{\alpha \E^{\I \chi}}$ and $\ket{\alpha \E^{-\I \chi}}$ as given by Eq.~\eqref{eq:hsvg11}. This is again an instance of decoherence, but it is now the atomic state that decoheres due to the environment provided by the photon field.

\subsection{\label{sec:matt-wave-interf}Matter-wave interferometry}

\begin{figure}
\centering
\includegraphics[scale=1]{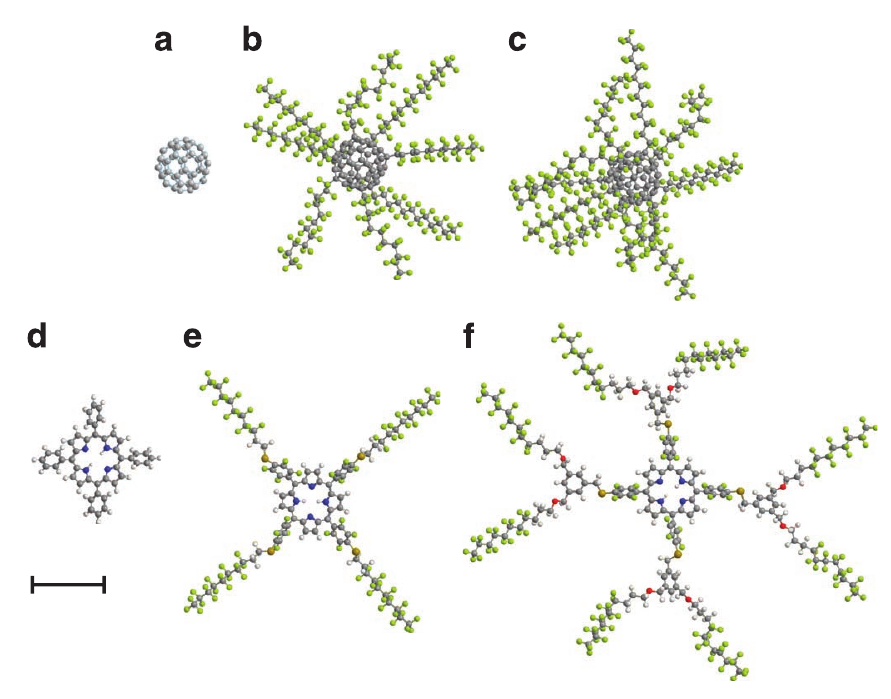}
\caption{Examples of molecular clusters used in matter-wave interferometry experiments, drawn to scale (the scale bar represents \unit[10]{\AA}). \emph{(a)} Fullerene C$_{60}$ ($m = \unit[720]{amu}$, 60 atoms). \emph{(b)} Perfluoroalkylated nanosphere PFNS8 ($m = \unit[5672]{amu}$, 356 atoms). \emph{(c)} PFNS10 ($m = \unit[6910]{amu}$, 430 atoms). \emph{(d)} Tetraphenylporphyrin TPP ($m = \unit[614]{amu}$, 78 atoms). \emph{(e)} TPPF84 ($m = \unit[2814]{amu}$, 202 atoms). \emph{(f)} TPPF152 ($m = \unit[5310]{amu}$, 430 atoms). Figure reproduced with permission from Ref.~\cite{Gerlich:2011:aa}. }
\label{fig:molecules}
\end{figure}

\begin{figure}
\centering
\includegraphics[scale=1]{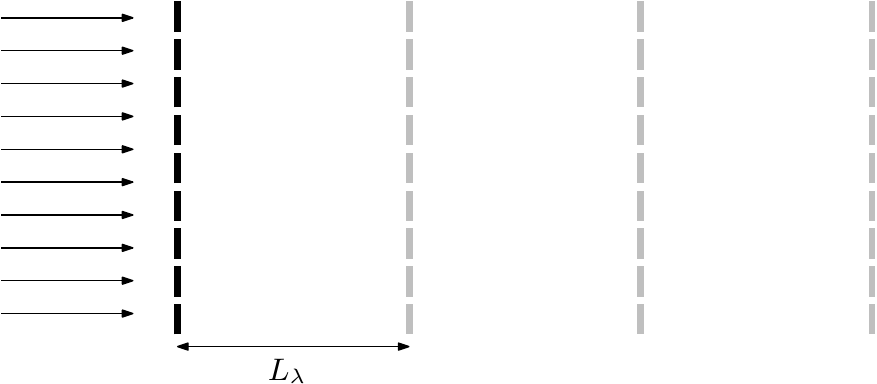}
\caption{Schematic illustration of the Talbot effect. When a plane wave of wavelength $\lambda$ is incident from the left on a diffraction grating, a repeating image of the grating is generated at distances equal to multiples of the Talbot length $L_\lambda = d^2/\lambda$ behind the grating, where $d$ is the spacing of the slits in the diffraction grating.}
\label{fig:tbeff}
\end{figure}

In matter-wave interferometry experiments with molecules (see Ref.~\cite{Hornberger:2012:ii} for a review), spatial interference fringes are demonstrated for mesoscopic molecules ranging from C$_{60}$ and C$_{70}$ fullerenes \cite{Arndt:1999:rc} to large molecular clusters (Fig.~\ref{fig:molecules}) \cite{Gerlich:2011:aa,Eibenberger:2013:az}. Because the de~Broglie wavelength of such molecules is on the order of picometers, one cannot use ordinary double-slit interferometry as one would do for photons or particles such as electrons \cite{Jonsson:1961:rz,Jonsson:1974:rz,Tonomura:1989:as}. Instead, the experiments are based on the Talbot effect familiar from classical optics \cite{Mansuripur:2009:zz}, a genuine interference phenomenon in which a plane wave incident on a diffraction grating creates images of the grating at multiples of the Talbot length $L_\lambda = d^2/\lambda$ behind the grating, where $d$ is the slit spacing and $\lambda$ is the wavelength of the incident wave (see Fig.~\ref{fig:tbeff}) \cite{Mansuripur:2009:zz,Hackermuller:2003:uu}. 

\begin{figure}
\centering
\includegraphics[scale=0.75]{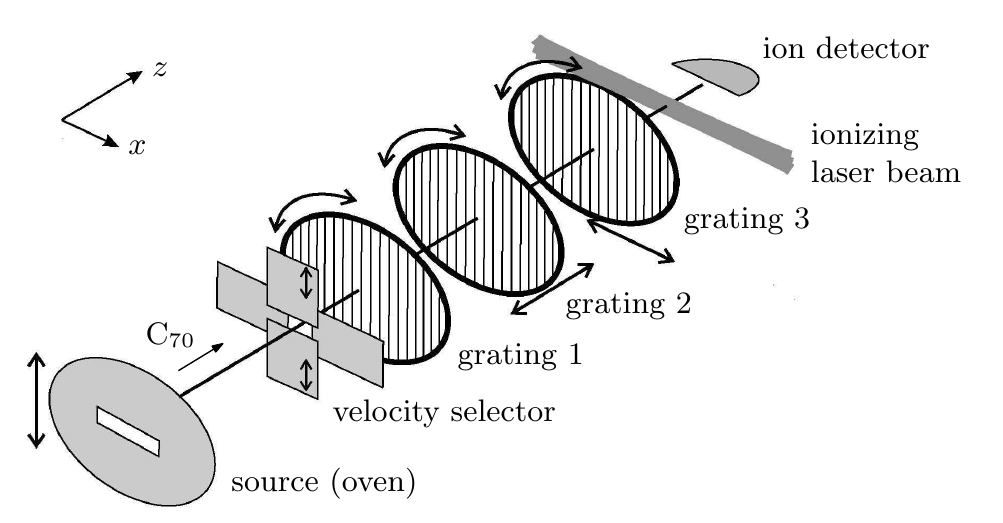}
\caption{Schematic illustration of the Talbot--Lau interferometer used in Refs.~\cite{Arndt:1999:rc,Brezger:2002:mu,Hornberger:2003:tv,Hackermueller:2002:wb,Hackermuller:2004:rd} for demonstrating interference patterns for C$_{60}$ and C$_{70}$ fullerene molecules, and for studying their decoherence. Molecules emitted from a source are velocity-selected and pass through the first grating to produce sufficient beam coherence. The second grating is a diffraction grating implementing the Talbot effect (see Fig.~\ref{fig:tbeff}). The third grating acts as a scanning mask for the molecular density pattern subsequently recorded by ionizing and detecting the molecules.  Figure adapted with permission from Ref.~\cite{Brezger:2002:mu}.}
\label{fig:c70-setup}
\end{figure}

In the matter-wave interferometry experiments with C$_{60}$ and C$_{70}$  molecules \cite{Brezger:2002:mu,Hornberger:2003:tv,Hackermueller:2002:wb,Hackermuller:2003:uu,Hackermuller:2004:rd}, one does not, however, work with an incident plane (i.e., coherent) wave as required for the Talbot effect, but rather with an uncollimated, incoherent molecular beam to allow for sufficiently high intensity. To accommodate such beams, the experiments make use of the three-grating setup shown in Fig.~\ref{fig:c70-setup}, realizing a so-called  Talbot--Lau interferometer \cite{Brezger:2002:mu,Hornberger:2003:tv,Hackermuller:2003:uu,Hackermueller:2002:wb,Hackermuller:2004:rd}. The first grating is used to produce sufficient transverse coherence of the molecular beam (on the order of 2--3 grating periods) at the location of the second grating, which plays the role of the diffraction grating in the Talbot effect (see again Fig.~\ref{fig:tbeff}) and is placed at the Talbot length $L_\lambda$ behind the first grating. (For C$_{70}$ molecules, the wavelength is a few picometer, and given the experimental slit spacing $d$ of about one micrometer, one finds a macroscopic Talbot length on the order of one meter; the exact Talbot length, and thus grating separation, in the fullerene experiments of Refs.~\cite{Brezger:2002:mu,Hornberger:2003:tv,Hackermueller:2002:wb,Hackermuller:2003:uu,Hackermuller:2004:rd} is $L_\lambda = \unit[38]{cm}$.) 

If there is indeed coherence between the different possible paths through the second grating, then the Talbot image of the diffraction grating will manifest itself as an oscillatory variation of the transverse molecular density at multiples of the Talbot length. To image this density pattern, a third grating, placed at a distance equal to the Talbot length behind the second grating, is scanned across the pattern (in the $x$ direction shown in Fig.~\ref{fig:c70-setup}), thus serving as a detection mask. The molecules that have passed through all three gratings are then ionized by a laser beam and detected. A sinusoidal variation of the number of detected molecules as a function of the position of the third grating indicates the presence of interference fringes. These fringes confirm the delocalization of the spatial wavefunction of the molecule due to the presence of the diffraction grating. Since the grating period is roughly one micrometer, the fringes therefore announce quantum coherence between two spatial locations one micrometer apart.\footnote{Such fringes could, in principle, also be explained in terms of the classical Moir\'e effect, which is a consequence of the blocking of rays by the grating. This effect, however, is independent of the velocity (and thus de Broglie wavelength) of the molecules. Therefore, a variation of the fringe visibility with molecular velocity indicates the presence of quantum interference arising from the Talbot effect. This variation was indeed observed in the fullerene experiment of Ref.~\cite{Brezger:2002:mu}, confirming the quantum nature of the observed interference fringes.}  

\begin{figure}
{\footnotesize \hspace{.4cm} \emph{(a)} \hspace{6.7cm} \emph{(b)} }

\vspace{.2cm}

\centering
\includegraphics[scale=0.9]{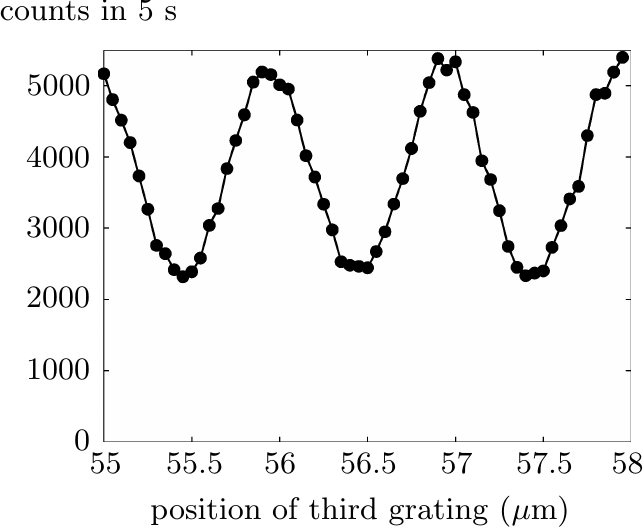} \hspace{.8cm} \includegraphics[scale=.9]{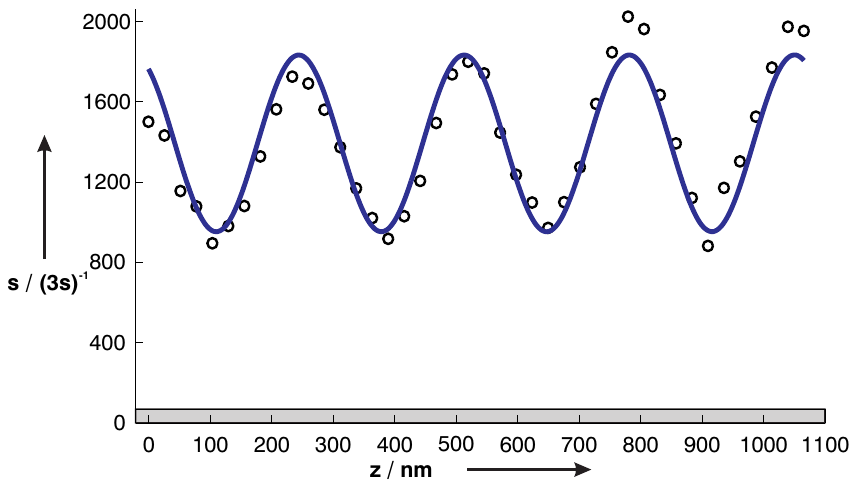}
\caption{Interference fringes observed in molecular interferometry experiments. \emph{(a)} Fringes for C$_{70}$ molecules, as reported by Brezger et al.\ \cite{Brezger:2002:mu}.  Figure adapted with permission from Ref.~\cite{Brezger:2002:mu}. \emph{(b)} Fringes for TPPF20 molecules, as observed by Eibenberger et al.\ \cite{Eibenberger:2013:az}. Measured fringe visibilities were $V=\unit[38]{\%}$ for the C$_{70}$ molecules and $V=\unit[33]{\%}$ for the TPPF20 molecules.  Solid lines are sinusoidal fits. Figure reproduced with permission from Ref.~\cite{Eibenberger:2013:az}.}
\label{fig:pattern}
\end{figure}

In an improved version \cite{Gerlich:2007:om} of this original Talbot--Lau setup, the mechanical diffraction grating is replaced by a standing laser light wave, which eliminates perturbations arising from van der Waals interactions between the grating walls and the molecules. An all-optical implementation using optical ionization gratings has also been realized \cite{Haslinger:2013:ii}. Interference fringes observed for C$_{70}$ molecules and the much larger TPPF20 molecules \cite{Eibenberger:2013:az} are shown in Fig.~\ref{fig:pattern}.
 
\begin{figure}
{\footnotesize  \hspace{.7cm} \emph{(a)} \hspace{7.3cm} \emph{(b)} }

\vspace{.2cm}

\centering
\includegraphics[scale=1]{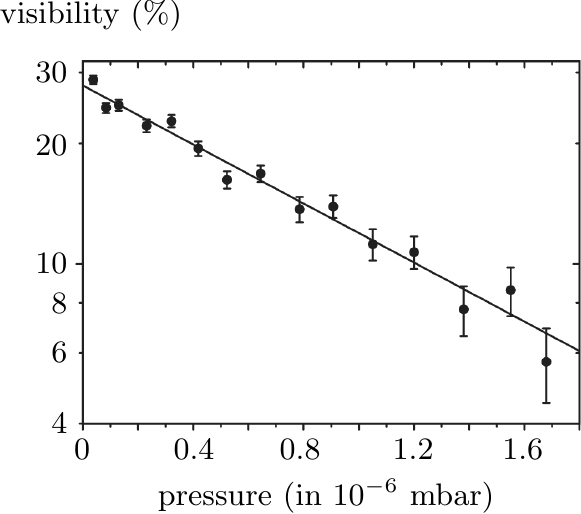} \hspace{1.5cm} \includegraphics [scale=.7]{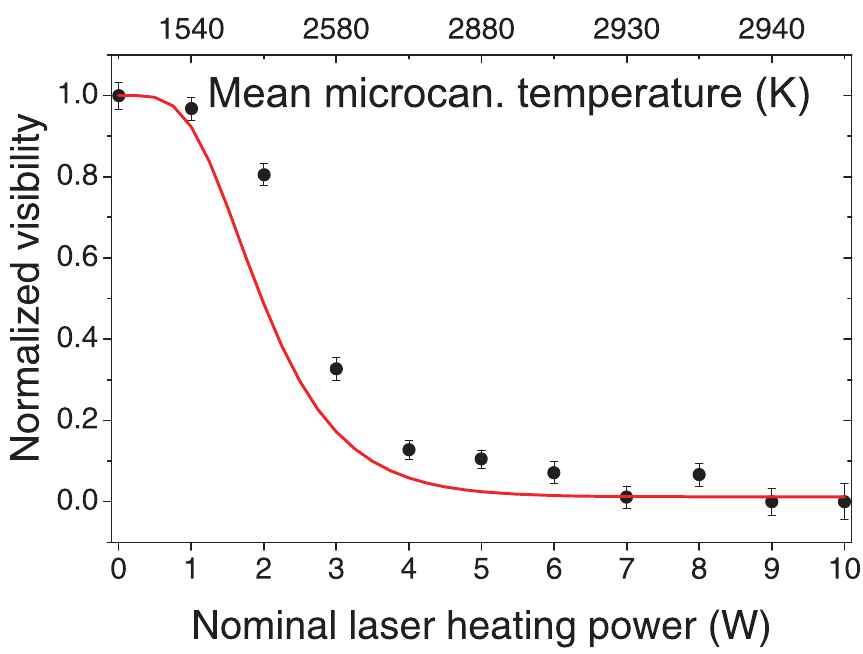}
\caption{Influence of different sources of decoherence on the visibility of interference fringes for C$_{70}$ fullerenes. \emph{(a)} Decoherence due to scattering of background gas molecules in the interferometer, as demonstrated in the experiments by Hornberger et al.\ \cite{,Hornberger:2003:tv} and Hackerm\"uller et al.\ \cite{Hackermuller:2003:uu}. The plot shows the fringe visibility as a function of the pressure of the background gas. Experimental data are represented by circles, and the theoretical prediction is shown as a solid line \cite{Hornberger:2003:un,Hornberger:2004:bb,Hornberger:2005:mo}.  Figure adapted with permission from  Ref.~\cite{Hackermuller:2003:uu}. \emph{(b)} Reduction in fringe visibility as a result of emission of thermal radiation from heated fullerenes, shown for different
  laser heating powers and corresponding mean microcanonical molecular temperatures, as demonstrated in the experiment by Hackerm\"uller et al.\
  \cite{Hackermuller:2004:rd}.  The theoretical prediction \cite{Hackermuller:2004:rd}  (see also Refs.~\cite{Hornberger:2005:mo,Hornberger:2006:tx}) is shown as solid line.  Figure reproduced with permission from Ref.~\cite{Hornberger:2012:ii}. }
\label{fig:c70-vis}
\end{figure}

Two important sources of decoherence ubiquitous in nature were studied in a controlled fashion using fullerene interferometry experiments: collisional decoherence (see Sec.~\ref{sec:collisionaldecoherence}) \cite{Hornberger:2003:tv,Hackermuller:2003:uu} and thermal decoherence \cite{Hackermuller:2004:rd}. To induce controlled collisional decoherence, the experiments by Hornberger et al.\ \cite{,Hornberger:2003:tv} and Hackerm\"uller et al.\ \cite{Hackermuller:2003:uu} introduced a background gas of adjustable pressure into the interferometer. Scattering of gas molecules by the fullerenes creates entanglement, and which-path information about the fullerene is carried away by the gas molecules. The higher the gas pressure, the greater the likelihood for a fullerene to collide with a gas particle, and thus the stronger the resulting decoherence effect. Figure~\ref{fig:c70-vis}\emph{a} shows the experimentally observed decrease of fringe visibility as a function of gas pressure, which was found to be in excellent agreement with theoretical models for collisional decoherence \cite{Hornberger:2003:un,Hornberger:2004:bb}. The experimental data were also used to confirm the predictions of more realistic collisional-decoherence models based on the quantum linear Boltzmann equation \cite{Hornberger:2006:tb,Hornberger:2006:tc,Hornberger:2008:ii,Busse:2009:aa,Vacchini:2009:pp,Busse:2010:aa,Busse:2010:oo,Hornberger:2012:ii}. Moreover, as already mentioned in Sec.~\ref{sec:collisionaldecoherence}, Talbot--Lau interferometry is sensitive to molecular rotations \cite{Gring:2010:aa,Stickler:2015:zz}, and this observation has inspired the development of collisional-decoherence models that take into account both spatial and orientational decoherence of anisotropic molecules \cite{Walter:2016:zz,Stickler:2016:yy,Papendell:2017:yy, Stickler:2018:oo,Stickler:2018:uu}; predictions derived from these models are in good agreement with experimental data \cite{Stickler:2018:oo}.

To study decoherence of fullerenes due to emission of thermal radiation, Hackerm\"uller et al.\ \cite{Hackermuller:2004:rd} used laser beams to heat the fullerene molecules to temperatures up to 3,000~K. Because the photons emitted from the heated molecules  carry away which-path information, spatial coherence---and thus fringe visibility---is reduced. By changing the molecular temperature, the strength of the resulting thermal decoherence can be adjusted. Figure~\ref{fig:c70-vis}\emph{b} shows the experimentally observed visibility of the interference fringes as a function of the laser heating power, with the corresponding (mean microcanonical) temperature of the molecular beam shown as well. We see that at temperatures below around \unit[1,500]{K} (the source temperature is \unit[900]{K}), thermal decoherence is still relatively weak and the fringe visibility is only mildly affected. Around \unit[2,000]{K}, the decoherence becomes strong enough for interference fringes to start being visibly reduced, while around \unit[2,500]{K} the visibility has been halved. Above \unit[3,000]{K}, decoherence is complete and fringes are no longer discernible. 

The observed dependence of the fringe visibility on molecular temperature is in good agreement with theoretical predictions \cite{Hackermuller:2004:rd}; see also Refs.~\cite{Hornberger:2005:mo,Hornberger:2006:tx} for a detailed theoretical analysis of the experiment. To explain the temperature dependence \cite{Hornberger:2005:mo}, one notes that only above \unit[2,000]{K} there is a non-negligible probability for a heated fullerene to emit a photon with a wavelength comparable to the extent of the spatial delocalization of the fullerene state (as given by the slit spacing $d$) such that appreciable which-path information can be obtained. Thus, this temperature marks the onset of decoherence. The emission of several photons of such wavelength is needed to cause a 50\% decrease in fringe visibility, which requires the higher temperature regime of \unit[2,500]{K}. Around \unit[3,000]{K}, the average number of emitted photons per fullerene becomes large enough to produce full decoherence with no remaining fringe visibility. While for fullerene molecules substantial thermal decoherence obtains at only relatively high temperatures, for much larger molecules it becomes a significant source of decoherence even at room temperature, showing the importance of thermal decoherence for understanding the quantum-to-classical transition on macroscopic scales \cite{Joos:2003:jh,Hornberger:2006:tx,Schlosshauer:2007:un}. For example, observation of interference fringes for molecules on the order of $10^9$ amu (which is orders of magnitude beyond current experiments) would likely necessitate cooling the molecules to their vibrational ground state at below \unit[77]{K} \cite{Hornberger:2012:ii, Kaltenbaek:2016:pp}.

\subsection{\label{sec:superc-syst}Superconducting systems}

Superconducting qubit systems, such as superconducting quantum interference devices (SQUIDs) and Cooper-pair boxes, play a prominent role in the exploration of coherence and decoherence in macroscopic systems. Their importance to fundamental studies of macroscopic quantum behavior was spelled out early by Leggett \cite{Leggett:1980:yt}, who suggested, in 1980, that such systems may become ideal vehicles for the creation of cat-like superpositions of macroscopically distinct states. Superconducting qubits are also considered promising candidates for the realization of a quantum computer \cite{Devoret:2013:pp}. 

Superconductivity is a phenomenon in which pairs of electrons of opposite spin condense into a boson-like particle, known as a Cooper pair. Each Cooper pair is in a low-energy ground  state. Provided the thermal vibrational energy of the crystal lattice of the material is lower than the energy gap between the ground and first excited states of the Cooper pair, interactions with the lattice cannot excite the Cooper pairs. The Cooper pairs can therefore freely move around the lattice, forming a resistance-free, persistent ``supercurrent'' whose collective center-of-mass motion may be described quantum-mechanically by a single, macroscopically extended wave function. When a thin insulating barrier between two pieces of superconducting material is inserted (known as a Josephson junction), Cooper pairs will tunnel through the barrier, leading to a flow of supercurrent even if no voltage is applied. This \emph{Josephson effect} is a purely quantum-mechanical phenomenon (for reviews, see, e.g., Refs.~\cite{Likharev:1979:ii,Makhlin:2001:oo}). 

\begin{figure}
  \centering
\includegraphics[scale=0.65]{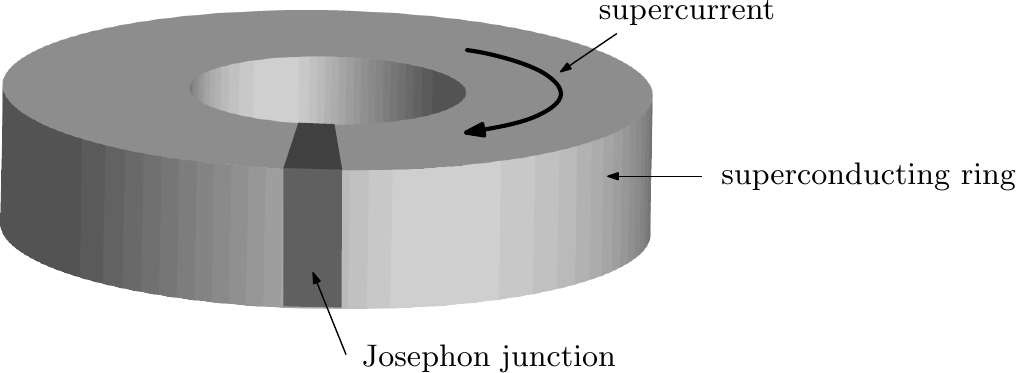}
  \caption{Schematic illustration of a SQUID. A ring of superconducting material is interrupted by a thin insulating barrier (Josephson junction), which causes a dissipation-free current (``supercurrent'') consisting of Cooper pairs to flow in the loop.}
\label{fig:squidscheme} 
\end{figure}

Figure~\ref{fig:squidscheme} schematically shows a SQUID with a single Josephson junction, a setup referred to as an rf-SQUID. The supercurrent creates an intrinsic magnetic flux $\Phi_\text{int}$ threading the loop, and in addition an external magnetic field is applied to provide an adjustable external magnetic flux $\Phi_\text{ext}$. The requirement that the macroscopic wave function around the loop must be continuous translates into a condition for the total trapped flux $\Phi=\Phi_\text{int}+ \Phi_\text{ext}$, which must obey $\Delta \phi_\text{J} + 2\pi \Phi/\Phi_0 = 2\pi k$, $k=1,2,\hdots$, where $\Delta \phi_\text{J}$ is the phase shift introduced by the Josephson junction and $\Phi_0=h/2e$ is the flux quantum. In this way, the total flux $\Phi$ becomes quantized and serves as the single macroscopic variable representing the collective quantum-mechanical evolution of the Cooper pairs. 

\begin{figure}
{\footnotesize \hspace{1cm} \emph{(a)} \hspace{6.8cm} \emph{(b)} }

\vspace{.2cm}

  \centering
\includegraphics[scale=0.85]{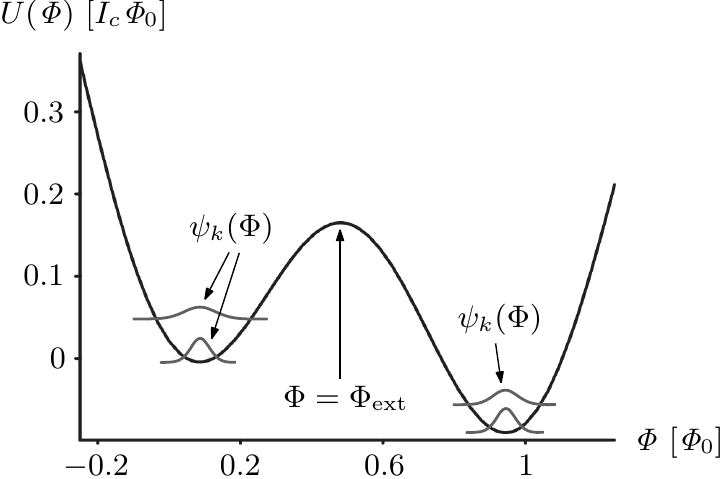} \hspace{1cm}\includegraphics[scale=0.85]{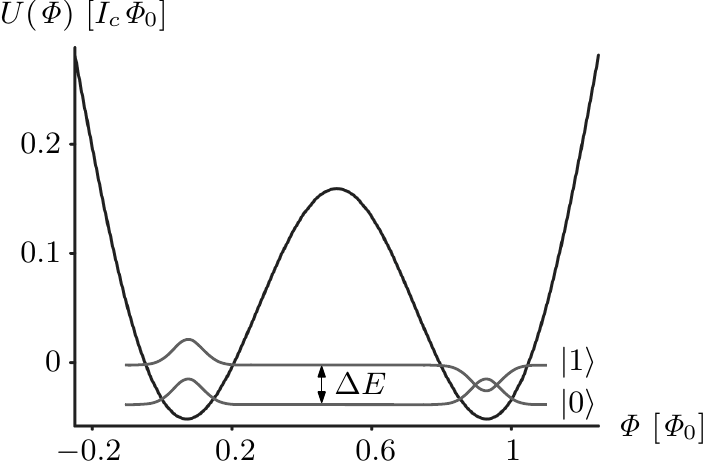}
\caption{Double-well potential $U(\Phi)$ governing the evolution of the macroscopic flux variable $\Phi$ in an rf-SQUID. \emph{(a)} Away from the bias point $\Phi_\text{ext} = \Phi_0/2$, the well is tilted. Low-lying energy eigenstates $\psi_k(\Phi) =\braket{\Phi}{k}$ are tightly localized in each well, representing approximate flux eigenstates. They also approximately correspond to a persistent supercurrent flowing in a fixed direction (clockwise or counterclockwise) around the SQUID loop. The potential $U(\Phi)$ is shown in units of $I_c \Phi_0$, where $I_c$ is the critical current of the Josephson junction and $\Phi_0$ is the flux quantum. \emph{(b)} At the bias point $\Phi_\text{ext} = \Phi_0/2$, the double well is symmetric. The two lowest-lying energy eigenstates become delocalized across the wells and consist of coherent superpositions of two macroscopic supercurrents flowing in opposite directions around the loop. These superpositions were observed in several experiments \cite{Friedman:2000:rr,Wal:2000:om,Chiorescu:2003:ta,Ilichev:2003:tv}.}
\label{fig:squidpot0} 
\end{figure}

The evolution of $\Phi$ is effectively governed by a tilted double-well potential $U(\Phi)$ in flux space (Fig.~\ref{fig:squidpot0}\emph{a}) \cite{Weiss:1999:tv}, with the amount of tilt determined by the applied external flux, and resonant quantum tunneling between the wells may occur \cite{Silvestrini:1996:ii,Rouse:1998:om}. Away from the bias point $\Phi_\text{ext}=\Phi_0/2$, each well contains low-lying energy eigenstates $\ket{k}$ that are well-localized within each well, corresponding to approximate eigenstates of the flux operator. These localized eigenstates in a given well also approximately correspond to a macroscopic supercurrent flowing in a definite direction (clockwise or counterclockwise) around the loop. The two lowest-lying energy eigenstates are well-separated from higher-energy states, such that the SQUID effectively acts as a two-state system, forming a superconducting flux qubit.

At the bias point $\Phi_\text{ext}=\Phi_0/2$, the double-well potential $U(\Phi)$ becomes symmetric (Fig.~\ref{fig:squidpot0}\emph{b}). The presence of the tunneling barrier leads to a level anticrossing that produces an energy ground state $\ket{0}$ and a first excited state $\ket{1}$ that are delocalized across the two wells. They are given by coherent superpositions of the states $\ket{\circlearrowright}$ and $\ket{\circlearrowleft}$ (which are localized in each well) representing ``classical'' clockwise and counterclockwise supercurrents, 
\begin{subequations}
\begin{align} 
\ket{0} &= \frac{1}{\sqrt{2}} \left(  \ket{\circlearrowright} + 
\ket{\circlearrowleft} \right), \label{eq:scqqqq3}\\
\ket{1} &= \frac{1}{\sqrt{2}} \left( - \ket{\circlearrowright} + 
\ket{\circlearrowleft} \right). \label{eq:scqqqq4}
\end{align}
\end{subequations}

\begin{figure}
{\footnotesize \hspace{1cm} \emph{(a)} \hspace{7.5cm} \emph{(b)} }

\centering
\begin{minipage}[c]{0.45\linewidth}
\centering
\vspace{0pt}
\includegraphics[width=0.6\textwidth]{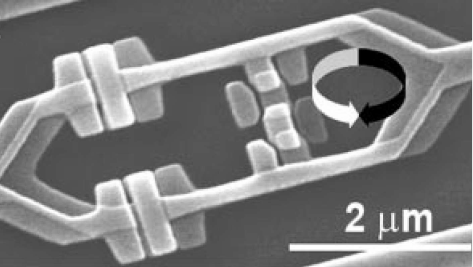}
\end{minipage}
\hspace{0.5cm}
\begin{minipage}[c]{0.5\linewidth}
\vspace{0pt}
\centering
\includegraphics[width=0.6\textwidth]{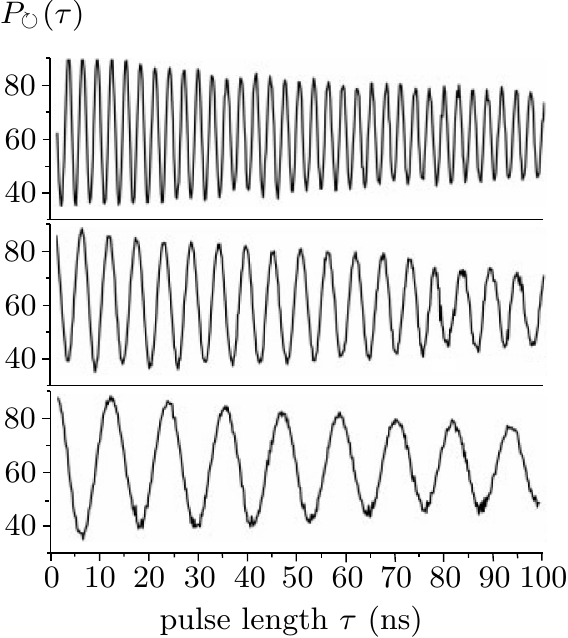}
\end{minipage}
\caption{\emph{(a)} Superconducting flux qubit used in the experiment by
  Chiorescu et al.\ \cite{Chiorescu:2003:ta}. The micrometer-sized superconducting loop is interrupted by three Josephson junctions (the use of three junctions, rather than one, enables easier tuning of the SQUID), and the flux in the loop is measured by coupling it to a second SQUID. The black and white arrows indicate the clockwise and counterclockwise directions of the supercurrent. \emph{(b)} Observation of superpositions of macroscopic supercurrents flowing in  opposite directions, as evidenced by measurements of Rabi oscillations and reported by Chiorescu et al.\ \cite{Chiorescu:2003:ta}. The plots show the occupation probability $P_\circlearrowright(\tau)$ for the clockwise supercurrent state as a function of the length $\tau$ of the microwave pulse. From top to bottom, the three data sets correspond to decreasing amplitude of the microwave pulse, with the Rabi frequency decreasing as also predicted by theory.  Figures adapted with permission from Ref.~\cite{Chiorescu:2003:ta}.}
\label{fig:chio}
\end{figure}

Such superposition states of macroscopic supercurrents flowing in opposite directions were first observed in several experiments in the early 2000s \cite{Friedman:2000:rr,Wal:2000:om,Chiorescu:2003:ta,Ilichev:2003:tv}. Friedman et al.\ \cite{Friedman:2000:rr} confirmed their existence indirectly through spectroscopic measurements of the energy splitting between them and found excellent agreement with theoretical predictions. The supercurrent in the experiment was several microampere. In experiments by Chiorescu et al.\ \cite{Chiorescu:2003:ta} (see Fig.~\ref{fig:chio}\emph{a}) and Ilichev et al.\ \cite{Ilichev:2003:tv}, the existence of the supercurrent superpositions was confirmed through the observation of Rabi oscillations between the states $\ket{\circlearrowright}$ and $\ket{\circlearrowleft}$ (Fig.~\ref{fig:chio}\emph{b}).

\begin{figure}
\centering
\includegraphics[scale=0.9]{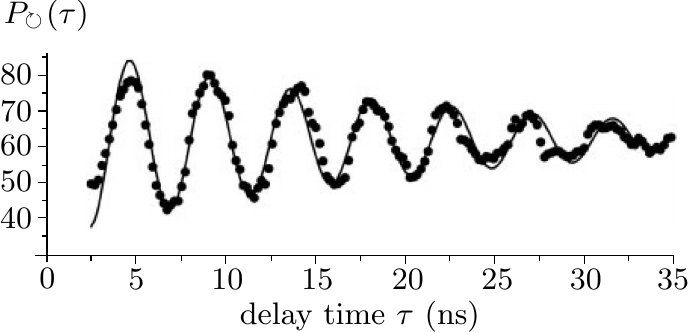}
\caption{Loss of coherence of a superposition of two supercurrents flowing in opposite directions in a SQUID, as measured in the experiment by Chiorescu et al.\
\cite{Chiorescu:2003:ta} using Ramsey interferometry. Progressive dephasing is observed as the decay of the oscillation amplitude over time. The oscillation represents the probability $P_{\circlearrowright}(\tau)$ of measuring a clockwise supercurrent as a function of the delay time $\tau$.  Figure adapted with permission from Ref.~\cite{Chiorescu:2003:ta}.}
\label{fig:chiorescu2}
\end{figure}

The gradual loss of coherence from these supercurrent superpositions was first measured by Chiorescu et al.\ \cite{Chiorescu:2003:ta} using Ramsey interferometry. The SQUID was tuned close to the bias point and initialized in the ground state $\ket{0}$. A $\pi/2$ microwave pulse was applied to transform this state into an equal-weight superposition of $\ket{0}$ and $\ket{1}$. The state was then allowed to evolve freely for a duration $\tau$, followed by the application of a second $\pi/2$ microwave pulse. In the resulting state, occupation probabilities for the supercurrent states $\ket{\circlearrowright}$ and $\ket{\circlearrowleft}$ will exhibit an oscillatory dependence on the delay time $\tau$. This dependence was experimentally observed (see Fig.~\ref{fig:chiorescu2}), with a measured frequency in excellent agreement with theoretical predictions. The characteristic time for the loss of phase coherence was obtained from the decay envelope of the oscillation and found to be around \unit[20]{ns}. In subsequent experiments with superconducting flux qubits, various influences that limit coherence time were studied, including flux noise \cite{Yoshihara:2006:ii,Bialczak:2007:uu} and photon noise \cite{Bertet:2005:un}; in the latter experiment, relatively long coherence times of several microseconds were observed \cite{Bertet:2005:un}. 

Loss of coherence has also been observed for superconducting charge qubits (Cooper-pair boxes \cite{Bouchiat:1998:ii}) and phase qubits. In a charge qubit, Cooper pairs tunnel through a Josephson junction onto a superconducting island, and the two qubit basis states are formed by states differing by the amount of charge on the island. Superpositions of such charge states were experimentally observed through Rabi oscillations \cite{Nakamura:1999:ub} and coherent oscillations with a decay time of $\unit[0.5]{\mu s}$ were measured \cite{Vion:2002:oo}. For phase qubits, where the variable of interest is the phase difference between the electrodes of the Josephson junction, superpositions of macroscopically distinct phase states with coherence times up to several $\mu$s have been observed \cite{Yu:2002:yb,Martinis:2002:qq}. Since then, several improved designs of superconducting qubits, such as quantronium, transmon, and fluxonium qubits, have further enhanced the coherence time \cite{Devoret:2013:pp}. For example, a 3D transmon has achieved coherence times on the order of $\unit[100]{\mu s}$ \cite{Rigetti:2012:aa,Sears:2012:ee}.

Drawing on experimental data, a number of theoretical studies have investigated and modeled loss-of-coherence processes in superconducting qubits, focusing on sources such as intrinsic quasiparticle tunneling \cite{Catelani:2012:zz},  the coupling to electromagnetic circuitry for SQUID readout \cite{Wal:2003:pp}, two-level defects in the Josephson junction \cite{Martinis:2005:zz}, and fluctuations in the bias current of the Josephson junction \cite{Martinis:2003:bz}. Such investigations have indicated that many sources of an intrinsic loss of coherence in superconducting qubits may be modeled in terms of an environment composed of effective two-level systems  \cite{Martinis:2002:qq,Martinis:2005:zz}, i.e., by means of a spin--spin model (compare Sec.~\ref{sec:spin-envir-models}) \cite{Dube:2001:zz,Prokofev:2000:zz}. 

It should be noted here that in the aforementioned experimental and theoretical studies of superconducting qubits, the loss of coherence is typically due to ensemble dephasing induced by fluctuations (noise), rather than by an entanglement-based transfer of information to an environment. While phenomenologically the effect on the system's density matrix may be similar for both processes, the physical differences between the two processes should be remembered; see the discussion in Sec.~\ref{sec:decoh-vers-diss}. 

\subsection{\label{sec:trapped}Ion traps}

Ion traps are one of the most promising and advanced platforms for the implementation of a quantum computer \cite{Haffner:2008:pp}. The idea of using trapped ions for quantum computation goes back to the groundbreaking papers by Cirac and Zoller \cite{Cirac:1995:tt} and Monroe et al.\ \cite{Monroe:1995:oo}. The dynamics of trapped ions have been studied extensively \cite{Leibfried:2003:om}. Several quantum-computational tasks have been experimentally realized using ion-trap qubits, including demonstrations of the Deutsch--Jozsa algorithm \cite{Deutsch:1989:mm}, quantum teleportation \cite{Barrett:2004:oo,Riebe:2004:qq}, quantum error correction \cite{Chiaverini:2004:aa}, decoherence-free subspaces \cite{Kielpinski:2001:uu,Roos:204:pp,Haffner:2005:zz,Langer:2005:uu},  entanglement between several ions \cite{Haffner:2005:sc}, and entanglement purification \cite{Reichle:2006:ii}; see Ref.~\cite{Haffner:2008:pp} for a comprehensive review. In ion-trap qubits, ions are bound by a time-dependent potential (Paul trap \cite{Paul:1990:oo,Leibfried:2003:om}), and the qubit states are formed by a pair of long-lived internal states of the ions \cite{Leibfried:2003:om,Haffner:2008:pp}. States of individual ions are typically initialized using optical-pumping techniques and are manipulated by laser pulses. Two-qubit operations may be carried out through coupling of collective motional degrees of freedom \cite{Cirac:1995:tt} or other means \cite{Haffner:2008:pp}.

Many studies of the loss of coherence in ion traps have focused on dephasing caused by noise and fluctuations in the physical (and often classical) parameters describing the trapping and control of the ions (see, e.g., Refs.~\cite{Turchette:2000:aa,Turchette:2000:oa,Brouard:2004:in,Grotz:2006:km,Stick:2006:aa,Seidelin:2006:rz,Haffner:2008:pp}). In fact, the influence of an environment is often \emph{simulated} by actively driving fluctuations in parameters such as the trap frequency or by applying external noise \cite{Myatt:2000:yy,Turchette:2000:aa}.  Since such loss of coherence is due to classical noise processes (manifesting their effect only in an ensemble average for which phase relations become smeared out) rather than entanglement-induced information transfer, the points discussed in Sec.~\ref{sec:decoh-vers-diss} regarding the distinction between noise-induced and entanglement-induced decoherence should be kept in mind. A frequently used technique for measuring dephasing in ion traps is Ramsey interferometry, in much the same way as we have already discussed for photons in a cavity (Sec.~\ref{sec:atoms-cavity}) and superconducting qubits (Sec.~\ref{sec:superc-syst}).

\begin{figure}
\centering
\includegraphics[scale=0.9]{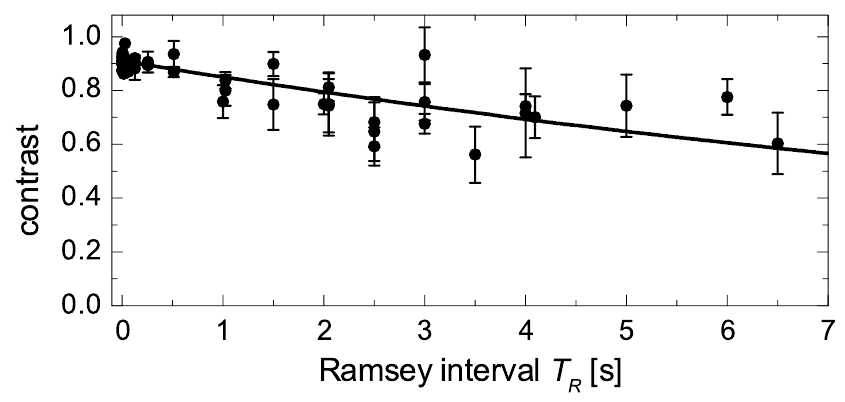} 
\caption{Loss of coherence of a superposition of two hyperfine levels in a single trapped $^9$Be$^+$ ion, as reported by Langer et al.\ \cite{Langer:2005:uu}. The loss of coherence is quantified by the Ramsey fringe contrast as a function of the wait time $T_R$ between the two $\pi/2$ pulses. The solid line is an exponential fit, from which a coherence time of $\unit[(14.7 \pm 1.6)]{s}$ is obtained.  Figure reproduced with permission from Ref.~\cite{Langer:2005:uu}.}
\label{fig:iondec}
\end{figure}

Two different kinds of superposition states in ion traps should be distinguished: superpositions of the internal atomic levels that represent the qubit basis states, and superpositions of the motional states of the ions. Since superpositions of the internal qubit levels are merely microscopic, they are not in the territory of the mesoscopic and macroscopic ``cat-like'' superposition states we have discussed previously in the context of photon fields, matter-wave interferometry, and SQUIDs. For such trapped-ion qubit states, a dominant source of loss of coherence are fluctuations in the magnetic trapping field (see, e.g., Refs.~\cite{SchmidtKaler:2003:pp,Brouard:2004:in,Grotz:2006:km, Haffner:2008:pp}). A superposition state of the ion will be sensitive to such fluctuations if its components differ in magnetic moment. Because the loss of coherence due to magnetic-field fluctuations has substantially limited achievable coherence times \cite{Haffner:2008:pp}, several ion-trap experiments have used qubit states that have the same magnetic moment and that are therefore insensitive to fluctuations of the magnetic field (see, e.g., Refs.~\cite{Haljan:2005:oo,Langer:2005:uu,Benhelm:2008:oo}). In this way, coherence times of \unit[10]{s} for a superposition of two hyperfine levels in a single $^9$Be$^+$ ion have been achieved \cite{Langer:2005:uu} (see Fig.~\ref{fig:iondec}). Other fluctuations in the experimental parameters that lead to a loss of coherence arise in the context of the control of the ion, for example, in the form of fluctuations in the intensity \cite{Schneider:1998:yz} and duration \cite{Miquel:1997:zz} of the laser beam, off-resonant excitations \cite{Steane:2000:ii}, AC-Stark shifts \cite{Haffner:2003:oo}, and detuning errors \cite{Leibfried:2003:mm}.

Let us now turn to the second kind of superposition states relevant to ion traps, namely, superpositions of motional states. Provided proper tuning of the trap parameters, the motion of a trapped ion is equivalent to that of a quasi-one-dimensional harmonically bound particle, and therefore the motional state of the ion may be represented by a harmonic oscillator \cite{Leibfried:2003:om,Wineland:2013:pp}. From a practical point of view, motional states of trapped ions are important because in certain implementations of ion-trap qubits, two-qubit gate operations are implemented by storing quantum information in motional states \cite{Cirac:1995:tt}. Different motional superposition states have been realized experimentally, including superpositions of coherent states \cite{Monroe:1996:tv,Myatt:2000:yy,Turchette:2000:aa,Wineland:2013:pp} and Fock states (i.e., number eigenstates) \cite{Myatt:2000:yy,Turchette:2000:aa}, and their dephasing has been observed \cite{Myatt:2000:yy,Turchette:2000:aa,SchmidtKaler:2003:pp}. In 1996, Monroe et al.\ \cite{Monroe:1996:tv} reported the observation of an ion for which the internal spin states $\ket{\uparrow}$ and $\ket{\downarrow}$ were quantum-correlated (entangled) with two coherent motional states $\ket{\alpha_\uparrow}$ and $\ket{\alpha_\downarrow}$ representing wave packets oscillating back and forth in the trap potential, 
\begin{equation}\label{eq:lidvg2}
\ket{\Psi} = \frac{1}{\sqrt{2}} \left( \ket{\uparrow}\ket{\alpha_\uparrow} + \ket{\downarrow} \ket{\alpha_\downarrow} \right).
\end{equation}
In the experiment, $\alpha_\downarrow=-\alpha_\uparrow$, so the motions of the wave packets were $180^\circ$ out of phase with each other; the maximum separation between the two wave packets at the turning points was \unit[83]{nm}, with a wave-packet size of \unit[7.1]{nm}. 

\begin{figure}
\centering
\includegraphics[scale=0.98]{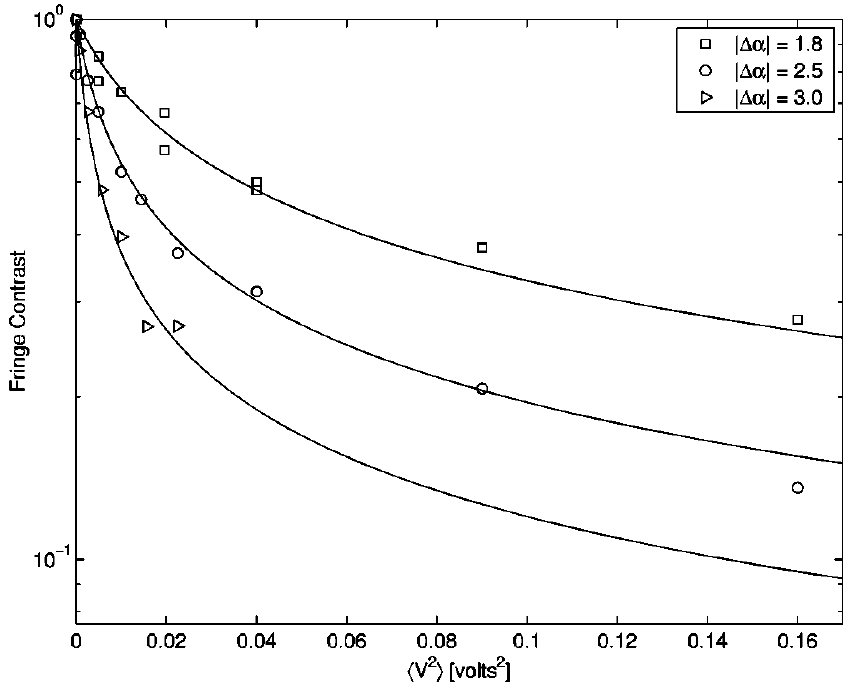} 
\caption{Dephasing of an entangled superposition $\ket{\Psi} = \frac{1}{\sqrt{2}} \left( \ket{\uparrow}\ket{\alpha_\uparrow} + \ket{\downarrow} \ket{\alpha_\downarrow} \right)$ of coherent states $\ket{\alpha_\uparrow}$ and $\ket{\alpha_\downarrow}$ for different sizes $\abs{\Delta \alpha} = \abs{\alpha_\uparrow-\alpha_\downarrow}$ of the superposition, as reported by Turchette et al.\ \cite{Turchette:2000:aa}. The loss of coherence is due to a simulated phase-damping environment, produced by varying the voltage $V$ that controls the frequency of the ion trap and then averaging over many such noisy realizations. The plot shows the Ramsey fringe contrast as a function of the applied mean-squared voltage noise $\langle V^2 \rangle$. Solid lines are based on a theoretical model; see Eq.~(44) of Ref.~\cite{Turchette:2000:aa}.  Figure reproduced with permission from Ref.~\cite{Turchette:2000:aa}.}
\label{fig:iondec2}
\end{figure}
 
Dephasing of such states was experimentally studied by Turchette et al.\ \cite{Turchette:2000:aa} using Ramsey interferometry  (see Fig.~\ref{fig:iondec2}). The authors simulated a dephasing environment by varying the trap frequency during the wait time between the two Ramsey $\pi/2$ pulses, which introduced a relative phase shift between the components in the superposition. The frequency was changed adiabatically to avoid energy transfer to and from the ion. The loss of coherence then appears as the result of an averaging over many different instances of the random noise process. It is therefore to be understood as a consequence of ensemble averaging, rather than entanglement with an environment (see again Sec.~\ref{sec:decoh-vers-diss} for comments on this distinction). Superpositions of Fock states, with number differences up to $\Delta n = 3$, and their dephasing were also observed. In a different experiment, Schmidt-Kaler et al.\ \cite{SchmidtKaler:2003:pp} measured motional center-of-mass-mode coherence times on the order of \unit[100]{ms} for a trapped $^{40}$Ca$^+$ ion described by a superposition of two vibrational states.

\begin{figure}
{\footnotesize \hspace{1.4cm}\emph{(a)} \hspace{3.95cm} \emph{(b)} \hspace{3.85cm} \emph{(c)} }

\centering
\includegraphics[scale=1.1]{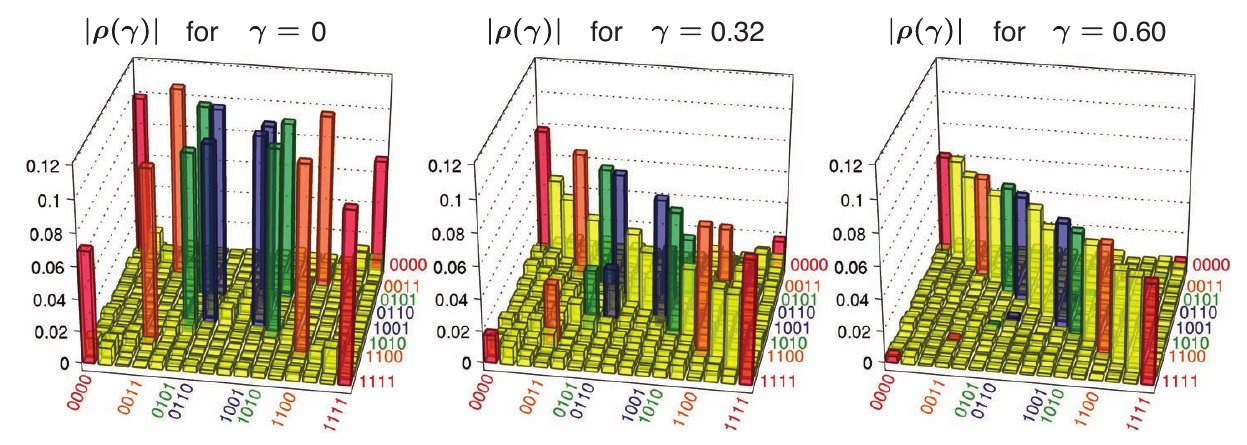}
\caption{Dynamics of multiparticle entanglement under the influence of an engineered dephasing environment, as studied experimentally by Barreiro et al.\ \cite{Barreiro:2010:aa}. The amount of dephasing, quantified by the parameter $\gamma$, increases from left to right. The plots show the absolute values of the tomographically reconstructed density matrix for four trapped-ion qubits subject to engineered dephasing. \emph{(a)} Without dephasing  ($\gamma=0$), the state violates a Bell--CHSH inequality. \emph{(b)} For modest dephasing ($\gamma=0.32$), the entanglement becomes bound. \emph{(c)} For stronger dephasing ($\gamma=0.60$), the entanglement disappears and the state becomes separable.  Figure reproduced with permission from Ref.~\cite{Barreiro:2010:aa}. }
\label{fig:entan}
\end{figure}

Ion traps have also been used to experimentally explore the dynamics of entanglement under the influence of an environment. For example, Barreiro et al.\ \cite{Barreiro:2010:aa} reported an experiment in which four entangled trapped-ion qubits were coupled to an engineered, tunable environment. By varying the amount of dephasing (represented by a parameter $\gamma$) introduced into the multiparticle system and then tomographically reconstructing the resulting state, crossovers between different entanglement regimes were observed (see Fig.~\ref{fig:entan}). In the absence of dephasing ($\gamma=0$), the entangled multiparticle state was shown to violate a Bell--CHSH inequality. With even a relatively small amount of dephasing ($\gamma=0.06$), the state no longer violates the inequality. Around $\gamma=0.3$, the state crosses over into bound entanglement \cite{Horodecki:1998:oo}, i.e., it becomes a state that is entangled but not distillable. Around $\gamma=0.6$, the state becomes completely separable, indicating that all entanglement initially present in the multiparticle state has been lost to dephasing. 

As already discussed in Sec.~\ref{sec:dfs}, trapped ions have also been used in experimental studies of decoherence-free subspaces \cite{Kielpinski:2001:uu,Roos:204:pp,Haffner:2005:zz,Langer:2005:uu}, reservoir engineering \cite{Poyatos:1996:um,Myatt:2000:yy,Turchette:2000:aa,Carvalho:2001:ua} (of which the aforementioned studies by Turchette et al.\ \cite{Turchette:2000:aa} are an example), and quantum state engineering \cite{Barreiro:2011:oo,Lin:2013:pp,Kienzler:2015:oo}.

\subsection{\label{sec:other}Other experimental areas}

We shall very briefly list a few other existing and prospective areas for the observation of decoherence.

\paragraph{Quantum dots} Decoherence of electron spins in quantum dots \cite{Hanson:2007:pp} has been studied for a number of sources of decoherence, including electrostatic fluctuations \cite{Kuhlmann:2013:aa,Arnold:2014:oo}, spin environments \cite{Fischer:2009:ii,Kuhlmann:2013:aa,Urbaszek:2013:pp,Delteil:2014:aa} and phonon environments representing acoustic vibrations of the crystal lattice \cite{Tighineanu:2018:ii}. 

\paragraph{Mechanical quantum resonators} Mechanical quantum resonators \cite{Aspelmeyer:2013:aa,Poot:2012:aa,Greenberg:2012:zz,Blencowe:2004:mm}, coupled either to electronic transducers (``quantum electromechanical systems'' \cite{Blencowe:2004:mm,Poot:2012:aa,Greenberg:2012:zz}) or photon fields (``cavity optomechanical systems'' \cite{Aspelmeyer:2013:aa}), are promising candidates for the generation of spatial macro-superpositions \cite{Arndt:2014:oo}. Such resonators may be effectively treated, under the right conditions, as a one-dimensional quantum harmonic oscillator, which represents the fundamental flexural mode.  Several potential decoherence mechanisms have been explored. For example, the role of intrinsic tunneling two-level defects (i.e., spin-$\frac{1}{2}$ particles) as a decohering (and dissipative) environment has been studied \cite{Mohanty:2002:mm,Ahn:2003:mt,Blencowe:2004:mm,Blencowe:2005:cc,Zolfagharkhani:2005:tv,Seoanez:2006:yb,Seoanez:2007:um,Remus:2009:im,Schlosshauer:2008:os,Aspelmeyer:2013:aa}, including consideration of decoherence models in which the resonator interacts with a collection of two-level systems that are in turn subject to a decohering bosonic bath \cite{Schlosshauer:2008:os,Remus:2009:im}. While decoherence in quantum resonators has often been investigated in the context of dissipative processes such as heating (see, e.g., Refs.~\cite{Mohanty:2002:mm,Ahn:2003:mt,Zolfagharkhani:2005:tv,Seoanez:2006:yb,Seoanez:2007:um,Aspelmeyer:2013:aa}), pure dephasing has also been observed and analyzed \cite{Fong:2012:aa,Zhang:2014:oo,Miao:2014:ii,Moser:2014:uu,Maillet:2016:zz}. For example, Ref.~\cite{Maillet:2016:zz} experimentally studied dephasing of a resonator due to a simulated phase reservoir, realized by inducing fluctuations in the resonator frequency through applied voltage noise. 
 
\paragraph{Bose--Einstein condensates} Different kinds of superposition states and nonclassical phenomena have been observed in Bose--Einstein condensates, including: interference fringes between independent, overlapping condensates arising from the indistinguishability of bosons \cite{Andrews:1997:um}; interference between single atoms in a coherently split condensate involving either spatial or internal degrees of freedom (see, e.g., Refs.~\cite{Shin:2004:lo,Gross:2010:gg}); and many-particle entanglement \cite{Tura:2014:oo,Schmied:2016:ll,Pezze:2018:uu}. For Bose--Einstein condensates described by a superposition of macroscopically different particle numbers (see, e.g., Refs.~\cite{Cirac:1998:mm,Ruostekoski:1998:mm,Gordon:1999:mh,Dunningham:2001:da,Calsamiglia:2001:tt,Louis:2001:mu,Micheli:2003:jn}), collisional decoherence  due to scattering processes between condensate and noncondensate atoms has been studied theoretically \cite{Dalvit:2000:bb}. Decoherence of phonons representing the collective quantum excitations of the condensate atoms in an isolated Bose--Einstein condensate has also been modeled \cite{Howl:2017:aa}. Ref.~\cite{Schrinski:2017:yy} considered the coherent splitting of a Bose--Einstein condensate into two distinct momentum modes traversing a Mach--Zehnder-like interferometer, such that the accumulated phase difference between the two arms of the interferometer leads to an interference signal at the output. The authors investigated the susceptibility of this interference signal to decoherence processes, as well as to hypothetical collapse theories such as continuous spontaneous localization models \cite{Bassi:2003:yb,Adler:2007:um,Bassi:2010:aa}.

\subsection{\label{sec:exper-tests-quant}Prospective tests of quantum mechanics}

Decoherence experiments are also useful for testing the universal validity of quantum mechanics \cite{Leggett:2002:uy,Marshall:2003:om,Bassi:2005:om,Pikovski:2012:aa,Arndt:2014:oo,Wan:2016:oo,Kaltenbaek:2016:pp,Stickler:2018:ii}, most notably with respect to the hypothetical presence of a novel nonunitary mechanism in nature that would break the linearity of the Schr\"odinger time evolution and lead to wave-function collapse. Such mechanisms that modify the linear structure of quantum mechanics are known under the headings of collapse theories, dynamical reduction models, and continuous spontaneous localization models; see Ref.~\cite{Bassi:2003:yb} for a comprehensive review. As long as the observable effect of such nonlinearities is to effectively destroy or prevent quantum interferences, the challenge is to distinguish such effects from those produced by ordinary quantum decoherence \cite{Adler:2007:um,Bassi:2010:aa}. One would need to sufficiently shield the system from decoherence in order to unambiguously isolate the postulated reduction effect. This goal is difficult to achieve, because for the collapse mechanism to become appreciable, the size of the system must be sufficiently large, in which case decoherence will in general be strong as well \cite{Tegmark:1993:uz,Nimmrichter:2013:aa}. 

The superpositions realized in current experiments are still not sufficiently macroscopic to rule out collapse theories, although it has been demonstrated \cite{Nimmrichter:2011:pr} that matter-wave interferometry with large molecular clusters (in the mass range between $10^6$ and $\unit[10^8]{amu}$) would be able to test the collapse theories proposed in Refs.~\cite{Adler:2007:um,Bassi:2010:aa}; such experiments may soon become technologically feasible \cite{Hornberger:2012:ii,Arndt:2014:oo}. Other promising avenues for testing quantum mechanics include motional superposition states of micromechanical oscillators \cite{Marshall:2003:om,Pikovski:2012:aa}, interference of free nanoparticles \cite{Romero:2011:aa,Wan:2016:oo} (an approach that offers the prospect of spatial superpositions separated by \unit[100]{nm} for particles on the order of $\unit[10^9]{amu}$ \cite{Wan:2016:oo}), and molecular nanorotors \cite{Stickler:2018:ii}. Ultimately, experiments carried out in space rather than on Earth might be able to push the limit for macroscopic superpositions to objects involving on the order of $10^{10}$ atoms \cite{Kaltenbaek:2016:pp}. In such space-based experiments, low background gas pressures ($\lesssim \unit[10^{-13}]{Pa}$) would minimize collisional decoherence, low temperatures ($\lesssim \unit[20]{K}$) would minimize thermal decoherence, and microgravity ($\lesssim \unit[10^{-9}]{g}$) would minimize decoherence induced by gravitational time dilation \cite{Pikovski:2015:oo}, potentially enabling tests of quantum gravity models \cite{Kaltenbaek:2016:pp}.

\begin{figure}
\centering
\includegraphics[scale=0.76]{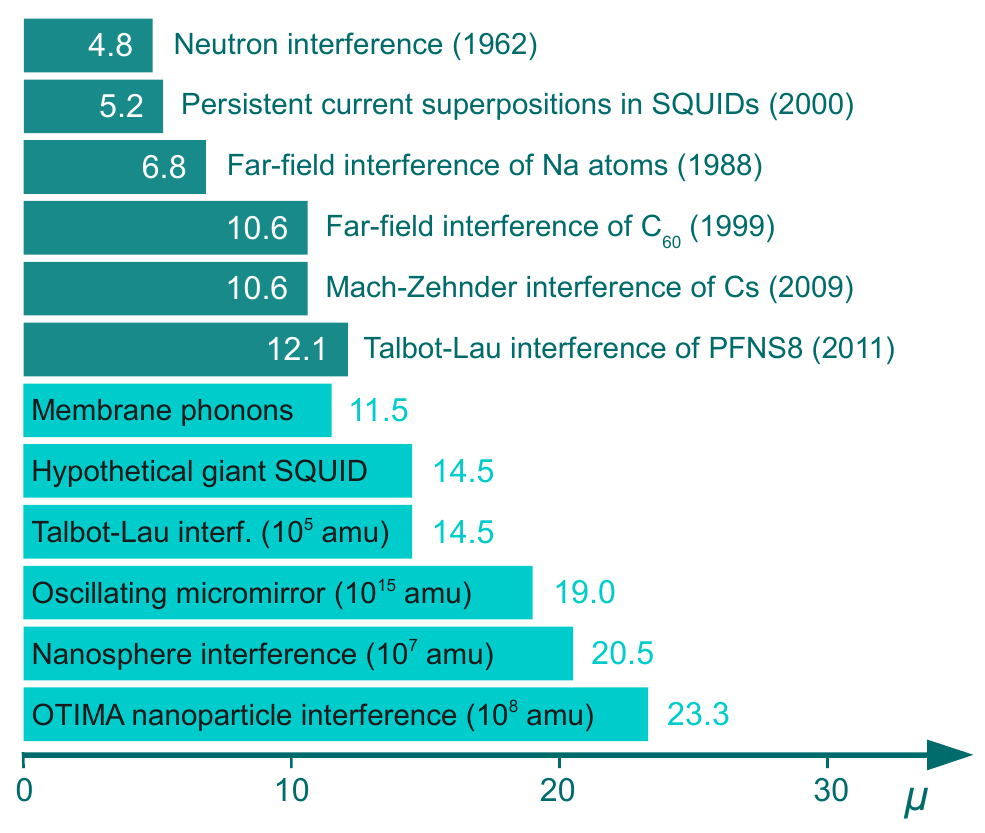}
\caption{Macroscopicity of mechanical superpositions reached in existing (top) and prospective (bottom) experiments, as reported in Refs.~\cite{Nimmrichter:2013:aa, Arndt:2014:oo}. The macroscopicity is quantified by a parameter $\mu$ defined by Nimmrichter and Hornberger \cite{Nimmrichter:2013:aa}. It represents the susceptibility of the superposition states to minimal modifications of quantum mechanics that would induce a dynamical reduction of the density operator to a classical mixture. References to existing experiments are as follows: neutron interference \cite{Zeilinger:1982:oo}; persistent current superpositions in SQUIDs \cite{Friedman:2000:rr} (see Sec.~\ref{sec:superc-syst}); far-field interference of Na atoms \cite{Keith:1988:uu}; far-field interference of C$_{60}$ \cite{Arndt:1999:rc} (see Sec.~\ref{sec:matt-wave-interf}); Mach--Zehnder interference of Cs \cite{Chung:2009:oo}; Talbot--Lau interference of PFNS8 \cite{Gerlich:2011:aa}. References to prospective experiments are as follows: 
membrane phonons refer to the experiment of Ref.~\cite{Teufel:2011:oo} extended in such a way that more than 1,000 oscillation cycles between the zero-phonon and one-phonon states become observable;  the hypothetical giant SQUID refers to a loop of length \unit[20]{mm}, a wire cross-section of $\unit[100]{\mu m^2}$, and a coherence time of \unit[1]{ms}; Talbot--Lau interferometry at $\unit[10^5]{amu}$ \cite{Nimmrichter:2011:pr}; oscillating micromirror \cite{Marshall:2003:om}; nanosphere interference \cite{Romero:2011:aa}; and OTIMA nanoparticle interference refers to an all-optical matter-wave interferometer in the time domain using pulsed ionization gratings \cite{Nimmrichter:2011:pr}.  Figure reproduced with permission from Ref.~\cite{Arndt:2014:oo}.}
\label{fig:macro}
\end{figure}

A related issue of interest is the question of how one may best quantify the ``catness''---i.e., the macroscopicity or ``size''---of a given superposition state. Various measures of macroscopicity have been suggested \cite{Leggett:1980:yt,Leggett:2002:uy,Dur:2002:pp,Bjork:2004:pp,Korsbakken:2007:pp,Marquardt:2008:ii,Lee:2011:oo,Frowis:2012:zz,Nimmrichter:2013:aa}. Most are focused on analyzing particular representations of quantum states and rely on quantum-information-theoretic measures such as the quantum Fisher information \cite{Frowis:2012:zz}. Such approaches tend to depend on particular basis choices for the decomposition of the wave function and can make it difficult to compare macroscopicities of states between physically different systems. Nimmrichter and Hornberger \cite{Nimmrichter:2013:aa} have introduced an alternative measure of macroscopicity that quantifies the extent to which a given superposition state would be capable of ruling out small modifications of quantum mechanics. To represent such a modification in the most general and model-independent way, the authors considered a dynamical-semigroup framework in which a dynamical generator obeying certain invariance and symmetry conditions is added to the evolution equation for the density operator of an $N$-particle system such that superpositions of macroscopically distinct states are dynamically transformed into classical mixtures. Specific collapse models, such as continuous spontaneous localization models \cite{Bassi:2003:yb}, may be recovered as special cases within this framework. The effect of the modification can be quantified in terms of the resulting coherence time $\tau$ of the superposition, leading to the definition of a corresponding macroscopicity parameter $\mu$ \cite{Nimmrichter:2013:aa}. Figure~\ref{fig:macro} shows estimates of $\mu$ for several existing and prospective experiments. One sees that, by this measure, the spatial superpositions involved in matter-wave interferometry  (see Sec.~\ref{sec:matt-wave-interf}) exhibit some of the largest macroscopicity; the aforementioned proposed experiments on free nanoparticles \cite{Romero:2011:aa,Wan:2016:oo} and micromechanical oscillators \cite{Marshall:2003:om,Pikovski:2012:aa} would also rank high on the macroscopicity scale. 

\section{\label{sec:impl-found-quant}Decoherence and the foundations of quantum mechanics}

Since the early days of quantum mechanics, the interpretation of the quantum formalism and its attending foundational questions have been the subject of much debate (see, for example, Bacciagaluppi and Valentini's analysis of the 1927 Solvay conference \cite{Bacciagaluppi:2006:yq}).  Especially given that decoherence theory was ``discovered'' only relatively recently, it is natural to ask what role decoherence may play in addressing foundational problems and informing the existing interpretations of quantum mechanics. One of the central topics in the foundations of quantum mechanics is known as the quantum measurement problem \cite{Bub:1997:iq,Wigner:1963:yt,Fine:1970:iq,Schlosshauer:2003:tv,Wallace:2008:ii,Schlosshauer:2011:ee}, and in Sec.~\ref{sec:mmtprob} we will discuss whether decoherence has anything of substance to say about it. In Sec.~\ref{sec:interp}, we will then briefly review the role that decoherence plays, or may play, in the various interpretations of quantum mechanics. In Sec.~\ref{sec:niels-bohrs-views}, we will comment on Niels Bohr's views on the primacy of classical concepts and their relationship to the quantum--classical correspondence described by decoherence. For in-depth discussions of the connections between decoherence and the foundations of quantum mechanics, see, for example, Refs.~\cite{Bacciagaluppi:2003:yz,Schlosshauer:2003:tv,Schlosshauer:2006:rw,Schlosshauer:2007:un}. Looking beyond the subject of decoherence, the interviews collected in Ref.~\cite{Schlosshauer:2011:ee} provide an overview of contemporary attitudes toward the interpretation of quantum mechanics.

\subsection{Decoherence and the measurement problem\label{sec:mmtprob}}

Application of the unitary Schr\"odinger evolution to a measuring apparatus interacting with a system prepared in a quantum superposition state cannot dynamically describe the stochastic selection of a particular term in the superposition as the measurement outcome (the ``collapse of the wave function''); rather, system and apparatus end up in an entangled state, with all terms of the original superposition still present and quantum-correlated with different apparatus states. This is the measurement problem: the question of how to reconcile the linear, deterministic evolution described by the Schr\"odinger equation with our observation of the occurrence of random measurement outcomes. Whether one considers the measurement problem a genuine difficulty depends strongly on one's interpretation of quantum states (see Ref.~\cite{Schlosshauer:2011:ee} for a representation of different views on the issue). Generally, the need to supply a dynamical account of the reduction of the superposition to a single outcome in the course of a measurement is much more acute when the quantum state is construed as a real, physical entity, rather than as encapsulating an observer's information or beliefs. (The latter, ``epistemic'' view is most radically, and consistently, realized in the QBist interpretation of quantum mechanics \cite{Fuchs:2014:pp}, in which quantum states represent an observer's beliefs---his probabilistic expectations---about his future experiences resulting from his interactions with the system.)

The measurement problem as just defined cannot be solved by decoherence \cite{Schlosshauer:2003:tv,Schlosshauer:2007:un}.  This is so for two reasons. First, the dynamics of decoherence processes are based entirely on the standard, unitary Schr\"odinger evolution. Second, the predictively relevant part of decoherence theory relies on reduced density matrices, which are derived from the requirement that they encode the correct quantum statistics (expectation values) for all measurements pertaining to only a subset of degrees of freedom of a composite system in a multipartite (and in general entangled) state. This derivation \emph{presumes} the existence and validity of the usual measurement axioms of quantum mechanics, in particular, the collapse postulate and Born's rule. In other words, for the kinds of entangled quantum states produced by decoherence-type interactions to be interpreted as describing a situation in which the system becomes ``classical,'' we need to take the existence of measurement outcomes as \emph{a priori} given, or otherwise give an account outside of decoherence of how measurement outcomes are produced, because the property of classicality is ultimately a statement about measurement statistics. Thus decoherence, by itself, cannot address the measurement problem in any substantial way.

Of course, to say that decoherence has no bearing on the measurement problem---or on any of the ``big'' foundational problems in general---is not to suggest that decoherence and its underlying ideas cannot be of relevance in the investigation of fundamental questions. In fact, to give just one example, further explorations of the role of the environment, such as those undertaken in the development of quantum Darwinism (see Sec.~\ref{sec:prol-inform-quant}), have already shed valuable light on deeper issues concerning information transfer, amplification, irreversibility, and communication in the quantum setting \cite{Zurek:2003:pl,Ollivier:2003:za,Ollivier:2004:im,Blume:2004:oo,Blume:2005:oo,Zurek:2009:om,Riedel:2010:un,Riedel:2011:un,Riedel:2012:un,Streltsov:2013:oo,Zurek:2013:xx,Zurek:2018:om,Zurek:2018:on} (see Ref.~\cite{Zurek:2009:om} for an overview of some of the relevant ideas).
 
If one takes the quantum measurement problem to include the preferred-basis problem \cite{Schlosshauer:2007:un}, and if the preferred-basis problem is understood in the sense defined in Sec.~\ref{sec:envir-induc-supers}, then decoherence solves it, as discussed there. Indeed, the ability of decoherence to dynamically define preferred bases is exploited in certain interpretations of quantum mechanics (see the following Sec.~\ref{sec:interp}). 

\subsection{Decoherence in interpretations of quantum mechanics\label{sec:interp}}

The interplay between decoherence and the interpretation of quantum mechanics goes back to the birth of decoherence. As mentioned in the Introduction, decoherence theory itself initially came about all but as a by-product of Zeh's development of an interpretation in the mold of Everett's many-worlds interpretation \cite{Zeh:1970:yt}. Since then, various interpretations have been assessed and refined in light of the insights and results brought about by the decoherence program. Most notably, decoherence has been used to define certain structural elements in interpretations, as well as identify internal consistency issues. Below, we shall give just give a few examples; the interested reader is pointed to Refs.~\cite{Schlosshauer:2003:tv, Bacciagaluppi:2003:yz} for in-depth discussions of the interplay between decoherence and interpretations. 

In Everett-style ``many worlds'' interpretations of quantum mechanics \cite{Everett:1957:rw,Wallace:2010:im}, the quantum state is interpreted realistically and never collapses; our observation of definite measurement outcomes is then explained as a continuous ``branching'' of universes, worlds, observers, and minds in the course of measurement-like interactions. Another version, which arguably includes Everett's own conception, interprets the global entangled state as merely describing relations between states (``relative-state interpretation''); see also Refs.~\cite{Rovelli:1996:rq,Mermin:1998:ii}. The preferred-basis problem is particularly acute in such interpretations, since the particular decomposition of the global quantum state (representing, in principle, the entire universe) defines the ``worlds'' (or ``relations'') and thus their properties at each instant in time; those worlds must also be appropriately connected in time. Here, the pointer states defined by the stability criterion of decoherence (see Sec.~\ref{sec:envir-induc-supers}) provide a ready-made solution, and in this way decoherence theory has played a critical role in defining the branching structure in Everett-style interpretations \cite{Zurek:1998:re,Butterfield:2001:ua,Wallace:2003:iq,Wallace:2003:iz,Wallace:2010:im}. Such an approach does not need to define the worlds \emph{a priori} or by means of an external rule; instead, the worlds are defined dynamically through the standard Schr\"odinger evolution, and since they are dynamically stable, they lead to robust, temporally extended trajectory-like branches. A relative-state interpretation that draws heavily from the insights and structures provided by decoherence theory is the ``existential interpretation'' of Zurek \cite{Zurek:1993:pu,Zurek:1998:re,Zurek:2004:yb}. This interpretation was later extended to include the results of quantum Darwinism \cite{Zurek:2009:om}, as well as a decoherence-inspired account of the origin of Born's rule based on symmetry and invariance properties of entangled system--environment states (``environment-assisted invariance'') \cite{Zurek:2002:ii,Zurek:2003:rv,Zurek:2003:pl,Zurek:2004:yb,Zurek:2009:om,Zurek:2018:on,Schlosshauer:2003:ms,Barnum:2003:yb,Mohrhoff:2004:tv}.

In modal interpretations of quantum mechanics \cite{Clifton:1996:op}, the physical quantity represented by an observable may be assigned a definite value even if the system is not in an eigenstate of that observable. The assignment of such definite values (corresponding to well-defined physical properties) must be in agreement with the prediction of quantum mechanics; in particular, the proper Born probabilities and time evolution must be recovered. Also, at least on macroscopic scales the modally assigned definite values ought to correspond to the definite ``classical'' quantities of our experience, such as well-localized positions. Recognizing the importance of environmental interactions  highlighted by decoherence theory, some modal interpretations have derived their value assignments from states obtained from an orthogonal decomposition of the decohered reduced density matrix \cite{Bacciagaluppi:1996:po,Bene:2001:po}. For finite-dimensional state spaces, the resulting states are found to be very close to the states that would be dynamically selected by the stability criterion of decoherence, ensuring proper classicality of the modally assigned properties \cite{Bacciagaluppi:1996:po,Bene:2001:po}. For infinite-dimensional state spaces, however, this agreement often breaks down; for example, for an environmental-scattering model it was shown that the modal properties obtained from the orthogonal decomposition of the decohered reduced density matrix were significantly delocalized while the pointer states indicated tight localization \cite{Bacciagaluppi:2000:yz}. Such inconsistencies can pinpoint limitations and empirical inadequacies of certain types of modal interpretations.

The consistent-histories interpretation of quantum mechanics \cite{Griffiths:1984:tr,Omnes:1994:pz,Griffiths:2002:tr} dispenses with the usual notions of measurement and instead defines time-ordered sequences of events (``histories'') for a closed system and assigns appropriate probabilities to these sequences. As a minimal requirement, such sets of histories must fulfill a consistency condition \cite{Griffiths:1984:tr,Omnes:1994:pz,Griffiths:2002:tr} to ensure the applicability of Boolean logic in the form of the additivity of probabilities. This, however, is not enough, as most consistent histories do not exhibit appropriate quasiclassicality for macroscopic systems \cite{GellMann:1990:uz,GellMann:1991:pp,Zurek:1993:pu,Paz:1993:ww,Albrecht:1993:pq,Dowker:1995:pa,Dowker:1996:ch}. To address this issue, the pointer bases obtained from decoherence have frequently been used to dynamically yield consistent, quasiclassical histories 
\cite{Zurek:1993:pu,Paz:1993:ww,Albrecht:1992:rz,Albrecht:1993:pq,Twamley:1993:bz}, and the importance of ``records'' (represented by stable
system--environment correlations) for the definition of quasiclassical histories  has been emphasized repeatedly \cite{Albrecht:1992:rz,Albrecht:1993:pq,Paz:1993:ww,Zurek:1993:pu,Zurek:2002:ii,GellMann:1998:xy}. In particular, the redundant environmental encoding of such records, as described by quantum Darwinism, has been identified as a key mechanism for ensuring consistent, stable, quasiclassical, objective histories \cite{Riedel:2016:oo} (see also Refs.~\cite{Zurek:1993:pu,Paz:1993:ww,Zurek:2002:ii,Zurek:2003:pl}).

\subsection{\label{sec:niels-bohrs-views}Bohr's views on the primacy of classical concepts}

In Niels Bohr's writings on quantum mechanics, the indispensability and primacy of ``classical concepts'' (such as position and momentum) is widely emphasized (see, for example, Refs.~\cite{Bohr:1949:mz,Bohr:1931:ii,Bohr:1935:re,Bohr:1996:mn,Bohr:1987:oo,Bohr:1958:lu}). Indeed, Howard has stated that ``the doctrine of classical concepts turns out to be more fundamental to Bohr's philosophy of physics than are better-known doctrines, like complementarity'' \cite[p.~202]{Howard:1994:lm}. Given that decoherence theory describes a dynamical emergence of classicality, it is not surprising that decoherence has sometimes been suggested to make Bohr's insistence on fundamental classical concepts superfluous. For example, Joos has traced the birth of the ideas of decoherence theory to a dissatisfaction with the ``orthodoxy of the Copenhagen school'' \cite[p.~54]{Joos:2006:yy}. He has argued that ``the message of decoherence'' is that ``we do not need to take classical notions as the starting point for physics,'' because ``these emerge through the dynamical process of decoherence from the quantum substrate'' \cite[p.~77]{Joos:2006:yy}. Similarly, Zeh \cite{Zeh:2000:rr} has asserted that
\begin{quote}
the Heisenberg--Bohr picture of quantum mechanics can now be claimed dead. Neither classical concepts, nor any uncertainty relations, complementarity, observables, quantum logic, quantum statistics, or quantum jumps have to be introduced on a fundamental level.
\end{quote}
Any analysis of such claims is complicated by the variety of meanings of the term ``classical''---referring variously to concepts, dynamical properties, statistics, phenomena, laws, or theories. A closer reading of Bohr's views reveals that his insistence on the primacy of classical concepts is chiefly grounded in epistemological concerns, and that it pertains to his understanding of the functional role of experiments \cite{Camilleri:2015:oo,Schlosshauer:2017:oo}. For Bohr, classical concepts are indispensable because without them, it would impossible to acquire empirical knowledge of the world through experiments. Therefore, according to Bohr, any interpretation of quantum mechanics must in the end fall back on the use of classical concepts, rendering circular any attempt to derive such concepts from the quantum formalism. This suggests that we must clearly distinguish between Bohr's \emph{epistemological} thesis of the primacy of classical concepts based on his view of the \emph{functional} role of an experiment, and the \emph{dynamical} problem of the quantum--classical transition. It is the latter that is addressed by decoherence theory, not the former  \cite{Camilleri:2015:oo,Schlosshauer:2017:oo}. 

While Bohr repeatedly emphasized the epistemological necessity of classical concepts, he offered only a few oblique comments on the problem of why, physically and dynamically, macroscopic systems and measurement apparatuses may be described in classical terms. These comments mostly involved the ``heaviness'' and large size of macroscopic systems, though they frequently also referred to irreversible amplification effects (see, e.g., Refs.~\cite{Bohr:1958:lu,Bohr:1958:mj}); for example, with regard to measurement apparatuses Bohr stated that they ``concern bodies sufficiently heavy to permit the quantum [effects] to be neglected in their description'' \cite[p.~170]{Bohr:1958:lu}.

Heisenberg, in a tantalizing passage, wrote that \cite[pp.~121--2]{Heisenberg:1989:zb}
\begin{quote}
  the system which is treated by the methods of quantum mechanics is in fact a part of a much bigger system (eventually the whole world); it is interacting with
  this bigger system; and one must add that the microscopic properties of the bigger system are (at least to a large extent) unknown. \dots The interaction with the bigger system with its undefined microscopic properties then introduces a new statistical element into the description \dots\ of the system under
  consideration. In the limiting case of the large dimensions this statistical element destroys the effects of the ``interference of probabilities'' in such a manner that the quantum-mechanical scheme really approaches the classical one in the limit.
\end{quote}
In a similar vein, elsewhere Heisenberg suggested that ``the interference terms are \dots\ removed by the partly undefined interactions of the measuring apparatus, with the system and with the rest of the world'' \cite[p.~23]{Heisenberg:1955:lm}. Even though one might identify a faint hint of the later ideas of the decoherence program in Heisenberg's pronouncements, there is no mention of the critical ingredient: entanglement. 

In the 1950s and 1960s, several of Bohr's disciples, including Weizs\"acker and Rosenfeld, attempted to develop a physical account of the quantum-to-classical transition based on notions of irreversibility, a strategy they saw as serving as a dynamical justification of Bohr's classical concepts \cite{Camilleri:2015:oo}. The thermodynamic theory of Daneri, Loinger, and Prosperi \cite{Daneri:1962:om}, which built on Bohr's hints concerning the role of irreversibility and amplification, is perhaps the best known (though circular \cite{Wigner:1995:jm, Bub:1971:ll}) approach. Yet, none of these early efforts recognized the role of entanglement in a dynamical explanation of quantum statistics turning into classical-looking distributions.

\section{\label{sec:concluding-remarks}Concluding remarks}

Schr\"odinger called entanglement ``\emph{the} characteristic trait of quantum mechanics, the one that enforces its entire departure from classical lines of thought'' \cite[p.~555]{Schrodinger:1935:jn}. He used his eponymous cat paradox to argue how entanglement amplified to macroscopic scales demonstrates the apparent irreconcilability of quantum mechanics with our ``classical'' experience of the everyday world. On this view, entanglement was perceived to be a peculiar quantum feature that would have to be tamed in order to bridge the gap between quantum and classical descriptions. So it is perhaps ironic that entanglement turned out to be the key to a dynamical explanation of the emergence of classicality in quantum mechanics. 

Without a doubt, future experiments will realize ever-larger Schr\"odinger cat--like states, and we will continue our journey toward the realization of a quantum computer. A key role in all such endeavors will be played by a deep understanding of decoherence and an ongoing development of decoherence models of increasing complexity and detail. It is indeed remarkable how the basic idea of decoherence---that entanglement of a quantum system with an environment has a dramatic influence on what is observable at the level of the system, an idea already spelled out in the very first paper by Zeh \cite{Zeh:1970:yt}---has enriched so thoroughly our theoretical understanding and experimental control of the quantum-to-classical transition. 

We shall close with a quote by Zeh himself, who not only was a pioneer of decoherence theory but remained, for the rest of his life, a steadfast, thoughtful advocate of the inseparability of his discovery from the interpretation of quantum mechanics. In 1996, he humbly observed \cite{Zeh:1996:gy} that decoherence is 
\begin{quote}
a normal consequence of interacting quantum mechanical systems. It can hardly be denied to occur---but it cannot explain anything that could not have been explained before. Remarkable is only its quantitative (realistic) aspect that seems to have been overlooked for long. 
\end{quote}

\section*{}
\addcontentsline{toc}{section}{References}


\begin{thebibliography}{100}
\expandafter\ifx\csname url\endcsname\relax
  \def\url#1{\texttt{#1}}\fi
\expandafter\ifx\csname urlprefix\endcsname\relax\def\urlprefix{URL }\fi
\expandafter\ifx\csname href\endcsname\relax
  \def\href#1#2{#2} \def\path#1{#1}\fi

\bibitem{Zeh:1970:yt}
H.~D. Zeh, On the interpretation of measurement in quantum theory, Found. Phys.
  1 (1970) 69--76.

\bibitem{Zurek:1981:dd}
W.~H. Zurek, Pointer basis of quantum apparatus: {I}nto what mixture does the
  wave packet collapse?, Phys. Rev. D 24 (1981) 1516--1525.

\bibitem{Zurek:1982:tv}
W.~H. Zurek, Environment-induced superselection rules, Phys. Rev. D 26 (1982)
  1862--1880.

\bibitem{Paz:2001:aa}
J.~P. Paz, W.~H. Zurek, Environment-induced decoherence and the transition from
  quantum to classical, in: R.~Kaiser, C.~Westbrook, F.~David (Eds.), Coherent
  Atomic Matter Waves, Les Houches Session LXXII, Vol.~72 of Les Houches Summer
  School Series, Springer, Berlin, 2001, pp. 533--614.

\bibitem{Zurek:2002:ii}
W.~H. Zurek, Decoherence, einselection, and the quantum origins of the
  classical, Rev. Mod. Phys. 75 (2003) 715--775.

\bibitem{Schlosshauer:2003:tv}
M.~Schlosshauer, Decoherence, the measurement problem, and interpretations of
  quantum mechanics, Rev. Mod. Phys. 76 (2004) 1267--1305.

\bibitem{Bacciagaluppi:2003:yz}
G.~Bacciagaluppi, The role of decoherence in quantum mechanics, in: E.~N. Zalta
  (Ed.), The Stanford Encyclopedia of Philosophy, 2012, online at \href{http://plato.stanford.edu/archives/win2012/entries/qm-decoherence}{\path{http://plato.stanford.edu/archives/win2012/entries/qm-decoherence}}.

\bibitem{Joos:2003:jh}
E.~Joos, H.~D. Zeh, C.~Kiefer, D.~Giulini, J.~Kupsch, I.-O. Stamatescu,
  Decoherence and the Appearance of a Classical World in Quantum Theory, 2nd
  Edition, Springer, New York, 2003.

\bibitem{Schlosshauer:2007:un}
M.~Schlosshauer, Decoherence and the Quantum-to-Classical Transition, Springer,
  Berlin/Heidelberg, 2007.

\bibitem{Schrodinger:1935:gs}
E.~Schr{\"o}dinger, Die gegenw{\"a}rtige {S}ituation in der {Q}uantenmechanik,
  Naturwissenschaften 23 (1935) 807--812, 823--828, 844--849.

\bibitem{Raimond:2001:aa}
J.~M. Raimond, M.~Brune, S.~Haroche, Manipulating quantum entanglement with
  atoms and photons in a cavity, Rev. Mod. Phys. 73 (2001) 565--582.

\bibitem{Hornberger:2012:ii}
K.~Hornberger, S.~Gerlich, S.~Nimmrichter, P.~Haslinger, M.~Arndt, Colloquium:
  Quantum interference with clusters and molecules, Rev. Mod. Phys. 84 (2012)
  157--173.

\bibitem{Leggett:2002:uy}
A.~J. Leggett, Testing the limits of quantum mechanics: motivation, state of
  play, prospects, J. Phys.: Condens. Matter 14 (2002) R415--R451.

\bibitem{Leibfried:2003:om}
D.~Leibfried, R.~Blatt, C.~Monroe, D.~Wineland, Quantum dynamics of single
  trapped ions, Rev. Mod Phys. 75 (2003) 281--324.

\bibitem{Haffner:2008:pp}
H.~H{\"a}ffner, C.~F. Roos, R.~Blatt, Quantum computing with trapped ions,
  Phys. Rep. 469 (2008) 155--203.

\bibitem{Lidar:2013:pp}
D.~A. Lidar, T.~A. Brun (Eds.), Quantum Error Correction, Cambridge University
  Press, 2013.

\bibitem{Camilleri:2009:aq}
K.~Camilleri, A history of entanglement: {D}ecoherence and the interpretation
  problem, Stud. Hist. Phil. Mod. Phys. 40 (2009) 290--302.

\bibitem{Joos:1999:po}
E.~Joos, Elements of environmental decoherence, in: P.~Blanchard, D.~Giulini,
  E.~Joos, C.~Kiefer, I.-O. Stamatescu (Eds.), Decoherence: Theoretical,
  Experimental, and Conceptual Problems, Springer, Berlin, 2000, pp. 1--17.

\bibitem{Camilleri:2015:oo}
K.~Camilleri, M.~Schlosshauer, Niels {B}ohr as philosopher of experiment:
  {D}oes decoherence theory challenge {B}ohr's doctrine of classical concepts?,
  Stud. Hist. Phil. Mod. Phys. 49 (2015) 73--83.

\bibitem{Kubler:1973:ux}
O.~K{\"u}bler, H.~D. Zeh, Dynamics of quantum correlations, Ann. Phys. (N.Y.)
  76 (1973) 405--418.

\bibitem{Joos:1985:iu}
E.~Joos, H.~D. Zeh, The emergence of classical properties through interaction
  with the environment, Z. Phys. B: Condens. Matter 59 (1985) 223--243.

\bibitem{Zurek:1986:uz}
W.~H. Zurek, Reduction of the wavepacket: {H}ow long does it take?, in: G.~T.
  Moore, M.~O. Scully (Eds.), Frontiers of Nonequilibrium Statistical
  Mechanics, Plenum Press, New York, 1986, pp. 145--149, first published in
  1984 as Los Alamos report LAUR 84-2750.

\bibitem{Walls:1985:pp}
D.~F. Walls, G.~J. Milburn, Effect of dissipation on quantum coherence, Phys.
  Rev. A 31 (1985) 2403--2408.

\bibitem{Walls:1985:lm}
D.~F. Walls, M.~J. Collett, G.~J. Milburn, Analysis of quantum measurement,
  Phys. Rev. D 32 (1985) 3208--3215.

\bibitem{Caldeira:1985:tt}
A.~O. Caldeira, A.~J. Leggett, Influence of damping on quantum interference:
  {A}n exactly soluble model, Phys. Rev. A 31 (1985) 1059--1066.

\bibitem{Zurek:1991:vv}
W.~H. Zurek, Decoherence and the transition from quantum to classical, Phys.
  Today 44 (1991) 36--44, see also the updated version available as eprint
  quant-ph/0306072.

\bibitem{Feynman:1982:yy}
R.~P. Feynman, Simulating physics with computers, Int. J. Theor. Phys. 21
  (1982) 467--488.

\bibitem{Deutsch:1985:ym}
D.~Deutsch, Quantum theory, the {C}hurch--{T}uring principle and the universal
  quantum computer, Proc. R. Soc. Lond. A 400 (1985) 97--117.

\bibitem{Deutsch:1992:tv}
D.~Deutsch, R.~Jozsa, Rapid solution of problems by quantum computation, Proc.
  R. Soc. Lond. A 439 (1992) 553--558.

\bibitem{Berthiaume:1992:lk}
A.~Berthiaume, G.~Brassard, Oracle quantum computing, in: Proc. Workshop on
  Physics of Computation: PhysComp '92, IEEE Computer Society Press, Los
  Alamitos, CA, 1992, pp. 60--62.

\bibitem{Berthiaume:1992:lm}
A.~Berthiaume, G.~Brassard, The quantum challenge to structural complexity
  theory, in: Proc. 7th Annual Structure in Complexity Theory Conf., IEEE
  Computer Society Press, Los Alamitos, CA, 1992, pp. 132--137.

\bibitem{Bernstein:1993:yy}
E.~Bernstein, U.~Vazirani, Quantum complexity theory, in: Proceedings of the
  25th Annual ACM Symposium on Theory of Computing, ACM, New York, 1993, pp.
  11--20.

\bibitem{Simon:1994:lk}
D.~Simon, On the power of quantum computation, in: Proc. 35th Annual Symp. on
  Foundations of Computer Science, IEEE Computer Society Press, Los Alamitos,
  CA, 1994, pp. 124--134.

\bibitem{Shor:1994:om}
P.~W. Shor, Polynomial-time algorithms for prime factorization and discrete
  logarithms on a quantum computer, in: Proceedings of the 35th Annual
  Symposium on the Foundations of Computer Science, IEEE Computer Society
  Press, Los Alamitos, CA, 1994, pp. 124--134.

\bibitem{Shor:1997:tt}
P.~W. Shor, Polynomial-time algorithms for prime factorization and discrete
  logarithms on a quantum computer, SIAM J. Comp. 26 (1997) 1484--1509.

\bibitem{Grover:1996:rr}
L.~K. Grover, A fast quantum mechanical algorithm for database search, in:
  Proc. of the 28th Annual ACM Symposium on the Theory of Computing, ACM, New
  York, 1996, pp. 212--219.

\bibitem{Grover:1997:mm}
L.~K. Grover, Quantum mechanics helps in search for a needle in a haystack,
  Phys. Rev. Lett. 79 (1997) 325--328.

\bibitem{Brune:1996:om}
M.~Brune, E.~Hagley, J.~Dreyer, X.~Ma{\^i}tre, A.~Maali, C.~Wunderlich, J.~M.
  Raimond, S.~Haroche, Observing the progressive decoherence of the ``meter''
  in a quantum measurement, Phys. Rev. Lett. 77 (1996) 4887--4890.

\bibitem{Arndt:1999:rc}
M.~Arndt, O.~Nairz, J.~Vos-Andreae, C.~Keller, G.~van~der Zouw, A.~Zeilinger,
  Wave--particle duality of {C$_{60}$} molecules, Nature 401 (1999) 680--682.

\bibitem{Friedman:2000:rr}
J.~R. Friedman, V.~Patel, W.~Chen, S.~K. Yolpygo, J.~E. Lukens, Quantum
  superposition of distinct macroscopic states, Nature 406 (2000) 43--46.

\bibitem{Wal:2000:om}
C.~H. van~der Wal, A.~C.~J. ter Haar, F.~K. Wilhelm, R.~N. Schouten, C.~J.
  P.~M. Harmans, T.~P. Orlando, S.~Lloyd, J.~E. Mooij, Quantum superposition of
  macroscopic persistent-current states, Science 290 (2000) 773--777.

\bibitem{Hornberger:2009:aq}
K.~Hornberger, Introduction to decoherence theory, in: A.~Buchleitner,
  C.~Viviescas, M.~Tiersch (Eds.), Entanglement and Decoherence: Foundations
  and Modern Trends, Vol. 768 of Lecture Notes in Physics, Springer, Berlin,
  2009, pp. 221--276.

\bibitem{Breuer:2002:oq}
H.-P. Breuer, F.~Petruccione, The Theory of Open Quantum Systems, Oxford
  University Press, Oxford, 2002.

\bibitem{Wooters:1979:az}
W.~K. Wootters, W.~H. Zurek, Complementarity in the double-slit experiment:
  {Q}uantum nonseparability and a quantitative statement of {B}ohr's principle,
  Phys. Rev. D 19 (1979) 473--484.

\bibitem{Englert:1996:km}
B.-G. Englert, Fringe visibility and which-way information: An inequality,
  Phys. Rev. Lett. 77 (1996) 2154--2157.

\bibitem{Landau:1927:uy}
L.~D. Landau, The damping problem in wave mechanics, Z. Phys. 45 (1927)
  430--441.

\bibitem{Neumann:1932:gq}
J.~von Neumann, Mathematische {G}rundlagen der {Q}uantenmechanik, Springer,
  Berlin, 1932.

\bibitem{Furry:1936:pp}
W.~H. Furry, Note on the quantum mechanical theory of measurement, Phys. Rev.
  49 (1936) 393--399.

\bibitem{Everett:1957:rw}
H.~Everett, ``{R}elative state'' formulation of quantum mechanics, Rev. Mod.
  Phys. 29 (1957) 454--462.

\bibitem{Einstein:1935:dr}
A.~Einstein, B.~Podolsky, N.~Rosen, Can quantum-mechanical description of
  physical reality be considered complete?, Phys. Rev. 47 (1935) 777--780.

\bibitem{Bell:1964:ep}
J.~S. Bell, On the {E}instein--{P}odolsky--{R}osen paradox, Physics 1 (1964)
  195--200.

\bibitem{Bell:1966:ph}
J.~S. Bell, On the problem of hidden variables in quantum mechanics, Rev. Mod.
  Phys. 38 (1966) 447--452.

\bibitem{Peres:1978:aa}
A.~Peres, Unperformed experiments have no results, Am. J. Phys. 46.

\bibitem{Paz:1993:ta}
J.~P. Paz, S.~Habib, W.~H. Zurek, Reduction of the wave packet: {P}referred
  observable and decoherence time scale, Phys. Rev. D 47 (1993) 488--501.

\bibitem{Leggett:1987:pm}
A.~J. Leggett, S.~Chakravarty, A.~T. Dorsey, M.~P.~A. Fisher, A.~Garg, Dynamics
  of the dissipative two-state system, Rev. Mod. Phys. 59 (1987) 1--85.

\bibitem{Mokarzel:2002:za}
S.~G. Mokarzel, A.~N. Salgueiro, M.~C. Nemes, Modeling the reversible
  decoherence of mesoscopic superpositions in dissipative environments, Phys.
  Rev. A 65 (2002) 044101.

\bibitem{Hornberger:2003:un}
K.~Hornberger, J.~E. Sipe, Collisional decoherence reexamined, Phys. Rev. A 68
  (2003) 012105.

\bibitem{Kraus:1971:ii}
K.~Kraus, General state changes in quantum theory, Ann. Phys. 64 (1971)
  311--355.

\bibitem{Kraus:1983:ee}
K.~Kraus, States, Effects, and Operations, Springer, Berlin, 1983.

\bibitem{Alicki:2007:uu}
R.~Alicki, K.~Lendi, Quantum Dynamical Semigroups and Applications, 2nd
  Edition, Vol. 717 of Lect. Notes Phys., Springer, Berlin/Heidelberg, 2007.

\bibitem{VonNeumann:1926:tv}
J.~von Neumann, Thermodynamik quantummechanischer {G}esamheiten, G{\"o}tt.
  Nach. 1 (1927) 273--291.

\bibitem{Espagnat:1966:mf}
B.~{d'E}spagnat, Two remarks on the theory of measurement, Nuovo Cimento Suppl.
  1 (1966) 828--838.

\bibitem{Espagnat:1988:cf}
B.~{d'E}spagnat, Conceptual Foundations of Quantum Mechanics, 2nd Edition,
  Benjamin, Reading, Massachusetts, 1976.

\bibitem{Espagnat:1995:ma}
B.~{d'E}spagnat, Veiled Reality, an Analysis of Present-Day Quantum Mechanical
  Concepts, Addison-Wesley, Reading, Massachusetts, 1995.

\bibitem{Fuchs:2014:pp}
C.~A. Fuchs, N.~D. Mermin, R.~Schack, An introduction to {QBism} with an
  application to the locality of quantum mechanics, Am. J. Phys. 82 (2014)
  749--754.

\bibitem{Wigner:1932:un}
E.~Wigner, On the quantum correction for thermodynamic equilibrium, Phys. Rev.
  40 (1932) 749--759.

\bibitem{Hillery:1984:tv}
M.~Hillery, R.~F. {O'C}onnell, M.~O. Scully, E.~P. Wigner, Distribution
  functions in physics: {F}undamentals, Phys. Rep. 106 (1984) 121--167.

\bibitem{Hudson:1974:ra}
R.~L. Hudson, When is the {W}igner quasi-probability density non-negative?,
  Rep. Math. Phys. 6 (1974) 249--252.

\bibitem{Gallis:1990:un}
M.~R. Gallis, G.~N. Fleming, Environmental and spontaneous localization, Phys.
  Rev. A 42 (1990) 38--48.

\bibitem{Diosi:1995:um}
L.~Di{\'o}si, Quantum master equation of a particle in a gas environment,
  Europhys. Lett. 30 (1995) 63--68.

\bibitem{Hornberger:2006:tb}
K.~Hornberger, Master equation for a quantum particle in a gas, Phys. Rev.
  Lett. 97 (2006) 060601.

\bibitem{Hornberger:2008:ii}
K.~Hornberger, B.~Vacchini, Monitoring derivation of the quantum linear
  {B}oltzmann equation, Phys. Rev. A 77 (2008) 022112.

\bibitem{Busse:2009:aa}
M.~Busse, K.~Hornberger, Emergence of pointer states in a non-perturbative
  environment, J. Phys. A: Math. Theor. 42 (2009) 362001.

\bibitem{Busse:2010:aa}
M.~Busse, K.~Hornberger, Pointer basis induced by collisional decoherence, J.
  Phys. A: Math. Theor. 43 (2010) 015303.

\bibitem{Harris:1981:rc}
R.~A. Harris, L.~Stodolsky, On the time dependence of optical activity, J.
  Chem. Phys. 74 (1981) 2145--2155.

\bibitem{Zeh:1999:qr}
H.~D. Zeh, The meaning of decoherence, in: P.~Blanchard, D.~Giulini, E.~Joos,
  C.~Kiefer, I.~Stamatescu (Eds.), Decoherence: {T}heoretical, Experimental,
  and Conceptual Problems, Lecture Notes in Physics {No.\ 538}, Springer,
  Berlin, 2000, pp. 19--42.

\bibitem{Trost:2009:ll}
J.~Trost, K.~Hornberger, Hund's paradox and the collisional stabilization of
  chiral molecules, Phys. Rev. Lett. 103 (2009) 023202.
\newblock \href {http://dx.doi.org/10.1103/PhysRevLett.103.023202}
  {\path{doi:10.1103/PhysRevLett.103.023202}}.

\bibitem{Bahrami:2012:oo}
M.~Bahrami, A.~Shafiee, A.~Bassi, Decoherence effects on superpositions of
  chiral states in a chiral molecule, Phys. Chem. Chem. Phys. 14 (2012)
  9214--9218.

\bibitem{Giulini:1995:zh}
D.~Giulini, C.~Kiefer, H.~D. Zeh, Symmetries, superselection rules, and
  decoherence, Phys. Lett. A 199 (1995) 291--298.

\bibitem{Giulini:2000:ry}
D.~Giulini, Decoherence: A dynamical approach to superselection rules?, Lect.
  Notes Phys. 559 (2000) 67--92.

\bibitem{Paz:1999:vv}
J.~P. Paz, W.~H. Zurek, Quantum limit of decoherence: Environment induced
  superselection of energy eigenstates, Phys. Rev. Lett. 82 (1999) 5181--5185.

\bibitem{Zurek:1993:pu}
W.~H. Zurek, Preferred states, predictabilty, classicality, and the
  environment-induced decoherence, Prog. Theor. Phys. 89 (1993) 281--312.

\bibitem{Zurek:1993:qq}
W.~H. Zurek, S.~Habib, J.~P. Paz, Coherent states via decoherence, Phys. Rev.
  Lett. 70 (1993) 1187--1190.

\bibitem{Zurek:1998:re}
W.~H. Zurek, Decoherence, einselection, and the existential interpretation
  ({T}he {R}ough {G}uide), Philos. Trans. R. Soc. London, Ser. A 356 (1998)
  1793--1821.

\bibitem{Diosi:2000:yn}
L.~Di{\'o}si, C.~Kiefer, Robustness and diffusion of pointer states, Phys. Rev.
  Lett. 85 (2000) 3552--3555.

\bibitem{Eisert:2003:ib}
J.~Eisert, Exact decoherence to pointer states in free open quantum systems is
  universal, Phys. Rev. Lett. 92 (2004) 210401.

\bibitem{Zurek:2003:pl}
W.~H. Zurek, Quantum {D}arwinism and envariance, in: J.~D. Barrow, P.~C.~W.
  Davies, C.~H. Harper (Eds.), Science and Ultimate Reality, Cambridge
  University Press, Cambridge, England, 2004, pp. 121--137.

\bibitem{Ollivier:2003:za}
H.~Ollivier, D.~Poulin, W.~H. Zurek, Emergence of objective properties from
  subjective quantum states: {E}nvironment as a witness, Phys. Rev. Lett. 93
  (2004) 220401.

\bibitem{Ollivier:2004:im}
H.~Ollivier, D.~Poulin, W.~H. Zurek, Environment as a witness: selective
  proliferation of information and emergence of objectivity, Phys. Rev. A 72
  (2005) 042113.

\bibitem{Blume:2004:oo}
R.~Blume-Kohout, W.~H. Zurek, A simple example of ``{Q}uantum {D}arwinism'':
  {R}edundant information storage in many-spin environments, Found. Phys. 35
  (2005) 1857--1876.

\bibitem{Blume:2005:oo}
R.~Blume-Kohout, W.~H. Zurek, Quantum {D}arwinism: {E}ntanglement, branches,
  and the emergent classicality of redundantly stored quantum information,
  Phys. Rev. A 73 (2006) 062310.

\bibitem{Zurek:2009:om}
W.~H. Zurek, Quantum {D}arwinism, Nature Phys. 5 (2009) 181--188.

\bibitem{Riedel:2010:un}
C.~J. Riedel, W.~H. Zurek, Quantum {D}arwinism in an everyday environment: Huge
  redundancy in scattered photons, Phys. Rev. Lett. 105 (2010) 020404.

\bibitem{Riedel:2011:un}
C.~J. Riedel, W.~H. Zurek, Redundant information from thermal illumination:
  quantum {D}arwinism in scattered photons, New J. Phys. 13 (2011) 073038.

\bibitem{Riedel:2012:un}
C.~J. Riedel, W.~H. Zurek, M.~Zwolak, The rise and fall of redundancy in
  decoherence and quantum {D}arwinism, New J. Phys. 14 (2012) 083010.

\bibitem{Zurek:2014:xx}
W.~H. Zurek, Quantum {D}arwinism, classical reality, and the randomness of
  quantum jumps, Phys. Today 67 (2014) 44--50.

\bibitem{Zwolak:2016:zz}
M.~Zwolak, C.~J. Riedel, W.~H. Zurek, Amplification, decoherence, and the
  acquisition of information by spin environments, Sci. Rep. 6 (2016) 25277.

\bibitem{Zwolak:2017:mm}
M.~Zwolak, W.~H. Zurek, Redundancy of einselected information in quantum
  {D}arwinism: {T}he irrelevance of irrelevant environment bits, Phys. Rev. A
  95 (2017) 030101(R).

\bibitem{Zurek:2018:on}
W.~H. Zurek, Quantum theory of the classical: quantum jumps, {B}orn's rule and
  objective classical reality via quantum {D}arwinism, Phil. Trans. R. Soc. A
  376 (2018) 20180107.

\bibitem{Unden:2018:ia}
T.~Unden, D.~Louzon, M.~Zwolak, W.~H. Zurek, F.~Jelezko, Revealing the
  emergence of classicality using nitrogen-vacancy centers. Phys. Rev. Lett. 
123 (2019) 140402. 

\bibitem{Zwolak:2014:tt}
M.~Zwolak, C.~J. Riedel, W.~H. Zurek, Amplification, redundancy, and the
  quantum {C}hernoff information, Phys. Rev. Lett. 112 (2014) 140406.

\bibitem{Blume:2007:oo}
R.~Blume-Kohout, W.~H. Zurek, Quantum {D}arwinism in quantum {B}rownian motion,
  Phys. Rev. Lett. 101 (2008) 240405.

\bibitem{Ollivier:2001:az}
H.~Ollivier, W.~H. Zurek, Quantum discord: {A} measure of the quantumness of
  correlations, Phys. Rev. Lett. 88 (2002) 017901.

\bibitem{Streltsov:2013:oo}
A.~Streltsov, W.~H. Zurek, Quantum discord cannot be shared, Phys. Rev. Lett.
  111 (2013) 040401.

\bibitem{Galve:2016:oo}
F.~Galve, R.~Zambrini, S.~Maniscalco, Non-{M}arkovianity hinders {Q}uantum
  {D}arwinism, Sci. Rep. 6 (2016) 19607.

\bibitem{Pleasance:2017:oo}
G.~Pleasance, B.~M. Garraway, Application of quantum {D}arwinism to a
  structured environment, Phys. Rev. A 96 (2017) 062105.

\bibitem{Ciampini:2018:ii}
M.~A. Ciampini, G.~Pinna, P.~Mataloni, M.~Paternostro, Experimental signature
  of quantum {D}arwinism in photonic cluster states, Phys. Rev. A 98 (2018)
  020101(R).

\bibitem{Cucchietti:2005:om}
F.~M. Cucchietti, J.~P. Paz, W.~H. Zurek, Gaussian decoherence from random spin
  environments, Phys. Rev. A 72 (2005) 052113.

\bibitem{Schneider:1998:yz}
S.~Schneider, G.~J. Milburn, Decoherence in ion traps due to laser intensity
  and phase fluctuations, Phys. Rev. A 57 (1998) 3748--3752.

\bibitem{Miquel:1997:zz}
C.~Miquel, J.~P. Paz, W.~H. Zurek, Quantum computation with phase drift errors,
  Phys. Rev. Lett. 78 (1997) 3971--3974.

\bibitem{Martinis:2003:bz}
J.~M. Martinis, S.~Nam, J.~Aumentado, K.~M. Lang, C.~Urbina, Decoherence of a
  superconducting qubit due to bias noise, Phys. Rev. B 67 (2003) 094510.

\bibitem{Vandersypen:2004:ra}
L.~M.~K. Vandersypen, I.~L. Chuang, {NMR} techniques for quantum control and
  computation, Rev. Mod. Phys. 76 (2004) 1037--1069.

\bibitem{Myatt:2000:yy}
C.~J. Myatt, B.~E. King, Q.~A. Turchette, C.~A. Sackett, D.~Kielpinski, W.~M.
  Itano, C.~Monroe, D.~J. Wineland, Decoherence of quantum superpositions
  through coupling to engineered reservoirs, Nature 403 (2000) 269--273.

\bibitem{Jaynes:1980:lm}
E.~Jaynes, Quantum beats, in: A.~O. Barut (Ed.), Foundations of Radiation
  Theory and Quantum Electrodynamics, Plenum Press, New York, 1980, pp. 37--43.

\bibitem{Peres:1980:im}
A.~Peres, Can we undo quantum measurements?, Phys. Rev. D 22 (1980) 879--883.

\bibitem{Scully:1982:yb}
M.~O. Scully, K.~Dr{\"u}hl, Quantum eraser: {A} proposed photon correlation
  experiment concerning observation and {``}delayed choice{''} in quantum
  mechanics, Phys. Rev. A 25 (1982) 2208--2213.

\bibitem{Scully:1991:yb}
M.~O. Scully, B.~G. Englert, H.~Walther, Quantum optical tests of
  complementarity, Nature (London) 351 (1991) 111--116.

\bibitem{Englert:1999:aq}
B.-G. Englert, M.~O. Scully, H.~Walther, Quantum erasure in double-slit
  interferometers with which-way detectors, Am. J. Phys. 67 (1999) 325--329.

\bibitem{Ashby:2016:pp}
J.~M. Ashby, P.~D. Schwarz, M.~Schlosshauer, Delayed-choice quantum eraser for
  the undergraduate laboratory, Am. J. Phys. 84 (2016) 95--105.

\bibitem{Schneider:1999:tt}
S.~Schneider, G.~J. Milburn, Decoherence and fidelity in ion traps with
  fluctuating trap parameters, Phys. Rev. A 59 (1999) 3766--3774.

\bibitem{Turchette:2000:aa}
Q.~A. Turchette, C.~J. Myatt, B.~E. King, C.~A. Sackett, D.~Kielpinski, W.~M.
  Itano, C.~Monroe, D.~J. Wineland, Decoherence and decay of motional quantum
  states of a trapped atom coupled to engineered reservoirs, Phys. Rev. A 62
  (2000) 053807.

\bibitem{Lindblad:1976:um}
G.~Lindblad, On the generators of quantum dynamical semigroups, Commun. Math.
  Phys. 48 (1976) 119--130.

\bibitem{Gorini:1976:tt}
V.~Gorini, A.~Kossakowski, E.~C.~G. Sudarshan, Completely positive dynamical
  semigroups of {$N$}-level systems, J. Math. Phys. 17 (1976) 821--825.

\bibitem{Alicki:2001:aa}
R.~Alicki, M.~Fannes, Quantum Dynamical Systems, Oxford University Press,
  Oxford, 2001.

\bibitem{Benatti:2005:ii}
F.~Benatti, R.~Floreanini, Open quantum dynamics: Complete positivity and
  entanglement, Int. J. Mod. Phys. B 19 (2005) 3063--3139.
\newblock \href {http://dx.doi.org/10.1142/S0217979205032097}
  {\path{doi:10.1142/S0217979205032097}}.

\bibitem{Horodecki:1995:oo}
R.~Horodecki, P.~Horodecki, M.~Horodecki, Violating {B}ell inequality by mixed
  spin 1/2 states: necessary and sufficient condition, Phys. Lett. A 200 (1995)
  340--344.

\bibitem{Peres:1996:oo}
A.~Peres, Separability criterion for density matrices, Phys. Rev. Lett. 77
  (1996) 1413--1415.
\newblock \href {http://dx.doi.org/10.1103/PhysRevLett.77.1413}
  {\path{doi:10.1103/PhysRevLett.77.1413}}.

\bibitem{Benatti:2002:oo}
F.~Benatti, R.~Floreanini, R.~Romano, Complete positivity and dissipative
  factorized dynamics, J. Phys. A: Math. Gen. 35 (2002) L551--L556.
\newblock \href {http://dx.doi.org/10.1088/0305-4470/35/39/101}
  {\path{doi:10.1088/0305-4470/35/39/101}}.

\bibitem{Gorini:1978:uf}
V.~Gorini, A.~Frigerio, M.~Verri, A.~Kossakowski, E.~C.~G. Sudarshan,
  Properties of quantum {M}arkovian master equations, Rep. Math. Phys. 13
  (1978) 149--173.

\bibitem{Davies:1974:tw}
E.~B. Davies, Markovian master equations, Commun. Math. Phys. 39 (1974)
  91--110.

\bibitem{Kossakowski:1972:tf}
A.~Kossakowski, On quantum statistical mechanics of non-{H}amiltonian systems,
  Rep. Math. Phys. 3 (1972) 247--288.

\bibitem{Hornberger:2006:tc}
K.~Hornberger, Monitoring approach to open quantum dynamics using scattering
  theory, EPL 77 (2007) 50007.

\bibitem{Davies:1976:uu}
E.~B. Davies, Quantum Theory of Open Systems, Academic Press, London, 1976.

\bibitem{Holevo:1996:ll}
A.~S. Holevo, Covariant quantum {M}arkovian evolutions, J. Math. Phys. 37~(4)
  (1996) 1812--1832.
\newblock \href {http://dx.doi.org/10.1063/1.531481}
  {\path{doi:10.1063/1.531481}}.

\bibitem{Redfield:1957:im}
A.~G. Redfield, On the theory of relaxation processes, IBM J. Res. Develop. 1
  (1957) 19--31.

\bibitem{Blum:1981:qq}
K.~Blum, Density Matrix Theory and Applications, Plenum Press, New York,
  London, 1981.

\bibitem{Caldeira:1983:on}
A.~O. Caldeira, A.~J. Leggett, Path integral approach to quantum {B}rownian
  motion, Physica A 121 (1983) 587--616.

\bibitem{Dumke:1979:ia}
R.~D{\"u}mcke, H.~Spohn, The proper form of the generator in the weak-coupling
  limit, Z. Phys. B 34 (1979) 419--422.

\bibitem{Davies:1976:oo}
E.~B. Davies, Markovian master equations. {II}, Mathematische Annalen 219
  (1976) 147--158.

\bibitem{Davies:1978:uu}
E.~B. Davies, A model of atomic radiation, Ann. Inst. H. Poincar{\'e} 38 (1978)
  91--110.

\bibitem{Dodin:2018:zz}
A.~Dodin, T.~Tscherbul, R.~Alicki, A.~Vutha, P.~Brumer, Secular versus
  nonsecular Redfield dynamics and Fano coherences in incoherent excitation: An
  experimental proposal, Phys. Rev. A 97 (2018) 013421.
\newblock \href {http://dx.doi.org/10.1103/PhysRevA.97.013421}
  {\path{doi:10.1103/PhysRevA.97.013421}}.

\bibitem{Barchielli:1991:fv}
A.~Barchielli, V.~P. Belavkin, Measurements continuous in time and a posteriori
  states in quantum mechanics, J. Phys. A: Math. Gen. 24 (1991) 1495--1514.

\bibitem{Belavkin:1989:an}
V.~P. Belavkin, Non-demolition measurements, nonlinear filtering and dynamic
  programming of quantum stochastic processes, in: Lecture Notes in Control and
  Information Sciences, Vol. 121, Springer, Berlin, 1989, pp. 245--265.

\bibitem{Belavkin:1989:am}
V.~P. Belavkin, A continuous counting observation and posterior quantum
  dynamics, J. Phys. A: Math. Gen. 22 (1989) L1109--L1114.

\bibitem{Belavkin:1989:um}
V.~P. Belavkin, A new wave equation for a continuous non-demolition
  measurement, Phys. Lett. A 140 (1989) 355--358.

\bibitem{Belavkin:1995:tt}
V.~P. Belavkin, The interplay of classical and quantum stochastics: Diffusion,
  measurement and filtering, in: P.~Garbaczewksi, M.~Wolf, A.~Veron (Eds.),
  Chaos: The Interplay Between Stochastic and Deterministic Behaviour, Lecture
  Notes in Physics, Springer, 1995, pp. 21--41.

\bibitem{Diosi:1988:wx}
L.~Di{\'o}si, Continuous quantum measurement and {I}t{\^o} formalism, Phys.
  Lett. A 129 (1988) 419--423.

\bibitem{Diosi:1988:hn}
L.~Di{\'o}si, Localized solution of a simple nonlinear quantum {L}angevin
  equation, Phys. Lett. A 132 (1988) 233--236.

\bibitem{Diosi:1988:bv}
L.~Di{\'o}si, Quantum stochastic processes as models for state vector
  reduction, J. Phys. A 21 (1988) 2885--2898.

\bibitem{Gisin:1984:qs}
N.~Gisin, Quantum measurements and stochastic processes, Phys. Rev. Lett. 52
  (1984) 1657--1660.

\bibitem{Gisin:1989:jn}
N.~Gisin, Stochastic quantum dynamics and relativity, Helv. Phys. Acta 62
  (1989) 363--371.

\bibitem{Wiseman:1994:qq}
H.~M. Wiseman, Quantum theory of continuous feedback, Phys. Rev. A 49 (1994)
  2133--2150.

\bibitem{Goan:2001:rz}
H.-S. Goan, G.~J. Milburn, H.~M. Wiseman, H.~B. Sun, Continuous quantum
  measurement of two coupled quantum dots using a point contact: A quantum
  trajectory approach, Phys. Rev. B 63 (2001) 125326.

\bibitem{Plenio:1998:bb}
M.~B. Plenio, P.~L. Knight, The quantum-jump approach to dissipative dynamics
  in quantum optics, Rev. Mod. Phys. 70 (1998) 101--144.

\bibitem{Sorgel:2015:pp}
L.~S{\"o}rgel, K.~Hornberger, Unraveling quantum {B}rownian motion: {P}ointer
  states and their classical trajectories, Phys. Rev. A 92 (2015) 062112.

\bibitem{Prokofev:2000:zz}
N.~V. Prokof'ev, P.~C.~E. Stamp, Theory of the spin bath, Rep. Prog. Phys. 63
  (2000) 669--726.

\bibitem{Dube:2001:zz}
M.~Dub{\'e}, P.~C.~E. Stamp, Mechanisms of decoherence at low temperatures,
  Chem. Phys. 268 (2001) 257--272.

\bibitem{Groeblacher:2013:im}
S.~Gr{\"o}blacher, A.~Trubarov, N.~Prigge, M.~Aspelmeyer, J.~Eisert,
  Observation of non-{M}arkovian micro-mechanical {B}rownian motion, Nature
  Comm. 6 (2015) 7606.

\bibitem{Nakajima:1958:im}
S.~Nakajima, On quantum theory of transport phenomena, Prog. Theor. Phys. 20
  (1958) 948--959.

\bibitem{Zwanzig:1960:om}
R.~Zwanzig, Statistical mechanics of irreversibility, in: Boulder Lecture Notes
  in Theoretical Physics, Vol. III, Interscience, New York, 1960, pp. 106--141.

\bibitem{Zwanzig:1960:mo}
R.~Zwanzig, Ensemble method in the theory of irreversibility, J. Chem. Phys. 33
  (1960) 1338--1341.

\bibitem{Chaturvedi:1979:pm}
S.~Chaturvedi, F.~Shibata, Time-convolutionless projection operator formalism
  for elimination of fast variables. {A}pplications to {B}rownian motion, Z.
  Phys. B 35 (1979) 297--308.

\bibitem{Shibata:1980:ma}
F.~Shibata, T.~Arimitsu, Expansion formulas and nonequilibrium statistical
  mechanics, J. Phys. Soc. Jpn. 49 (1980) 891--897.

\bibitem{Royer:1972:um}
A.~Royer, Cumulant expansions and pressure broadening as an example of
  relaxation, Phys. Rev. A 6 (1972) 1741--1760.

\bibitem{Royer:2003:za}
A.~Royer, Combining projection superoperators and cumulant expansions in open
  quantum dynamics with initial correlations and fluctuating {H}amiltonians and
  environments, Phys. Lett. A 315 (2003) 335--351.

\bibitem{Feynman:1963:jj}
R.~Feynman, F.~L. Vernon, The theory of a general quantum system interacting
  with a linear dissipative system, Ann. Phys. (N.Y.) 24 (1963) 118--173.

\bibitem{Caldeira:1983:gv}
A.~Caldeira, A.~Leggett, Quantum tunneling in a dissipative system, Ann. Phys.
  (N.Y.) 149 (1983) 374--456.

\bibitem{Lounasmaa:1974:yb}
O.~V. Lounasmaa, Experimental Principles and Methods below 1 K, Academic Press,
  New York, 1974.

\bibitem{Adler:2006:yb}
S.~L. Adler, Normalization of collisional decoherence: Squaring the delta
  function, and an independent cross-check, J. Phys. A: Math. Gen. 39 (2006)
  14067--14074.

\bibitem{Hackermuller:2003:uu}
L.~Hackerm{\"u}ller, K.~Hornberger, B.~Brezger, A.~Zeilinger, M.~Arndt,
  Decoherence in a {T}albot--{L}au interferometer: the influence of molecular
  scattering, Appl. Phys. B 77 (2003) 781--787.

\bibitem{Hornberger:2003:tv}
K.~Hornberger, S.~Uttenthaler, B.~Brezger, L.~Hackerm{\"u}ller, M.~Arndt,
  A.~Zeilinger, Collisional decoherence observed in matter wave interferometry,
  Phys. Rev. Lett. 90 (2003) 160401.

\bibitem{Vacchini:2009:pp}
B.~Vacchini, K.~Hornberger, Quantum linear {B}oltzmann equation, Phys. Rep. 478
  (2009) 71--120.

\bibitem{Busse:2010:oo}
M.~Busse, P.~Pietrulewicz, H.-P. Breuer, K.~Hornberger, Stochastic simulation
  algorithm for the quantum linear {B}oltzmann equation, Phys. Rev. E 82 (2010)
  026706.

\bibitem{Hornberger:2004:bb}
K.~Hornberger, J.~E. Sipe, M.~Arndt, Theory of decoherence in a matter wave
  {T}albot--{L}au interferometer, Phys. Rev. A 70 (2004) 053608.

\bibitem{Nimmrichter:2011:pr}
S.~Nimmrichter, K.~Hornberger, P.~Haslinger, M.~Arndt, Testing spontaneous
  localization theories with matter-wave interferometry, Phys. Rev. A 83 (2011)
  043621.

\bibitem{Kokorowski:2001:ub}
D.~A. Kokorowski, A.~D. Cronin, T.~D. Roberts, D.~E. Pritchard, From single- to
  multiple-photon decoherence in an atom interferometer, Phys. Rev. Lett. 86
  (2001) 2191--2195.

\bibitem{Uys:2005:yb}
H.~Uys, J.~D. Perreault, A.~D. Cronin, Matter-wave decoherence due to a gas
  environment in an atom interferometer, Phys. Rev. Lett. 95 (2005) 150403.

\bibitem{Tegmark:1993:uz}
M.~Tegmark, Apparent wave function collapse caused by scattering, Found. Phys.
  Lett. 6 (1993) 571--590.

\bibitem{Gring:2010:aa}
M.~Gring, S.~Gerlich, S.~Eibenberger, S.~Nimmrichter, T.~Berrada, M.~Arndt,
  H.~Ulbricht, K.~Hornberger, M.~M{\"u}ri, M.~Mayor, M.~B{\"o}ckmann, , N.~L.
  Doltsinis, Influence of conformational molecular dynamics on matter wave
  interferometry, Phys. Rev. A 81 (2010) 031604(R).

\bibitem{Stickler:2015:zz}
B.~A. Stickler, K.~Hornberger, Molecular rotations in matter-wave
  interferometry, Phys. Rev. A 92 (2015) 023619.

\bibitem{Walter:2016:zz}
K.~Walter, B.~A. Stickler, K.~Hornberger, Collisional decoherence of polar
  molecules, Phys. Rev. A 93 (2016) 063612.

\bibitem{Stickler:2016:yy}
B.~A. Stickler, B.~Papendell, K.~Hornberger, Spatio-orientational decoherence
  of nanoparticles, Phys. Rev. A 94 (2016) 033828.

\bibitem{Papendell:2017:yy}
B.~Papendell, B.~A. Stickler, K.~Hornberger, Quantum angular momentum diffusion
  of rigid bodies, New J. Phys. 19~(12) (2017) 122001.

\bibitem{Stickler:2018:oo}
B.~A. Stickler, F.~T. Ghahramani, K.~Hornberger, Rotational alignment decay and
  decoherence of molecular superrotors, Phys. Rev. Lett. 121 (2018) 243402.

\bibitem{Stickler:2018:uu}
B.~A. Stickler, B.~Schrinski, K.~Hornberger, Rotational friction and diffusion
  of quantum rotors, Phys. Rev. Lett. 121 (2018) 040401.

\bibitem{Hu:1992:om}
B.~L. Hu, J.~P. Paz, Y.~Zhang, Quantum {B}rownian motion in a general
  environment: {E}xact master equation with nonlocal dissipation and colored
  noise, Phys. Rev. D 45 (1992) 2843--2861.

\bibitem{Weiss:1999:tv}
U.~Weiss, Quantum Dissipative Systems, World Scientific, Singapore, 1999.

\bibitem{Unruh:1989:rc}
W.~G. Unruh, W.~H. Zurek, Reduction of a wavepacket in quantum {B}rownian
  motion, Phys. Rev. D 40 (1989) 1071--1094.

\bibitem{Lombardo:2005:ia}
F.~C. Lombardo, P.~I. Villar, Decoherence induced by zero point fluctuations in
  quantum {B}rownian motion, Phys. Lett. A 336 (2005) 16--24.

\bibitem{Haake:1932:tt}
F.~Haake, R.~Reibold, Strong damping and low-temperature anomalies for the
  harmonic oscillator, Phys. Rev. A 32 (1985) 2462--2475.

\bibitem{Grabert:1988:bf}
H.~Grabert, P.~Schramm, G.-L. Ingold, Quantum {B}rownian motion: The functional
  integral approach, Phys. Rep. 168 (1988) 115--207.

\bibitem{Gallis:1992:im}
M.~R. Gallis, Spatial correlations of random potentials and the dynamics of
  quantum coherence, Phys. Rev. A 45 (1992) 47--53.

\bibitem{Anglin:1997:za}
J.~R. Anglin, J.~P. Paz, W.~H. Zurek, Deconstructing decoherence, Phys. Rev. A
  55 (1997) 4041--4053.

\bibitem{Unruh:1995:uy}
W.~G. Unruh, Maintaining coherence in quantum computers, Phys. Rev. A 51 (1995)
  992--997.

\bibitem{Palma:1996:yy}
G.~M. Palma, K.-A. Suominen, A.~K. Ekert, Quantum computers and dissipation,
  Proc. R. Soc. Lond. A 452 (1996) 567--584.

\bibitem{Schlosshauer:2008:os}
M.~Schlosshauer, A.~P. Hines, G.~J. Milburn, Decoherence and dissipation of a
  quantum harmonic oscillator coupled to two-level systems, Phys. Rev. A 77
  (2008) 022111.

\bibitem{Stamp:1998:im}
P.~C.~E. Stamp, Quantum environments: {S}pin baths, oscillator baths, and
  applications to quantum magnetism, in: S.~Tomsovic (Ed.), Tunnelling in
  Complex Systems, World Scientific, Singapore, 1998, pp. 101--197.

\bibitem{Prokofev:1995:ab}
N.~V. Prokof{'}ev, P.~C.~E. Stamp, Decoherence in the quantum dynamics of a
  {``}central spin{''} coupled to a spin environment. Eprint \href
  {http://arxiv.org/abs/cond-mat/9511011} {\path{arXiv:cond-mat/9511011}}.

\bibitem{Prokofev:1993:aa}
N.~V. Prokof{'}ev, P.~C.~E. Stamp, Giant spins and topological decoherence: a
  {H}amiltonian approach, J. Phys. Chem. Lett. 5 (1993) L663--L670.

\bibitem{Dobrovitski:2003:az}
V.~V. Dobrovitski, H.~A.~D. Raedt, M.~I. Katsnelson, B.~N. Harmon, Quantum
  oscillations without quantum coherence, Phys. Rev. Lett. 90 (2003) 210401.

\bibitem{Caldeira:1993:bz}
A.~O. Caldeira, A.~H. {Castro Neto}, T.~O. de~Carvalho, Dissipative quantum
  systems modeled by a two-level-reservoir coupling, Phys. Rev. B 48 (1993)
  13974--13976.

\bibitem{Dowling:2003:tv}
J.~P. Dowling, G.~J. Milburn, Quantum technology: The second quantum
  revolution, Proc. R. Soc. Lond. A 361 (2003) 1655--1674.

\bibitem{Lidar:1998:uu}
D.~A. Lidar, I.~L. Chuang, K.~B. Whaley, Decoherence-free subspaces for quantum
  computation, Phys. Rev. Lett. 81 (1998) 2594--2597.

\bibitem{Zanardi:1997:yy}
P.~Zanardi, M.~Rasetti, Noiseless quantum codes, Phys. Rev. Lett. 79 (1997)
  3306--3309.

\bibitem{Zanardi:1997:tv}
P.~Zanardi, M.~Rasetti, Error avoiding quantum codes, Mod. Phys. Lett. B 11
  (1997) 1085--1093.

\bibitem{Zanardi:1998:oo}
P.~Zanardi, Dissipation and decoherence in a quantum register, Phys. Rev. A 57
  (1998) 3276--3284.

\bibitem{Lidar:1999:fa}
D.~A. Lidar, D.~Bacon, K.~B. Whaley, Concatenating decoherence-free subspaces
  with quantum error correcting codes, Phys. Rev. Lett. 82 (1999) 4556--4559.

\bibitem{Bacon:2000:yy}
D.~Bacon, J.~Kempe, D.~A. Lidar, K.~B. Whaley, Universal fault-tolerant quantum
  computation on decoherence-free subspaces, Phys. Rev. Lett. 85 (2000)
  1758--1761.

\bibitem{Duan:1998:yb}
L.-M. Duan, G.-C. Guo, Reducing decoherence in quantum-computer memory with all
  quantum bits coupling to the same environment, Phys. Rev. A 57 (1998)
  737--741.

\bibitem{Zanardi:2001:oo}
P.~Zanardi, Stabilizing quantum information, Phys. Rev. A 63 (2001) 012301.

\bibitem{Knill:2000:aa}
E.~Knill, R.~Laflamme, L.~Viola, Theory of quantum error correction for general
  noise, Phys. Rev. Lett. 82 (2000) 2525--2528.

\bibitem{Lidar:2003:aa}
D.~A. Lidar, K.~B. Whaley, Decoherence-free subspaces and subsystems, in:
  F.~Benatti, R.~Floreanini (Eds.), Irreversible Quantum Dynamics, Vol. 622 of
  Springer Lecture Notes in Physics, Springer, Berlin, 2003, pp. 83--120, also
  available as eprint quant-ph/0301032.

\bibitem{Lidar:2014:pp}
D.~A. Lidar, Review of decoherence-free subspaces, noiseless subsystems, and
  dynamical decoupling, Adv. Chem. Phys. 154 (2014) 295--354.

\bibitem{Kempe:2001:oo}
J.~Kempe, D.~Bacon, D.~A. Lidar, K.~B. Whaley, Theory of decoherence-free
  fault-tolerant universal quantum computation, Phys. Rev. A 63 (2001) 042307.

\bibitem{Choi:2006:tt}
M.-D. Choi, D.~W. Kribs, Method to find quantum noiseless subsystems, Phys.
  Rev. Lett. 96 (2006) 050501.

\bibitem{Beny:2007:pp}
C.~B{\'e}ny, A.~Kempf, D.~W. Kribs, Generalization of quantum error correction
  via the {H}eisenberg picture, Phys. Rev. Lett. 98 (2007) 100502.

\bibitem{BlumeKohout:2010:pp}
R.~Blume-Kohout, H.~K. Ng, D.~Poulin, L.~Viola, Information-preserving
  structures: {A} general framework for quantum zero-error information, Phys.
  Rev. A 82 (2010) 062306.

\bibitem{Bacon:1999:aq}
D.~Bacon, D.~A. Lidar, K.~B. Whaley, Robustness of decoherence-free subspaces
  for quantum computation, Phys. Rev. A 60 (1999) 1944--1955.

\bibitem{Kattemolle:2018:ii}
J.~Kattem\"olle, J.~van Wezel, Dynamical fidelity susceptibility of
  decoherence-free subspaces, Phys. Rev. A 99 (2019) 062340.
\newblock \href {http://dx.doi.org/10.1103/PhysRevA.99.062340}
  {\path{doi:10.1103/PhysRevA.99.062340}}.

\bibitem{Altepeter:2004:ll}
J.~B. Altepeter, P.~G. Hadley, S.~M. Wendelken, A.~J. Berglund, P.~G. Kwiat,
  Experimental investigation of a two-qubit decoherence-free subspace, Phys.
  Rev. Lett. 92 (2004) 147901.

\bibitem{Steane:1996:cd}
A.~M. Steane, Error correcting codes in quantum theory, Phys. Rev. Lett. 77
  (1996) 793--797.

\bibitem{Shor:1995:rx}
P.~W. Shor, Scheme for reducing decoherence in quantum computer memory, Phys.
  Rev. A 52 (1995) R2493--R2496.

\bibitem{Steane:2001:dx}
A.~M. Steane, Quantum computing and error correction, in: P.~Turchi, A.~Gonis
  (Eds.), Decoherence and Its Implications in Quantum Computation and
  Information Transfer, IOS Press, Amsterdam, 2001, pp. 284--298, also
  available as eprint quant-ph/0304016.

\bibitem{Knill:2002:rx}
E.~Knill, R.~Laflamme, A.~Ashikhmin, H.~Barnum, L.~Viola, W.~Zurek,
  Introduction to quantum error correction, LA Science 27 (2002) 188--225.

\bibitem{Nielsen:2000:tt}
M.~A. Nielsen, I.~L. Chuang, Quantum Computation and Quantum Information,
  Cambridge University Press, Cambridge, 2000.

\bibitem{Kwiat:2000:kv}
P.~G. Kwiat, A.~J. Berglund, J.~B. Altepeter, A.~G. White, Experimental
  verification of decoherence-free subspaces, Science 290 (2000) 498--501.

\bibitem{Kielpinski:2001:uu}
D.~Kielpinski, V.~Meyer, M.~A. Rowe, C.~A. Sackett, W.~M. Itano, C.~Monroe,
  D.~J. Wineland, A decoherence-free quantum memory using trapped ions, Science
  291 (2001) 1013--1015.

\bibitem{Viola:2001:ra}
L.~Viola, E.~M. Fortunato, M.~A. Pravia, E.~Knill, R.~Laflamme, D.~G. Cory,
  Experimental realization of noiseless subsystems for quantum information
  processing, Science 293 (2001) 2059--2063.

\bibitem{Roos:204:pp}
C.~F. Roos, G.~P.~T. Lancaster, M.~Riebe, H.~H\"affner, W.~H\"ansel, S.~Gulde,
  C.~Becher, J.~Eschner, F.~Schmidt-Kaler, R.~Blatt, Bell states of atoms with
  ultralong lifetimes and their tomographic state analysis, Phys. Rev. Lett. 92
  (2004) 220402.
\newblock \href {http://dx.doi.org/10.1103/PhysRevLett.92.220402}
  {\path{doi:10.1103/PhysRevLett.92.220402}}.

\bibitem{Haffner:2005:zz}
H.~H{\"a}ffner, F.~Schmidt-Kaler, W.~H{\"a}nsel, C.~F. Roos, T.~K{\"o}rber,
  M.~Chwalla, M.~Riebe, J.~Benhelm, U.~D. Rapol, C.~Becher, R.~Blatt, Robust
  entanglement, Appl. Phys. B 81 (2005) 151--153.

\bibitem{Langer:2005:uu}
C.~Langer, R.~Ozeri, J.~D. Jost, J.~Chiaverini, B.~DeMarco, A.~Ben-Kish, R.~B.
  Blakestad, J.~Britton, D.~B. Hume, W.~M. Itano, D.~Leibfried, R.~Reichle,
  T.~Rosenband, T.~Schaetz, P.~O. Schmidt, D.~J. Wineland, Long-lived qubit
  memory using atomic ions, Phys. Rev. Lett. 95 (2005) 060502.
\newblock \href {http://dx.doi.org/10.1103/PhysRevLett.95.060502}
  {\path{doi:10.1103/PhysRevLett.95.060502}}.

\bibitem{Mohseni:2003:pp}
M.~Mohseni, J.~S. Lundeen, K.~J. Resch, A.~M. Steinberg, Experimental
  application of decoherence-free subspaces in an optical quantum-computing
  algorithm, Phys. Rev. Lett. 91 (2003) 187903.

\bibitem{Deutsch:1989:mm}
D.~Deutsch, Quantum computational networks, Proc. R. Soc. Lond. A 425 (1989)
  73--90.

\bibitem{Zhang:2006:zz}
Q.~Zhang, J.~Yin, T.-Y. Chen, S.~Lu, J.~Zhang, X.-Q. Li, T.~Yang, X.-B. Wang,
  J.-W. Pan, Experimental fault-tolerant quantum cryptography in a
  decoherence-free subspace, Phys. Rev. A 73 (2006) 020301(R).

\bibitem{Pushin:2011:zz}
D.~A. Pushin, M.~G. Huber, M.~Arif, D.~G. Cory, Experimental realization of
  decoherence-free subspace in neutron interferometry, Phys. Rev. Lett. 107
  (2011) 150401.

\bibitem{Dalvit:2000:bb}
D.~A.~R. Dalvit, J.~Dziarmaga, W.~H. Zurek, Decoherence in {B}ose--{E}instein
  condensates: Towards bigger and better {S}chr{\"o}dinger cats, Phys. Rev. A
  62 (2000) 013607.

\bibitem{Poyatos:1996:um}
J.~F. Poyatos, J.~I. Cirac, P.~Zoller, Quantum reservoir engineering with laser
  cooled trapped ions, Phys. Rev. Lett. 77 (1996) 4728--4731.

\bibitem{Carvalho:2001:ua}
A.~R.~R. Carvalho, P.~Milman, R.~L. de~Matos~Filho, L.~Davidovich, Decoherence,
  pointer engineering, and quantum state protection, Phys. Rev. Lett. 86 (2001)
  4988--4991.

\bibitem{Barreiro:2011:oo}
J.~T. Barreiro, M.~M{\"u}ller, P.~Schindler, D.~Nigg, T.~Monz, M.~Chwalla,
  M.~Hennrich, C.~F. Roos, P.~Zoller, R.~Blatt, An open-system quantum
  simulator with trapped ions, Nature 470 (2011) 486--491.

\bibitem{Lin:2013:pp}
Y.~Lin, J.~P. Gaebler, F.~Reiter, T.~R. Tan, R.~Bowler, A.~S. S{\o}rensen,
  D.~Leibfried, D.~J. Wineland, Dissipative production of a maximally entangled
  steady state of two quantum bits, Nature 504 (2013) 415--418.

\bibitem{Kienzler:2015:oo}
D.~Kienzler, H.-Y. Lo, B.~Keitch, L.~de~Clercq, F.~Leupold, F.~Lindenfelser,
  M.~Marinelli, V.~Negnevitsky, J.~P. Home, Quantum harmonic oscillator state
  synthesis by reservoir engineering, Science 347 (2015) 53--56.

\bibitem{Shankar:2013:pp}
S.~Shankar, M.~Hatridge, Z.~Leghtas, K.~M. Sliwa, A.~Narla, U.~Vool, S.~M.
  Girvin, L.~Frunzio, M.~Mirrahimi, M.~H. Devoret, Autonomously stabilized
  entanglement between two superconducting quantum bits, Nature 504 (2013)
  419--422.

\bibitem{Krauter:2011:ll}
H.~Krauter, C.~A. Muschik, K.~Jensen, W.~Wasilewski, J.~M. Petersen, J.~I.
  Cirac, E.~S. Polzik, Entanglement generated by dissipation and steady state
  entanglement of two macroscopic objects, Phys. Rev. Lett. 107 (2011) 080503.

\bibitem{Verstraete:2009:ii}
F.~Verstraete, M.~M. Wolf, J.~I. Cirac, Quantum computation, quantum state
  engineering, and quantum phase transitions driven by dissipation, Nature
  Phys. 5 (2009) 633--636.

\bibitem{Braun:2002:aa}
D.~Braun, Creation of entanglement by interaction with a common heat bath,
  Phys. Rev. Lett. 89 (2002) 277901.
\newblock \href {http://dx.doi.org/10.1103/PhysRevLett.89.277901}
  {\path{doi:10.1103/PhysRevLett.89.277901}}.

\bibitem{Benatti:2003:aa}
F.~Benatti, R.~Floreanini, M.~Piani, Environment induced entanglement in
  {M}arkovian dissipative dynamics, Phys. Rev. Lett. 91 (2003) 070402.
\newblock \href {http://dx.doi.org/10.1103/PhysRevLett.91.070402}
  {\path{doi:10.1103/PhysRevLett.91.070402}}.

\bibitem{Kim:2002:oo}
M.~S. Kim, J.~Lee, D.~Ahn, P.~L. Knight, Entanglement induced by a single-mode
  heat environment, Phys. Rev. A 65 (2002) 040101.
\newblock \href {http://dx.doi.org/10.1103/PhysRevA.65.040101}
  {\path{doi:10.1103/PhysRevA.65.040101}}.

\bibitem{Jakobczyk:2002:oo}
L.~Jak{\'o}bczyk, Entangling two qubits by
  dissipation, J. Phys. A: Math. Theor. 35~(30) (2002) 6383--6392.
\newblock \href {http://dx.doi.org/10.1088/0305-4470/35/30/313}
  {\path{doi:10.1088/0305-4470/35/30/313}}.

\bibitem{Zyczkowski:2001:ii}
K.~\ifmmode~\dot{Z}\else \.{Z}\fi{}yczkowski, P.~Horodecki, M.~Horodecki,
  R.~Horodecki, Dynamics of quantum entanglement, Phys. Rev. A 65 (2001)
  012101.
\newblock \href {http://dx.doi.org/10.1103/PhysRevA.65.012101}
  {\path{doi:10.1103/PhysRevA.65.012101}}.

\bibitem{Lee:2004:uu}
J.~Lee, I.~Kim, D.~Ahn, H.~McAneney, M.~S. Kim, Completely positive
  non-{M}arkovian decoherence, Phys. Rev. A 70 (2004) 024301.
\newblock \href {http://dx.doi.org/10.1103/PhysRevA.70.024301}
  {\path{doi:10.1103/PhysRevA.70.024301}}.

\bibitem{Barreiro:2010:aa}
J.~T. Barreiro, P.~Schindler, O.~G{\"u}hne, T.~Monz, M.~Chwalla, C.~F. Roos,
  M.~Hennrich, R.~Blatt, Experimental multiparticle entanglement dynamics
  induced by decoherence, Nature Phys. 6 (2010) 943--946.

\bibitem{Viola:1998:uu}
L.~Viola, S.~Lloyd, Dynamical suppression of decoherence in two-state quantum
  systems, Phys. Rev. A 58 (1998) 2733--2744.

\bibitem{Viola:1999:zp}
L.~Viola, E.~Knill, S.~Lloyd, Dynamical decoupling of open quantum systems,
  Phys. Rev. Lett. 82 (1999) 2417--2421.

\bibitem{Zanardi:1999:oo}
P.~Zanardi, Symmetrizing evolutions, Phys. Lett. A 258 (1999) 77--91.

\bibitem{Viola:2000:pp}
L.~Viola, E.~Knill, S.~Lloyd, Dynamical generation of noiseless quantum
  subsystems, Phys. Rev. Lett. 85 (2000) 3520--3523.

\bibitem{Wu:2002:aa}
L.-A. Wu, D.~A. Lidar, Creating decoherence-free subspaces using strong and
  fast pulses, Phys. Rev. Lett. 88 (2002) 207902.

\bibitem{Wu:2002:bb}
L.-A. Wu, M.~S. Byrd, D.~A. Lidar, Efficient universal leakage elimination for
  physical and encoded qubits, Phys. Rev. Lett. 89 (2002) 127901.

\bibitem{Viola:1999:pp}
L.~Viola, S.~Lloyd, E.~Knill, Universal control of decoupled quantum systems,
  Phys. Rev. Lett. 83 (1999) 4888--4891.

\bibitem{West:2010:oo}
J.~R. West, D.~A. Lidar, B.~H. Fong, M.~F. Gyure, High fidelity quantum gates
  via dynamical decoupling, Phys. Rev. Lett. 105 (2010) 230503.

\bibitem{Gaitan:2008:uu}
F.~Gaitan, Quantum Error Correction and Fault Tolerant Quantum Computing, CRC
  Press, Boca Raton, 2008.

\bibitem{Cory:1998:uu}
D.~G. Cory, M.~D. Price, W.~Maas, E.~Knill, R.~Laflamme, W.~H. Zurek, T.~F.
  Havel, S.~S. Somaroo, Experimental quantum error correction, Phys. Rev. Lett.
  81 (1998) 2152--2155.

\bibitem{Lidar:2001:oo}
D.~A. Lidar, D.~Bacon, J.~Kempe, K.~B. Whaley, Decoherence-free subspaces for
  multiple-qubit errors. {II.} {U}niversal, fault-tolerant quantum computation,
  Phys. Rev. A 63 (2001) 022307.

\bibitem{Arndt:2014:oo}
M.~Arndt, K.~Hornberger, Testing the limits of quantum mechanical
  superpositions, Nature Phys. 10 (2014) 271--277.

\bibitem{Haroche:2006:hh}
S.~Haroche, J.-M. Raimond, Exploring the Quantum: Atoms, Cavities, and Photons,
  Oxford University Press, Oxford, 2006.

\bibitem{Kaiser:2001:tm}
R.~Kaiser, C.~Westbrook, F.~David (Eds.), Coherent Atomic Matter Waves, Les
  Houches Session LXXII, Les Houches Summer School Series, Springer, Berlin,
  2001.

\bibitem{Davidovich:1996:sa}
L.~Davidovich, M.~Brune, J.~M. Raimond, S.~Haroche, Mesoscopic quantum
  coherences in cavity {QED}: Preparation and decoherence monitoring schemes,
  Phys. Rev. A 53 (1996) 1295--1309.

\bibitem{Kuhr:2007:aa}
S.~Kuhr, S.~Gleyzes, C.~Guerlin, J.~Bernu, U.~B. Hoff, S.~Del{\'e}glise,
  S.~O.~M. Brune, J.-M. Raimond, Ultrahigh finesse {F}abry-{P}{\'e}rot
  superconducting resonator, Appl. Phys. Lett. 90 (2007) 164101.

\bibitem{Deleglise:2008:oo}
S.~Del{\'e}glise, I.~Dotsenko, C.~Sayrin, J.~Bernu, M.~Brune, J.-M. Raimond,
  S.~Haroche, Reconstruction of non-classical cavity field states with
  snapshots of their decoherence, Nature 455 (2008) 510--514.

\bibitem{Auffeves:2003:za}
A.~Auffeves, P.~Maioli, T.~Meunier, S.~Gleyzes, G.~Nogues, M.~Brune, J.~M.
  Raimond, S.~Haroche, Entanglement of a mesoscopic field with an atom induced
  by photon graininess in a cavity, Phys. Rev. Lett. 91 (2003) 230405.

\bibitem{Vlastakis:2013:pp}
B.~Vlastakis, G.~Kirchmair, Z.~Leghtas, S.~E. Nigg, L.~Frunzio, S.~M. Girvin,
  M.~Mirrahimi, M.~H. Devoret, R.~J. Schoelkopf, Deterministically encoding
  quantum information using 100-photon {S}chr{\"o}dinger cat states, Science
  342~(6158) (2013) 607--610.
\newblock \href {http://dx.doi.org/10.1126/science.1243289}
  {\path{doi:10.1126/science.1243289}}.

\bibitem{Hermann-Avigliano:2015:tt}
C.~Hermann-Avigliano, N.~Cisternas, M.~Brune, J.-M. Raimond, C.~Saavedra,
  Scheme for efficient generation of mesoscopic field-state superposition in
  cavity {QED}, Phys. Rev. A 91 (2015) 013815.

\bibitem{Maitre:1997:tv}
X.~Ma{\^i}tre, E.~Hagley, J.~Dreyer, A.~Maali, C.~W.~M. Brune, J.~M. Raimond,
  S.~Haroche, An experimental study of a {S}chr{\"o}dinger cat decoherence with
  atoms and cavities, J. Mod. Opt. 44 (1997) 2023--2032.

\bibitem{Brune:1992:zz}
M.~Brune, S.~Haroche, J.~M. Raimond, L.~Davidovich, N.~Zagury, Manipulation of
  photons in a cavity by dispersive atom--field coupling: Quantum-nondemolition
  measurements and generation of {``}{S}chr{\"o}dinger cat{''} states, Phys.
  Rev. A 45 (1992) 5193--5214.

\bibitem{Ramsey:1950:pp}
N.~F. Ramsey, A molecular beam resonance method with separated oscillating
  fields, Phys. Rev. 78 (1950) 695--699.

\bibitem{Gerlich:2011:aa}
S.~Gerlich, S.~Eibenberger, M.~Tomandl, S.~Nimmrichter, K.~Hornberger, P.~J.
  Fagan, J.~T{\"u}xen, M.~Mayor, M.~Arndt, Quantum interference of large
  organic molecules, Nature Comm. 2 (2012) 263.

\bibitem{Eibenberger:2013:az}
S.~Eibenberger, S.~Gerlich, M.~Arndt, M.~Mayor, J.~T{\"u}xen, Matter-wave
  interference with particles selected from a molecular library with masses
  exceeding 10,000 amu, Phys. Chem. Chem. Phys. 15 (2013) 14696--14700.

\bibitem{Jonsson:1961:rz}
C.~J{\"o}nsson, Elektroneninterferenzen an mehreren k{\"u}nstlich hergestellten
  {F}einspalten, Z. Physik 161 (1961) 454--474.

\bibitem{Jonsson:1974:rz}
C.~J{\"o}nsson, Electron diffraction at multiple slits, Am. J. Phys. 42 (1974)
  4--11.

\bibitem{Tonomura:1989:as}
A.~Tonomura, J.~Endo, T.~Matsuda, T.~Kawasaki, H.~Ezawa, Demonstration of
  single-electron buildup of an interference pattern, Am. J. Phys. 57 (1989)
  117--120.

\bibitem{Mansuripur:2009:zz}
M.~Mansuripur, Classical Optics and its Applications, 2nd Edition, Cambridge
  University Press, Cambridge, 2009.

\bibitem{Brezger:2002:mu}
B.~Brezger, L.~Hackerm{\"u}ller, S.~Uttenthaler, J.~Petschinka, M.~Arndt,
  A.~Zeilinger, Matter-wave interferometer for large molecules, Phys. Rev.
  Lett. 88 (2002) 100404.

\bibitem{Hackermueller:2002:wb}
L.~Hackerm{\"u}ller, S.~Uttenthaler, K.~Hornberger, E.~Reiger, B.~Brezger,
  A.~Zeilinger, M.~Arndt, Wave nature of biomolecules and fluorofullerenes,
  Phys. Rev. Lett. 91 (2003) 090408.

\bibitem{Hackermuller:2004:rd}
L.~Hackerm{\"u}ller, K.~Hornberger, B.~Brezger, A.~Zeilinger, M.~Arndt,
  Decoherence of matter waves by thermal emission of radiation, Nature 427
  (2004) 711--714.

\bibitem{Gerlich:2007:om}
S.~Gerlich, L.~Hackerm{\"u}ller, K.~Hornberger, A.~Stibor, H.~Ulbricht,
  F.~Goldfarb, T.~Savas, M.~M{\"u}ri, M.~Mayor, M.~Arndt, A
  {K}apitza--{D}irac--{T}albot--{L}au interferometer for highly polarizable
  molecules, Nature Phys. 3 (2007) 711.

\bibitem{Haslinger:2013:ii}
P.~Haslinger, N.~D{\"o}rre, P.~Geyer, J.~Rodewald, S.~Nimmrichter, M.~Arndt, A
  universal matter-wave interferometer with optical ionization gratings in the
  time domain, Nature Phys. 9 (2013) 144--148.

\bibitem{Hornberger:2005:mo}
K.~Hornberger, L.~Hackerm{\"u}ller, M.~Arndt, Influence of molecular
  temperature on the coherence of fullerenes in a near-field interferometer,
  Phys. Rev. A 71 (2005) 023601.

\bibitem{Hornberger:2006:tx}
K.~Hornberger, Thermal limitation of far-field matter-wave interference, Phys.
  Rev. A 73 (2006) 052102.

\bibitem{Kaltenbaek:2016:pp}
R.~Kaltenbaek, M.~Aspelmeyer, P.~F. Barker, A.~Bassi, J.~Bateman, K.~Bongs,
  S.~Bose, C.~Braxmaier, {\v{C}}.~Brukner, B.~Christophe, M.~Chwalla, P.-F.
  Cohadon, A.~M. Cruise, C.~Curceanu, K.~Dholakia, L.~Di{\'o}si,
  K.~D{\"o}ringshoff, W.~Ertmer, J.~Gieseler, N.~G{\"u}rlebeck,
  G.~Hechenblaikner, A.~Heidmann, S.~Herrmann, S.~Hossenfelder, U.~Johann,
  N.~Kiesel, M.~Kim, C.~L{\"a}mmerzahl, A.~Lambrecht, M.~Mazilu, G.~J. Milburn,
  H.~M{\"u}ller, L.~Novotny, M.~Paternostro, A.~Peters, I.~Pikovski, A.~{Pilan
  Zanoni}, E.~M. Rasel, S.~Reynaud, C.~J. Riedel, M.~Rodrigues, L.~Rondin,
  A.~Roura, W.~P. Schleich, J.~Schmiedmayer, T.~Schuldt, K.~C. Schwab,
  M.~Tajmar, G.~M. Tino, H.~Ulbricht, R.~Ursin, V.~Vedral, Macroscopic quantum
  resonators ({MAQRO}): 2015 update, EPJ Quantum Technology 3 (2016) 5.
\newblock \href {http://dx.doi.org/10.1140/epjqt/s40507-016-0043-7}
  {\path{doi:10.1140/epjqt/s40507-016-0043-7}}.

\bibitem{Leggett:1980:yt}
A.~J. Leggett, Macroscopic quantum systems and the quantum theory of
  measurement, Suppl. Prog. Theor. Phys. 69 (1980) 80--100.

\bibitem{Devoret:2013:pp}
M.~H. Devoret, R.~J. Schoelkopf, Superconducting circuits for quantum
  information: An outlook, Science 339 (2013) 1169--1174.

\bibitem{Likharev:1979:ii}
K.~K. Likharev, Superconducting weak links, Rev. Mod. Phys. 51 (1979) 101--159.

\bibitem{Makhlin:2001:oo}
Y.~Makhlin, G.~Sch{\"o}n, A.~Shnirman, Quantum-state engineering with
  {J}osephson-junction devices, Rev. Mod. Phys. 73 (2001) 357--400.

\bibitem{Chiorescu:2003:ta}
I.~Chiorescu, Y.~Nakamura, C.~J. P.~M. Harmans, J.~E. Mooij, Coherent quantum
  dynamics of a superconducting flux qubit, Science 21 (2003) 1869--1871.

\bibitem{Ilichev:2003:tv}
E.~Il'ichev, N.~Oukhanski, A.~Izmalkov, T.~Wagner, M.~Grajcar, H.-G. Meyer,
  A.~Y. Smirnov, A.~M. van~den Brink, M.~H.~S. Amin, A.~M. Zagoskin, Continuous
  monitoring of {R}abi oscillations in a {J}osephson flux qubit, Phys. Rev.
  Lett. 91 (2003) 097906.

\bibitem{Silvestrini:1996:ii}
P.~Silvestrini, B.~Ruggiero, Y.~N. Ovchinnikov, Resonant macroscopic quantum
  tunneling in {SQUID} systems, Phys. Rev. B 54 (1996) 1246--1250.

\bibitem{Rouse:1998:om}
R.~Rouse, S.~Han, J.~E. Lukens, Resonant tunneling between macroscopically
  distinct levels of a {SQUID}, in: G.~D. Palazzi, C.~Cosmelli, L.~Zanello
  (Eds.), Phenomenology of Unification from Present to Future, World Scientic,
  Singapore, 1998, pp. 207--209.

\bibitem{Yoshihara:2006:ii}
F.~Yoshihara, K.~Harrabi, A.~O. Niskanen, Y.~Nakamura, J.~S. Tsai, Decoherence
  of flux qubits due to $1/f$ flux noise, Phys. Rev. Lett. 97 (2006) 167001.
\newblock \href {http://dx.doi.org/10.1103/PhysRevLett.97.167001}
  {\path{doi:10.1103/PhysRevLett.97.167001}}.

\bibitem{Bialczak:2007:uu}
R.~C. Bialczak, R.~McDermott, M.~Ansmann, M.~Hofheinz, N.~Katz, E.~Lucero,
  M.~Neeley, A.~D. O'Connell, H.~Wang, A.~N. Cleland, J.~M. Martinis, $1/f$
  flux noise in {J}osephson phase qubits, Phys. Rev. Lett. 99 (2007) 187006.
\newblock \href {http://dx.doi.org/10.1103/PhysRevLett.99.187006}
  {\path{doi:10.1103/PhysRevLett.99.187006}}.

\bibitem{Bertet:2005:un}
P.~Bertet, I.~Chiorescu, G.~Burkard, K.~Semba, C.~J. P.~M. Harmans, D.~P.
  DiVincenzo, J.~E. Mooij, Dephasing of a superconducting qubit induced by
  photon noise, Phys. Rev. Lett. 95 (2005) 257002.

\bibitem{Bouchiat:1998:ii}
V.~Bouchiat, D.~Vion, P.~Joyez, D.~Esteve, M.~H. Devoret, Quantum coherence
  with a single {C}ooper pair, Phys. Scr. 1998 (1998) 165--170.

\bibitem{Nakamura:1999:ub}
Y.~Nakamura, Y.~A. Pashkin, J.~S. Tsai, Coherent control of macroscopic quantum
  states in a single-{C}ooper-pair box, Nature 398 (1999) 786--788.

\bibitem{Vion:2002:oo}
D.~Vion, A.~Aassime, A.~Cottet, P.~Joyez, H.~Pothier, C.~Urbina, D.~Esteve,
  M.~H. Devoret, Manipulating the quantum state of an electrical circuit,
  Science 296 (2002) 886--889.

\bibitem{Yu:2002:yb}
Y.~Yu, S.~Han, X.~Chu, S.-I. Chu, Z.~Wang, Coherent temporal oscillations of
  macroscopic quantum states in a {J}osephson junction, Science 296 (2002)
  889--892.

\bibitem{Martinis:2002:qq}
J.~M. Martinis, S.~Nam, J.~Aumentado, C.~Urbina, Rabi oscillations in a large
  {J}osephson-junction qubit, Phys. Rev. Lett. 89 (2002) 117901.

\bibitem{Rigetti:2012:aa}
C.~Rigetti, J.~M. Gambetta, S.~Poletto, B.~L.~T. Plourde, J.~M. Chow, A.~D.
  C\'orcoles, J.~A. Smolin, S.~T. Merkel, J.~R. Rozen, G.~A. Keefe, M.~B.
  Rothwell, M.~B. Ketchen, M.~Steffen, Superconducting qubit in a waveguide
  cavity with a coherence time approaching 0.1 ms, Phys. Rev. B 86 (2012)
  100506.
\newblock \href {http://dx.doi.org/10.1103/PhysRevB.86.100506}
  {\path{doi:10.1103/PhysRevB.86.100506}}.

\bibitem{Sears:2012:ee}
A.~P. Sears, A.~Petrenko, G.~Catelani, L.~Sun, H.~Paik, G.~Kirchmair,
  L.~Frunzio, L.~I. Glazman, S.~M. Girvin, R.~J. Schoelkopf, Photon shot noise
  dephasing in the strong-dispersive limit of circuit {QED}, Phys. Rev. B 86
  (2012) 180504.
\newblock \href {http://dx.doi.org/10.1103/PhysRevB.86.180504}
  {\path{doi:10.1103/PhysRevB.86.180504}}.

\bibitem{Catelani:2012:zz}
G.~Catelani, S.~E. Nigg, S.~M. Girvin, R.~J. Schoelkopf, L.~I. Glazman,
  Decoherence of superconducting qubits caused by quasiparticle tunneling,
  Phys. Rev. B 86 (2012) 184514.
\newblock \href {http://dx.doi.org/10.1103/PhysRevB.86.184514}
  {\path{doi:10.1103/PhysRevB.86.184514}}.

\bibitem{Wal:2003:pp}
C.~H. {van der Wal}, F.~K. Wilhelm, C.~J. P.~M. Harmans, J.~E. Mooij,
  Engineering decoherence in {J}osephson persistent-current qubits, Eur. Phys.
  J. B 31 (2003) 111--124.

\bibitem{Martinis:2005:zz}
J.~M. Martinis, K.~B. Cooper, R.~McDermott, M.~Steffen, M.~Ansmann, K.~Osborn,
  K.~Cicak, S.~Oh, D.~P. Pappas, R.~W. Simmonds, C.~C. Yu, Decoherence in
  {J}osephson qubits from dielectric loss, Phys. Rev. Lett. 95 (2005) 210503.

\bibitem{Cirac:1995:tt}
J.~I. Cirac, P.~Zoller, Quantum computations with cold trapped ions, Phys. Rev.
  Lett. 74 (1995) 4091--4094.

\bibitem{Monroe:1995:oo}
C.~Monroe, D.~M. Meekhof, B.~E. King, W.~M. Itano, D.~J. Wineland,
  Demonstration of a fundamental quantum logic gate, Phys. Rev. Lett. 75 (1995)
  4714--4717.
\newblock \href {http://dx.doi.org/10.1103/PhysRevLett.75.4714}
  {\path{doi:10.1103/PhysRevLett.75.4714}}.

\bibitem{Barrett:2004:oo}
M.~D. Barrett, J.~Chiaverini, T.~Schaetz, J.~Britton, W.~M. Itano, J.~D. Jost,
  E.~Knill, C.~Langer, D.~Leibfried, R.~Ozeri, D.~J. Wineland, Deterministic
  quantum teleportation of atomic qubits, Nature 429 (2004) 737--739.

\bibitem{Riebe:2004:qq}
M.~Riebe, H.~H{\"a}ffner, C.~F. Roos, W.~H{\"a}nsel, J.~Benhelm, G.~P.~T.
  Lancaster, T.~W. K{\"o}rber, C.~Becher, F.~Schmidt-Kaler, D.~F.~V. James,
  R.~Blatt, Deterministic quantum teleportation with atoms, Nature 429 (2004)
  734--737.

\bibitem{Chiaverini:2004:aa}
J.~Chiaverini, D.~Leibfried, T.~Schaetz, M.~D. Barrett, R.~B. Blakestad,
  J.~Britton, W.~M. Itano, J.~D. Jost, E.~Knill, C.~Langer, R.~Ozeri, D.~J.
  Wineland, Realization of quantum error correction, Nature 432 (2004)
  602--605.

\bibitem{Haffner:2005:sc}
H.~H{\"a}ffner, W.~H{\"a}nsel, C.~F. Roos, J.~Benhelm, D.~{Chek-al-kar},
  M.~Chwalla, T.~K{\"o}rber, U.~D. Rapol, M.~Riebe, P.~O. Schmidt, C.~Becher,
  O.~G{\"u}hne, W.~D{\"u}rr, R.~Blatt, Scalable multiparticle entanglement of
  trapped ions, Nature 438 (2005) 643--646.

\bibitem{Reichle:2006:ii}
R.~Reichle, D.~Leibfried, E.~Knill, J.~Britton, R.~B. Blakestad, J.~D. Jost,
  C.~Langer, R.~Ozeri, S.~Seidelin, D.~J. Wineland, Experimental purification
  of two-atom entanglement, Nature 443 (2006) 838--841.

\bibitem{Paul:1990:oo}
W.~Paul, Electromagnetic traps for charged and neutral particles, Rev. Mod.
  Phys. 62 (1990) 531--540.
\newblock \href {http://dx.doi.org/10.1103/RevModPhys.62.531}
  {\path{doi:10.1103/RevModPhys.62.531}}.

\bibitem{Turchette:2000:oa}
Q.~A. Turchette, D.~Kielpinski, B.~E. King, D.~Leibfried, D.~M. Meekhof, C.~J.
  Myatt, M.~A. Rowe, C.~A. Sackett, C.~S. Wood, W.~M. Itano, C.~Monroe, D.~J.
  Wineland, Heating of trapped ions from the quantum ground state, Phys. Rev. A
  61 (2000) 063418.

\bibitem{Brouard:2004:in}
S.~Brouard, J.~Plata, Decoherence of trapped-ion internal and vibrational
  modes: {T}he effect of fluctuating magnetic fields, Phys. Rev. A 70 (2004)
  013413.

\bibitem{Grotz:2006:km}
T.~Grotz, L.~Heaney, W.~T. Strunz, Quantum dynamics in fluctuating traps:
  {M}aster equation, decoherence, and heating, Phys. Rev. A 74 (2006) 022102.

\bibitem{Stick:2006:aa}
D.~Stick, W.~K. Hensinger, S.~Olmschenk, M.~J. Madsen, K.~Schwab, C.~Monroe,
  Ion trap in a semiconductor chip, Nature Phys. 2 (2006) 36--39.

\bibitem{Seidelin:2006:rz}
S.~Seidelin, J.~Chiaverini, R.~Reichle, J.~J. Bollinger, D.~Leibfried,
  J.~Britton, J.~H. Wesenberg, R.~B. Blakestad, R.~J. Epstein, D.~B. Hume,
  W.~M. Itano, J.~D. Jost, C.~Langer, R.~Ozeri, N.~Shiga, D.~J. Wineland,
  Microfabricated surface-electrode ion trap for scalable quantum information
  processing, Phys. Rev. Lett. 96 (2006) 253003.

\bibitem{SchmidtKaler:2003:pp}
F.~Schmidt-Kaler, S.~Gulde, M.~Riebe, T.~Deuschle, A.~Kreuter, G.~Lancaster,
  C.~Becher, J.~Eschner, H.~H{\"a}ffner, R.~Blatt, The coherence of qubits
  based on single {Ca} ions, J. Phys. B 36~(3) (2003) 623--636.

\bibitem{Haljan:2005:oo}
P.~C. Haljan, P.~J. Lee, K.-A. Brickman, M.~Acton, L.~Deslauriers, C.~Monroe,
  Entanglement of trapped-ion clock states, Phys. Rev. A 72 (2005) 062316.
\newblock \href {http://dx.doi.org/10.1103/PhysRevA.72.062316}
  {\path{doi:10.1103/PhysRevA.72.062316}}.

\bibitem{Benhelm:2008:oo}
J.~Benhelm, G.~Kirchmair, C.~F. Roos, R.~Blatt, Experimental
  quantum-information processing with ${^{43}\text{C}\text{a}}^{+}$ ions, Phys.
  Rev. A 77 (2008) 062306.
\newblock \href {http://dx.doi.org/10.1103/PhysRevA.77.062306}
  {\path{doi:10.1103/PhysRevA.77.062306}}.

\bibitem{Steane:2000:ii}
A.~Steane, C.~F. Roos, D.~Stevens, A.~Mundt, D.~Leibfried, F.~Schmidt-Kaler,
  R.~Blatt, Speed of ion-trap quantum-information processors, Phys. Rev. A 62
  (2000) 042305.
\newblock \href {http://dx.doi.org/10.1103/PhysRevA.62.042305}
  {\path{doi:10.1103/PhysRevA.62.042305}}.

\bibitem{Haffner:2003:oo}
H.~H\"affner, S.~Gulde, M.~Riebe, G.~Lancaster, C.~Becher, J.~Eschner,
  F.~Schmidt-Kaler, R.~Blatt, Precision measurement and compensation of optical
  {S}tark shifts for an ion-trap quantum processor, Phys. Rev. Lett. 90 (2003)
  143602.
\newblock \href {http://dx.doi.org/10.1103/PhysRevLett.90.143602}
  {\path{doi:10.1103/PhysRevLett.90.143602}}.

\bibitem{Leibfried:2003:mm}
D.~Leibfried, B.~DeMarco, V.~Meyer, D.~Lucas, M.~Barrett, J.~Britton, W.~M.
  Itano, B.~Jelenkovi{\'c}, C.~Langer, T.~Rosenband, D.~J. Wineland,
  Experimental demonstration of a robust, high-fidelity geometric two ion-qubit
  phase gate, Nature 422 (2003) 412--415.

\bibitem{Wineland:2013:pp}
D.~J. Wineland, Nobel lecture: Superposition, entanglement, and raising
  {S}chr{\"o}dinger's cat, Rev. Mod. Phys. 85 (2013) 1103--1114.
\newblock \href {http://dx.doi.org/10.1103/RevModPhys.85.1103}
  {\path{doi:10.1103/RevModPhys.85.1103}}.

\bibitem{Monroe:1996:tv}
C.~Monroe, D.~M. Meekhof, B.~E. King, D.~J.~A. Wineland, {``}{S}chr{\"o}dinger
  cat{''} superposition state of an atom, Science 272 (1996) 1131--1136.

\bibitem{Horodecki:1998:oo}
M.~Horodecki, P.~Horodecki, R.~Horodecki, Mixed-state entanglement and
  distillation: Is there a ``bound'' entanglement in nature?, Phys. Rev. Lett.
  80 (1998) 5239--5242.
\newblock \href {http://dx.doi.org/10.1103/PhysRevLett.80.5239}
  {\path{doi:10.1103/PhysRevLett.80.5239}}.

\bibitem{Hanson:2007:pp}
R.~Hanson, L.~P. Kouwenhoven, J.~R. Petta, S.~Tarucha, L.~M.~K. Vandersypen,
  Spins in few-electron quantum dots, Rev. Mod. Phys. 79 (2007) 1217--1265.

\bibitem{Kuhlmann:2013:aa}
A.~V. Kuhlmann, J.~Houel, A.~Ludwig, L.~Greuter, D.~Reuter, A.~D. Wieck,
  M.~Poggio, R.~J. Warburton, Charge noise and spin noise in a semiconductor
  quantum device, Nature Phys. 9 (2013) 570--575.

\bibitem{Arnold:2014:oo}
C.~Arnold, V.~Loo, A.~Lema\^{\i}tre, I.~Sagnes, O.~Krebs, P.~Voisin,
  P.~Senellart, L.~Lanco, Cavity-enhanced real-time monitoring of single-charge
  jumps at the microsecond time scale, Phys. Rev. X 4 (2014) 021004.
\newblock \href {http://dx.doi.org/10.1103/PhysRevX.4.021004}
  {\path{doi:10.1103/PhysRevX.4.021004}}.

\bibitem{Fischer:2009:ii}
J.~Fischer, M.~Trif, W.~A. Coish, D.~Loss, Spin interactions, relaxation and
  decoherence in quantum dots, Solid State Comm. 149 (2009) 1443--1450.

\bibitem{Urbaszek:2013:pp}
B.~Urbaszek, X.~Marie, T.~Amand, O.~Krebs, P.~Voisin, P.~Maletinsky,
  A.~H\"ogele, A.~Imamoglu, Nuclear spin physics in quantum dots: An optical
  investigation, Rev. Mod. Phys. 85 (2013) 79--133.

\bibitem{Delteil:2014:aa}
A.~Delteil, W.-b. Gao, P.~Fallahi, J.~Miguel-Sanchez,
  A.~Imamo\ifmmode~\breve{g}\else \u{g}\fi{}lu, Observation of quantum jumps of
  a single quantum dot spin using submicrosecond single-shot optical readout,
  Phys. Rev. Lett. 112 (2014) 116802.
\newblock \href {http://dx.doi.org/10.1103/PhysRevLett.112.116802}
  {\path{doi:10.1103/PhysRevLett.112.116802}}.

\bibitem{Tighineanu:2018:ii}
P.~Tighineanu, C.~L. Dree\ss{}en, C.~Flindt, P.~Lodahl, A.~S. S\o{}rensen,
  Phonon decoherence of quantum dots in photonic structures: Broadening of the
  zero-phonon line and the role of dimensionality, Phys. Rev. Lett. 120 (2018)
  257401.

\bibitem{Aspelmeyer:2013:aa}
M.~Aspelmeyer, T.~J. Kippenberg, F.~Marquardt, Cavity optomechanics, Rev. Mod.
  Phys. 86 (2014) 1391--1452.

\bibitem{Poot:2012:aa}
M.~Poot, H.~S.~J. {van der Zant}, Mechanical systems in the quantum regime,
  Phys. Rep. 511 (2012) 273--335.

\bibitem{Greenberg:2012:zz}
Y.~Greenberg, Y.~A. Pashkin, E.~Il{'}ichev, Nanomechanical resonators,
  Physics-Uspekhi 55~(4) (2012) 382--407.

\bibitem{Blencowe:2004:mm}
M.~Blencowe, Quantum electromechanical systems, Phys. Rep. 395 (2004) 159--222.

\bibitem{Mohanty:2002:mm}
P.~Mohanty, D.~A. Harrington, K.~L. Ekinci, Y.~T. Yang, M.~J. Murphy, M.~L.
  Roukes, Intrinsic dissipation in high-frequency micromechanical resonators,
  Phys. Rev. B 66 (2002) 085416.

\bibitem{Ahn:2003:mt}
K.-H. Ahn, P.~Mohanty, Quantum friction of micromechanical resonators at low
  temperatures, Phys. Rev. Lett. 90 (2003) 085504.

\bibitem{Blencowe:2005:cc}
M.~P. Blencowe, Nanoelectromechanical systems, Contemporary Physics 46 (2005)
  249--264.

\bibitem{Zolfagharkhani:2005:tv}
G.~Zolfagharkhani, A.~Gaidarzhy, S.-B. Shim, R.~L. Badzey, P.~Mohanty, Quantum
  friction in nanomechanical oscillators at millikelvin temperatures, Phys.
  Rev. B 72 (2005) 224101.

\bibitem{Seoanez:2006:yb}
C.~Seo{\'a}nez, F.~Guinea, A.~H. {Castro Neto}, Dissipation due to two-level
  systems in nano-mechanical devices, Europhys. Lett. 78 (2007) 60002.

\bibitem{Seoanez:2007:um}
C.~Seo{\'a}nez, F.~Guinea, A.~H. {Castro Neto}, Surface dissipation in
  nanoelectromechanical systems: {U}nified description with the standard
  tunneling model and effects of metallic electrodes, Phys. Rev. B 77 (2008)
  125107.

\bibitem{Remus:2009:im}
L.~G. Remus, M.~P. Blencowe, Y.~Tanaka, Damping and decoherence of a
  nanomechanical resonator due to a few two-level systems, Phys. Rev. B 80
  (2009) 174103.

\bibitem{Fong:2012:aa}
K.~Y. Fong, W.~H.~P. Pernice, H.~X. Tang, Frequency and phase noise of
  ultrahigh $q$ silicon nitride nanomechanical resonators, Phys. Rev. B 85
  (2012) 161410.
\newblock \href {http://dx.doi.org/10.1103/PhysRevB.85.161410}
  {\path{doi:10.1103/PhysRevB.85.161410}}.

\bibitem{Zhang:2014:oo}
Y.~Zhang, J.~Moser, J.~G\"uttinger, A.~Bachtold, M.~I. Dykman, Interplay of
  driving and frequency noise in the spectra of vibrational systems, Phys. Rev.
  Lett. 113 (2014) 255502.
\newblock \href {http://dx.doi.org/10.1103/PhysRevLett.113.255502}
  {\path{doi:10.1103/PhysRevLett.113.255502}}.

\bibitem{Miao:2014:ii}
T.~Miao, S.~Yeom, P.~Wang, B.~Standley, M.~Bockrath, Graphene
  nanoelectromechanical systems as stochastic-frequency oscillators, Nano Lett.
  14~(6) (2014) 2982--2987.

\bibitem{Moser:2014:uu}
J.~Moser, A.~Eichler, J.~G{\"u}ttinger, M.~I. Dykman, A.~Bachtold, Nanotube
  mechanical resonators with quality factors of up to 5 million, Nature
  Nanotechnology 9 (2014) 1007--1011.

\bibitem{Maillet:2016:zz}
O.~Maillet, F.~Vavrek, A.~D. Fefferman, O.~Bourgeois, E.~Collin, Classical
  decoherence in a nanomechanical resonator, New J. Phys. 18 (2016) 073022.

\bibitem{Andrews:1997:um}
M.~R. Andrews, C.~G. Townsend, H.-J. Miesner, D.~S. Durfee, D.~M. Kurn,
  W.~Ketterle, Observation of interference between two {B}ose condensates,
  Science 275 (1997) 637--641.

\bibitem{Shin:2004:lo}
Y.~Shin, M.~Saba, T.~A. Pasquini, W.~Ketterle, D.~E. Pritchard, A.~E.
  Leanhardt, Atom interferometry with {B}ose--{E}instein condensates in a
  double-well potential, Phys. Rev. Lett. 92 (2004) 050405.

\bibitem{Gross:2010:gg}
C.~Gross, T.~Zibold, E.~Nicklas, J.~Est{\`e}ve, M.~K. Oberthaler, Nonlinear
  atom interferometer surpasses classical precision limit, Nature 464 (2010)
  1165--1169.

\bibitem{Tura:2014:oo}
J.~Tura, R.~Augusiak, A.~B. Sainz, T.~V{\'e}rtesi, M.~Lewenstein, A.~Ac{\'i}n,
  Detecting nonlocality in many-body quantum states, Science 344 (2014)
  1256--1258.

\bibitem{Schmied:2016:ll}
R.~Schmied, J.-D. Bancal, B.~Allard, M.~Fadel, V.~Scarani, P.~Treutlein,
  N.~Sangouard, Bell correlations in a {B}ose--{E}instein condensate, Science
  352 (2016) 441--444.

\bibitem{Pezze:2018:uu}
L.~Pezz\`e, A.~Smerzi, M.~K. Oberthaler, R.~Schmied, P.~Treutlein, Quantum
  metrology with nonclassical states of atomic ensembles, Rev. Mod. Phys. 90
  (2018) 035005.
\newblock \href {http://dx.doi.org/10.1103/RevModPhys.90.035005}
  {\path{doi:10.1103/RevModPhys.90.035005}}.

\bibitem{Cirac:1998:mm}
J.~I. Cirac, M.~Lewenstein, K.~M{\o}lmer, P.~Zoller, Quantum superposition
  states of {B}ose--{E}instein condensates, Phys. Rev. A 57 (1998) 1208--1218.

\bibitem{Ruostekoski:1998:mm}
J.~Ruostekoski, M.~J. Collett, R.~Graham, D.~F. Walls, Macroscopic
  superpositions of {B}ose--{E}instein condensates, Phys. Rev. A 57 (1998)
  511--517.

\bibitem{Gordon:1999:mh}
D.~Gordon, C.~M. Savage, Creating macroscopic quantum superpositions with
  {B}ose--{E}instein condensates, Phys. Rev. A 59 (1999) 4623--4629.

\bibitem{Dunningham:2001:da}
J.~A. Dunningham, K.~Burnett, Proposals for creating {S}chr{\"o}dinger cat
  states in {B}ose--{E}instein condensates, J. Mod. Opt. 48 (2001) 1837--1853.

\bibitem{Calsamiglia:2001:tt}
J.~Calsamiglia, M.~Mackie, K.-A. Suominen, Superposition of macroscopic numbers
  of atoms and molecules, Phys. Rev. Lett. 87 (2001) 160403.

\bibitem{Louis:2001:mu}
P.~J.~Y. Louis, P.~M.~R. Brydon, C.~M. Savage, Macroscopic quantum
  superposition states in {B}ose--{E}instein condensates: Decoherence and many
  modes, Phys. Rev. A 64 (2001) 053613.

\bibitem{Micheli:2003:jn}
A.~Micheli, D.~Jaksch, J.~I. Cirac, P.~Zoller, Many particle entanglement in
  two-component {B}ose--{E}instein condensates, Phys. Rev. A 67 (2003) 013601.

\bibitem{Howl:2017:aa}
R.~Howl, C.~Sab{\'{\i}}n, L.~Hackerm{\"u}ller, I.~Fuentes, Quantum decoherence
  of phonons in {B}ose--{E}instein condensates, J. Phys. B 51~(1) (2017)
  015303.
\newblock \href {http://dx.doi.org/10.1088/1361-6455/aa9622}
  {\path{doi:10.1088/1361-6455/aa9622}}.

\bibitem{Schrinski:2017:yy}
B.~Schrinski, K.~Hornberger, S.~Nimmrichter, Sensing spontaneous collapse and
  decoherence with interfering {B}ose--{E}instein condensates, Quantum Sci.
  Technol. 2 (2017) 044010.

\bibitem{Bassi:2003:yb}
A.~Bassi, G.~C. Ghirardi, Dynamical reduction models, Phys. Rep. 379 (2003)
  257--426.

\bibitem{Adler:2007:um}
S.~L. Adler, Lower and upper bounds on {CSL} parameters from latent image
  formation and {IGM} heating, J. Phys. A 40 (2007) 2935--2957.

\bibitem{Bassi:2010:aa}
A.~Bassi, D.-A. Deckert, L.~Ferialdi, Breaking quantum linearity: {C}onstraints
  from human perception and cosmological implications, EPL 92 (2010) 50006.

\bibitem{Marshall:2003:om}
W.~Marshall, C.~Simon, R.~Penrose, D.~Bouwmeester, Towards quantum
  superpositions of a mirror, Phys. Rev. Lett 91 (2003) 130401.

\bibitem{Bassi:2005:om}
A.~Bassi, E.~Ippoliti, S.~L. Adler, Towards quantum superpositions of a mirror:
  An exact open systems analysis, Phys. Rev. Lett. 94 (2005) 030401.

\bibitem{Pikovski:2012:aa}
I.~Pikovski, M.~R. Vanner, M.~Aspelmeyer, M.~S. Kim, {\v C}.~Brukner, Probing
  {P}lanck-scale physics with quantum optics, Nature Phys. 8 (2012) 393--397.

\bibitem{Wan:2016:oo}
C.~Wan, M.~Scala, G.~W. Morley, A.~A. Rahman, H.~Ulbricht, J.~Bateman, P.~F.
  Barker, S.~Bose, M.~S. Kim, Free nano-object {R}amsey interferometry for
  large quantum superpositions, Phys. Rev. Lett. 117 (2016) 143003.

\bibitem{Stickler:2018:ii}
B.~A. Stickler, B.~Papendell, S.~Kuhn, B.~Schrinski, J.~Millen, M.~Arndt,
  K.~Hornberger, Probing macroscopic quantum superpositions with nanorotors,
  New J. Phys. 20 (2018) 122001.

\bibitem{Nimmrichter:2013:aa}
S.~Nimmrichter, K.~Hornberger, Macroscopicity of mechanical quantum
  superposition states, Phys. Rev. Lett. 110 (2013) 160403.

\bibitem{Romero:2011:aa}
O.~Romero-Isart, A.~C. Pflanzer, F.~Blaser, R.~Kaltenbaek, N.~Kiesel,
  M.~Aspelmeyer, J.~I. Cirac, Large quantum superpositions and interference of
  massive nanometer-sized objects, Phys. Rev. Lett. 107 (2011) 020405.

\bibitem{Pikovski:2015:oo}
I.~Pikovski, M.~Zych, F.~Costa, {\v C}.~Brukner, Universal decoherence due to
  gravitational time dilation, Nature Phys. (2015) 668--672.

\bibitem{Zeilinger:1982:oo}
A.~Zeilinger, R.~Gaehler, C.~Shull, W.~Treimer, Experimental status and recent
  results of neutron interference optics, AIP Conf. Proc. 89 (1982) 93--99.

\bibitem{Keith:1988:uu}
D.~W. Keith, M.~L. Schattenburg, H.~I. Smith, D.~E. Pritchard, Diffraction of
  atoms by a transmission grating, Phys. Rev. Lett. 61 (1988) 1580--1583.
\newblock \href {http://dx.doi.org/10.1103/PhysRevLett.61.1580}
  {\path{doi:10.1103/PhysRevLett.61.1580}}.

\bibitem{Chung:2009:oo}
K.-Y. Chung, S.-w. Chiow, S.~Herrmann, S.~Chu, H.~M\"uller, Atom interferometry
  tests of local {L}orentz invariance in gravity and electrodynamics, Phys.
  Rev. D 80 (2009) 016002.
\newblock \href {http://dx.doi.org/10.1103/PhysRevD.80.016002}
  {\path{doi:10.1103/PhysRevD.80.016002}}.

\bibitem{Teufel:2011:oo}
J.~D. Teufel, T.~Donner, D.~Li, J.~W. Harlow, M.~S. Allman, K.~Cicak, A.~J.
  Sirois, J.~D. Whittaker, K.~W. Lehnert, R.~W. Simmonds, Sideband cooling of
  micromechanical motion to the quantum ground state, Nature 475 (2011)
  359--363.

\bibitem{Dur:2002:pp}
W.~D\"ur, C.~Simon, J.~I. Cirac, Effective size of certain macroscopic quantum
  superpositions, Phys. Rev. Lett. 89 (2002) 210402.
\newblock \href {http://dx.doi.org/10.1103/PhysRevLett.89.210402}
  {\path{doi:10.1103/PhysRevLett.89.210402}}.

\bibitem{Bjork:2004:pp}
G.~Bj{\"o}rk, P.~G.~L. Mana, A size criterion for macroscopic superposition
  states, J. Optics B 6~(11) (2004) 429--436.

\bibitem{Korsbakken:2007:pp}
J.~I. Korsbakken, K.~B. Whaley, J.~Dubois, J.~I. Cirac, Measurement-based
  measure of the size of macroscopic quantum superpositions, Phys. Rev. A 75
  (2007) 042106.
\newblock \href {http://dx.doi.org/10.1103/PhysRevA.75.042106}
  {\path{doi:10.1103/PhysRevA.75.042106}}.

\bibitem{Marquardt:2008:ii}
F.~Marquardt, B.~Abel, J.~von Delft, Measuring the size of a quantum
  superposition of many-body states, Phys. Rev. A 78 (2008) 012109.
\newblock \href {http://dx.doi.org/10.1103/PhysRevA.78.012109}
  {\path{doi:10.1103/PhysRevA.78.012109}}.

\bibitem{Lee:2011:oo}
C.-W. Lee, H.~Jeong, Quantification of macroscopic quantum superpositions
  within phase space, Phys. Rev. Lett. 106 (2011) 220401.
\newblock \href {http://dx.doi.org/10.1103/PhysRevLett.106.220401}
  {\path{doi:10.1103/PhysRevLett.106.220401}}.

\bibitem{Frowis:2012:zz}
F.~Fr{\"o}wis, W.~D{\"u}r, Measures of macroscopicity for quantum spin systems,
  New J. Phys. 14~(9) (2012) 093039.

\bibitem{Bacciagaluppi:2006:yq}
G.~Bacciagaluppi, A.~Valentini, Quantum Theory at the Crossroads: Reconsidering
  the 1927 Solvay Conference, Cambridge University Press, Cambridge, 2007, also
  available as eprint quant-ph/0609184.

\bibitem{Bub:1997:iq}
J.~Bub, Interpreting the Quantum World, 1st Edition, Cambridge University
  Press, Cambridge, England, 1997.

\bibitem{Wigner:1963:yt}
E.~P. Wigner, The problem of measurement, Am. J. Phys. 31 (1963) 6--15.

\bibitem{Fine:1970:iq}
A.~Fine, Insolubility of the quantum measurement problem, Phys. Rev. D 2 (1970)
  2783--2787.

\bibitem{Wallace:2008:ii}
D.~Wallace, Philosophy of quantum mechanics, in: D.~Rickles (Ed.), The Ashgate
  Companion to Contemporary Philosophy of Physics, Routledge, New York, NY,
  2008, pp. 16--98.

\bibitem{Schlosshauer:2011:ee}
M.~Schlosshauer, Elegance and Enigma: The Quantum Interviews, 1st Edition,
  Springer, Berlin/Heidelberg, 2011.

\bibitem{Schlosshauer:2006:rw}
M.~Schlosshauer, A.~Fine, Decoherence and the foundations of quantum mechanics,
  in: J.~Evans, A.~Thorndike (Eds.), Quantum Mechanics at the Crossroads: New
  Perspectives from History, Philosophy and Physics, Springer, Berlin, 2006,
  pp. 125--148.

\bibitem{Zurek:2013:xx}
W.~H. Zurek, Wave-packet collapse and the core quantum postulates:
  {D}iscreteness of quantum jumps from unitarity, repeatability, and actionable
  information, Phys. Rev. A 87 (2013) 052111.

\bibitem{Zurek:2018:om}
W.~H. Zurek, Quantum reversibility is relative, or does a quantum measurement
  reset initial conditions?, Phil. Trans. R. Soc. A 376 (2018) 20170315.

\bibitem{Wallace:2010:im}
D.~Wallace, Decoherence and ontology, in: S.~Saunders, J.~Barrett, A.~Kent,
  D.~Wallace (Eds.), Many Worlds? {E}verett, Quantum Theory and Reality, Oxford
  University Press, Oxford, 2010, pp. 53--72.

\bibitem{Rovelli:1996:rq}
C.~Rovelli, Relational quantum mechanics, Int. J. Theor. Phys. 35 (1996)
  1637--1678.

\bibitem{Mermin:1998:ii}
N.~D. Mermin, The {I}thaca interpretation of quantum mechanics, Pramana 51
  (1998) 549--565.

\bibitem{Butterfield:2001:ua}
J.~N. Butterfield, Some worlds of quantum theory, in: R.~J. Russell,
  P.~Clayton, K.~Wegter-McNelly, J.~Polkinghorne (Eds.), Quantum Mechanics:
  Scientific Perspectives on Divine Action, Vatican Observatory and The Center
  for Theology and the Natural Sciences, Vatican City State, 2001, pp.
  111--140, also available as an eprint from the Pittsburgh Philosophy of
  Science Archive at \href{http://philsci-archive.pitt.edu/archive/00000203}{\path{http://philsci-archive.pitt.edu/archive/00000203}}.

\bibitem{Wallace:2003:iq}
D.~Wallace, Worlds in the {E}verett interpretation, Stud. Hist. Philos. Mod.
  Phys. 33 (2002) 637--661.

\bibitem{Wallace:2003:iz}
D.~Wallace, Everett and structure, Stud. Hist. Philos. Mod. Phys. 34 (2003)
  87--105.

\bibitem{Zurek:2004:yb}
W.~H. Zurek, Probabilities from entanglement, {B}orn's rule $p_k= |\psi_k|^2$
  from envariance, Phys. Rev. A 71 (2005) 052105.

\bibitem{Zurek:2003:rv}
W.~H. Zurek, Environment-assisted invariance, entanglement, and probabilities
  in quantum physics, Phys. Rev. Lett. 90 (2003) 120404.

\bibitem{Schlosshauer:2003:ms}
M.~Schlosshauer, A.~Fine, On {Z}urek's derivation of the {B}orn rule, Found.
  Phys. 35 (2005) 197--213.

\bibitem{Barnum:2003:yb}
H.~Barnum, No-signalling-based version of {Z}ure{k's} derivation of quantum
  probabilities: {A} note on {`}{E}nvironment-assisted invariance,
  entanglement, and probabilities in quantum physics{'}. Eprint \href
  {http://arxiv.org/abs/quant-ph/0312150} {\path{arXiv:quant-ph/0312150}}.

\bibitem{Mohrhoff:2004:tv}
U.~Mohrhoff, Probabilities from envariance?, Int. J. Quantum Inf. 2 (2004)
  221--230.

\bibitem{Clifton:1996:op}
R.~Clifton, The properties of modal interpretations of quantum mechanics, Br.
  J. Philos. Sci. 47 (1996) 371--398.

\bibitem{Bacciagaluppi:1996:po}
G.~Bacciagaluppi, M.~Hemmo, Modal interpretations, decoherence and
  measurements, Stud. Hist. Philos. Mod. Phys. 27 (1996) 239--277.

\bibitem{Bene:2001:po}
G.~Bene, Quantum origin of classical properties within the modal
  interpretations. Eprint \href {http://arxiv.org/abs/quant-ph/0104112}
  {\path{arXiv:quant-ph/0104112}}.

\bibitem{Bacciagaluppi:2000:yz}
G.~Bacciagaluppi, Delocalized properties in the modal interpretation of a
  continuous model of decoherence, Found. Phys. 30 (2000) 1431--1444.

\bibitem{Griffiths:1984:tr}
R.~B. Griffiths, Consistent histories and the interpretation of quantum
  mechanics, J. Stat. Phys. 36 (1984) 219--272.

\bibitem{Omnes:1994:pz}
R.~Omn{\`e}s, The Interpretation of Quantum Mechanics, Princeton University
  Press, Princeton, 1994.

\bibitem{Griffiths:2002:tr}
R.~B. Griffiths, Consistent Quantum Theory, Cambridge University Press,
  Cambridge, 2002.

\bibitem{GellMann:1990:uz}
M.~Gell-Mann, J.~Hartle, Quantum mechanics in the light of quantum cosmology,
  in: S.~Kobayashi, H.~Ezawa, Y.~Murayama, S.~Nomura (Eds.), Proceedings of the
  3rd International Symposium on the Foundations of Quantum Mechanics in the
  Light of New Technology (Tokyo, Japan, August 1989), Physical Society of
  Japan, Tokio, 1990, pp. 321--343.

\bibitem{GellMann:1991:pp}
M.~Gell-Mann, J.~B. Hartle, Quantum mechanics in the light of quantum
  cosmology, in: W.~H. Zurek (Ed.), Complexity, Entropy, and the Physics of
  Information, Santa Fe Institute of Studies in the Science of Complexity,
  Addison-Wesley, Redwood City, 1991, pp. 425--458.

\bibitem{Paz:1993:ww}
J.~P. Paz, W.~H. Zurek, Environment-induced decoherence, classicality and
  consistency of quantum histories, Phys. Rev. D 48 (1993) 2728--2738.

\bibitem{Albrecht:1993:pq}
A.~Albrecht, Following a ``collapsing'' wave function, Phys. Rev. D 48 (1993)
  3768--3778.

\bibitem{Dowker:1995:pa}
F.~Dowker, A.~Kent, Properties of consistent histories, Phys. Rev. Lett. 75
  (1995) 3038--3041.

\bibitem{Dowker:1996:ch}
F.~Dowker, A.~Kent, On the consistent histories approach to quantum mechanics,
  J. Stat. Phys. 82 (1996) 1575--1646.

\bibitem{Albrecht:1992:rz}
A.~Albrecht, Investigating decoherence in a simple system, Phys. Rev. D 46
  (1992) 5504--5520.

\bibitem{Twamley:1993:bz}
J.~Twamley, Phase-space decoherence: {A} comparison between consistent
  histories and environment-induced superselection, Phys. Rev. D 48 (1993)
  5730--5745.

\bibitem{GellMann:1998:xy}
M.~Gell-Mann, J.~B. Hartle, Strong decoherence, in: D.~H. Feng, B.~L. Hu
  (Eds.), Quantum Classical Correspondence: The 4th Drexel Symposium on Quantum
  Nonintegrability, International Press, Cambridge, Massachussetts, 1998, pp.
  3--35.

\bibitem{Riedel:2016:oo}
C.~J. Riedel, W.~H. Zurek, M.~Zwolak, Objective past of a quantum universe:
  {R}edundant records of consistent histories, Phys. Rev. A 93 (2016) 032126.

\bibitem{Bohr:1949:mz}
N.~Bohr, Discussions with {E}instein on epistemological problems in atomic
  physics, in: P.~A. Schilpp (Ed.), Albert Einstein: Philosopher--Scientist,
  Vol.~7, Library of Living Philosophers, Evanston, Illinois, 1949, pp.
  201--241.

\bibitem{Bohr:1931:ii}
N.~Bohr, Maxwell and modern theoretical physics, Nature 128 (1931) 691--692.

\bibitem{Bohr:1935:re}
N.~Bohr, Can quantum-mechanical description of physical reality be considered
  complete?, Phys. Rev. 48 (1935) 696--702.

\bibitem{Bohr:1996:mn}
N.~Bohr, Collected {W}orks. {V}ol.~7. Foundations of Quantum Mechanics~II
  (1933--1958), North Holland, Amsterdam, 1996, edited by J. Kalckar.

\bibitem{Bohr:1987:oo}
N.~Bohr, Essays 1932--1957 on Atomic Physics and Human Knowledge, Vol.~2 of The
  Philosophical Writings of Niels Bohr, Ox Bow Press, Woodbridge, Conn., 1987.

\bibitem{Bohr:1958:lu}
N.~Bohr, On atoms and human knowledge, D{\ae}dalus 87 (1958) 164--175.

\bibitem{Howard:1994:lm}
D.~Howard, What makes a classical concept classical? {T}oward a reconstruction
  of {N}iels {B}ohr's philosophy of physics, in: Niels Bohr and Contemporary
  Philosophy, Vol. 158 of Boston Studies in the Philosophy of Science, Kluwer,
  Dordrecht, 1994, pp. 201--229.

\bibitem{Joos:2006:yy}
E.~Joos, The emergence of classicality from quantum theory, in: P.~Clayton,
  P.~Davies (Eds.), The Re-Emergence of Emergence: The Emergentist Hypothesis
  from Science to Religion, Oxford University Press, Oxford, 2006, pp. 53--77.

\bibitem{Zeh:2000:rr}
H.~D. Zeh, The problem of conscious observation in quantum mechanical
  description, Found. Phys. Lett. 13 (2000) 221--233.

\bibitem{Schlosshauer:2017:oo}
M.~Schlosshauer, K.~Camilleri, Bohr and the problem of the quantum-to-classical
  transition, in: J.~Faye, H.~Folse (Eds.), Niels Bohr and Philosophy of
  Physics: Twenty-First-Century Perspectives, Bloomsbury Publishing, London,
  UK, 2017, pp. 223--233.

\bibitem{Bohr:1958:mj}
N.~Bohr, Quantum physics and philosophy: Causality and complementarity, in:
  R.~Klibanksy (Ed.), Philosophy at Mid-Century: A Survey, La Nuova Italia
  Editrice, Florence, 1958, pp. 308--314.

\bibitem{Heisenberg:1989:zb}
W.~Heisenberg, Physics and Philosophy. The Revolution in Modern Science,
  Penguin, London, 1989.

\bibitem{Heisenberg:1955:lm}
W.~Heisenberg, The development of the interpretation of the quantum theory, in:
  W.~Pauli, L.~Rosenfeld, V.~Weisskopf (Eds.), Niels Bohr and the Development
  of Physics: Essays Dedicated to Niels Bohr on the Occasion of his Seventieth
  Birthday, McGraw Hill, New York, 1955, pp. 12--29.

\bibitem{Daneri:1962:om}
A.~Daneri, A.~Loinger, G.~M. Prosperi, Quantum theory of measurement and
  ergodicity conditions, Nucl. Phys. 33 (1962) 297--319.

\bibitem{Wigner:1995:jm}
E.~Wigner, The Collected Works of Eugene Wigner. Part B, Vol.~6 of
  Philosophical Reflections and Syntheses, Springer, Berlin, 1995.

\bibitem{Bub:1971:ll}
J.~Bub, Comment on the {D}aneri--{L}oinger--{P}rosperi quantum theory of
  measurement, in: T.~Bastin (Ed.), Quantum Theory and Beyond, Cambridge
  University, Cambridge, 1971, pp. 65--70.

\bibitem{Schrodinger:1935:jn}
E.~Schr{\"o}dinger, Discussion of probability relations between separated
  systems, Proc. Cambridge Philos. Soc. 31 (1935) 555--563.

\bibitem{Zeh:1996:gy}
H.~D. Zeh, What is achieved by decoherence?, in: M.~Ferrero, A.~van~der Merwe
  (Eds.), New Developments on Fundamental Problems in Quantum Physics (Oviedo
  II), Kluwer, Dordrecht, 1997, pp. 441--452.

\end{thebibliography}

\end{document}